%% file: theseshort.tex
\documentclass[12pt,twoside,openright]{report} 
\usepackage[T1]{fontenc}
\usepackage[latin1]{inputenc}
\usepackage[francais,english]{babel}

\usepackage{epsf}
\usepackage{amsmath,amsfonts}
\usepackage{amssymb,amsfonts}
\usepackage{makeidx}

\usepackage{fancyheadings}
\usepackage{these}
\makeindex

\include{macros}

\begin{document}
\pagenumbering{roman}
\include{front}

\renewcommand{\baselinestretch}{1.3} \normalsize
\selectlanguage{francais}
\tableofcontents
\cleardoublepage
\renewcommand{\thepage}{\arabic{page}}
\include{intro}
\include{chap2}
\include{chap3}

\include{chap4}

\include{chap5}

\appendix
\sloppy
\renewcommand{\baselinestretch}{1.1} \normalsize

\chapter[Higgs branch, HyperK{\"a}hler quotient and duality in N=2 SYM
theories]
{Higgs branch, HyperK{\"a}hler quotient and duality in SUSY N=2 
Yang-Mills theories\label{hk}}
\begin{publi}{Int. J. Mod. Phys. A12 (1997) 4907}
\end{publi}

\chapter{$R^2$ Corrections and Non-perturbative Dualities of 
        N=4 String ground states\label{tt}}
\begin{publi}{Nucl. Phys. B 510 (1998) 423}
\end{publi}

\chapter{On $R^4$ threshold corrections in IIB string theory and 
       (p,q) string instantons\label{pq}}
\begin{publi}{Nucl. Phys. B 508 (1997) 509}
\end{publi}

\chapter{U-duality and D-brane Combinatorics \label{dc}}
\begin{publi}{Phys. Lett. B 418 (1998) 61}
\end{publi}

\chapter[A note on non-perturbative $R^4$ couplings]
{A note on non-perturbative \\$R^4$ couplings\label{nr4}}
\begin{publi}{accept{\'e} pour publication dans Phys. Lett.}
\end{publi}

\chapter{Calculable $e^{-1/\lambda}$ Effects \label{dds}}
\begin{publi}{Nucl. Phys. B 512 (1998) 61}
\end{publi}

\chapter{M-Theory and U-duality on $T^d$ with Gauge Backgrounds\label{mu}}
\begin{publi}{accept{\'e} pour publication dans Nucl. Phys. B.}
\end{publi}


\renewcommand{\baselinestretch}{1} \normalsize

\bibliofront

\renewcommand\bibname{R{\'e}f{\'e}rences bibliographiques}
\clearemptydoublepage

\include{thesebbl}

{\pagestyle{empty}\cleardoublepage\newpage}
\eject
\include{resume}

\printindex
\end{document}

%% file: macros.tex
\newcommand{\Tr}{{\rm Tr}\ }

\newcommand{\axion}{\hbox{\Large $a$} }
\newcommand{\A}{{\cal A}}

\def\a{\alpha}

\newcommand{\be}{\begin{equation}}
\newcommand{\ee}{\end{equation}}
\newcommand{\ba}{\begin{eqnarray}}
\newcommand{\ea}{\end{eqnarray}}
\renewcommand{\sp}{\; , \; \; }

\def\tr{\mbox{tr}}

\newcommand{\CC}{\mathbb{C}}
\newcommand{\RR}{\mathbb{R}}
\newcommand{\II}{\mathbb{I}}
\newcommand{\NN}{\mathbb{N}}
\newcommand{\Zint}{\mathbb{Z}}
\newcommand{\Real}{\mathbb{R}}

\def\Tr{\,{\rm Tr}\, }
\def\det{\,{\rm det}\, }

\def\tr{\,{\rm tr}\, }

\def\Re{\,{\rm Re}\, }

\def\iS{S_2}

\def\a{\alpha}
\def\b{\beta}

\def\p{\pi}

\def\t{\tau}
\def\th{\vartheta}

\def\et{\eta}

\def\thb{\bar{\th}}

\def\bb{\bar{b}}

\def\F{{\cal F}}

\def\A{{\cal A}}

\newcommand{\taubar}{\bar{\tau}}

\newcommand{\Ftilde}{{\widetilde{\cal F}}}

\newcommand{\e}{e}

\def\ft1{\tilde{F}_1}

\def\ti{\times}
\def\ss{\ar{a}{b}}

\def\bss{\ar{\ba}{\bb}}
\def\hg{\ar{h}{g}}
\def\w{\wedge}

\def\sqr#1#2{{\vcenter{\vbox{\hrule height.#2pt
        \hbox{\vrule width.#2pt height#1pt \kern#1pt
           \vrule width.#2pt}
        \hrule height.#2pt}}}}

\def\lsim{{\displaystyle
{{\raise-8pt\hbox{$ <$}}
\atop{\raise5pt\hbox{$\sim$}}}}}
\def\gsim{{\displaystyle
{{\raise-8pt\hbox{$ >$}}
\atop{\raise5pt\hbox{$\sim$}}}}}
\def\slsim{{\displaystyle
{{\raise-8pt\hbox{$\scriptstyle <$}}
\atop{\raise5pt\hbox{$\scriptstyle \sim$}}}}}
\def\sgsim{{\displaystyle
{{\raise-8pt\hbox{$\scriptstyle  >$}}
\atop{\raise5pt\hbox{$\scriptstyle \sim$}}}}}
\newcommand{\ar}[2]{\genfrac{[}{]}{0pt}{}{#1}{#2}}

\newcommand{\sump}[0]{\sum_{(h,g)}\!{\raise 4pt \hbox{$'$}}\,}
\def\ti{\times}
\def\ss{\ar{a}{b}}
\def\bss{\ar{\ba}{\bb}}
\def\hg{\ar{h}{g}}

\def\w{\wedge}
\def\Tr{\,{\rm Tr}\, }
\def\det{\,{\rm det}\, }

\def\tr{\,{\rm tr}\, }

\def\Re{\,{\rm Re}\, }

\def\a{\alpha}
\def\b{\beta}

\def\e{\epsilon}

\def\t{\tau}
\def\p{\pi}

\def\th{\vartheta}

\def\et{\eta}

\def\bal{\bar{\alpha}}
\def\bbe{\bar{\beta}}
\def\thb{\bar{\th}}

\def\ba{\bar{a}}
\def\bb{\bar{b}}

\def\F{{\cal F}}

\def\sp{\; , \; \; }

\def\ed{\end{document}}

\def\II{\rm II}

\def\fig#1#2#3#4{
\begin{figure}
\begin{center}
\mbox{\epsfysize #1 \epsffile{#2}}
\end{center}
\caption{#3}
\label{#4}
\end{figure}}


%% file: front.tex
\topmargin=0in
\enlargethispage{3cm}
\thispagestyle{empty}
\vspace*{-2cm}
\centerline{\scshape 
\large Centre de Physique Th{\'e}orique -- {\'E}cole Polytechnique}
\vskip 1.5cm
\centerline{\Large \bfseries TH{\`E}SE DE DOCTORAT DE L'UNIVERSIT{\'E} PARIS VI}
\vskip .8cm
\centerline{\large Sp{\'e}cialit{\'e} : \bfseries\scshape PHYSIQUE TH{\'E}ORIQUE}
\vskip .5cm
\centerline{ \mbox{\epsfysize 3cm \epsffile{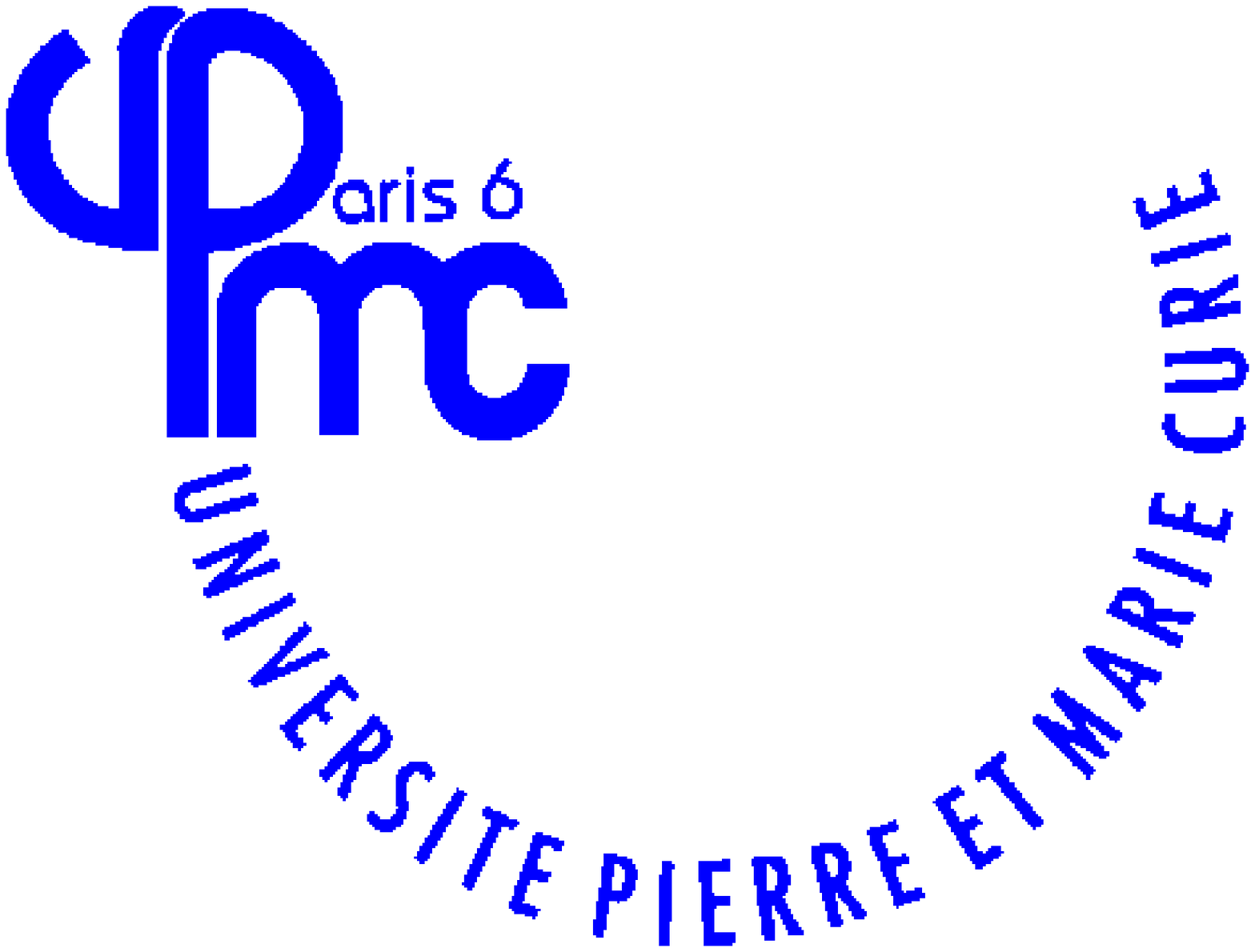}} }
\vskip 1.2cm
\centerline{pr{\'e}sent{\'e}e par}
\vskip .6cm
\centerline{\Large \bf Boris PIOLINE}
\vskip 1cm
\centerline{pour obtenir le grade de}
\vskip .6cm
\centerline{\Large \bf Docteur de l'Universit{\'e} Paris VI}
\vskip 1.5cm
\centerline{Sujet :}
\vskip 1cm
\centerline{\LARGE \bfseries \itshape Aspects non perturbatifs}
\vskip .3cm
\centerline{\LARGE \bfseries \itshape de la th{\'e}orie des supercordes}
\vskip 3cm
\noindent
Soutenue le 21 avril 1998 devant le jury compos{\'e} de~:
\vskip 1cm
\begin{tabular}{cll}
MM. & Luis {\'A}lvarez-Gaum{\'e},& rapporteur,\\
    & Ignatios Antoniadis,& directeur de th{\`e}se, \\
    & Jean-Loup Gervais, &\\
    & Costas Kounnas, & rapporteur,\\
    & Eliezer Rabinovici, &\\
    & Paul Windey, &pr{\'e}sident du jury,\\
\&  & Jean-Bernard Zuber.\\ 
\end{tabular}
\topmargin=2cm
\clearemptydoublepage

\thispagestyle{empty}
\vspace*{\stretch{1}}
\begin{flushright} {\scshape {\`A} mon fr{\`e}re et mes parents.}
\end{flushright}
\vspace*{\stretch{2}}
\clearemptydoublepage

\renewcommand{\baselinestretch}{1.3} \normalsize
\thispagestyle{empty}
\enlargethispage{5mm}
\centerline{\large \itshape Remerciements}
\vskip 1cm
Mon travail de th{\`e}se s'est d{\'e}roul{\'e} au Centre de Physique 
Th{\'e}orique de l'{\'E}cole Polytechnique du mois de septembre 1995 au
mois d'avril 1998~; je remercie sinc{\`e}rement 
les directeurs successifs Guy Laval et 
Marie-No{\"e}lle Bussac de m'avoir admis au sein 
de ce laboratoire, et ainsi 
permis de travailler dans un cadre scientifique exceptionnel.
Ignatios Antoniadis et Constantin Bachas ont bien voulu m'int{\'e}grer
dans leur {\'e}quipe de <<cordistes>> et je leur suis profond{\'e}ment
reconnaissant de m'avoir initi{\'e} {\`a} un domaine de recherche
d'un abord difficile mais r{\'e}ellement passionnant.  
Ignatios a dirig{\'e} mes recherches doctorales 
vers des directions fructueuses avec une grande
patience, gentillesse et disponibilit{\'e}~; non content de r{\'e}pondre {\`a}
mes interrogations purement scientifiques, il a su me communiquer sa
passion et sa pugnacit{\'e} tout en me laissant ma{\^\i}tre de d{\'e}terminer
l'orientation de mes recherches~; je tiens {\`a} lui exprimer ma plus
grande gratitude. Edouard Br{\'e}zin et Jean Iliopoulos m'ont 
fait go{\^u}ter aux d{\'e}lices de la th{\'e}orie quantique des champs, et 
ma vie en a {\'e}t{\'e} chang{\'e}e. Bient{\^o}t relay{\'e} par Paul Windey,
Costas Kounnas m'a le premier parl{\'e} avec la faconde qui le caract{\'e}rise
d'<<une th{\'e}orie of everything>>
qui m'a sembl{\'e} bien attirante~; je lui dois,
ainsi qu'{\`a} Jean Iliopoulos et Gabriele Veneziano, d'avoir pu
effectuer mon service national de la coop{\'e}ration {\`a} la division
th{\'e}orie du CERN, o{\`u} pendant seize mois, moyennant une modeste
assistance informatique aux membres de la division, j'ai pu interagir
avec d'illustres physiciens et faire progresser consid{\'e}rablement
mes recherches. Je leur en suis tr{\`e}s reconnaissant.

Luis {\'A}lvarez-Gaum{\'e} et Costas Kounnas ont accept{\'e} la 
lourde t{\^a}che d'{\^e}tre les rapporteurs de cette th{\`e}se
de poids cons{\'e}quent, je les en remercie chaleureusement.
Jean-Loup Gervais, Eliezer Rabinovici, Paul Windey
et Jean-Bernard Zuber ont accept{\'e} de participer au jury,
et c'est un honneur dont je leur suis tr{\`e}s reconnaissant.
Je remercie particuli\`erement Paul Windey pour avoir
accept\'e de prendre la pr\'esidence du jury, et
Jean-Bernard Zuber pour ses num\'ereuses et pr\'ecieuses remarques
stylistiques.

Costas Kounnas et Elias Kiritsis, aux enthousiasmes
batailleurs, m'ont permis d'entrer dans le jeu de la recherche 
en th{\'e}orie des cordes, par les s{\'e}ances initiatiques qu'ils
ont organis{\'e}es au CERN, et au cours de collaborations enflamm{\'e}es
(car il n'y a pas de fum{\'e}e sans feu) avec Andrea Gregori, 
Niels Obers et Marios Petropoulos~; Tom Taylor m'a aid{\'e} {\`a} 
contrer les vell{\'e}it{\'e}s somnambulatoires d'Ignatios au cours
de longues veill{\'e}es de travail~;
Niels Obers a conjugu{\'e} ses
efforts aux miens contre vents et mar{\'e}es alg{\'e}briques,
et avec Kristin F{\"o}rger m'a aid{\'e} {\`a} corriger les nombreuses
coquilles se cachant dans ce manuscrit~; Eliezer
Rabinovici a accept{\'e} de sacrifier sa vie conjugale aux exigences
de la M-th{\'e}orie~: qu'ils re{\c c}oivent tous l'expression de ma
gratitude et mon amiti{\'e}. 

La vie du laboratoire se partage souvent entre personnel scientifique
et personnel technique. Au cours de ma th{\`e}se j'ai pu faire
l'exp{\'e}rience inverse gr{\^a}ce au soutien et {\`a} l'amiti{\'e} 
des membres du secr{\'e}tariat de la division th{\'e}orie du CERN, 
Nanie, Suzy, Marie-No{\"e}lle et Jeanne, et d'Elena, ma coll{\`e}gue
informaticienne. Je remercie {\'e}galement Fran{\c c}oise Andalo
et Brigitte Oisline pour leur grande gentillesse et efficacit{\'e},
et Jean-Luc Bellon pour ses multiples conseils informatiques.

L'exp{\'e}rience de la th{\`e}se comporte
des p{\'e}riodes fastes et des p{\'e}riodes creuses~: les premi{\`e}res
ne compenseraient les secondes sans le soutien de 
mes corr{\'e}ligionnaires et concurrents Herv{\'e} Partouche et
Pierre Vanhove, des autres th{\'e}sards et post-docs du laboratoire, de mes
coll{\`e}gues et amis du CERN, {\'E}ric, Philippe, Niels, 
Emilian, Christophe et Christoph, et de quelques
co-coop{\'e}rants avec qui nous arpentions d'autres sommets.
Je remercie aussi chaleureusement mes amis
pour m'avoir conserv{\'e} leur amiti{\'e} pendant ces trois 
ann{\'e}es d'{\'e}loignement~; tenant {\`a} les garder, je pr{\'e}serverai
leur anonymat. Finalement, c'est {\`a} mes parents et
mon fr{\`e}re que je d{\'e}die ce m{\'e}moire, pour leur soutien et
leur compr{\'e}hension.

\clearpage
\thispagestyle{empty}
\vspace*{\stretch{1}}
\centerline{\large \itshape Avertissement}
\vskip 1cm
Ce m{\'e}moire a {\'e}t{\'e} {\'e}crit avec le souci de rendre les travaux
effectu{\'e}s dans le cadre de cette th{\`e}se accessibles {\`a} un public 
non expert. Au cours de ses trente ann{\'e}es d'existence, la th{\'e}orie 
des cordes a cependant suscit{\'e} un nombre consid{\'e}rable de contributions
{\`a} la base des d{\'e}veloppements les plus r{\'e}cents. J'ai
donc fait le choix d'introduire, autant que possible, toutes les
notions implicites dans les travaux pr{\'e}sent{\'e}s ici, 
en mettant l'accent plus sur
les concepts que sur les d{\'e}tails techniques. Le lecteur trouvera
dans la bibliographie les r{\'e}f{\'e}rences n{\'e}cessaires pour donner
une signification plus pr{\'e}cise {\`a} ces concepts. 
L'{\it invitation au voyage} s'adresse {\`a} un public large et
{\'e}vite r{\'e}solument toute {\'e}quation. Elle fournit une
pr{\'e}sentation synth{\'e}tique des arguments menant {\`a} la 
formulation de la M-th{\'e}orie, et r{\'e}sume mes contributions
{\`a} ce domaine. Celles-ci se regroupent en trois th{\`e}mes
principaux~: la dualit{\'e} {\'e}lectrique-magn{\'e}tique 
des th{\'e}ories de jauge (chapitre 2), les dualit{\'e}s des
th{\'e}ories de supercordes et leurs effets non perturbatifs
(chapitres 3 et 4), la th{\'e}orie des matrices (chapitre 5).
Chaque chapitre contient une pr{\'e}sentation p{\'e}dagogique
des concepts mis en jeu, et un r{\'e}sum{\'e} de mes contributions~;
les publications elles-m{\^e}mes sont reproduites en version 
originale dans les annexes A {\`a}
G. Les contributions aux actes de conf{\'e}rence n'ont pas {\'e}t{\'e}
reproduites dans ce m{\'e}moire.

\vspace*{\stretch{1}}
\clearemptydoublepage


%% file: intro.tex
\index{supergravit{\'e}!solitons de la|see{brane}}
\index{Dirichlet, membrane de|see{D-brane}}
\index{boucle, d{\'e}veloppement en boucle|see{genre, d{\'e}veloppement en}}
\index{couplage des cordes|see{dilaton}}
\index{mod{\`e}le standard|see{standard}}
\index{calcul semi-classique|see{semi-classique}}
\index{BPS, saturation|see{saturation BPS}}
\index{dualit{\'e}!d'espace-cible|see{T-dualit{\'e}}}
\index{Riemann, surface de|see{Riemann}}
\index{vari{\'e}t{\'e}!K_3@$K_3$ |see{$K_3$}}
\index{dualit{\'e}!de la th{\'e}orie IIB 10D|see{S-dualit{\'e}}}
\index{K3, surface@$K_3$, surface@homologie de|see{cycle d'homologie}}
\index{orbifold|see{vari{\'e}t{\'e} orbifold}}
\index{vari{\'e}t{\'e}!de compactification|see{compactification}}
\index{dualit{\'e}!des th{\'e}ories de supergravit{\'e}|see{sym{\'e}tries cach{\'e}es}}
\index{BPS, {\'e}tats|see{{\'e}tats BPS}}
\index{vari{\'e}t{\'e}!de Calabi-Yau|see{Calabi-Yau}}
\index{instanton!s{\'e}rie de|see{s{\'e}rie d'instantons}}
\index{conifold|see{singularit{\'e} de conifold}}
\index{NS5-brane|see{cinq-brane de Neveu-Schwarz}}
\index{dualit{\'e}!T-dualit{\'e}|see{T-dualit{\'e}}}
\index{soliton!des th{\'e}ories de supergravit{\'e}|see{brane}} 
\index{dualit{\'e}!U-dualit{\'e}|see{U-dualit{\'e}}}
\index{dualit{\'e}!S-dualit{\'e}|see{S-dualit{\'e}}}
\index{U-dualit{\'e}!de la th{\'e}orie IIB|see{S-dualit{\'e}}}
\index{moment, {\'e}tat de|see{Kaluza-Klein}}
\index{groupe!de T-dualit{\'e}|see{T-dualit{\'e}}}
\index{groupe!de U-dualit{\'e}|see{U-dualit{\'e}}}
\index{reduction dim@r{\'e}duction dimensionnelle!de Kaluza-Klein|see{Kaluza-Klein, r{\'e}duction de}}

\chapter{Invitation au voyage}
En l'espace de quelques ann{\'e}es, de l'article fondateur de Seiberg
et Witten obtenant la premi{\`e}re solution exacte d'une th{\'e}orie
de jauge supersym{\'e}trique fortement coupl{\'e}e en quatre dimensions 
gr{\^a}ce aux sym{\'e}tries de dualit{\'e},
{\`a} la proposition de Banks, Fischler, Shenker et Susskind de
d{\'e}finition explicite non perturbative de la th{\'e}orie des 
supercordes, le paysage de la th{\'e}orie des champs
et des cordes a subi un profond bouleversement, auquel j'ai
eu la chance d'assister, participer modestement quelquefois.
Les questions fondamentales de la th{\'e}orie des champs, 
structure du vide, spectre {\`a} basse {\'e}nergie, confinement, 
jusqu'alors hors de port{\'e}e de la th{\'e}orie de perturbations,
devenaient ainsi accessibles au traitement analytique, tout
au moins dans le cadre des th{\'e}ories supersym{\'e}triques.
En m{\^e}me temps, on d{\'e}couvrait dans le spectre 
des diff{\'e}rentes th{\'e}ories de supercordes
une richesse d'{\'e}tats non perturbatifs {\'e}tendus, dits $p$-branes,
identifi{\'e}s aux {\'e}tats fondamentaux d'une th{\'e}orie des
supercordes duale~; ses ph{\'e}nom{\`e}nes
non perturbatifs dans le cadre d'une th{\'e}orie de cordes
devenaient accessibles au calcul perturbatif dans la
th{\'e}orie duale. Les cinq th{\'e}ories des supercordes {\'e}taient ainsi
identifi{\'e}es comme d{\'e}veloppements de perturbation
en diff{\'e}rents r{\'e}gimes d'une th{\'e}orie ma{\^\i}tresse 
encore myst{\'e}rieuse, la M-th{\'e}orie, dans laquelle
les cordes ne semblent pas occuper de position
privil{\'e}gi{\'e}es. Par un retour remarquable,
ces d{\'e}veloppements offrent de nouvelles perspectives
sur les th{\'e}ories de jauge, d{\'e}crivant les degr{\'e}s de
libert{\'e} des solitons de $p$-branes de la th{\'e}orie
des cordes. Il me semblait n{\'e}cessaire d'ouvrir ce m{\'e}moire
par un panorama historique et synth{\'e}tique de ce paysage
en mouvement qui constitue l'arri{\`e}re-plan de mon travail
de th{\`e}se, et qui a d{\'e}termin{\'e} au jour le jour la
direction de mes recherches. 

\section{De la th{\'e}orie des champs quantiques...}

La th{\'e}orie quantique des champs 
et son {\it alter ego},
la th{\'e}orie statistique des champs, repr{\'e}sentent {\`a} l'heure actuelle
les fers de lance de l'arsenal de la physique th{\'e}orique moderne
pour traiter un large spectre de probl{\`e}mes s'{\'e}tendant de la
physique des particules {\'e}l{\'e}mentaires {\`a} la physique de la
mati{\`e}re condens{\'e}e et des milieux d{\'e}sordonn{\'e}s. Ces th{\'e}ories
prennent le relais de la m{\'e}canique quantique de Bohr, Heisenberg et Planck
et de la m{\'e}canique statistique de Boltzmann
lorsqu'il n'est plus possible d'isoler un petit nombre de degr{\'e}s
de libert{\'e}, comme c'est le cas en physique des particules
lorsque les effets de production de paires deviennent importants,
ou en physique des supraconducteurs lorsque les excitations coh{\'e}rentes
collectives du nuage {\'e}lectronique d{\'e}terminent les propri{\'e}t{\'e}s
de conduction. Elles rendent compte
de nombreux r{\'e}sultats exp{\'e}rimentaux avec une pr{\'e}cision in{\'e}gal{\'e}e,
dont l'illustration la plus frappante est sans doute donn{\'e}e par
l'accord {\`a} pr{\`e}s de $10^{-10}$
des calculs d'{\'e}lectrodynamique quantique
\index{electrodynamique@{\'e}lectrodynamique!quantique} avec les
mesures de la structure hyperfine des raies de r{\'e}sonance atomique.
\index{hyperfine, structure}
La th{\'e}orie des champs de jauge non ab{\'e}lienne 
de groupe de jauge $SU(3)\times SU(2)\times U(1)$
d{\'e}crivant le mod{\`e}le
standard \cite{Yang:1954ek,Weinberg:1967pk,Salam:1968rm}
\index{standard, mod{\`e}le} ne peut
encore {\^e}tre confront{\'e}e {\`a} des mesures d'une telle pr{\'e}cision,
mais reproduit jusqu'{\`a} pr{\'e}sent l'ensemble des r{\'e}sultats obtenus
dans les divers acc{\'e}l{\'e}rateurs de particules \cite{Barnett:1996hr}. 
\index{acc{\'e}lerateur} Les d{\'e}saccords
tant{\^o}t observ{\'e}s se sont jusqu'{\`a} pr{\'e}sent r{\'e}v{\'e}l{\'e}s
imputables {\`a} des erreurs de protocole exp{\'e}rimental,
et dans l'hypoth{\`e}se o{\`u} un tel
d{\'e}saccord viendrait {\`a} {\^e}tre confirm{\'e}, certaines particularit{\'e}s de
ces th{\'e}ories seraient susceptibles d'{\^e}tre revues
sans pour autant remettre en cause 
le sch{\'e}ma g{\'e}n{\'e}ral de la th{\'e}orie quantique des champs.

\subsection{Divergences et renormalisations}

Cet {\'e}tat de gr{\^a}ce de la th{\'e}orie des champs
contraste singuli{\`e}rement avec le scepticisme g{\'e}n{\'e}ral 
qui en a entour{\'e} la naissance dans les ann{\'e}es 1940, ent{\^a}ch{\'e}e
il est vrai d'incoh{\'e}rences math{\'e}matiques de mauvais augure.
Si les infinis apparaissant dans le calcul de quantit{\'e}s aussi
{\'e}l{\'e}mentaires que la masse de l'{\'e}lectron induite par les fluctuations
du vide ont {\'e}t{\'e} assez rapidement dompt{\'e}s par une 
proc{\'e}dure de {\it renormalisation}, 
la puissance pr{\'e}dictive de telles
th{\'e}ories {\'e}tait n{\'e}anmoins diminu{\'e}e, puisqu'il 
{\'e}tait n{\'e}cessaire de {\it normaliser} ces infinis
\index{renormalisation}
\index{divergence ultraviolette}
{\`a} la valeur finie donn{\'e}e par l'exp{\'e}rience. Sans compter l'aspect
inesth{\'e}tique et {\it ad hoc} de la proc{\'e}dure, la s{\'e}rie de
perturbation obtenue s'av{\'e}rait elle-m{\^e}me divergente, et pouvait
au mieux s'interpr{\'e}ter comme une {\it s{\'e}rie asymptotique}
\index{asymptotique, s{\'e}rie}
\index{serie de perturbation@s{\'e}rie de perturbation!en th{\'e}orie des champs}
{\`a} faible couplage. 
\index{serie de perturbation@s{\'e}rie de perturbation!en th{\'e}orie des champs}
La transposition des 
m{\'e}thodes du groupe de renormalisation de K. Wilson  en th{\'e}orie
\index{renormalisation!groupe de}
statistique des champs {\`a}
la th{\'e}orie quantique des champs 
a permis une compr{\'e}hension bien plus satisfaisante de ces infinis.
L'origine de ces divergences {\'e}tait alors
identifi{\'e}e comme le r{\'e}sultat de l'{\'e}volution des constantes
de couplage en fonction de
l'{\'e}chelle d'observation, apr{\`e}s int{\'e}gration
des degr{\'e}s de libert{\'e}s aux {\'e}chelles interm{\'e}diaires.
Les constantes de couplage <<nues>> de la th{\'e}orie aux {\'e}chelles infiniment
petites apparaissaient ainsi infinies {\`a} l'{\'e}chelle d'observation.
Les th{\'e}ories des
champs dites {\it renormalisables}
\index{renormalisation!renormalisabilit{\'e} des th{\'e}ories de jauge} 
ne pouvaient donc {\^e}tre d{\'e}finies sans 
r{\'e}f{\'e}rence {\`a} une {\'e}chelle de r{\'e}gularisation ultraviolette,
et seuls les couplages pr{\'e}servant la renormalisabilit{\'e} 
{\'e}taient pertinents pour la d{\'e}termination des interactions
{\`a} l'{\'e}chelle d'observation, les autres ne survivant pas au
flot de renormalisation.
\index{couplage effectif}
En particulier, le couplage des th{\'e}ories de jauge, sans dimension
et donc juste pertinent, pr{\'e}sentait un comportement tr{\`e}s diff{\'e}rent
sous le groupe de renormalisation, tendant {\`a} z{\'e}ro aux
grandes distances dans les th{\'e}ories
de jauge ab{\'e}liennes ou non ab{\'e}liennes avec un nombre
suffisant de champs de
mati{\`e}re, tandis qu'il augmentait dans l'infrarouge pour les
th{\'e}ories de jauge non ab{\'e}liennes avec suffisamment peu (ou pas)
de champs de mati{\`e}re, finalement quittant le r{\'e}gime perturbatif.
Inversement, aux petites distances, le couplage des premi{\`e}res 
divergeait au p{\^o}le dit {\it de Landau}
\index{Landau, p{\^o}le de}, tandis qu'il d{\'e}croissait
vers z{\'e}ro pour les secondes, dites pour cette raison
{\it asymptotiquement libres}\index{libert{\'e} asymptotique}. 
Cette derni{\`e}re cat{\'e}gorie,
dans laquelle entre le mod{\`e}le standard, devenait ainsi le
\index{standard, mod{\`e}le}
seul candidat {\`a} une th{\'e}orie fondamentale
susceptible de d{\'e}crire les interactions de jauge. 
La libert{\'e} asymptotique permettait en outre de calculer
de nombreuses pr{\'e}dictions {\`a} haute {\'e}nergie par un
d{\'e}veloppement perturbatif, aboutissant {\`a} l'accord
que l'on sait avec les r{\'e}sultats des exp{\'e}riences
de diffusion profond{\'e}ment in{\'e}lastique.\index{diffusion
profond{\'e}ment in{\'e}lastique}

\subsection{Au-del{\`a} de la th{\'e}orie de perturbation}

Si seules les th{\'e}ories de jauge asymptotiquement libres peuvent
pr{\'e}tendre {\`a} une description fondamentale de la mati{\`e}re,
on con{\c c}oit l'importance de d{\'e}finir
ces th{\'e}ories au niveau non perturbatif, sinon de d{\'e}velopper
des m{\'e}thodes de calcul dans ce r{\'e}gime.
L'{\'e}chelle d'{\'e}nergie du spectre des particules stables,
leptons, m{\'e}sons et baryons ($\sim 1 {\rm GeV}$), est en effet bien inf{\'e}rieure
{\`a} l'{\'e}chelle des quarks et gluons sond{\'e}e par les 
exp{\'e}riences de diffusion profond{\'e}ment in{\'e}lastique ($\sim 200 {\rm GeV}$),
\index{Lambda_{QCD}@$\Lambda_{QCD}$}
et correspond donc {\`a} un r{\'e}gime de fort couplage effectif~;
la d{\'e}monstration du confinement du nombre quantique
de couleur\index{confinement} et de la brisure de sym{\'e}trie chirale 
\index{sym{\'e}trie chirale, brisure de} est donc hors de port{\'e}e 
des m{\'e}thodes perturbatives ordinaires.

La d{\'e}finition d'une th{\'e}orie des champs hors
du r{\'e}gime perturbatif est en soit un probl{\`e}me majeur, et
les approches constructivistes ne sont pas encore en mesure
de d{\'e}finir rigoureusement les th{\'e}ories de jauge en dimension 4.
Inspir{\'e}e par les mod{\`e}les de physique statistique, la
r{\'e}gularisation sur r{\'e}seau des th{\'e}ories de jauge
\index{r{\'e}seau, th{\'e}orie de jauge sur}
offre une d{\'e}finition commode {\`a} tout couplage,
rel{\`e}guant ces complications au traitement de la limite
de grand volume. Elle
se pr{\^e}te particuli{\`e}rement bien au calcul
sur superordinateur, et des r{\'e}sultats tr{\`e}s encourageants
ont d{\'e}j{\`a} {\'e}t{\'e} obtenus \cite{Weingarten:1996ig}. Son efficacit{\'e} sur le plan
analytique n'est cependant pas aussi {\'e}vidente, et une
meilleure compr{\'e}hension, au moins qualitative, 
des ph{\'e}nom{\`e}nes non perturbatifs
peut {\^e}tre obtenue gr{\^a}ce aux m{\'e}thodes {\it semi-classiques},
bas{\'e}es sur la recherche des solutions classiques et leur 
quantification perturbative.
\index{semi-classique!calcul en th{\'e}orie des champs}

Le d{\'e}veloppement de ces m{\'e}thodes remonte {\`a} 
la d{\'e}couverte des {\it solitons} de
Polyakov et 't Hooft \cite{Polyakov:1974ek,'tHooft:1976fv}
et des {\it instantons} de 't Hooft et Belavin {\it et al.} 
\cite{'tHooft:1976fv,Belavin:1975rt}
\footnote{On se reportera avec profit {\`a}
\cite{Coleman:1977,Coleman:1985} pour une introduction
{\'e}l{\'e}mentaire {\`a} ces techniques.}.
Les solitons correspondent
\index{soliton!des th{\'e}ories de jauge}
\index{instanton!des th{\'e}ories de jauge}
{\`a} des solutions non triviales des {\'e}quations classiques
du mouvement, ind{\'e}pendantes du temps et 
localis{\'e}es dans l'espace~; leur stabilit{\'e} est
garantie classiquement par la topologie non triviale de
ces configurations de champs. Leur masse cro{\^\i}t comme $1/g^2$, et
ils {\'e}chappent donc au spectre perturbatif.
Ce sont des candidats {\`a} la
repr{\'e}sentation de particules
du spectre non perturbatif, et leurs propri{\'e}t{\'e}s peuvent en principe
{\^e}tre d{\'e}termin{\'e}es par quantification autour de ces solutions.

Les instantons sont au contraire localis{\'e}s dans le temps autant
que dans l'espace, et correspondent {\`a} des transitions 
par effet tunnel entre les vides classiques de la th{\'e}orie.
Ils entrent en principe sur un pied d'{\'e}galit{\'e}
avec le vide trivial dans la 
d{\'e}finition non perturbative de la th{\'e}orie, bien
que le calcul se limite dans la pratique {\`a} la prise en compte
d'un gaz {\it dilu{\'e}} d'instantons. 
Leur contribution n'est pertinente que dans le cas de processus
interdits perturbativement, sans quoi la contribution 
en $e^{-1/g^2}$ des instantons 
ne saurait {\'e}merger des contributions perturbatives
en $g^n$.

\subsection{Supersym{\'e}trie, dualit{\'e} et calculabilit{\'e}} 

Ces m{\'e}thodes non perturbatives ont tr{\`e}s r{\'e}cemment
subi un essor remarquable dans le
cadre des th{\'e}ories de jauge {\`a} supersym{\'e}trie
{\'e}tendue, pour lesquelles
les propri{\'e}t{\'e}s de supersym{\'e}trie garantissent l'absence
de corrections perturbatives pour certaines quantit{\'e}s physiques
au-del{\`a} d'un certain ordre, ainsi que la stabilit{\'e},
pour des valeurs g{\'e}n{\'e}riques du couplage,
d'{\'e}tats v{\'e}rifiant la propri{\'e}t{\'e} de Bogomolny, Prasad et Sommerfield (BPS),
\index{supersym{\'e}trie}
\index{etats BPS@{\'e}tats BPS}
c'est-{\`a}-dire annihil{\'e}s par certaines charges de la super-alg{\`e}bre
de Poincar{\'e}\footnote{Les d{\'e}sint{\'e}grations d'{\'e}tats BPS ont {\'e}t{\'e}
\index{desintegration@d{\'e}sint{\'e}gration d'{\'e}tats BPS}
mises en envidence dans les th{\'e}ories de supersym{\'e}trie $N=2$
\cite{seiberg/witten:1994.2,Ferrari:1997} mais n'interviendront
pas dans les cas de supersym{\'e}trie plus {\'e}lev{\'e}e consid{\'e}r{\'e}s 
dans ce m{\'e}moire.}.
\index{etats BPS@{\'e}tats BPS}
D{\`e}s 1978, une conjecture de 
Montonen et Olive \cite{Montonen:1977sn}
\index{Montonen-Olive, conjecture de}
raffin{\'e}e par Osborn
\cite{Osborn:1979tq} proposait que
les th{\'e}ories de jauge {\`a} supersym{\'e}trie $N=4$, finies {\`a} tout
ordre en perturbation, poss{\'e}daient au niveau quantique la sym{\'e}trie
de dualit{\'e} {\'e}lectrique-magn{\'e}tique, dite {\it S-dualit{\'e}}. 
\index{dualit{\'e}!{\'e}lectrique-magn{\'e}tique}
Cette sym{\'e}trie est bien
connue dans l'{\'e}lectrodynamique classique de Maxwell, o{\`u} elle
\index{electrodynamique@{\'e}lectrodynamique!de Maxwell}
{\'e}change champs {\'e}lectriques et champs
magn{\'e}tiques, charges {\'e}lectriques
et charges magn{\'e}tiques, {\it tout en inversant le couplage de jauge}.
Bien que l'absence
exp{\'e}rimentale de monop{\^o}les magn{\'e}tiques laisse supposer que
\index{monop{\^o}le magn{\'e}tique}
cette sym{\'e}trie n'est pas pr{\'e}serv{\'e}e dans la r{\'e}alit{\'e},
on conjecturait au contraire que cette transformation de dualit{\'e} restait
une sym{\'e}trie {\it quantique} des th{\'e}ories 
de jauge supersym{\'e}triques $N=4$.
\index{S-dualit{\'e}!de SYM $N=4$}
L'aspect le plus frappant de cette conjecture est qu'elle 
reliait un r{\'e}gime de faible couplage, dans lequel la th{\'e}orie
de perturbation donne des informations fiables, {\`a} un r{\'e}gime
de fort couplage hors d'atteinte jusqu'alors. La d{\'e}couverte
par Witten et Olive \cite{Witten:1978mh}
que la masse des {\'e}tats BPS
{\'e}tait prot{\'e}g{\'e}e de corrections quantiques
pour toute valeur du couplage, 
donnait les premi{\`e}res indications de la validit{\'e} de cette
conjecture, en permettant l'identification des {\it monop{\^o}les magn{\'e}tiques} de
't Hooft et Polyakov aux {\it bosons de jauge} fondamentaux $W^{\pm}$
dans la formulation duale.
Un pas d{\'e}cisif
dans la d{\'e}monstration de cette conjecture {\'e}tait effectu{\'e} par
Sen en 1994, qui montrait que l'extension 
de cette dualit{\'e} binaire
{\`a} un groupe discret infini $Sl(2,\Zint)$, 
dit de S-dualit{\'e}, agissant par {\it transformations modulaires} sur
le param{\`e}tre complexe
$S=\frac{\theta}{2\pi}+ i\frac{4\pi}{g^2}$, 
o{\`u} l'on a inclus l'angle $\theta$ couplant {\`a} la
densit{\'e} topologique $\int F\wedge F$, impliquait
l'existence d'{\'e}tats solitoniques de charge {\'e}lectrique
et magn{\'e}tique $(p,q)$ pour tous $p$ et $q$ entiers 
\index{pq, dyons de charge@$(p,q)$, dyons de charge}
premiers entre eux~; il reliait l'existence de ces {\'e}tats 
aux vides d'une m{\'e}canique quantique sur l'espace des
modules des monop{\^o}les de charge magn{\'e}tique $q$, et construisait
explicitement la fonction d'onde du monop{\^o}le de 
charge magn{\'e}tique 2 \cite{Sen:1994yi}. 
Ces r{\'e}sultats sont pour une grande part 
{\`a} l'origine de la ru{\'e}e vers l'El Dorado
\index{El Dorado}
des dualit{\'e}s.

La d{\'e}couverte du gisement revient cependant {\`a} Seiberg et Witten,
qui la m{\^e}me ann{\'e}e 1994, montraient
comment, en utilisant les contraintes de la supersym{\'e}trie
et de la dualit{\'e} {\'e}lectrique-magn{\'e}tique
sur la structure globale de l'espace des vides, on pouvait d{\'e}terminer
l'action effective de basse {\'e}nergie, exacte {\`a} toute valeur du
couplage, pour une large classe de th{\'e}ories de jauge 
{\`a} supersym{\'e}trie
{\'e}tendue $N=2$ \cite{seiberg/witten:1994.1,seiberg/witten:1994.2}. 
\index{Seiberg-Witten!solution de}
Ce tour de force 
donnait pour la premi{\`e}re fois la dynamique {\`a} grande distance,
ou fort couplage,
d'une th{\'e}orie des champs non triviale 
en quatre dimensions d'espace-temps, et
fournissait ainsi un terrain d'{\'e}tude privil{\'e}gi{\'e} pour les
ph{\'e}nom{\`e}nes non perturbatifs de confinement, brisure de sym{\'e}trie
\index{confinement}
\index{sym{\'e}trie chirale, brisure de}
chirale et condensation de monop{\^o}les. L'argument de Seiberg et Witten
utilisait de mani{\`e}re cruciale la dualit{\'e} {\'e}lectrique-magn{\'e}tique,
celle-ci reliant diff{\'e}rentes descriptions {\'e}quivalentes
d'un m{\^e}me point de l'espace des modules.

Malheureusement, ces th{\'e}ories {\`a} supersym{\'e}trie {\'e}tendue,
bien que traitables analytiquement, 
n'ont que peu de pertinence
ph{\'e}nom{\'e}nologique, la raison premi{\`e}re 
\index{phenomenologie@ph{\'e}nom{\'e}nologie}
en {\'e}tant l'impossibilit{\'e}
d'introduire de la mati{\`e}re chirale. Elles peuvent certes donner lieu {\`a}
des th{\'e}ories supersym{\'e}triques N=1 ou sans supersym{\'e}trie
par brisure douce, encore contr{\^o}lables math{\'e}matiquement,
mais toujours non chirales. Peu de temps avant son
article fondateur avec E. Witten, N. Seiberg proposait une 
version diff{\'e}rente de la dualit{\'e} {\'e}lectrique-magn{\'e}tique
\index{Seiberg, dualit{\'e} dans SYM $N=1$}
dans le cadre des th{\'e}ories de jauge de supersym{\'e}trie
$N=1$, dite {\it dualit{\'e} infrarouge}
\cite{Seiberg:1994bz}. Sur la base 
d'arguments de comptage de degr{\'e}s de libert{\'e}, 
de limites de d{\'e}couplage
et de correspondance d'anomalies chiralesde 't Hooft, Seiberg proposait
\index{anomalie!de 't Hooft}
que la th{\'e}orie $N=1$ de groupe de jauge $SU(N_c)$ et en pr{\'e}sence de  
$N_f$ saveurs de quarks 
soit identique {\it {\`a} grande distance} {\`a} une
th{\'e}orie de groupe de jauge diff{\'e}rent $SU(N_f-N_c)$, $N_f$ saveurs
de quarks et $N_f^2$ singlets de jauge~; 
les bosons de jauge {\it non ab{\'e}liens} d'une
th{\'e}orie apparaissaient ainsi comme des solitons dans la 
description duale. Si cette conjecture a {\'e}t{\'e} {\'e}tendue
{\`a} un grand nombre de situations et a consid{\'e}rablement
fait progresser notre compr{\'e}hension de la dynamique des
th{\'e}ories de jauge, sa puissance calculatoire est malheureusement
rest{\'e}e limit{\'e}e jusqu'{\`a} pr{\'e}sent.

\section{...aux th{\'e}ories des supercordes...}

Couronn{\'e}e de succ{\`e}s pour la description des trois interactions
{\'e}lectromagn{\'e}tique, forte et faible dans le cadre du mod{\`e}le
standard\index{mod{\`e}le standard}, la th{\'e}orie des champs doit cependant 
reconna{\^\i}tre son {\'e}chec {\`a} d{\'e}crire la force de gravitation
au niveau quantique. Renormalisables en dimension deux, les
\index{renormalisation!renormalisabilit{\'e} de la gravitation}
interactions gravitationnelles g{\'e}n{\`e}rent de s{\'e}v{\`e}res
divergences ultraviolettes en dimension sup{\'e}rieure, 
\index{divergence ultraviolette}
et de nouveaux contre-termes
et donc constantes de couplage doivent {\^e}tre introduits {\`a}
chaque ordre en perturbation. 
Ces difficult{\'e}s ne sont finalement pas surprenantes,
si l'on r{\'e}alise que la notion de gravit{\'e} quantique remet
en cause les fondements de la structure de l'espace-temps
aux distances inf{\'e}rieures {\`a} l'{\'e}chelle de Planck
\index{Planck, {\'e}chelle de}
($10^{-35}$ m, ou $10^{19}$ GeV),
et donc la notion m{\^e}me de champ d{\'e}fini en tout point de l'espace.
Des efforts importants ont {\'e}t{\'e}
consacr{\'e}s {\`a} transposer les m{\'e}thodes des th{\'e}ories de
jauge {\`a} la gravitation dans le cadre de la th{\'e}orie
de gravitation de boucles \cite{Rovelli:1997yv}, 
mais leur succ{\`e}s est encore
incertain. Face {\`a} ces difficult{\'e}s, la th{\'e}orie des
supercordes appara{\^\i}{}t {\`a} l'heure actuelle comme le seul
candidat viable 
{\`a} l'unification quantique de toutes les interactions fondamentales.

\subsection{Th{\'e}ories perturbatives de cordes}
\index{bosonique, th{\'e}orie des cordes}
Initialement d{\'e}velopp{\'e}e sous le nom de {\it mod{\`e}les duaux}
\index{duaux, mod{\`e}les}
{\`a} la fin des ann{\'e}es soixante afin de rendre compte des
trajectoires de Regge observ{\'e}es sur le spectre des r{\'e}sonances
\index{Regge, trajectoire de}
hadroniques de spin {\'e}lev{\'e}, la th{\'e}orie des cordes a d{\'e}couvert
sa v{\'e}ritable vocation lorsqu'il a {\'e}t{\'e} r{\'e}alis{\'e} que la 
pr{\'e}sence d'une particule de spin 2 dans son spectre de masse
nulle en faisait un candidat naturel {\`a} la description de
la gravitation quantique\cite{Scherk:1974ca}~;
ce candidat {\'e}tait d'autant plus attirant que contrairement
aux th{\'e}ories des champs, toutes les amplitudes y {\'e}taient
finies.  La raison de ce miracle tient {\`a} 
la structure {\'e}tendue que la th{\'e}orie conf{\`e}re aux objets
{\'e}l{\'e}mentaires, de petites cordes relativistes oscillantes 
d{\'e}crivant une
{\it surface d'univers} au cours de leur propagation dans l'espace-temps,
et interagissant de mani{\`e}re g{\'e}om{\'e}trique par coupure et
recollement ({\it splitting and joining}), correspondant aux
\index{splitting and joining}
bifurcations de cette surface (figure \ref{corde}). 
La tension de la corde $1/\alpha'$ d{\'e}finit l'{\'e}chelle de masse
\index{tension!de la corde fondamentale}
de la th{\'e}orie, dite {\it {\'e}chelle des cordes}, 
tandis que les interactions sont pond{\'e}r{\'e}es
par le {\it couplage des cordes} sans dimension $g$. Vues de loin, ces cordes
\index{dilaton}
se comportent comme des particules ponctuelles dont la nature, la masse
et les interactions sont d{\'e}termin{\'e}es par l'{\'e}tat 
d'oscillation interne. Tout comme la ligne d'univers d'une particule
\index{ligne d'univers}
ordinaire supporte une th{\'e}orie des champs unidimensionelle, la
surface d'univers de la corde supporte une {\it th{\'e}orie des champs
\index{conforme, th{\'e}orie des champs!sur la surface d'univers}
bidimensionnelle conforme}\footnote{Cette derni{\`e}re condition,
n{\'e}cessaire au d{\'e}couplage du champ de Liouville sur la surface
\index{Liouville, champ de}
d'univers, peut {\^e}tre relax{\'e}e dans le cadre des {\it th{\'e}ories
des cordes non critiques}. 
\index{non critiques, th{\'e}orie des cordes}
Ces derni{\`e}res seront ignor{\'e}es dans
le cadre de cette th{\`e}se, bien que de r{\'e}cents progr{\`e}s permettent
d'esp{\'e}rer en leur quantification.}. 
Les contraintes d'invariance conforme restreignent
la dimensionalit{\'e} de l'espace-temps {\`a} la {\it
dimension critique} $D=26$, et lui imposent en outre de v{\'e}rifier
\index{critique, dimension}
une extension des {\'e}quations de la relativit{\'e} g{\'e}n{\'e}rale,
corrig{\'e}es {\`a} tous les ordres en $\alpha'$.
La gravitation {\'e}merge donc naturellement, et cette correspondance 
fixe la taille caract{\'e}ristique des cordes {\`a}
l'{\'e}chelle de Planck. 
\index{Planck, {\'e}chelle de}
\fig{5cm}{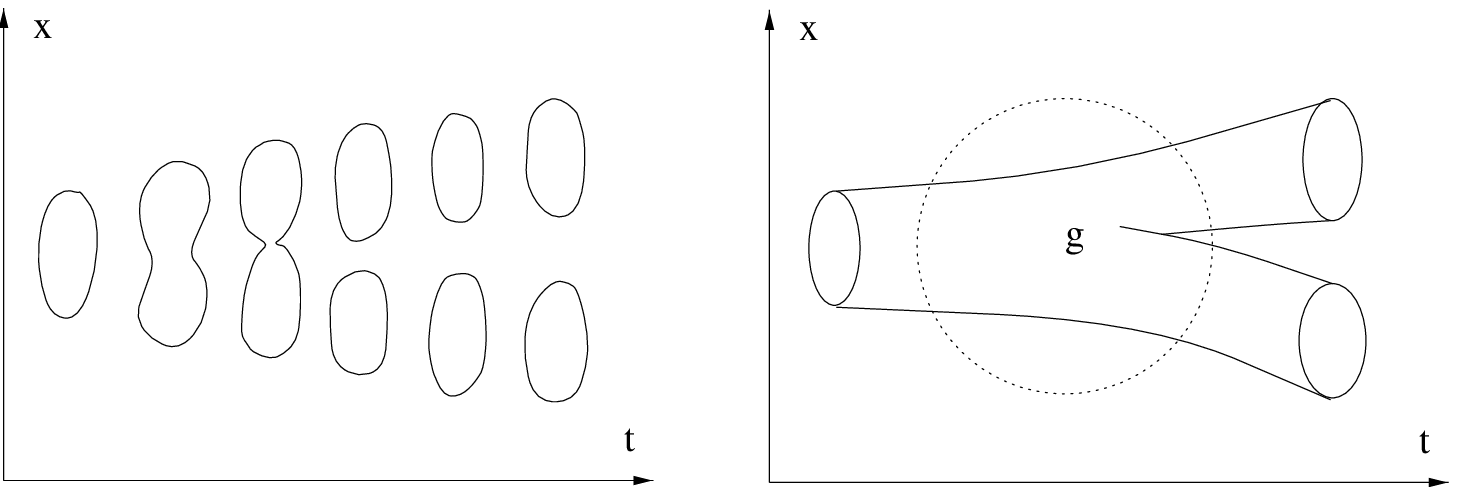}{Interaction de {\it splitting} et surface d'univers
de la corde}{corde}

Cette version simpliste, dite {\it corde bosonique}
n'est cependant pas viable en raison de la pr{\'e}sence d'une particule
de masse carr{\'e}e n{\'e}gative, le tachyon, r{\'e}v{\'e}latrice de
\index{tachyon}
l'instabilit{\'e} de cette th{\'e}orie, ainsi qu'en raison de l'absence de
degr{\'e}s de libert{\'e} fermioniques dans le spectre. On peut
cependant remplacer la th{\'e}orie conforme bidimensionnelle
\index{conforme, th{\'e}orie des champs!superconforme}
en une th{\'e}orie conforme localement supersym{\'e}trique,
ou {\it superconforme}, pour obtenir, apr{\`e}s une projection
judicieuse \cite{Gliozzi:1976jf}, une th{\'e}orie sans
\index{GSO, projection}
tachyon de dimension critique $D=10$\footnote{Les th{\'e}ories
superconformes {\'e}tendues conduisent {\`a} une dimension
critique $D=2$ ($N=2$) ou $D=-2$ ($N=4$) dont la pertinence
n'est pas {\'e}vidente. Certaines constructions h{\'e}t{\'e}rotiques
permettent d'obtenir $D=4$ pour une supersym{\'e}trie de surface
d'univers $N=(2,1)$ \cite{Ooguri:1991fp}, mais ne seront pas consid{\'e}r{\'e}es dans
ce m{\'e}moire.}. L'invariance superconforme {\it sur la surface 
d'univers} conduit alors {\`a} une th{\'e}orie supersym{\'e}trique
{\it dans l'espace ambiant}. Ces th{\'e}ories sont en g{\'e}n{\'e}ral
\index{chiralit{\'e}}
\index{anomalie!gravitationnelle}
chirales, et les contraintes de compensation
d'anomalies gravitationnelles s{\'e}lectionnent une dizaine de mod{\`e}les
\cite{Kawai:1986vd}, dont seuls six mod{\`e}les ne pr{\'e}sentent
pas de tachyon~: on construit
\index{supercordes, th{\'e}orie des}
ainsi les {\it cordes ouvertes de type I} avec supersym{\'e}trie 
d'espace-temps $N=1$ {\`a} dix dimensions et groupe
de jauge SO(32)~; les {\it cordes ferm{\'e}es de
type IIA et IIB} avec supersym{\'e}trie $N=2$ non chirale (IIA)
ou chirale (IIB) {\`a} dix dimensions~; 
les {\it cordes ferm{\'e}es h{\'e}t{\'e}rotiques}
avec supersym{\'e}trie $N=1$ {\`a} dix dimensions et groupes de
jauge $SO(32)$ {\it ou} $E_8\times E_8$~; 
et finalement la corde h{\'e}t{\'e}rotique
non supersym{\'e}trique de groupe de jauge $SO(16) \times SO(16)$
\cite{Alvarez-Gaume:1986jb},
que nous omettrons dans la suite en nous restreignant aux {\it th{\'e}ories
des supercordes critiques supersym{\'e}triques}.

Le spectre de ces th{\'e}ories de supercordes
se compose donc d'un {\'e}tage de masse nulle comprenant
une particule scalaire $\phi$
dite {\it dilaton}, dont la valeur moyenne $g=e^\phi$
\index{dilaton|textit}
d{\'e}finit le couplage de la th{\'e}orie, d'une particule de
spin 2 identifi{\'e}e au graviton
\index{graviton}, et d'un certain nombre de 
{\it tenseurs antisym{\'e}triques  de jauge} 
\index{tenseur antisym{\'e}trique}
d{\'e}pendant de la th{\'e}orie consid{\'e}r{\'e}e~;
ces tenseurs g{\'e}n{\'e}ralisent la notion de potentiel vecteur $A_{\mu}$
{\`a} un tenseur antisym{\'e}trique {\`a} $p$ indices $B$, ou {\it $p$-forme},
invariant sous la transformation de jauge $B \rightarrow B+d\Lambda$
o{\`u} $\Lambda$ est une $(p-1)$-forme quelconque.
Le spectre inclut {\'e}galement une tour d'{\'e}tats massifs
de spin arbitrairement {\'e}lev{\'e} et de masses carr{\'e}es en
progression arithm{\'e}tique $\mathcal{M}^2=N/\alpha'$, 
\index{supermassifs, {\'e}tats}
tout comme la gamme pythagoricienne
\index{Pythagore}
des modes propres de la corde vibrante. Les modes $N\ne 0$
n'apparaissent qu'aux {\'e}nergies de l'ordre de l'{\'e}chelle de
Planck, et leurs effets peuvent donc {\^e}tre int{\'e}gr{\'e}s 
\index{Planck, {\'e}chelle de}
\index{action effective!des modes de masse nulle}
pour donner une th{\'e}orie des champs 
de basse {\'e}nergie r{\'e}gissant la dynamique des {\'e}tats de masse nulle~;
les sym{\'e}tries de diff{\'e}omorphismes et de jauge ajout{\'e}es
aux propri{\'e}t{\'e}s de supersym{\'e}trie d'espace-temps identifient cette
th{\'e}orie des champs
aux th{\'e}ories de supergravit{\'e} introduites par 
\index{supergravit{\'e}}
Freedman, van Nieuwenhuizen, Ferrara, Deser et Zumino
\cite{Freedman:1976xh,Deser:1976eh}.
Bien que ces th{\'e}ories soient non renormalisables au sens du 
\index{divergence ultraviolette!en supergravit{\'e}}
\index{renormalisation!renormalisabilit{\'e} de la gravitation}
comptage de puissances, les th{\'e}ories de supercordes en fournissent
une r{\'e}gularisation coh{\'e}rente et compatible avec toutes les
invariances de jauge.

\subsection{Interactions et th{\'e}orie de perturbations}
La th{\'e}orie des supercordes telle qu'elle {\'e}merge au milieu des
ann{\'e}es 1980 rassemble donc en r{\'e}alit{\'e} six th{\'e}ories 
des cordes critiques distinctes,
s{\'e}lectionn{\'e}es par un cahier des charges drastique: absence 
d'instabilit{\'e} tachyonique, et absence d'anomalies gravitationnelles
et de jauge. Ces th{\'e}ories sont d{\'e}finies par leur d{\'e}veloppement
\index{serie de perturbation@s{\'e}rie de perturbation!en th{\'e}orie des cordes}
perturbatif en somme sur les surfaces de Riemann de
\index{surface de Riemann}
genre arbitraire $n$,  rempla{\c c}ant les diagrammes de Feynman
\index{Feynman, diagramme de}
ordinaires {\`a} $n$ boucles, pond{\'e}r{\'e}es par une puissance du
couplage $g^n$ (figure \ref{genre}). Gr{\^a}ce {\`a} l'invariance conforme, les pattes 
externes des diagrammes peuvent {\^e}tre
incorpor{\'e}es comme des insertions d'{\it op{\'e}rateurs de vertex}
\index{amplitude de diffusion}
locaux dans la th{\'e}orie des champs bidimensionnelle.
Il faut encore sommer sur les diff{\'e}rentes {\it m{\'e}triques}
sur la surface d'univers, ou plus pr{\'e}cis{\'e}ment sur leurs
{\it classes d'{\'e}quivalence conforme}, lesquelles sont
d{\'e}termin{\'e}es par la {\it structure complexe} de la surface
de Riemann~:
cette op{\'e}ration est analogue {\`a} la sommation sur les moments
des particules se propageant dans les boucles, et prend {\'e}galement
\index{serie de perturbation@s{\'e}rie de perturbation!en th{\'e}orie des champs}
en compte les facteurs de sym{\'e}trie des diagrammes.
La d{\'e}finition des th{\'e}ories des supercordes
est donc intrins{\`e}quement perturbative en le param{\`e}tre $g$.
Par contraste, la th{\'e}orie conforme sur chaque surface 
est d{\'e}finie non perturbativement en $\alpha'$,
bien qu'il puisse s'av{\'e}rer n{\'e}cessaire 
d'effectuer un d{\'e}veloppement perturbatif en $\alpha'$
lorsque cette th{\'e}orie n'est pas int{\'e}grable.
\fig{1.7cm}{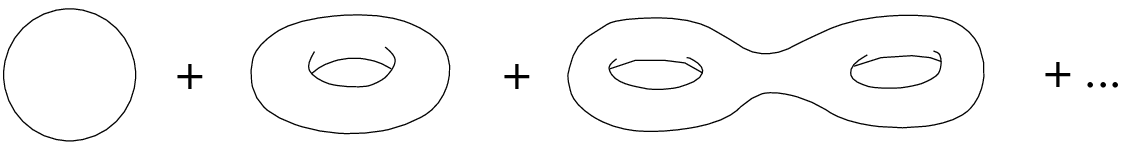}{D{\'e}veloppement en surface de Riemann de genre
  arbitraire~: contribution {\`a} l'ordre des arbres, une
boucle, deux boucles,...}{genre}

\subsection{Compactification et d{\'e}g{\'e}n{\'e}rescence du vide}
\index{compactification}
La restriction sur la dimensionnalit{\'e} $D=10$ de l'espace
peut {\^e}tre lev{\'e}e en supposant certaines d'entre elles
compactes de rayon tr{\`e}s faible. La vari{\'e}t{\'e}
correspondant {\`a} ces directions compactes est astreinte
{\`a} v{\'e}rifier les {\'e}quations d'Einstein
g{\'e}n{\'e}ralis{\'e}es~; la pr{\'e}servation de la supersym{\'e}trie
d'espace-temps restreint la vari{\'e}t{\'e} de compactification
aux espaces de Calabi-Yau, mais laisse encore un nombre 
\index{Calabi-Yau, vari{\'e}t{\'e} de}
important de possibilit{\'e}s~: ainsi, dans le cas de la
compactification $N=2$ {\`a} 4 dimensions, quelques 11000
espaces de Calabi-Yau topologiquement distincts sont connus,
avec pour chacun plusieurs param{\`e}tres de d{\'e}formations 
continues ! Notons de plus que l'on peut construire 
des th{\'e}ories de supercordes directement {\`a} quatre dimensions
en rempla{\c c}ant les dimensions internes par une th{\'e}orie
\index{conforme, th{\'e}orie des champs}
superconforme de charge centrale {\'e}quivalente
\index{fermionique, construction}
\cite{Kawai:1986va,Antoniadis:1987rn}. On obtient dans certains
cas des descriptions conformes {\it exactes} 
de compactification g{\'e}om{\'e}trique sur des espaces de Calabi-Yau
\index{Calabi-Yau, vari{\'e}t{\'e} de}
\cite{Gepner:1987vz} ou sur des solutions cosmologiques
\cite{Kounnas:1992wc}, ou bien des descriptions sans 
{\'e}quivalent g{\'e}om{\'e}trique connu.

Cette abondance de mod{\`e}les pour une th{\'e}orie d'unification
\index{vide!d{\'e}g{\'e}n{\'e}rescence en th{\'e}orie des cordes}
ne justifie cependant pas que l'on se d{\'e}sint{\'e}resse 
de ces th{\'e}ories~:
ces mod{\`e}les doivent en effet {\^e}tre consid{\'e}r{\'e}s comme 
des {\it {\'e}tats du vide diff{\'e}rents} des quelques th{\'e}ories
des supercordes fondamentales {\`a} dix dimensions, d{\'e}termin{\'e}s
par les valeurs moyennes de champs scalaires de masse
nulle, dits champs de module, param{\'e}trant la vari{\'e}t{\'e} de compactification.
\index{module, champ de}
Ces scalaires prennent en g{\'e}n{\'e}ral leurs valeurs dans une 
vari{\'e}t{\'e} abstraite courbe, dite {\it espace des modules}. 
\index{espace des modules}
La s{\'e}lection du vide physique,
ou en d'autres termes la g{\'e}n{\'e}ration de masse pour les
champs de modules, reste incomprise {\`a} l'heure actuelle.

\subsection{Sym{\'e}tries perturbatives et sym{\'e}tries cach{\'e}es}
La d{\'e}couverte des {\it dualit{\'e}s d'espace cible}
({\it target space dualities}, ou T-dualit{\'e}s)
\index{T-dualit{\'e}}
apporte une confirmation de ce point de vue. 
Ces dualit{\'e}s apparaissent dans l'exemple le plus simple
de la compactification sur un cercle de rayon $R$~: le spectre comporte
alors des {\it {\'e}tats de moment} de masse $\mathcal{M}=m/R$ correspondant aux
\index{Kaluza-Klein!excitation de!textit}
excitations de la th{\'e}orie des champs 
portant un moment interne $P=m/R$ selon la direction compacte,
mais aussi des {\it {\'e}tats d'enroulement} sp{\'e}cifiques {\`a} la
\index{enroulement!etat d'@{\'e}tat d'}
th{\'e}orie des cordes, de masse 
$\mathcal{M}=n R/\alpha'$ correspondant {\`a} une corde enroul{\'e}e 
$n$ fois autour du cercle (figure \ref{tdual}). 
Ces deux {\'e}tats sont {\'e}chang{\'e}s sous
l'inversion du rayon $R\rightarrow \alpha'/R$, et on peut
montrer que cette transformation est une sym{\'e}trie exacte de la
th{\'e}orie conforme, donc valable {\`a} tous les ordres dans 
la th{\'e}orie de perturbation. Cette sym{\'e}trie $\Zint_2$ se
g{\'e}n{\'e}ralise en une sym{\'e}trie de T-dualit{\'e} $SO(d,d,\Zint)$ pour les
compactifications sur tore $T^d$, et dans le cas des compactifications
sur espaces de Calabi-Yau donnent lieu {\`a} la {\it sym{\'e}trie
miroir}, qui identifie deux espaces de topologies 
diff{\'e}rentes\index{miroir, sym{\'e}trie}. La T-dualit{\'e} agit comme 
{\it automorphisme} des cordes h{\'e}t{\'e}rotiques, mais comme
{\it isomorphisme} des cordes de type IIA et IIB, {\'e}chang{\'e}es
sous la T-dualit{\'e}. Elle aura {\'e}galement des cons{\'e}quences
\index{dualit{\'e}!des th{\'e}ories IIA et IIB}
inattendues dans le cadre des th{\'e}ories de cordes ouvertes, 
ainsi que nous le verrons incessamment.
\fig{3cm}{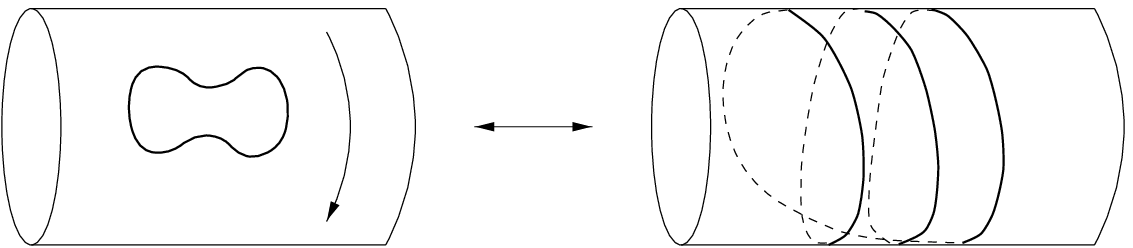}{La T-dualit{\'e} des cordes ferm{\'e}es
{\'e}change les {\'e}tats de moment
sur un cercle de rayon $R$ (gauche) et les {\'e}tats d'enroulement
sur un cercle de rayon $\alpha'/R$ (droite).}{tdual}

Les T-dualit{\'e}s agissent ordre par ordre en th{\'e}orie
des perturbations et apparaissent comme sym{\'e}tries {\it
continues}\footnote{Les {\'e}tats d'enroulement et de moment charg{\'e}s 
par rapport {\`a}
cette sym{\'e}trie continue sont en effet absents de la th{\'e}orie
effective {\`a} basse {\'e}nergie.} de
la th{\'e}orie de supergravit{\'e} d{\'e}crivant la dynamique
{\`a} basse {\'e}nergie. 
Les th{\'e}ories de supergravit{\'e} pr{\'e}sentent cependant, outre
ces sym{\'e}tries perturbatives, des {\it sym{\'e}tries cach{\'e}es}
\index{sym{\'e}tries cach{\'e}es}
reliant des r{\'e}gimes de couplages diff{\'e}rents,
analogues {\`a} la S-dualit{\'e} {\'e}lectrique-magn{\'e}tique des th{\'e}ories
de jauge \cite{Cremmer:1979up}. 
L'exemple le plus simple est sans doute la S-dualit{\'e}
$Sl(2,\Zint)$ 
\index{S-dualit{\'e}!de la th{\'e}orie IIB}
de la th{\'e}orie de type IIB, agissant par transformations modulaires
sur le param{\`e}tre complexe $S=\axion + i e^{-\phi}$, o{\`u}
$\axion$ correspond {\`a} la valeur moyenne d'un champ scalaire
de la th{\'e}orie de type IIB~; les deux tenseurs antisym{\'e}triques
$(B_{\mu\nu},\mathcal{B}_{\mu\nu})$ se transforment en outre comme
un doublet sous cette sym{\'e}trie. La th{\'e}orie de supergravit{\'e}
est invariante sous le groupe {\it continu} $Sl(2,\Real)$
\cite{Schwarz:1983qr}~; celui-ci est 
bris{\'e} en un sous-groupe discret
par l'existence de la corde fondamentale, charg{\'e}e
sous cette sym{\'e}trie. Par analogie avec la dualit{\'e} de Montonen-
Olive, on est donc amen{\'e} {\`a} conjecturer la validit{\'e}
{\it quantique} de cette sym{\'e}trie, et donc l'existence
d'un multiplet de {\it cordes solitoniques} de charges $(p,q)$
\index{pq, cordes de charge@$(p,q)$, cordes de charge}
par rapport aux champs $(B,\mathcal{B})$. Cette conjecture suppose
naturellement l'existence d'une {\it extension non perturbative}
de la th{\'e}orie des supercordes de type IIB.

De telles {\it sym{\'e}tries cach{\'e}es} sont visibles dans de nombreux
cas, et se conjuguent avec les dualit{\'e}s d'espace cible pour former
un groupe de dualit{\'e} plus large, dit de {\it U-dualit{\'e}},
\index{U-dualit{\'e}}
agissant dans l'extension non perturbative hypoth{\'e}tique
des th{\'e}ories de supercordes, ou {\it reliant diff{\'e}rentes
th{\'e}ories de supercordes} moyennant une red{\'e}finition des champs
\cite{Hull:1995ys}. 
L'existence de sym{\'e}trie de l'action
effective ne suffirait pas {\`a} conclure {\`a} l'existence de
ces dualit{\'e}s, mais l'{\'e}tude du spectre non perturbatif
des th{\'e}ories de supergravit{\'e}
en apporte une confirmation {\'e}clatante.

\subsection{Spectre de solitons, membranes et D-branes}
\index{brane}
\index{soliton!des th{\'e}ories de supergravit{\'e}}
\index{D-brane}
Bien que la formulation non perturbative des th{\'e}orie des supercordes
reste encore un probl{\`e}me ouvert {\`a} ce stade, l'{\'e}tude
du spectre semi-classique des solitons de la th{\'e}orie de supergravit{\'e}
donne une indication du spectre non perturbatif de cette th{\'e}orie
hypoth{\'e}tique. En particulier, pour autant que ces {\'e}tats soient
suffisamment {\'e}tendus pour que la description de basse {\'e}nergie
soit valide, les solitons v{\'e}rifiant la propri{\'e}t{\'e} BPS peuvent
{\^e}tre identifi{\'e}s avec des {\'e}tats BPS non perturbatifs de la
th{\'e}orie des cordes.
\index{etats BPS@{\'e}tats BPS}

La d{\'e}termination de ces {\'e}tats est probablement
{\`a} l'origine du bouleversement de notre compr{\'e}hension des th{\'e}ories
des supercordes. Elle r{\'e}v{\`e}le en effet l'existence d'une vari{\'e}t{\'e}
d'{\'e}tats BPS {\'e}tendus dits {\it $p$-branes}, 
g{\'e}n{\'e}ralisant la notion
de particule ponctuelle ( $p=0$ ) et de corde ( $p=1$ ) 
{\`a} des objets comportant $p\ge 0$ dimensions internes.
Ces {\'e}tats, infiniment lourds et localis{\'e}s {\`a} faible
couplage, correspondent {\`a} des champs de fond 
des th{\'e}ories de supergravit{\'e}
concentr{\'e}s au voisinage d'hypersurfaces de dimension $p+1$.
Ces objets poss{\`e}dent n{\'e}anmoins des modes propres de vibration leur
conf{\'e}rant une dynamique, qui peut en principe {\^e}tre calcul{\'e}e
dans l'approximation de l'espace des modules, comme dans le 
cas des monop{\^o}les magn{\'e}tiques de la th{\'e}orie des champs.

Ces {\'e}tats {\'e}tendus ne sont pas sans rappeler les D-branes
{\'e}tudi{\'e}es d{\`e}s le d{\'e}but des ann{\'e}es 1990 
dans le cadre de la T-dualit{\'e} des th{\'e}ories de cordes
ouvertes
\cite{Horava:1989ga,Dai:1989ua}. 
Contrairement aux th{\'e}ories de cordes ferm{\'e}es
\index{T-dualit{\'e}!en cordes ouvertes}
o{\`u} la transformation $R\rightarrow \alpha'/R$ agit de mani{\`e}re triviale,
la propagation libre des cordes ouvertes sur un cercle de rayon
$R$ est identifi{\'e}e dans l'image duale {\`a} la propagation sur un cercle
de rayon $\alpha'/R$ {\it en pr{\'e}sence d'un d{\'e}faut ponctuel} sur lequel
les extr{\'e}mit{\'e}s des cordes ouvertes sont contraintes de s'attacher
(figure \ref{tdualouvert}).
Ce d{\'e}faut, localis{\'e} dans la dimension compacte
dualis{\'e}e, d{\'e}finit une
hypersurface de dimension 9 dans l'espace minkovskien {\`a} 10 dimensions,
et donc une {\it 8-brane} (ou, dans le cas d'une compactification
sur un tore de dimension $d$, une $(9-d)$-brane). Ces objets,
baptis{\'e}s {\it D$p$-branes} (ou D-branes) 
par r{\'e}f{\'e}rence {\`a} la condition de bord
de Dirichlet satisfaite par les coordonn{\'e}es de plongement de la
\index{Dirichlet, condition de}
corde ouverte, ont {\'e}t{\'e} identifi{\'e}s avec les $p$-branes
de la supergravit{\'e} dans un article majeur de Polchinski, identifiant
leur couplage aux champs de jauge des supergravit{\'e}s
de type II \cite{Polchinski:1995mt}. 
Leur r{\^o}le s'est consid{\'e}rablement d{\'e}velopp{\'e}  lorsqu'il
a {\'e}t{\'e} r{\'e}alis{\'e} que la description de leurs interactions 
se ramenait {\`a} une th{\'e}orie de jauge usuelle sur leur volume
d'univers, permettant ainsi l'application fructueuse de r{\'e}sultats
obtenus dans le cadre de la th{\'e}orie des cordes au
domaine des th{\'e}ories
de jauge. Notons {\'e}galement l'existence dans les th{\'e}ories 
de type II et h{\'e}t{\'e}rotiques de la 
{\it 5-brane de Neveu-Schwarz}, ou {\it NS5-brane}, charg{\'e}e
magn{\'e}tiquement
par rapport au tenseur $B_{\mu\nu}$ du secteur gravitationnel universel.
\index{cinq-brane!de Neveu-Schwarz}
Contrairement aux D-branes, cet objet solitonique 
ne poss{\`e}de {\`a} l'heure actuelle pas de
description en termes de th{\'e}orie conforme bidimensionnelle.
\fig{3cm}{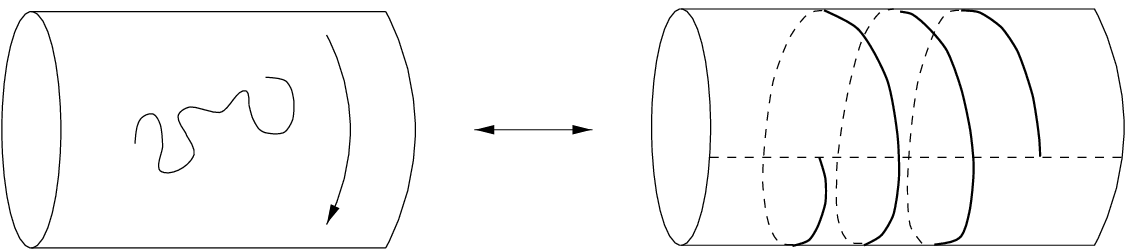}{La T-dualit{\'e} des cordes ouvertes
{\'e}change les cordes ouvertes libres (gauche) avec les cordes
ouvertes aux extr{\'e}mit{\'e}s attach{\'e}es sur la D-brane (droite).}
{tdualouvert}

\section{...{\`a} la M-th{\'e}orie}

\index{M-th{\'e}orie}
La d{\'e}couverte de ces solitons {\'e}tendus remettait en cause
la vision perturbative des th{\'e}ories de supercordes :
de la corde fondamentale et des solitons de membranes,
quels sont les objets fondamentaux d{\'e}finissant la th{\'e}orie
au niveau non perturbatif ? Quand il existe une corde
solitonique en plus de la corde fondamentale, existe-t-il
une description duale dans laquelle leurs r{\^o}les soient 
{\'e}chang{\'e}s ? On savait d{\'e}ja que les cordes de type IIA
et de type IIB {\'e}taient reli{\'e}es par la sym{\'e}trie miroir
perturbative, les autres cordes pourraient-elles {\^e}tre reli{\'e}es 
par des dualit{\'e}s non perturbatives analogues {\`a}
celles mises en {\'e}vidence en th{\'e}orie des champs ?

\subsection{Dualit{\'e} des th{\'e}ories de cordes}

Ces d{\'e}veloppements conduisirent ainsi en 1995 {\`a} la formulation
de l'hypoth{\`e}se, largement confirm{\'e}e par la suite, que
les cinq th{\'e}ories de supercordes supersym{\'e}triques
{\it ne formaient que diff{\'e}rentes facettes de la
m{\^e}me th{\'e}orie}, dans diff{\'e}rentes approximations de son espace
des modules.  
Cette id{\'e}e d{\'e}passe largement en puissance le 
concept de dualit{\'e} {\'e}lectrique-magn{\'e}tique
apparu en th{\'e}orie des champs, et s'inspire d'un
faisceau de co{\"\i}ncidences remarquables~; le lecteur
pourra s'orienter {\`a} l'aide de la M-appemonde figure
(\ref{map}).
\fig{8cm}{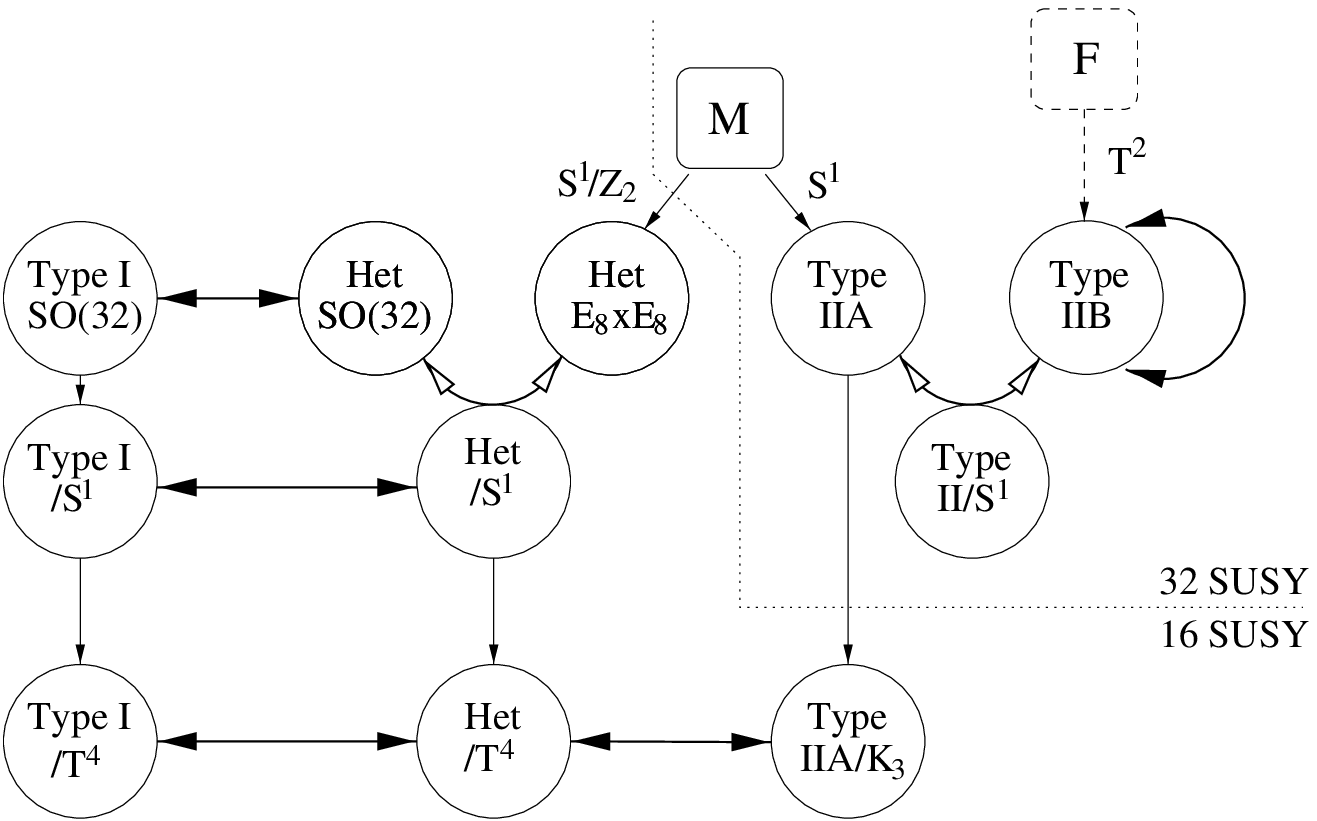}{Sch{\'e}ma synth{\'e}tique des dualit{\'e}s de
cordes {\`a} 16 et 32 charges supersym{\'e}triques. Les fl{\`e}ches verticales
repr{\'e}sentent les compactifications, les doubles fl{\`e}ches pleines les
dualit{\'e}s non perturbatives, les doubles fl{\`e}ches claires les
dualit{\'e}s perturbatives.}{map}

Ainsi, les cordes de type I
et h{\'e}t{\'e}rotiques de groupe de jauge SO(32) poss{\`e}dent la
m{\^e}me action moyennant la red{\'e}finition $g_{het}=1/g_{I}$.
\index{dualit{\'e}!h{\'e}t{\'e}rotique - type I}
La corde de type I
{\`a} dix dimensions poss{\`e}de un outre un soliton de corde, correspondant
{\`a} la 1-brane de Polchinski, dont la structure
n'est autre que celle de la corde fondamentale de la th{\'e}orie 
h{\'e}t{\'e}rotique. Ces observations conduisent {\`a} postuler que 
la th{\'e}orie des supercordes de type I et la th{\'e}orie des supercordes
h{\'e}t{\'e}rotiques SO(32) ne sont que deux d{\'e}veloppements perturbatifs,
l'un {\`a} $g_I$ faible, l'autre {\`a} $g_{I}$ grand, d'une m{\^e}me th{\'e}orie
d{\'e}finie {\`a} toute valeur de $g_{I}$. Les th{\'e}ories h{\'e}t{\'e}rotiques
de groupe de jauge $SO(32)$ et $E_8\times E_8$ {\'e}tant perturbativement
{\'e}quivalentes par T-dualit{\'e} apr{\`e}s compactification sur un cercle,
on voit que l'on ram{\`e}ne ainsi les trois th{\'e}ories de supercordes
avec supersym{\'e}trie $N=1$ {\`a} dix dimensions {\`a} une seule.

En r{\'e}alit{\'e}, le cas le plus clair de dualit{\'e} 
fut d'abord observ{\'e} dans les compactifications
{\`a} six dimensions de la corde h{\'e}t{\'e}rotique sur un tore
$T^4$, et de la corde de type IIA sur une vari{\'e}t{\'e} $K_3$.
\index{K3, surface@$K_3$, surface}
\index{compactification!sur $K_3$}
Ces deux mod{\`e}les, de supersym{\'e}trie $N=2$
en six dimensions, poss{\`e}dent le m{\^e}me espace des modules
$\Real^+ \times SO(4,20,\Real)/(SO(4)\times SO(20))$ correspondant
au dilaton et aux modules du r{\'e}seau pair autodual $\Gamma_{4,20}$
d{\'e}finissant la compactification de la th{\'e}orie h{\'e}t{\'e}rotique d'une
part~; au dilaton et aux modules de la compactification
de la corde de type IIA sur $K_3$ d'autre part. Les actions effectives
\index{espace des modules!de Het/$T^4$ - IIA/$K_3$}
de basse {\'e}nergie sont en correspondance sous l'identification
$g_{IIA}=1/g_{het}$, et les {\'e}tats du spectre perturbatif correspondant
au r{\'e}seau $\Gamma_{4,20}$ 
sont identifi{\'e}s avec les D-branes enroul{\'e}es
sur les 4 cycles auto-duaux et 20 cycles anti-auto-duaux de l'homologie
\index{cycle d'homologie!de $K_3$}
de $K_3$. Cette dualit{\'e} sera largement discut{\'e}e dans ce
m{\'e}moire, o{\`u} nous l'utiliserons pour d{\'e}duire des couplages
{\it exacts} dans une th{\'e}orie gr{\^a}ce {\`a} un calcul
{\it perturbatif} dans la th{\'e}orie duale.

\subsection{Le cha{\^\i}non manquant}

La conjecture de dualit{\'e} entre les th{\'e}ories de type IIA et 
h{\'e}t{\'e}rotiques
souffrait cependant de n'{\^e}tre valable qu'apr{\`e}s compactification
{\`a} six dimensions, et laissait myst{\'e}rieux le lien entre les th{\'e}ories
de type I (supersym{\'e}trie $N=1$)
et les th{\'e}ories de type II ($N=2$) {\`a} dix dimensions.
Le cha{\^\i}non manquant fut postul{\'e} par E. Witten
en 1995, et le myst{\`e}re qui l'entoure encore maintenant lui 
a valu l'appellation de {\it M-th{\'e}orie}. La justification
\index{M-th{\'e}orie}
\index{Kaluza-Klein!r{\'e}duction de SUGRA 11D en IIA}
de son existence repose principalement sur le lien entre la 
th{\'e}orie de supergravit{\'e}
$N=2$ {\`a} dix dimensions d{\'e}crivant la dynamique de basse {\'e}nergie des
supercordes de type IIA avec la th{\'e}orie de supergravit{\'e} $N=1$ {\`a}
\index{supergravit{\'e}!{\`a} onze dimensions}
onze dimensions construite par Cremmer, Julia et Scherk
\cite{Cremmer:1978km}. La supergravit{\'e} {\`a} 11 dimensions restitue 
en effet la supergravit{\'e}
de type IIA apr{\`e}s compactification {\`a} la Kaluza-Klein sur un cercle,
c'est-{\`a}-dire en omettant toute d{\'e}pendance sur la direction
interne\cite{Kaluza:1921,Klein:1926}.
Le rayon de la onzi{\`e}me dimension $R_{11}$ et l'{\'e}chelle de Planck {\`a}
onze dimensions $l_{11}$ se trouvent ainsi identifi{\'e}s avec
la constante de couplage $g_{IIA}$ de la th{\'e}orie des cordes de type IIA
et l'{\'e}chelle de corde $\alpha'$
selon\footnote{Les coefficients num{\'e}riques ont {\'e}t{\'e} omis dans
cette formule. Des relations {\'e}quivalentes mais utiles sont
$\alpha'=l_{11}^3/R_{11}$, $g_{IIA}=R_{11}/\sqrt{\alpha'}$.
}
\begin{equation}
R_{11}/l_{11}=g_{IIA}^{2/3}\ ,\quad l_{11}=\sqrt{\alpha'}g_{IIA}^{1/3}
\ .
\end{equation}
Cette identification entre th{\'e}ories de supergravit{\'e} peut {\^e}tre
{\'e}tendue {\`a} plus haute {\'e}nergie en identifiant les D0-branes
\index{D-brane!D0-branes}
\index{Kaluza-Klein!excitation de}
de la th{\'e}orie de type IIA avec
les modes de Kaluza-Klein du graviton de onze dimensions
\cite{Townsend:1995kk,Witten:1995ex}~;
les D2-branes et la corde fondamentale de la th{\'e}orie de type IIA
sugg{\`e}rent alors l'existence d'une
{\it membrane} de la M-th{\'e}orie, enroul{\'e}e ou non selon le cercle
\index{membrane, soliton de SUGRA 11D}
de rayon $R_{11}$, tandis les D4- et NS5-branes descendraient d'une
hypoth{\'e}tique {\it 5-brane}. Membranes et 5-branes apparaissent
\index{cinq-brane!soliton de SUGRA 11D}
du reste comme solitons de la supergravit{\'e} {\`a} onze dimensions.
Ces objets ne peuvent cependant {\^e}tre quantifi{\'e}s {\`a} ce jour, et 
la d{\'e}finition m{\^e}me de la M-th{\'e}orie est encore inconnue~;
nous reviendrons sur ce probl{\`e}me dans la suite.

L'existence de cette th{\'e}orie {\'e}tant postul{\'e}e, il est naturel
d'appliquer les m{\'e}thodes de construction de descendants habituelles
en th{\'e}orie des cordes. En particulier, tandis que la compactification sur
un cercle $S_1$ restitue la th{\'e}orie de type IIA, la compactification
sur un orbifold $S_1/\Zint_2$ (autrement dit, un segment), doit donner
une th{\'e}orie de supersym{\'e}trie moiti{\'e}, soit $N=1$. Les
constructions d'orbifold entra{\^\i}nent g{\'e}n{\'e}riquement  
l'existence d'{\'e}tats dits twist{\'e}s, localis{\'e}s sur les points fixes,
\index{twist{\'e}, {\'e}tat}
\index{compactification!de la M-th{\'e}orie sur un segment}
soit ici les deux 9-branes {\`a} chaque extr{\'e}mit{\'e} du segment.
Tandis que la th{\'e}orie des cordes offre une m{\'e}thode g{\'e}n{\'e}rale
pour d{\'e}terminer ces {\'e}tats, on doit ici se contenter d'arguments
indirects. Horava et Witten ont ainsi pu montrer que l'annulation
locale des anomalies gravitationnelles n{\'e}cessitait l'existence
\index{anomalie!gravitationnelle}
de bosons de jauge $E_8$ se propageant sur chaque bord
\cite{Horava:1996ma}. Dans la
limite o{\`u} la longueur du segment tend vers z{\'e}ro, on restitue
ainsi le contenu de la th{\'e}orie h{\'e}t{\'e}rotique $E_8\times E_8$,
dont la corde fondamentale est donn{\'e}e par la membrane (ou 2-brane)
de la M-th{\'e}orie suspendue entre les deux neuf-branes !

\subsection{La th{\'e}orie non perturbative des supercordes}
On voit donc que les cinq th{\'e}ories des cordes de supersym{\'e}trie
$N=1$ correspondent {\`a} cinq limites d'une th{\'e}orie ma{\^\i}tresse
qui admet un d{\'e}veloppement diff{\'e}rent en 
s{\'e}rie de cordes perturbative dans chaque limite, tandis que la supergravit{\'e}
{\`a} onze dimensions repr{\'e}sente la limite de basse {\'e}nergie
de cette th{\'e}orie. Plus pr{\'e}cis{\'e}ment,
nous avons discut{\'e} un certain nombre de conditions satisfaites
par ces cinq formulations limites qui autorisent l'existence d'un
prolongement dans l'int{\'e}rieur du domaine des param{\`e}tres.
Cette approche est assez similaire {\`a} celle adopt{\'e}e en g{\'e}om{\'e}trie
diff{\'e}rentielle, o{\`u} l'on d{\'e}finit une vari{\'e}t{\'e} par les cartes
\index{vari{\'e}t{\'e}!diff{\'e}rentielle}
locales sur un ensemble d'ouverts et par les fonctions de transitions.
L'analogie a ses limites, car dans le
cas pr{\'e}sent, sur chaque carte
l'information n'est que de nature asymptotique, en raison de la
non convergence de la s{\'e}rie asymptotique. On ne donne donc la
\index{asymptotique, s{\'e}rie}
th{\'e}orie que sur un ensemble de mesure nulle $g=0$, et il
faudrait un argument d'analyticit{\'e} pour prolonger cette information
{\`a} des valeurs finies des constantes de couplage.

Quoi qu'il en soit, cette th{\'e}orie non perturbative des supercordes,
que nous d{\'e}nommerons M-th{\'e}orie dans la suite de cet expos{\'e},
ne semble pas donner de place privil{\'e}gi{\'e}e aux cordes elles-m{\^e}mes,
mais plut{\^o}t aux membranes et cinq-branes. Elle doit {\'e}galement
pr{\'e}senter la propri{\'e}t{\'e} d'engendrer une alg{\`e}bre
de courants $E_8$ d{\`e}s lors qu'elle est d{\'e}finie en 
pr{\'e}sence de bord, ce qui l'apparente aux th{\'e}ories
de Chern-Simons \cite{Horava:1997dd}.
\index{Chern-Simons, th{\'e}orie de}
La th{\'e}orie des supermembranes semble en l'{\'e}tat actuel
incoh{\'e}rente et ne peut servir de d{\'e}finition {\`a} la M-th{\'e}orie.
Dans la suite, nous discuterons une proposition r{\'e}cente pour
d{\'e}finir la M-th{\'e}orie sur le front de lumi{\`e}re.
\index{lumi{\`e}re, front de}
Pour l'instant, nous adopterons une approche moins ambitieuse
mais de rapport plus imm{\'e}diat, et envisagerons dans quelle mesure
les techniques de calcul semi-classique en th{\'e}orie des champs
peuvent {\^e}tre transpos{\'e}es {\`a} la th{\'e}orie des cordes.

\subsection{Approche semi-classique {\`a} la th{\'e}orie des supercordes}
\index{semi-classique!calcul en th{\'e}orie des cordes}
Comme nous l'avons discut{\'e} plus haut, les m{\'e}thodes 
semi-classiques pour la d{\'e}termination non perturbative d'amplitudes
physiques sont particuli{\`e}rement efficaces en th{\'e}orie des champs
lorsque l'on s'int{\'e}resse {\`a} des processus interdits perturbativement.
La th{\'e}orie des supercordes, consid{\'e}r{\'e}e dans un
de ses vides supersym{\'e}triques, remplit pr{\'e}cis{\'e}ment cette
condition, du moins lorsque l'on s'int{\'e}resse aux interactions 
dominantes {\`a} basse {\'e}nergie. On peut donc chercher {\`a} calculer
les corrections non perturbatives en incluant les configurations
d'instantons de la th{\'e}orie des cordes.  
\index{instanton!des th{\'e}ories des cordes}
\index{enroulement!des solitons en instantons}
L'{\'e}tude des points-selles euclidiens de la th{\'e}orie de supergravit{\'e}
d{\'e}crivant la dynamique de basse {\'e}nergie permet de d{\'e}terminer
ces configurations, tout comme l'{\'e}tude des solutions classiques
minkovskiennes fournissait le spectre BPS non perturbatif des $p$-branes.
Tout naturellement, les solutions euclidiennes s'obtiennent
en enroulant la ligne d'univers de genre temps
\index{ligne d'univers}
des solutions minkovskiennes statiques autour d'un cercle de genre 
espace du continuum d'espace-temps. On obtient ainsi un ``spectre''
d'objets instantoniques, localis{\'e}s {\`a} un instant  donn{\'e},
mais {\'e}tendus dans $p+1$ directions spatiales, {\`a} la diff{\'e}rence des 
instantons ponctuels des th{\'e}ories de jauge en dimension 4.
Malheureusement, l'absence de formulation non perturbative de la
th{\'e}orie des supercordes ne nous permet pas d'obtenir les r{\`e}gles
de sommation sur ces configurations de mani{\`e}re d{\'e}ductive.
\index{instanton!mesure d'int{\'e}gration}
Cette approche serait donc vou{\'e}e {\`a} l'{\'e}chec, s'il n'existait
un certain nombre d'amplitudes physiques d{\'e}terminables exactement
gr{\^a}ce par des arguments de dualit{\'e}. L'examen de ces quantit{\'e}s,
exactes {\`a} toute valeur du couplage, dans la limite de faible
couplage, permet d'identifier les contributions non perturbatives
de ces instantons, et donne des indications pr{\'e}cieuses sur les
r{\`e}gles {\`a} appliquer dans des cas plus g{\'e}n{\'e}raux.
Cette strat{\'e}gie a {\'e}t{\'e} mise en \oe{}uvre dans le cadre de ce
travail de th{\`e}se, et fera l'objet d'une discussion approfondie
dans ce m{\'e}moire. 

\subsection{M comme Matrice ?}
\index{M-th{\'e}orie}
\index{matrices, th{\'e}orie des}
D{\'e}finie par le prolongement hypoth{\'e}tique de th{\'e}ories perturbatives, 
la th{\'e}orie non perturbative des supercordes gagnerait en cr{\'e}dibilit{\'e}
et pr{\'e}dictivit{\'e} {\`a} recevoir une formulation intrins{\`e}que qui
ne fasse appel {\`a} aucun d{\'e}veloppement perturbatif.
Bien que la r{\'e}ponse {\`a} cette question reste encore hors d'atteinte,
Banks, Fischler, Shenker et Susskind ont propos{\'e} une
\index{Banks, Fischler, Shenker et Susskind, conjecture de}
formulation \cite{Banks:1997vh,Susskind:1997cw} qui 
reproduit bon nombre des caract{\'e}ristiques suppos{\'e}es de la 
M-th{\'e}orie, dont l'amplitude de diffusion graviton-graviton
pr{\'e}dite par la th{\'e}orie de supergravit{\'e} {\`a} 11 dimensions
\cite{Becker:1997wh}.
Cette approche, connue sous le nom de th{\'e}orie des matrices
({\it M(atrix) Theory}) postule que la dynamique de la M-th{\'e}orie 
peut {\^e}tre d{\'e}crite {\it sur le  front de lumi{\`e}re} par la
{\it m{\'e}canique quantique supersym{\'e}trique}
de 9 matrices hermitiennes de $u(N)$ dans la limite de grand $N$,
obtenue par r{\'e}duction dimensionnelle de la th{\'e}orie de Yang-Mills
$U(N)$ supersym{\'e}trique {\`a} dix dimensions~; les 9 champs scalaires
$A_I$ issus de la r{\'e}duction du potentiel vecteur $A_{\mu}$
sont identifi{\'e}s aux 9 coordonn{\'e}es {\it non commutatives}
transverses au front de lumi{\`e}re. Ce mod{\`e}le n'est autre que 
celui d{\'e}crivant la dynamique de $N$ D0-branes, ainsi identifi{\'e}es
\index{D-brane!D0-branes}
\index{parton}
aux composants {\'e}l{\'e}mentaires, ou {\it partons}, de la M-th{\'e}orie.
Cette formulation rappelle les d{\'e}veloppements
encore r{\'e}cents des mod{\`e}les de matrices, qui avaient en effet
permis de faire le lien entre la m{\'e}canique statistique d'une
matrice al{\'e}atoire gaussienne dans la limite de {\it double scaling}
et le d{\'e}veloppement en genre d'une th{\'e}orie de cordes.
La connection avec les th{\'e}ories de supercordes critiques {\'e}tait
cependant rest{\'e}e hors de port{\'e}e, et le calcul limit{\'e} au
calcul de certains exposants critiques. L'analogie a pourtant ses
limites, car c'est ici les fluctuations d'un nombre continu de matrices
param{\'e}tr{\'e}es par le temps propre qu'il faut consid{\'e}rer. La
propri{\'e}t{\'e} de supersym{\'e}trie {\'e}tendue de la th{\'e}orie des
matrices permet cependant d'esp{\'e}rer que ce probl{\`e}me reste
traitable. Sa recevabilit{\'e} reste cependant subordonn{\'e}e {\`a}
la d{\'e}monstration de l'invariance de 
Lorentz $SO(1,10)$ {\`a} onze dimensions,
\index{Lorentz, invariance {\`a} 11D}
non manifeste sur le front de lumi{\`e}re.

Encore cette formulation ne traite-t-elle que la M-th{\'e}orie dans
l'espace plat {\`a} onze dimensions, soit une limite particuli{\`e}re
de la th{\'e}orie non perturbative des supercordes.
Pour pr{\'e}tendre au nom de M-th{\'e}orie, elle doit
pouvoir d{\'e}crire les configurations
de champs de fond les plus g{\'e}n{\'e}rales, et en particulier
pouvoir {\^e}tre compactifi{\'e}e
\index{compactification!de la th{\'e}orie des matrices}
\footnote{Plus pr{\'e}cisement, l'espace de Hilbert de la th{\'e}orie des
matrices devrait comporter des secteurs de supers{\'e}lection
d{\'e}crivant les diff{\'e}rentes compactifications possibles.}.
Le m{\'e}canisme de compactification 
sur des vari{\'e}t{\'e}s quelconques est bien loin d'{\^e}tre compris,
mais d{\'e}j{\`a}, la compactification sur des vari{\'e}t{\'e}s toro{\"\i}dales 
plates de dimension $d$
se r{\'e}v{\`e}le beaucoup plus complexe qu'en th{\'e}ories
des champs ou des cordes habituelles, puisque la m{\'e}canique quantique
de matrices $U(N)$ devient une th{\'e}orie de jauge
supersym{\'e}trique $U(N)$ en dimension $d+1$, dans la limite de grand $N$ !
Ce saut qualitatif compromet la pr{\'e}dictivit{\'e} de cette th{\'e}orie,
en particulier lorsque $d\ge 3$ puisqu'alors la th{\'e}orie de jauge
devient non renormalisable et donc mal d{\'e}finie {\`a} courte distance,
\index{renormalisation!renormalisabilit{\'e} des th{\'e}ories de jauge}
o{\`u} elle doit {\^e}tre compl{\'e}t{\'e}e par des degr{\'e}s de libert{\'e}
suppl{\'e}mentaires. Les U-dualit{\'e}s de la M-th{\'e}orie sont alors
\index{U-dualit{\'e}!et th{\'e}orie des matrices}
identifi{\'e}es aux dualit{\'e}s {\'e}lectrique-magn{\'e}tique et g{\'e}om{\'e}triques
de ces th{\'e}ories de jauge {\'e}tendues. Au cours de ce travail de th{\`e}se,
nous avons {\'e}galement montr{\'e} comment les modules de la
compactification pouvaient s'interpr{\'e}ter dans le cadre de la
th{\'e}orie des matrices, et {\'e}tudi{\'e} le spectre des
{\'e}tats BPS dans les deux formulations. L'identification de la th{\'e}orie
des matrices {\`a} la M-th{\'e}orie, et en particulier aux th{\'e}ories
de supercordes limites, n'a cependant {\'e}t{\'e} v{\'e}rifi{\'e}e 
que dans le secteur BPS, et on peut l{\'e}gitimement se
demander si cette {\'e}quivalence, comme les conjectures de dualit{\'e}
des cordes, s'{\'e}tend au spectre entier des th{\'e}ories de supercordes.

\subsection[Au del{\`a} du cadre de ce m{\'e}moire]{Au del{\`a} du cadre de ce m{\'e}moire...}
Le panorama historique que nous avons dessin{\'e} jusqu'{\`a} pr{\'e}sent
ne donne qu'un aper{\c c}u impressionniste et impr{\'e}cis
du bouleversement survenu en quelques ann{\'e}es dans les
th{\'e}ories des champs et des cordes. 
Nous aurons l'occasion
de rem{\'e}dier {\`a} une partie des insuffisances de cette
introduction dans la
suite de ce m{\'e}moire pour les sujets touchant de plus pr{\`e}s
{\`a} mes recherches doctorales, renvoyant
le lecteur aux articles
originaux et articles de revue cit{\'e}s en r{\'e}f{\'e}rence
pour plus de d{\'e}tails. Nous devrons {\'e}galement ignorer certains 
d{\'e}veloppements prometteurs que nous n'avons pu aborder
dans le cadre de cette th{\`e}se mais que nous 
esp{\'e}rons explorer dans un proche avenir~:
\begin{itemize}
\item La th{\'e}orie des supercordes de type IIB admet des sch{\'e}mas de
compactification non perturbative dans lesquels le dilaton n'est 
pas uniforme, et d{\'e}fini {\`a} une transformation de S-dualit{\'e} 
$Sl(2,\Zint)$ pr{\`e}s. Ces configurations peuvent naturellement
{\^e}tre interpr{\'e}t{\'e}es dans le cadre d'une th{\'e}orie en dimension 12,
introduite par C. Vafa sous le nom de F-th{\'e}orie \cite{Vafa:1996xn}. 
\index{F-th{\'e}orie}
Bien que cette derni{\`e}re soit encore mal d{\'e}finie
(voir \cite{Bars:1997bz} pour une tentative int{\'e}ressante), on peut ainsi 
construire une nouvelle classe de mod{\`e}les, dont, par
compactification sur un espace de Calabi-Yau de dimension complexe 4,
certains mod{\`e}les de supersym{\'e}trie $N=1$ {\`a} quatre dimensions 
d'int{\'e}r{\^e}t ph{\'e}nom{\'e}nologique. Cette proc{\'e}dure permet {\'e}galement
\index{phenomenologie@ph{\'e}nom{\'e}nologie}
de construire des th{\'e}ories de jauge de contenu arbitraire dans
la limite o{\`u} la gravit{\'e} se d{\'e}couple, et de r{\'e}soudre celles-ci
explicitement \cite{Katz:1997fh,Klemm:1997gg}.

\item La mod{\'e}lisation des trous noirs en termes d'{\'e}tats li{\'e}s
\index{trou noir, entropie des}
de D-branes permet une description des degr{\'e}s de libert{\'e}
microscopiques de ces objets. On a ainsi pu d{\'e}river la formule
de Bekenstein-Hawking pour l'entropie des trous noirs extr{\'e}maux 
et quasi extr{\'e}maux, correspondant aux intersections BPS et
quasi-BPS de D-branes
(pour une courte revue, voir par exemple \cite{Maldacena:1997tm}). 
Ce calcul a {\'e}t{\'e} r{\'e}cemment {\'e}tendu
aux trous noirs non extr{\'e}maux, dont le trou noir de Schwarzschild,
gr{\^a}ce {\`a} certaines transformations de U-dualit{\'e} non 
consid{\'e}r{\'e}es dans ce m{\'e}moire \cite{Sfetsos:1997xs}.

\item L'{\'e}tude de la dynamique de volume d'univers des
solitons de la M-th{\'e}orie permet de comprendre 
les dualit{\'e}s des th{\'e}ories de jauge comme action g{\'e}om{\'e}trique
sur la configuration solitonique
(voir \cite{Giveon:1998sr} pour une revue 
de ces d{\'e}veloppements)~; elle sugg{\`e}re {\'e}galement 
l'existence d'une classe de th{\'e}ories de jauge non triviales en dimension
sup{\'e}rieure {\`a} quatre, dont l'absence de formulation lagrangienne
ne diminue pas l'int{\'e}r{\^e}t. Elle permet enfin de relier les
excitations des th{\'e}ories de jauge conformes, et en particulier
de la th{\'e}orie de Yang-Mills N=4 {\`a} quatre dimensions, aux
excitations de supergravit{\'e} sur l'horizon de ces solitons,
\index{horizon, g{\'e}om{\'e}trie de l'}
tout au moins dans la limite de grand $N$
\cite{Maldacena:1997re}.
\end{itemize}

\section{Pr{\'e}sentation de mes recherches doctorales}

Au cours de cette th{\`e}se, je me suis tout d'abord int{\'e}ress{\'e}
aux propri{\'e}t{\'e}s de dualit{\'e} des th{\'e}ories de champs 
supersym{\'e}triques $N=2$. Dans leur article fondateur de 1994, 
Seiberg et Witten ont montr{\'e} comment les corrections
non perturbatives {\`a} la m{\'e}trique des scalaires
de la branche de Coulomb, o{\`u} la sym{\'e}trie de
jauge est bris{\'e}e en un sous-groupe ab{\'e}lien, pouvaient
{\^e}tre calcul{\'e}es gr{\^a}ce aux propri{\'e}t{\'e}s de dualit{\'e} 
{\'e}lectrique-magn{\'e}tique. En revanche, la branche de Higgs,
o{\`u} la valeur moyenne des scalaires des hypermultiplets
brise enti{\`e}rement la sym{\'e}trie de jauge, est prot{\'e}g{\'e}e 
de telles corrections par la supersym{\'e}trie globale.
Les corrections provenant de la th{\'e}orie des cordes ne
sont cependant pas exclues, ce qui a motiv{\'e} l'{\'e}tude
de cette branche sous la direction d'Ignatios Antoniadis
(annexe \ref{hk}) \cite{Antoniadis:1996ra}.
J'ai en particulier pu montrer l'existence d'une nouvelle dualit{\'e}
identifiant la branche de Higgs d'une th{\'e}orie de jauge 
avec $N_c$ couleurs  et $N_f$ saveurs avec celle d'une th{\'e}orie
de $N_f-N_c$ couleurs et $N_f$ saveurs. Cette dualit{\'e}
des branches de Higgs rappelle la dualit{\'e} infrarouge sugg{\'e}r{\'e}e
par Seiberg dans les th{\'e}ories des champs supersym{\'e}triques
$N=1$ correspondantes~; elle a par la suite  {\'e}t{\'e}
interpr{\'e}t{\'e}e en termes d'action g{\'e}om{\'e}trique sur les
configurations de D-branes. Je me suis ensuite pench{\'e}
avec Herv{\'e} Partouche sur le probl{\`e}me de la brisure
partielle de supersym{\'e}trie globale $N=2$ par
les termes de Fayet-Iliopoulos magn{\'e}tiques qu'il avait
mise en {\'e}vidence avec Ignatios Antoniadis et Tomasz
Taylor \cite{antoniadis:1996.1}~; 
cette {\'e}tude a conclu {\`a} l'impossibilit{\'e} d'{\'e}tendre 
ce processus en pr{\'e}sence d'hypermultiplets charg{\'e}s minimalement
coupl{\'e}s, et ne sera pas reprise dans ce m{\'e}moire
\cite{Partouche:1996yp}.

Mon int{\'e}r{\^e}t s'est ensuite port{\'e} vers les dualit{\'e}s
des th{\'e}ories de cordes, au contact d'Elias Kiritsis
et Costas Kounnas {\`a} la division Th{\'e}orie du CERN. 
Apr{\`e}s l'{\'e}tude d'une proposition r{\'e}cente de d{\'e}nombrement
des {\'e}tats BPS des th{\'e}ories de cordes h{\'e}t{\'e}rotiques compactifi{\'e}es
sur $T^6$ ou de type IIA compactifi{\'e}e sur $K_3 \times T^2$ 
\cite{Dijkgraaf:1997it}
qui s'est r{\'e}v{\'e}l{\'e}e incompl{\`e}te, nous
nous sommes int{\'e}ress{\'e}s, en collaboration avec
Andrea Gregori, Niels Obers et Marios Petropoulos,
aux corrections de seuil aux interactions gravitationnelles 
{\`a} quatre d{\'e}riv{\'e}es dans les th{\'e}ories $N=4$ de
rang non maximal (annexe \ref{tt})\cite{Gregori:1997hi}. 
Sous la dualit{\'e} h{\'e}t{\'e}rotique-type II,
la contribution {\`a} une boucle de type II donne lieu
{\`a} des corrections non perturbatives en $e^{-1/g^2}$ 
correspondant {\`a} des {\it cinq-branes} h{\'e}t{\'e}rotiques enroul{\'e}es
sur le tore $T^6$.
Cette g{\'e}n{\'e}ralisation des r{\'e}sultats
de Harvey et Moore \cite{Harvey:1996ir} a mis en {\'e}vidence la restriction 
de la S-dualit{\'e} {\`a} un sous-groupe de $Sl(2,\Zint)$ dans ces
mod{\`e}les d'orbifolds sans point fixes, ainsi que certains
comportements inattendus de th{\'e}ories h{\'e}t{\'e}rotiques
{\`a} fort couplage. 

Cette premi{\`e}re confrontation avec les effets non perturbatifs
de la th{\'e}orie des cordes m'a conduit {\`a} {\'e}tudier,
en collaboration avec Ignatios Antoniadis et Tomasz Taylor,
d'autres couplages {\`a} quatre d{\'e}riv{\'e}es dans ces th{\'e}ories
$N=4$, qui ont cette fois la propri{\'e}t{\'e} d'{\^e}tre donn{\'e}s
exactement {\`a} une boucle du cot{\'e} h{\'e}t{\'e}rotique 
(annexe \ref{dds})\cite{Antoniadis:1997zt}. Leur
interpr{\'e}tation du point de vue des cordes de type II correspond
alors {\`a} des effets non perturbatifs en $e^{-1/g}$ de D-branes
s'enroulant sur les cycles d'homologie de $K_3\times T^2$.
Nous avons ainsi obtenu les premiers exemples 
explicites d'effets d'instantons
en espace courbe, et retrouv{\'e} le d{\'e}nombrement des cycles alg{\'e}briques
de $K_3$ de genre donn{\'e}, conjectur{\'e} ind{\'e}pendamment par Berschadsky,
Sadov et Vafa \cite{Bershadsky:1996qy}.

Je me suis simultan{\'e}ment 
tourn{\'e} avec Elias Kiritsis vers l'{\'e}tude de situations
plus pures o{\`u} les contributions de ces instantons pouvaient
{\^e}tre comprises plus pr{\'e}cis{\'e}ment. Nous nous sommes ainsi int{\'e}ress{\'e}s
aux couplages en $R^4$ de  la th{\'e}orie de type IIB, pour lesquels
Green et Gutperle avaient conjectur{\'e} une expression non perturbative
en termes d'une fonction non holomorphe modulaire de $Sl(2,\Zint)$
\cite{Green:1997tv}. Nous
avons {\'e}tendu leur conjecture aux compactifications de la corde
de type IIB sur $T^2$ et $T^3$, pour lesquelles nous avons montr{\'e}
que les s{\'e}ries d'Eisenstein invariantes sous les groupes 
de U-dualit{\'e} $Sl(3,\Zint)$
et $Sl(5,\Zint)$ reproduisaient les contributions perturbatives
{\`a} l'ordre des arbres et {\`a} une boucle, ainsi que des sommes
d'instantons interpr{\'e}tables en termes de cordes de type $(p,q)$,
images de la corde fondamentale sous la sym{\'e}trie $Sl(2,\Zint)_B$,
enroul{\'e}es sur les deux-cycles du tore de compactification
(annexe \ref{pq})\cite{Kiritsis:1997em}.
Nous avons ensuite {\'e}tendu ces r{\'e}sultats aux compactifications
toro{\"\i}dales quelconques de type IIA ou IIB, obtenant les
contributions de D$p$-branes de dimensionalit{\'e} arbitraire 
{\`a} ces couplages (annexe \ref{dc})\cite{Pioline:1997pu}. 
En m{\^e}me temps, nous remarquions que ces
contributions ne suffisaient pas {\`a} reproduire un r{\'e}sultat
invariant sous la dualit{\'e} en dimension $D\le 6$, et
qu'elles devaient {\^e}tre compl{\'e}t{\'e}es par des contributions
en $e^{-1/g^2}$ dont l'interpr{\'e}tation reste obscure
{\`a} ce jour. Tout r{\'e}cemment, j'ai pu d{\'e}montr{\'e} l'absence
de contributions des formes cuspo{\"\i}{}dales aux couplages en
$R^4$, fournissant ainsi la d{\'e}monstration d{\'e}finitive
de la conjecture de Green et Gutperle (annexe \ref{nr4})\cite{Pioline:1998}.
Ce r{\'e}sultat appara{\^\i}{}t {\'e}galement comme cons{\'e}quence
de l'{\'e}tude des couplages {\`a} quatre d{\'e}riv{\'e}es cit{\'e}
plus haut. 

La th{\'e}orie des matrices, pr{\'e}tendante {\`a} une d{\'e}finition
non perturbative de la M-th{\'e}\-orie et donc de la th{\'e}orie
des supercordes, ayant remport{\'e} des succ{\`e}s remarquables,
je me suis pench{\'e} sur ces d{\'e}veloppements en collaboration
avec Eliezer Rabinovici et Niels Obers, dans l'espoir de
clarifier ces contributions instantoniques (annexe \ref{mu})
\cite{Obers:1997kk}. 
Nous avons ainsi
pu faire le lien avec des r{\'e}sultats obtenus peu avant
par Eliezer en collaboration avec Amit Giveon, Shlyamon Elitzur  
et David Kutasov sur l'impl{\'e}mentation de la sym{\'e}trie de
U-dualit{\'e} en th{\'e}orie des matrices \cite{Elitzur:1997zn}. Nous avons {\'e}galement
pu {\'e}tendre ces r{\'e}sultats, en d{\'e}montrant comment la 
th{\'e}orie des matrices devait {\^e}tre modifi{\'e}e pour
d{\'e}crire des compactifications en pr{\'e}sence de champs de
fond de jauge, n{\'e}cessaires {\`a} la r{\'e}alisation de la
U-dualit{\'e}.


%% file: chap2.tex
\chapter{Dualit{\'e}s non perturbatives des th{\'e}ories des champs
supersym{\'e}triques}
Comme nous l'avons d{\'e}crit dans notre introduction, la notion 
de dualit{\'e} {\'e}lectrique-magn{\'e}tique mise en {\'e}vidence dans
les th{\'e}ories de jauges non ab{\'e}liennes {\`a} supersym{\'e}trie
{\'e}tendue a jou{\'e} un r{\^o}le essentiel dans l'{\'e}mergence de
la conjecture de dualit{\'e} des th{\'e}ories de supercordes.
Nous en donnerons ici une pr{\'e}sentation sommaire, qui
nous permettra d'introduire de nombreux concepts de base 
implicites dans les chapitres ult{\'e}rieurs. Nous suivrons ainsi
le cheminement de mon apprentissage, et introduirons le lecteur {\`a} la 
publication en appendice \ref{hk}. Ces id{\'e}es ne prendront leur pleine
dimension qu'une fois transpos{\'e}es aux th{\'e}ories
de supergravit{\'e} qui d{\'e}crivent la dynamique {\`a} basse
{\'e}nergie des th{\'e}ories de supercordes, et pr{\'e}sentent
des sym{\'e}tries <<cach{\'e}es>> non perturbatives analogues
{\`a} la S-dualit{\'e} des th{\'e}ories de jauge. La br{\`e}ve
description que nous en donnerons nous introduira aux
dualit{\'e}s des th{\'e}ories de supercordes, qui feront l'objet
du chapitre suivant.

\section{Des monop{\^o}les magn{\'e}tiques {\`a} la S-dualit{\'e}}

\subsection{Dualit{\'e} et monop{\^o}les magn{\'e}tiques}
\index{dualit{\'e}!{\'e}lectrique-magn{\'e}tique}
L'id{\'e}e de dualit{\'e} {\'e}lectrique-magn{\'e}tique remonte sans 
doute aux premiers jours de l'{\'e}lectrodynamique classique
maxwellienne, dont les {\'e}quations dans le vide pr{\'e}sentent
une sym{\'e}trie globale $U(1)$ m{\'e}langeant champs 
{\'e}lectrique et champs magn{\'e}tique selon
$E+i B \rightarrow e^{i\alpha} (E+i B)$. 
La validit{\'e} en pr{\'e}sence de mati{\`e}re de cette sym{\'e}trie 
impliquerait l'existence de particules
ponctuelles charg{\'e}es magn{\'e}tiquement, dites
{\it monop{\^o}les magn{\'e}tiques}, qui aurait l'avantage
\index{monop{\^o}le magn{\'e}tique!de Dirac}
d'expliquer la quantification de la charge~: en effet,
l'inobservabilit{\'e} de la singularit{\'e} de Dirac dans
le potentiel vecteur en pr{\'e}sence de deux 
{\it dyons} de charges {\'e}lectriques et magn{\'e}tiques 
$(q_e,q_m)$ et $(q'_e,q'_m)$ impose la condition 
de Dirac-Schwinger-Zwanziger
\cite{Dirac:1931,Schwinger:1966nj,Zwanziger:1968}
\index{Dirac, condition de quantification de}
\begin{equation}
q_e q_m' - q'_e q_m \in 4\pi \hbar \Zint \ ;
\end{equation}
la charge {\'e}lectrique $q_e$ et la charge magn{\'e}tique $q_m$ 
sont donc restreintes {\`a} prendre leurs valeurs dans le r{\'e}seau
\begin{equation}
q_e = \hbar g~e\ ,\quad q_m = \frac{4\pi}{g} m\ , \quad 
(e,m)\in\Zint\times \Zint;
\end{equation}
o{\`u} la premi{\`e}re relation d{\'e}finit notre convention pour
le couplage de jauge $g$. On voit ainsi que la dualit{\'e} 
{\'e}lectrique-magn{\'e}tique {\'e}changeant charges {\'e}lectrique
et magn{\'e}tique s'accompagne d'une {\it inversion 
du couplage de jauge}\footnote{Nous choisissons d{\`e}s {\`a} pr{\'e}sent
les unit{\'e}s de mesure $c=\hbar=1$.}
\begin{equation}
g \rightarrow \frac{4\pi}{g}\ , e \leftrightarrow m
\end{equation}
{\'e}changeant faible couplage et fort couplage.
Les monop{\^o}les de la th{\'e}orie de Maxwell correspondent
cependant {\`a} des configurations singuli{\`e}res du
champ de jauge, et leur existence est malheureusement 
fortement exclue par les donn{\'e}es
exp{\'e}rimentales. 

\subsection{Monop{\^o}le de 't Hooft-Polyakov et conjecture
de Montonen-Olive}
L'int{\'e}r{\^e}t pour ces particules a {\'e}t{\'e} relanc{\'e} avec la
d{\'e}couverte de solutions classiques
de th{\'e}ories de jauges non ab{\'e}liennes {\`a} 
sym{\'e}trie spontan{\'e}ment bris{\'e}e, charg{\'e}es magn{\'e}tiquement
sous le groupe de jauge ab{\'e}lien r{\'e}siduel
\cite{'tHooft:1974qc,Polyakov:1974ek}. 
\index{monop{\^o}le magn{\'e}tique!de 't Hooft-Polyakov}
La charge magn{\'e}tique $m$ est donn{\'e}e par un invariant
topologique, et ces {\'e}tats sont donc stables classiquement.
Leur masse est donn{\'e}e classiquement dans le cas du mod{\`e}le
de Georgi-Glashow par
\index{Georgi-Glashow, mod{\`e}le de}
\begin{equation}
\mathcal{M}_{m}= \frac{4\pi |a|}{g}\cdot |m|\ , \quad m\in \Zint\ ,
\end{equation}
o{\`u} $a^2=\frac{1}{2}\tr\phi^2$ d{\'e}signe la valeur moyenne du scalaire de
Higgs \index{Higgs!scalaire de}
dans la repr{\'e}sentation adjointe du groupe de jauge $SU(2)$. Ceci
est {\`a} comparer {\`a} la masse des bosons $W^{\pm}$ du spectre perturbatif,
\begin{equation}
\mathcal{M}_{e}= |a|\ g\cdot  |e|\ , \quad e=\pm 1\ .
\end{equation}
Ces deux objets saturent donc la {\it borne de Bogomolny}  
\cite{Bogomolny:1976de}\index{Bogomolny, borne de}
apparaissant dans la limite de Prasad et Sommerfield
\index{Prasad-Sommerfield, limite de}
\cite{Prasad:1975kr} du mod{\`e}le de Georgi-Glashow, o{\`u} la longueur du
champ de Higgs est gel{\'e}e~:
\begin{equation}
\label{mbogo}
\mathcal{M}\ge |a| \sqrt{ q^2_e + q^2_m}\ .
\end{equation}
Cette relation, {\it invariante sous la dualit{\'e}
{\'e}lectrique-magn{\'e}tique}, recevra plus loin une interpr{\'e}tation 
dans le cadre des th{\'e}ories supersym{\'e}triques.
Un calcul semi-classique de
diffusion montre {\'e}galement 
que la force statique entre deux monop{\^o}les s'annule,
tout comme l'interaction entre deux particules $W^{\pm}$ de m{\^e}me charge,
en raison de la compensation entre l'{\'e}change r{\'e}pulsif de bosons
vecteurs et l'{\'e}change attractif de particules de Higgs
\index{Higgs!particule de}.
Ces constatations conduisirent en  1977 Montonen et Olive {\`a} formuler
\index{Montonen-Olive, conjecture de}
l'hypoth{\`e}se de dualit{\'e} {\'e}lectrique-magn{\'e}tique 
des th{\'e}ories de jauge de groupe $SU(2)$, g{\'e}n{\'e}ralis{\'e}e peu apr{\`e}s
{\`a} tous les groupes compacts par Goddard, Nuyts et Olive,
postulant l'{\'e}quivalence des th{\'e}ories de jauge de 
couplage $g$ et $4\pi/g$, les monop{\^o}les magn{\'e}tiques de 't Hooft-
Polyakov jouant le r{\^o}le des champs fondamentaux dans
la formulation duale
\cite{Montonen:1977sn,Goddard:1977qe}.

\subsection{Supersym{\'e}trie et propri{\'e}t{\'e} BPS}
L'id{\'e}e de dualit{\'e} dans les th{\'e}ories quantiques des champs
en tant que telle n'est pas neuve : la dualit{\'e} de Kramers-Wannier
\index{dualit{\'e}!de Kramers-Wannier}
entre les phases de haute temp{\'e}rature et de basse temp{\'e}rature
du mod{\`e}le d'Ising-Onsager en est sans doute le premier exemple ;
le mod{\`e}le de {\it sine}-Gordon {\`a} 1+1 dimensions
s'est {\'e}galement r{\'e}v{\'e}l{\'e} dual {\`a} fort couplage
au mod{\`e}le de Thirring, les solitons d'une th{\'e}orie
s'identifiant aux champs fondamentaux de la th{\'e}orie duale
\cite{Coleman:1975bu}
\index{dualit{\'e}!de sine Gordon-Thirring}.
L'importance particuli{\`e}re de la dualit{\'e} de Montonen-Olive
tient au fait qu'elle correspond
{\`a} une th{\'e}orie des champs {\`a} 3+1 dimensions non int{\'e}grable,
et candidate {\`a} une description ph{\'e}nom{\`e}nologique des
interactions fortes et {\'e}lectrofaibles. Dans sa
formulation premi{\`e}re de Montonen et Olive, cette conjecture
n'est cependant pas tenable, pour deux raisons apparemment
distinctes. Premi{\`e}rement, le couplage de jauge, par suite des corrections
quantiques, d{\'e}pend de l'{\'e}chelle d'observation, et l'identification
$g\leftrightarrow 4\pi/g$ n'a pas de sens~; en outre la masse des
solitons en fonction de ce couplage est sujette {\`a} des corrections
quantiques et la validit{\'e} de la formule (\ref{mbogo}) n'est pas assur{\'e}e~;
enfin, les monop{\^o}les du mod{\`e}le de Georgi-Glashow
sont des particules scalaires qui ne sauraient
{\^e}tre identifi{\'e}es aux particules fondamentales de spin 1.

Ces trois probl{\`e}mes peuvent {\^e}tre r{\'e}solus simultan{\'e}ment au
prix d'une perte de g{\'e}n{\'e}ralit{\'e} et de pertinence
ph{\'e}nom{\`e}nologique, en consid{\'e}rant des th{\'e}ories de jauge
{\`a} supersym{\'e}trique {\'e}tendue. La supersym{\'e}trie donne
lieu {\`a} des th{\'e}or{\`e}mes de non renormalisation permettant
de contr{\^o}ler les corrections quantiques. 
\index{non renormalisation!de l'action {\`a} 2 d{\'e}riv{\'e}es dans SYM $N=4$}
En particulier, les th{\'e}ories $N=4$ sont finies
{\`a} tout ordre en perturbation, et l'action effective {\`a} deux
d{\'e}riv{\'e}es ne re{\c c}oit aucune correction quantique
\index{action effective!{\`a} deux d{\'e}riv{\'e}es}
\footnote{Les th{\'e}ories $N=4$ sont en effet des exemples
de th{\'e}ories conformes {\`a} quatre dimensions. La sym{\'e}trie
\index{conforme, th{\'e}orie des champs!SYM $N=4$}
conforme est spontan{\'e}ment bris{\'e}e par la valeur
moyenne $a$ du champ de Higgs.}. Le couplage
de jauge $g$ est donc {\'e}gale {\`a} sa valeur <<nue>>, et parler
de son inversion ne pose pas de difficult{\'e}.
La supersym{\'e}trie organise aussi
le spectre en repr{\'e}sentations d{\'e}pendant des 
{\it charges centrales} $Z^{ij}$ de
l'alg{\`e}bre de supersym{\'e}trie {\'e}tendue
\index{charge centrale}
\index{supersym{\'e}trie!charge centrale}
\footnote{L'ouvrage \cite{wess/bagger:1992} rassemble les informations
essentielles sur la supersym{\'e}trie et la supergravit{\'e}.}
\begin{subequations}
\begin{align}
\{ Q^i_\alpha, Q^j_{\bar \beta} \} &= \sigma^\mu_{\alpha\bar\beta}~P_\mu \\
\{Q_\alpha^i,Q_\beta^j\} &=2 \epsilon_{\alpha\beta} Z^{ij}\ .
\end{align}
\end{subequations}
\index{supersym{\'e}trie!alg{\'e}bre de}
\index{supersym{\'e}trie!repr{\'e}sentations de la}
o{\`u} les indices grecs correspondent aux indices spinoriels
et les indices $i=1\dots N$ aux $N$ charges supersym{\'e}triques.
La repr{\'e}sentation irr{\'e}ductible g{\'e}n{\'e}rique, de dimensions $2^{2N}$ 
v{\'e}rifie l'in{\'e}galit{\'e}
\begin{equation}
\label{mz}
\mathcal{M} \ge |Z_{\lambda}|
\end{equation}
o{\`u} $Z_{\lambda}$ d{\'e}signe toute valeur propre de la matrice
antisym{\'e}trique $Z^{ij}$. Lorsque l'in{\'e}galit{\'e} (\ref{mz})
est satur{\'e}e pour {\it une} valeur propre $Z_{\lambda}$, la
repr{\'e}sentation g{\'e}n{\'e}rique se r{\'e}duit alors en deux
repr{\'e}sentations irr{\'e}ductibles de dimension moiti{\'e}
{\it annihil{\'e}es par la moiti{\'e} des charges supersym{\'e}triques}.
Ce processus dichotomique se poursuit lorsque l'{\'e}galit{\'e}
dans l'{\'e}quation (\ref{mz}) se produit pour plusieurs 
valeurs propres\footnote{La dimension est encore divis{\'e}e par deux
lorsque toutes les charges centrales s'annulent.}.
L'{\'e}galit{\'e} dans la formule de masse (\ref{mz}) 
est alors {\it exacte {\`a} tout couplage}, sans quoi la
dimension de la repr{\'e}sentation changerait de mani{\`e}re
discontinue. Dans le cas de la th{\'e}orie de Yang-Mills $SU(2)$
de supersym{\'e}trie $N=4$, la charge centrale s'{\'e}crit
\begin{equation}
Z^{ij} = |a| ~( Q_e + i Q_m )~ \epsilon^{ij} 
\end{equation}
o{\`u} $\epsilon^{ij}$ est la forme symplectique standard,
et la borne de Bogomolny (\ref{mbogo}) appara{\^\i}{}t comme
\index{Bogomolny, borne de}
une cons{\'e}quence de l'alg{\`e}bre de supersym{\'e}trie~; les
monop{\^o}les magn{\'e}tiques, comme les bosons $W^{\pm}$
appartiennent ainsi {\`a} des repr{\'e}sentations courtes
de l'alg{\`e}bre de supersym{\'e}trie $N=4$
et comprennent donc un {\'e}tat de spin 1
\footnote{Le spin des monop{\^o}les provient de la
quantification des modes z{\'e}ro fermioniques en leur
pr{\'e}sence.}. Plus
g{\'e}n{\'e}ralement, on appelle {\it borne de Bogomolny-Prasad-
Sommerfield}  (ou {\it borne BPS}) l'in{\'e}galit{\'e} (\ref{mz}), et 
{\it {\'e}tats BPS} les {\'e}tats la saturant.  Ces {\'e}tats
sont centraux dans l'{\'e}tude des dualit{\'e}s en th{\'e}ories
des champs et th{\'e}ories de cordes.

\subsection{S-dualit{\'e} de la th{\'e}orie de Yang-Mills $N=4$}

Ignor{\'e}e pendant une quinzaine d'ann{\'e}es, la conjecture de
dualit{\'e} {\'e}lectrique-magn{\'e}\-tique a de nouveau concentr{\'e}
l'int{\'e}r{\^e}t en 1994 gr{\^a}ce au travail de Sen qui en a donn{\'e}
une v{\'e}rification essentielle. En pr{\'e}sence d'un couplage
topologique $\theta \int \Tr F\wedge F$, la charge {\'e}lectrique 
effective subie l'<<effet Witten>> 
\index{Witten, effet}
\index{angle $\theta$}
$e\rightarrow e+\frac{\theta}{2\pi} m$ \cite{Witten:1979ey}, 
de sorte que la masse
d'un {\'e}tat BPS de charges {\'e}lectrique $e$ et magn{\'e}tique $m$
\index{masse, formule de!des dyons de SYM $N=4$}
(\ref{mbogo})
devient 
\begin{equation}
\label{msl2} \mathcal{M}= |a| \sqrt{ \frac{|e + S m|^2}{\iS} }
\mbox{ o{\`u} } S= \frac{\theta}{2\pi} + i\frac{4\pi}{g^2} 
:= S_1 + i S_2\ .
\end{equation}
Cette relation pr{\'e}sente la propri{\'e}t{\'e} d'invariance sous 
le groupe modulaire $Sl(2,\Zint)$ des matrices $2\times 2$
\index{groupe!modulaire $Sl(2,\Zint)$}
de d{\'e}terminant 1 et coefficients entiers agissant par
\begin{equation}
S \rightarrow \frac{aS + b}{cS + d} \sp
\begin{pmatrix} e \\ m \end{pmatrix}
\rightarrow 
\begin{pmatrix} a & b \\ c & d \end{pmatrix}
\begin{pmatrix} e\\ m \end{pmatrix}
\sp 
a,b,c,d\in \Zint,\ ad-bc=1
\end{equation} 
On est ainsi amen{\'e} {\`a} conjecturer l'invariance de la th{\'e}orie
quantique elle-m{\^e}me sous cette sym{\'e}trie, baptis{\'e}e S-dualit{\'e}.
\index{S-dualit{\'e}!de SYM $N=4$}
Ce groupe est engendr{\'e} par la transformation de dualit{\'e}
{\'e}lectrique-magn{\'e}tique $S\rightarrow -1/S$, qui g{\'e}n{\'e}ralise
la dualit{\'e} de Montonen-Olive {\`a} une valeur non nulle
de l'angle $\theta$, ainsi que par le
flot spectral $S \rightarrow S+1$, qui ne change pas
les amplitudes mais agit non trivialement sur le spectre.
\index{flot spectral!dans SYM $N=4$}
L'existence de bosons de jauge $W^{\pm}$ 
dans le spectre implique alors {\it l'existence de l'orbite sous 
$Sl(2,\Zint)$ de ces {\'e}tats}, soit tous les {\'e}tats de charge
$(p,q)$, o{\`u} $p$ et $q$ sont deux entiers premiers entre eux.
\index{pq, dyons de charge@$(p,q)$, dyons de charge}
Ces {\'e}tats BPS doivent correspondre
{\`a} l'existence d'un {\'e}tat fondamental supersym{\'e}trique pour la m{\'e}canique 
quantique d{\'e}finie sur l'espace des modules classiques des
monop{\^o}les de charge $q$
\index{espace des modules!des monop{\^o}les magn{\'e}tiques}
\index{monop{\^o}le magn{\'e}tique!espace des modules}
\footnote{Cet espace a fait l'objet d'une {\'e}tude math{\'e}matique
approfondie, r{\'e}sum{\'e}e dans l'ouvrage \cite{Atiyah:1988}.}, 
et Sen a pu construire explicitement la
fonction d'onde dans le cas 
du monop{\^o}le magn{\'e}tique de charge $(1,2)$ \cite{Sen:1994yi}.
 
La conjecture de S-dualit{\'e} des th{\'e}ories de jauge supersym{\'e}triques
$N=4$ a depuis fait l'objet de nombreux tests. En particulier, il a
{\'e}t{\'e} prouv{\'e} que la fonction de partition de la 
\index{partition, fonction de!des th{\'e}ories SYM}
version topologiquement <<twist{\'e}e>> de ces th{\'e}ories de jauge sur une vari{\'e}t{\'e}
arbitraire {\'e}tait invariante sous la S-dualit{\'e}
\cite{Vafa:1994tf}, tout comme l'{\'e}tait la divergence infrarouge de
\index{divergence infrarouge}
l'{\'e}nergie libre en pr{\'e}sence de flux {\'e}lectriques et magn{\'e}tiques,
\cite{Girardello:1995gf}. La d{\'e}monstration 
rigoureuse de l'invariance de la th{\'e}orie de jauge $N=4$ originale 
est cependant rest{\'e}e hors d'atteinte
\footnote{La fonction
de partition de la th{\'e}orie de jauge {\it ab{\'e}lienne} non supersym{\'e}trique
de Maxwell sur une vari{\'e}t{\'e} de dimension 4 quelconque a {\'e}t{\'e} calcul{\'e}e
ind{\'e}pendamment par E. Verlinde et E. Witten
\cite{Verlinde:1995mz,Witten:1995gf} et 
pr{\'e}sente un poids modulaire non trivial
$((\chi+\sigma)/4,(\chi-\sigma/4))$ sous les transformations 
de S-dualit{\'e}, o{\`u} $\chi$ est la caract{\'e}ristique d'Euler et $\sigma$ la
signature de la vari{\'e}t{\'e}.}. Elle appara{\^\i}t
maintenant comme une cons{\'e}quence {\'e}l{\'e}mentaire  de la conjecture de
dualit{\'e} des th{\'e}ories de supercordes. Cette conjecture
n'a cependant pas encore
men{\'e} au calcul de quantit{\'e}s non-triviales, comme par exemple
des couplages {\`a} plus de deux d{\'e}riv{\'e}es, principalement en 
raison de la m{\'e}connaissance des contraintes de
la supersym{\'e}trie sur ces quantit{\'e}s. Ce probl{\`e}me sera
abord{\'e} dans le chapitre 3 dans le cadre  des
th{\'e}ories de supercordes. En revanche, l'extension
par  Seiberg et Witten de la conjecture
de S-dualit{\'e} $N=4$  aux th{\'e}ories supersym{\'e}triques $N=2$
a d{\'e}clench{\'e} un grand retentissement tant en physique
des particules qu'en math{\'e}matiques.

\section{Dualit{\'e} dans les th{\'e}ories de jauge de supersym{\'e}trie N=2}
En regard des th{\'e}ories de jauge {\`a} supersym{\'e}trie $N=4$
bien dociles, les th{\'e}ories de jauge 
$N=2$ pr{\'e}sentent une richesse {\'e}tourdissante. En particulier,
l'action effective de basse {\'e}nergie n'est plus astreinte au
th{\'e}or{\`e}me de non renormalisation et la fonction beta ne s'annule
en g{\'e}n{\'e}ral plus, de sorte que la dynamique
peut r{\'e}server des surprises inattendues. Les contraintes
de la supersym{\'e}trie restent cependant encore assez fortes
pour permettre la d{\'e}termination exacte de l'action effective
{\`a} deux d{\'e}riv{\'e}es, comme l'ont montr{\'e} Seiberg et Witten.

\subsection{Action microscopique, action effective et supersym{\'e}trie N=1}
Comme nous l'avons discut{\'e} dans l'introduction, les degr{\'e}s
de libert{\'e} des th{\'e}ories de jauge asymptotiquement libres 
{\`a} grande distance, correspondant aux particules observables
{\`a} basse {\'e}nergie, n'ont que peu {\`a} voir avec les degr{\'e}s
de libert{\'e} microscopiques en termes desquels la th{\'e}orie
est formul{\'e}e, ceci en raison de la croissance rapide du couplage
de jauge aux grandes {\'e}chelles. On cherche donc {\`a} retenir les
degr{\'e}s de libert{\'e} de masse minimale, et caract{\'e}riser leurs
interactions {\`a} basse {\'e}nergie en termes d'une action effective,
\index{action effective}
obtenue apr{\`e}s int{\'e}gration de degr{\'e}s de libert{\'e} de masse
sup{\'e}rieure. Cette action effective peut {\^e}tre d{\'e}velopp{\'e}e 
en puissance des moments ou d{\'e}riv{\'e}es, et l'on ne retient
d'ordinaire que les termes cin{\'e}tiques {\`a} deux d{\'e}riv{\'e}es au plus.
Dans les th{\'e}ories supersym{\'e}triques, ceux-ci peuvent {\^e}tre
caract{\'e}ris{\'e}s par le {\it superpotentiel} $W$, 
\index{superpotentiel}
fonction {\it holomorphe} des composantes scalaires $z^i$
des multiplets chiraux, le {\it potentiel de K{\"a}hler}
$K$,
\index{Kahler, potentiel de@K{\"a}hler, potentiel de}
fonction {\it r{\'e}elle} des m{\^e}mes champs, et la matrice des
{\it couplages de jauge } $\tau_{ij}$
et $\theta_{ij}$
\index{action effective!{\`a} deux d{\'e}riv{\'e}es}
\index{angle $\theta$}
L'action {\`a} deux d{\'e}riv{\'e}es
des champs bosoniques s'{\'e}crit alors
\begin{equation}
\mathcal{S}= \int d^4x \left( |W(z)|^2 + 
\frac{\partial^2 K}{\partial z^i \partial z^{\bar j} }
Dz^i Dz^{\bar j}
-\frac{1}{4g^2_{ij}} F^{(i)} \wedge * F^{(j)}
+ \theta_{ij} F^{(i)} \wedge F^{(j)} \right)
\end{equation}
o{\`u} nous n'avons retenu que les termes purement bosoniques,
suffisants pour d{\'e}terminer l'action compl{\`e}te par supersym{\'e}trie.
Une formulation plus g{\'e}om{\'e}trique d{\'e}finirait les champs $z^i$ 
comme application de l'espace-temps dans
une vari{\'e}t{\'e} k{\"a}hlerienne $\mathcal{M}$, et le superpotentiel,
dont la phase est sans signification, comme section 
holomorphe d'un fibr{\'e} en ligne sur cette vari{\'e}t{\'e}.

\subsection{Supersym{\'e}trie $N=2$, g{\'e}om{\'e}tries sp{\'e}ciales
et hyper\-k{\"a}h\-le\-riennes} 
En pr{\'e}sence de supersym{\'e}trie {\'e}tendue $N=2$, la structure
de l'action effective devient plus contrainte. Les multiplets
de l'alg{\`e}bre de supersym{\'e}trie $N=1$
se combinent pour former des multiplets de l'alg{\`e}bre $N=2$ :
\index{supersym{\'e}trie!globale $N=2$}
\begin{itemize}
\item un multiplet chiral et un multiplet vectoriel $N=1$ forment
un {\it multiplet vectoriel} N=2, correspondant {\`a}
un scalaire complexe $z$, deux fermions de Weyl\footnote{Les propri{\'e}t{\'e}s de chiralit{\'e} (Weyl) et r{\'e}alit{\'e}
(Majorana) des repr{\'e}sentations spinorielles sont discut{\'e}es
\index{spinorielle, repr{\'e}sentation}
en d{\'e}tail dans la r{\'e}f{\'e}rence \cite{Kugo:1983bn}.}
et un boson vecteur~;
\index{vectoriel, multiplet|textit}
\item deux multiplets chiraux N=1 forment un {\it hypermultiplet}
N=2, correspondant {\`a} deux scalaires complexes $Q$ et $\tilde Q$
\index{hypermultiplet|textit}
et deux fermions de Weyl.
\end{itemize}
La vari{\'e}t{\'e} k{\"a}hlerienne $\mathcal{M}$ se s{\'e}pare alors,
au moins localement, en un produit $\mathcal{M}_V \otimes \mathcal{M}_H$
d'une vari{\'e}t{\'e} de {\it g{\'e}om{\'e}trie sp{\'e}ciale} $\mathcal{M}$ 
\index{geometrie sp{\'e}ciale, vari{\'e}t{\'e} de@g{\'e}om{\'e}trie sp{\'e}ciale, vari{\'e}t{\'e} de}
\index{vari{\'e}t{\'e}!de g{\'e}om{\'e}trie sp{\'e}ciale},
de dimension $2N_V$
param{\'e}tr{\'e}e par les scalaires des $N_V$ multiplets vectoriels,
et d'une vari{\'e}t{\'e} {\it hyperk{\"a}hlerienne} $\mathcal{M}_H$,
\index{hyperk{\"a}hlerienne, vari{\'e}t{\'e}}
\index{vari{\'e}t{\'e}!hyperk{\"a}hlerienne}
de dimension $4 N_H$
correspondant aux $N_H$ hypermultiplets. 
Le potentiel de K{\"a}hler s'{\'e}crit donc comme une somme de deux
termes d{\'e}pendant de chaque facteur s{\'e}par{\'e}ment, et
les couplages et angles de jauge ne d{\'e}pendent que des scalaires
des multiplets vectoriels. 
Les transformations de jauge correspondant aux interactions
{\'e}lectro-magn{\'e}tiques sont d{\'e}finies par les {\it isom{\'e}tries}
communes aux deux vari{\'e}t{\'e}s, et leur {\it jaugement}
introduit le seul couplage entre les deux facteurs sous la forme
d'un superpotentiel non-trivial (et bien entendu, de d{\'e}riv{\'e}es
covariantes)\footnote{La r{\'e}f{\'e}rence \cite{Andrianopoli:1996vr}
offre une description g{\'e}n{\'e}rale des couplages de la
supergravit{\'e} et supersym{\'e}trie $N=2$.}. 
En particulier, {\it les hypermultiplets neutres
et les multiplets vectoriels sont d{\'e}coupl{\'e}s} en l'absence
de la gravitation. Ce point sera {\`a} la base d'arguments de
non renormalisation dans la suite de l'expos{\'e}.
\index{decouplage@d{\'e}couplage!des hypers et vecteurs}

Du c{\^o}t{\'e} des multiplets vectoriels, potentiel de K{\"a}hler,
couplages et angles de jauge 
se trouvent unifi{\'e}s en terme d'une {\it section 
holomorphe} $(a_{D;i}(z),a^i(z))$
d'un fibr{\'e} vectoriel de groupe symplectique 
\index{groupe!symplectique}
$Sp(N_V,\Zint)$ sur la vari{\'e}t{\'e}
$\mathcal{M}_V$ selon
\begin{equation}
\label{prepo}
K=a_{D,i} \overline{a^i} - \overline{a_{D,i}} a^i
\sp
\tau_{ij}=\frac{\theta_{ij}}{2\pi} + i\frac{4\pi}{g^2_{ij}}
=\frac{\partial a_{D;i}}{\partial a^j}
\end{equation}
Cette description incorpore pr{\'e}cis{\'e}ment les transformations
de dualit{\'e} {\'e}lectrique-ma\-gn{\'e}tique correspondant aux
rotations symplectiques $Sp(N_V,\Zint)$ sur le vecteur 
{\`a} $2N_V$ composantes $(F_i,*F^i)$, g{\'e}n{\'e}ralisant
les transformations de dualit{\'e} $Sl(2,\Zint)=Sp(1,\Zint)$ 
au cas $N_V\ge 1$. Sous cette rotation, la section
holomorphe se transforme comme un vecteur, laissant le
potentiel de K{\"a}hler invariant. La matrice des couplages
de jauge $\tau_{ij}$ se transforme {\'e}galement en accord
avec la dualit{\'e} {\'e}lectrique-magn{\'e}tique. Lorsqu'elles sont
ind{\'e}pendantes, on peut 
choisir les quantit{\'e}s $a^i$ elles-m{\^e}mes comme coordonn{\'e}es
$z$ sur la vari{\'e}t{\'e} de g{\'e}om{\'e}trie sp{\'e}ciale\footnote{
On se prive ainsi de cas fort int{\'e}ressants
o{\`u} la supersym{\'e}trie globale $N=2$ se trouve spontan{\'e}ment 
\index{supersym{\'e}trie!brisure partielle de la}
bris{\'e}e en $N=1$ \cite{antoniadis:1996.1}.}. 
Les fonctions $a_{D;i}$ peuvent alors {\^e}tre
obtenues en termes d'une fonction holomorphe $\mathcal{F}(a_i)$  
dite {\it pr{\'e}potentiel}
\index{pr{\'e}potentiel}, selon 
$a_{D;i}=\partial\mathcal{F}/\partial a^i$. Ce choix
obscurcit cependant le r{\^o}le des transformations de dualit{\'e}.

La vari{\'e}t{\'e} hyperk{\"a}hlerienne d{\'e}finissant les interactions
des hypermultiplets est elle aussi tr{\`e}s contrainte. Elle ne
poss{\`e}de malheureusement pas une caract{\'e}risation aussi
explicite que la vari{\'e}t{\'e} de g{\'e}om{\'e}trie sp{\'e}ciale,
la raison profonde en {\'e}tant l'absence de description 
en termes de superchamps hors de la couche de masse des
hypermultiplets
\index{hypermultiplet}
\footnote{Il existe bien une description en termes de {\it 
super-espace harmonique}
\index{harmonique, super-espace}, mais elle introduit une infinit{\'e} de
champs auxiliaires qui ne peuvent en g{\'e}n{\'e}ral {\^e}tre
{\'e}limin{\'e}s \cite{Galperin:1984av}. 
Il existe {\'e}galement une classe de vari{\'e}t{\'e}s
hyperk{\"a}hleriennes dites {\it sp{\'e}ciales}
\cite{Cecotti:1989qn}, correspondant
aux fibr{\'e}s cotangents des vari{\'e}t{\'e}s de g{\'e}om{\'e}trie
sp{\'e}ciale, qui joue un r{\^o}le particulier en th{\'e}orie
des cordes, mais leur {\'e}tude n'a pas {\'e}t{\'e} d{\'e}velopp{\'e}e
\index{hyperk{\"a}hlerienne sp{\'e}ciale, vari{\'e}t{\'e}}
\index{vari{\'e}t{\'e}!hyperk{\"a}hlerienne sp{\'e}ciale}
{\`a} sa pleine mesure.}. 
Par d{\'e}finition, ces vari{\'e}t{\'e}s admettent trois
structures complexes covariantement constantes r{\'e}alisant
l'alg{\`e}bre des unit{\'e}s des quaternions
\index{quaternions}
\footnote{On trouvera en appendice \ref{hk} une description plus compl{\`e}te
des vari{\'e}t{\'e}s hyperk{\"a}hleriennes.}. Le choix d'une structure
complexe particuli{\`e}re {\'e}quivaut au choix d'une sous-alg{\`e}bre
$N=1$ dans l'alg{\`e}bre de supersym{\'e}trie $N=2$. La forme
de K{\"a}hler covariantement constante 
associ{\'e}e en fait une vari{\'e}t{\'e} symplectique
\index{vari{\'e}t{\'e}!symplectique}
holomorphe dont le groupe d'holonomie est au plus
\index{holonomie!de la m{\'e}trique}
compris dans $Sp(N_H)\subset SO(4,N)$. En particulier, 
la courbure de Ricci est nulle, et ces vari{\'e}t{\'e}s
(en dimension 4) d{\'e}finissent ainsi des instantons
\index{instanton!gravitationnel}
gravitationnels. Elles interviennent {\'e}galement comme espace
des modules des monop{\^o}les et instantons, et les
constructions de ADHM \cite{atiyah:1978} et de 
Kronheimer-Nakajima \cite{kronheimer:1989,nakajima:1994} en donnent des
r{\'e}alisations explicites, loin cependant d'en {\'e}puiser
\index{espace des modules!des instantons}
\index{espace des modules!des monop{\^o}les magn{\'e}tiques}
\index{monop{\^o}le magn{\'e}tique!espace des modules}
la vari{\'e}t{\'e}. L'{\'e}tude de leur r{\'e}alisation 
comme espace des modules de th{\'e}ories de jauge
$N=2$ renormalisables fait l'objet de la publication
en appendice \ref{hk}.

Cette caract{\'e}risation des actions {\`a} l'ordre de deux
d{\'e}riv{\'e}es vaut aussi bien pour l'action microscopique
d{\'e}finissant la th{\'e}orie que pour l'action effective
en donnant la limite {\`a} grande distance. Dans le premier
cas, on se restreint {\`a} une
action microscopique renormalisable, ce qui implique que
les vari{\'e}t{\'e}s des multiplets vectoriels et des
hypermultiplets sont triviales.
Les transformations de jauge agissent alors par
des repr{\'e}sentations {\it lin{\'e}aires} sur les
divers multiplets. En particulier, 
les multiplets vectoriels 
se transforment dans la repr{\'e}sentation adjointe,
tandis que les deux scalaires complexes des hypermultiplets 
se transforment dans deux repr{\'e}sentations conjugu{\'e}es,
soit au total une repr{\'e}sentation {\it r{\'e}elle}.
Cette absence de chiralit{\'e} est en fait la principale
\index{chiralit{\'e}}
limitation des th{\'e}ories de jauge de supersym{\'e}trie
$N\ge 2$. Le second cas est bien plus int{\'e}ressant,
puisqu'il d{\'e}crit la dynamique {\`a} basse {\'e}nergie, une fois
que l'effet des modes massifs a {\'e}t{\'e} int{\'e}gr{\'e}.

\subsection{Phase de Coulomb, phase de Higgs, et associ{\'e}es}
\index{phase!de Coulomb}
\index{phase!de Higgs}
\index{Higgs!phase de}
Le comportement {\`a} grande distance des th{\'e}ories de champs
montre en g{\'e}n{\'e}ral une vari{\'e}t{\'e} de phases distingu{\'e}es
par des param{\`e}tres d'ordre correspondant aux valeurs 
moyennes de certains op{\'e}rateurs. Dans le cas des th{\'e}ories
de jauge $N=2$ asymptotiquement libre, ces param{\`e}tres
d'ordre correspondent aux valeurs moyennes des champs
scalaires des multiplets vectoriels et des hypermultiplets,
autoris{\'e}es par les directions plates du potentiel
\index{directions plates!des th{\'e}ories de SYM}
scalaire
\footnote{Des valeurs moyennes correspondant {\`a} des
minima locaux du potentiel effectif 
correspondraient {\`a} des vides non supersym{\'e}triques.},
et ces phases correspondent {\`a} des branches distinctes
de l'espace des modules.
Les valeurs
moyennes des scalaires des multiplets vectoriels brisent
g{\'e}n{\'e}riquement la sym{\'e}trie de jauge en un
sous-groupe ab{\'e}lien de m{\^e}me rang (le tore de Cartan)
et correspondent {\`a} la {\it phase de Coulomb}. 
Les valeurs moyennes des hypermultiplets dans la
repr{\'e}sentation fondamentale peuvent en revanche
briser le groupe de jauge compl{\`e}tement, r{\'e}sultant
en la phase dite {\it de Higgs}. Dans le cas d'une th{\'e}orie
de jauge $SU(2)$ avec $N_f$ doublets d'hypermultiplets,
ces deux phases sont exclusives l'une de l'autre, et se
connectent classiquement en un point o{\`u} tous les scalaires
s'annulent et la sym{\'e}trie de jauge est restaur{\'e}e. 
Dans des cas plus g{\'e}n{\'e}raux, on peut avoir des branches
mixtes o{\`u} hypers et vecteurs condensent, et des restaurations
de sym{\'e}tries partielles sur des sous-vari{\'e}t{\'e}s de l'espace
des modules
\index{sym{\'e}trie de jauge!restauration}. 

Cette structure de l'espace des modules est bas{\'e}e sur l'analyse
des directions plates du potentiel scalaire microscopique,
et peut fort bien {\^e}tre modifi{\'e}e par les corrections quantiques.
L'absence de superpotentiel sur la vari{\'e}t{\'e} de g{\'e}om{\'e}trie
\index{superpotentiel}
sp{\'e}ciale garantit que les directions plates de la phase
de Coulomb ne seront pas lev{\'e}es, mais leur g{\'e}om{\'e}trie
a toutes les raisons d'{\^e}tre modifi{\'e}e, comme le montre
le calcul explicite de Seiberg et Witten
\cite{seiberg/witten:1994.1,seiberg/witten:1994.2}. Nous en donnons
ici une pr{\'e}sentation sommaire, renvoyant aux cours et articles de revue
\cite{Lerche:1996xu,Alvarez-Gaume:1997mv} pour plus de d{\'e}tails.

\subsection{G{\'e}om{\'e}trie de la phase de Coulomb}
\index{Seiberg-Witten!solution de}
La phase de Coulomb appara{\^\i}t dans le cas le plus simple
d'une th{\'e}orie de jauge $SU(2)$ sans hypermultiplets.
Le potentiel montre des directions plates correspondant
{\`a} la valeur moyenne complexe $a$ du triplet de Higgs
\index{Higgs!scalaire de}
$\phi=a\sigma_3$, brisant la sym{\'e}trie de
jauge en un sous-groupe ab{\'e}lien $U(1)$. L'espace des modules classique
correspond donc au plan complexe $\CC$ quotient{\'e} par la sym{\'e}trie
$\Zint_2: a \rightarrow -a$ correspondant {\`a} une transformation
de jauge r{\'e}siduelle, et param{\'e}tr{\'e} de mani{\`e}re univoque
par $u=a^2/2$. La singularit{\'e} conique {\`a} l'origine
correspond {\`a} la restauration classique de la sym{\'e}trie de 
jauge lorsque $a=0$
\index{sym{\'e}trie de jauge!restauration}. 
D'une mani{\`e}re g{\'e}n{\'e}rale, on attend
que {\it chaque singularit{\'e} corresponde {\`a} la divergence
\index{divergence infrarouge}
infrarouge associ{\'e}e {\`a} l'apparition d'un {\'e}tat de masse nulle.}
Cette image de l'espace des modules est certainement correcte
pour $|u|$ grand, car $u$ a la dimension d'une {\'e}nergie carr{\'e}e,
mesur{\'e}e par rapport {\`a} l'{\'e}chelle d'{\'e}nergie 
dynamiquement g{\'e}n{\'e}r{\'e}e
\begin{equation} 
\label{lambda}
\left(\frac{\Lambda}{\mu}\right)^{\beta_0} =
  e^{-\frac{8\pi^2}{g^2(\mu)}}\ ,
\end{equation}
o{\`u} $\beta_0=4$ est le coefficient de la fonction beta {\`a} une boucle~;
\index{beta, fonction!en th{\'e}ories de jauge}
ce r{\'e}gime correspond aux petites distances pour lesquelles
la th{\'e}orie est asymptotiquement libre. La section symplectique
\index{libert{\'e} asymptotique}
$(a_D,a)$ y est donn{\'e}e par un calcul perturbatif, {\`a}
une boucle en raison d'un th{\'e}or{\`e}me de non renormalisation
\index{non renormalisation!de la phase de Coulomb pour SYM $N=2$}
perturbative~:
\begin{equation}
a_D(u) \sim \frac{i}{\pi} \sqrt{2u}\log(u)\ ,\quad a(u) \sim \sqrt{2u}
\end{equation}
et pr{\'e}sente une monodromie non triviale sous
les rotations $u \rightarrow e^{2i\pi} u$:
\index{monodromie}
\begin{equation}
\label{monoinf}
\begin{pmatrix} a_D \\ a \end{pmatrix}
\rightarrow
\begin{pmatrix} -1 & 2 \\ 0 & -1 \end{pmatrix}
\begin{pmatrix} a_D \\ a \end{pmatrix}
\end{equation}
Par compacit{\'e} de l'espace des modules, cette monodromie
doit {\^e}tre {\'e}gale au produit des monodromies autour
des singularit{\'e}s apparaissant {\`a} $|u|$ fini, soit en 
r{\'e}gion de fort couplage. La monodromie associ{\'e}e 
{\`a} l'apparition de $d$ particules de charges $(e,m)$
et de masse nulle peut {\^e}tre calcul{\'e}e dans une description
duale o{\`u} sa charge est purement {\'e}lectrique, et s'{\'e}crit
\begin{equation}
\label{monosing}
\begin{pmatrix} 1-2 d~ em & 2d~ e^2\\ -2d~m^2 & 1+2d~em \end{pmatrix}
\in Sl(2,\Zint)\ .
\end{equation}
Le probl{\`e}me de la factorisation de la monodromie {\`a} l'infini
(\ref{monoinf}) en un produit de monodromies (\ref{monosing})
pour des valeurs $(e_i,m_i)$ des charges {\'e}lectriques est bien
d{\'e}fini, et admet une unique solution sous la forme de {\it deux}
singularit{\'e}s de charges $(0,1)$ et $(1,1)$ correspondant
respectivement {\`a} un {\it monop{\^o}le magn{\'e}tique} 
\index{monop{\^o}le magn{\'e}tique!solution de Seiberg-Witten}
et un {\it dyon} devenant de masse nulle en ces deux points.
\index{singularit{\'e}!de l'espace des modules}
\index{espace des modules!singularit{\'e}}
La singularit{\'e} classique de l'espace des modules est
ainsi r{\'e}solue en deux points distincts, $u=\pm \Lambda ^2$,
o{\`u} la th{\'e}orie admet une description duale faiblement
coupl{\'e}e. 
L'espace des modules quantique $\mathcal{M}_Q$
correspond donc {\`a} la sph{\`e}re de Riemann 
priv{\'e}e de ces deux points
et du point {\`a} l'infini, et  
il ne reste plus qu'{\`a} trouver une section holomorphe
$(a,a_D)$ sur $\mathcal{M}_Q$ poss{\'e}dant les monodromies
requises au voisinage des trois singularit{\'e}s.
\fig{3cm}{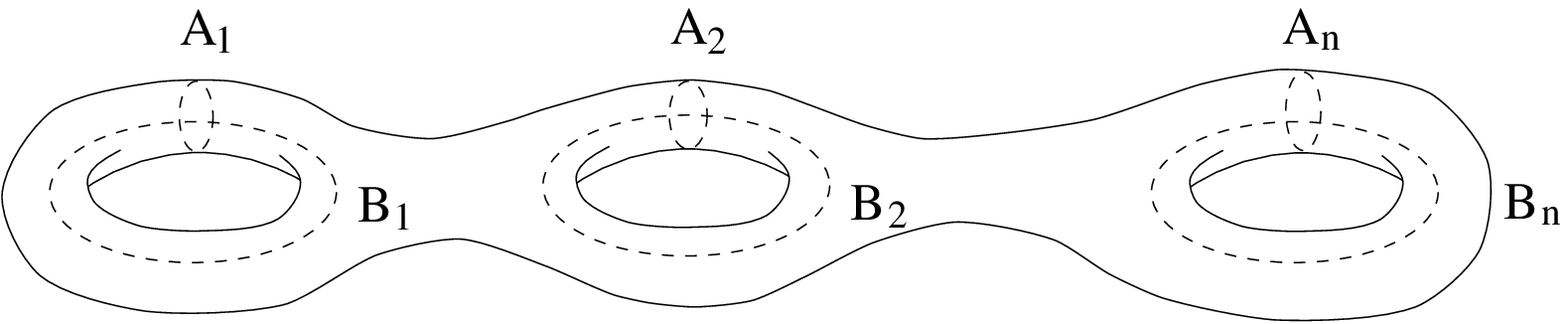}{Base symplectique $(A_i,B_i)$ des cycles
d'une surface de Riemann de genre $n$.}{riemann}

Ce probl{\`e}me peut {\^e}tre rattach{\'e} {\`a} l'{\'e}tude des
surfaces de Riemann en remarquant que le couplage de jauge
\index{surface de Riemann!solution de Seiberg-Witten}
$\tau$ peut {\^e}tre vu comme le param{\`e}tre modulaire
d'un tore auxiliaire associ{\'e} {\`a} chaque point de 
l'espace des modules. Plus g{\'e}n{\'e}ralement, pour un
groupe de rang $N_V$ spontan{\'e}ment bris{\'e} en $U(1)^{N_V}$,
les composantes $(a^i,_{D;i})$ peuvent {\^e}tre interpr{\'e}t{\'e}es
comme les p{\'e}riodes $(\oint_{A_i}\lambda_{\rm SW},\oint_{B_i}
\lambda_{\rm SW})$ 
d'une certaine forme m{\'e}romorphe $\lambda_{\rm SW}$
sur les cycles $A_i,B_i$ d'une surface de Riemann de genre $N_V$ et
de matrice de p{\'e}riodes $\tau_{ij}$
(figure \ref{riemann})
\footnote{La matrice des p{\'e}riodes $\tau_{ij}$ d{\'e}termine la \
structure complexe de la surface de Riemann, et correspond aux
p{\'e}riodes $\oint_{A_i} \omega_j$ de la 1-forme holomorphe
duale au cycle $B_j$, voir par example \cite{D'Hoker:1988ta}.}
. Les monodromies symplectiques
$Sp(N_V,\Zint)$ sur les p{\'e}riodes correspondent alors simplement
aux monodromies affectant la base symplectique des cycles 
d'homologie $A_i,B_i$
\index{surface de Riemann!homologie des}
lorsque les modules de la surface de Riemann sont vari{\'e}s. 
Les p{\'e}riodes peuvent alors {\^e}tre d{\'e}termin{\'e}es en r{\'e}solvant
les {\'e}quations de Picard-Fuchs avec les conditions asymptotiques
appropri{\'e}es. On peut ainsi obtenir
une solution en termes de fonctions hyperg{\'e}om{\'e}triques
pour la section $(a_D,a)(u)$, et remonter ainsi au pr{\'e}potentiel
{\it exact} $\mathcal{F}(a)$. Le d{\'e}veloppement {\`a} faible couplage
\begin{equation}
\label{finst}
\mathcal{F}(a)=\frac{i}{\pi}a^2\ln(a/\Lambda)+ a^2
\sum_{n=1}^{\infty} c_n\left(\frac{\Lambda}{a}\right)^{4n}\ .
\end{equation}
montre alors une {\it s{\'e}rie d'effets non perturbatifs}
d'ordre 
\begin{equation}
\Lambda^{4n} = \mu^{4n} e^{-\frac{8\pi^2 n}{g^2(\mu)}}
\end{equation}
que l'on peut identifier aux contributions {\`a} $n$
instantons. Un calcul explicite permet du reste de v{\'e}rifier
\index{serie d'instantons@s{\'e}rie d'instantons!pr{\'e}potentiel de SYM $N=2$}
les premiers termes de cette s{\'e}rie. La m{\^e}me m{\'e}thode permet d'obtenir
l'action effective de basse {\'e}nergie pour des th{\'e}ories
de jauge $N=2$ asymptotiquement libres de groupe de jauge quelconque,
en pr{\'e}sence d'hypermultiplets charg{\'e}s.
Nous reviendrons sur l'{\'e}tude de s{\'e}ries d'instantons
analogues {\`a} l'{\'e}quation (\ref{finst}), mais cette
fois dans le contexte de la th{\'e}orie des supercordes au
chapitre 4. 

La solution de Seiberg et Witten constitue un tour de force
unanimement salu{\'e}, et donne le premier exemple 
non-trivial de calcul exact {\`a} toute valeur du couplage.
De nombreux enseignements ont pu en {\^e}tre tir{\'e}s
pour la compr{\'e}hension des ph{\'e}nom{\`e}nes de confinement,
brisure de sym{\'e}trie chirale et condensation de monop{\^o}les.
Dans sa version <<twist{\'e}e>>, elle a {\'e}galement permis
une avanc{\'e}e majeure dans la classification des
\index{vari{\'e}t{\'e}!diff{\'e}rentielles de dimension 4}
\index{Seiberg-Witten!invariants de}
vari{\'e}t{\'e}s diff{\'e}rentielles de dimension 4, gr{\^a}ce au
calcul des invariants topologiques de Seiberg et Witten.

\subsection{G{\'e}om{\'e}trie de la phase de Higgs}
\index{phase!de Higgs}
\index{Higgs!phase de}
Contrairement {\`a} la branche de Coulomb, la g{\'e}om{\'e}trie de la
branche de Higgs est prot{\'e}g{\'e}e de toute correction quantique.
Le couplage de jauge $g$, ou de mani{\`e}re {\'e}quivalente 
l'{\'e}chelle $\Lambda$, peut en effet {\^e}tre assimil{\'e} {\`a} la
valeur moyenne dans le vide du champ scalaire d'un multiplet
vectoriel gel{\'e} 
\footnote{tout comme la masse des hypermultiplets peut s'interpr{\'e}ter
comme la valeur moyenne d'un multiplet vectoriel gel{\'e} correspondant
au jaugement de la sym{\'e}trie de saveur.}. 
Le d{\'e}couplage des multiplets vectoriels
et des hypermultiplets garantit alors l'ind{\'e}pendance de la
m{\'e}trique sur l'espace des modules des hypermultiplets en
la constante de couplage.  
\index{decouplage@d{\'e}couplage!des hypers et vecteurs}
\index{non renormalisation!de la phase de Higgs dans SYM $N=2$}
Ceci n'est cependant plus le cas en th{\'e}orie des supercordes de
type II supersym{\'e}triques $N=2$, o{\`u} la constante de couplage 
des cordes est un hypermultiplet dont peut donc d{\'e}pendre la
m{\'e}trique de l'espace des hypermultiplets
\footnote{L'espace des modules des hypermultiplets en th{\'e}orie
des cordes est une vari{\'e}t{\'e} quaternionique et non plus 
\index{vari{\'e}t{\'e}!quaternionique}
hyperk{\"a}hlerienne. Il est cependant possible de d{\'e}finir
une limite <<plate>> dans laquelle les effets gravitationnels
disparaissent, et les corrections de th{\'e}ories des cordes
subsistent.}. Cette remarque est {\`a} l'origine de notre int{\'e}r{\^e}t
pour ce sujet, bien qu'elle n'est pas {\'e}t{\'e} approfondie 
dans le cadre de ce travail de th{\`e}se. 
L'absence de corrections quantiques, pour d{\'e}cevante qu'elle soit,
ne diminue pour autant pas l'int{\'e}r{\^e}t de l'{\'e}tude de la phase
de Higgs dans les th{\'e}ories de jauge $N=2$. 
Elle permet en effet de donner des v{\'e}rifications non triviales
des conjectures de dualit{\'e} {\'e}lectrique-magn{\'e}tique, comme
nous l'avons montr{\'e} en collaboration avec Ignatios Antoniadis
dans la publication en appendice \ref{hk}.
Nous en donnons maintenant une pr{\'e}sentation sommaire.

Comme dans le cas de la branche de Coulomb classique, la structure 
de la branche de Higgs s'obtient en d{\'e}terminant les directions
plates du potentiel scalaire, mais pour une valeur moyenne nulle des
scalaires des multiplets vectoriels, et en identifiant les
solutions reli{\'e}es par la sym{\'e}trie de jauge r{\'e}siduelle. 
Le potentiel scalaire est 
g{\'e}n{\'e}r{\'e} par le jaugement des isom{\'e}tries de la vari{\'e}t{\'e}
hyperk{\"a}hlerienne~; dans le langage de la supersym{\'e}trie
$N=1$, il est la somme des {\it F-termes} des multiplets chiraux,
{\'e}gaux aux d{\'e}riv{\'e}es du superpotentiel, et des 
\index{superpotentiel}\index{D-terme}
{\it D-termes} des multiplets vectoriels~:
\begin{equation}
\label{scalpot}
V= \sum_{I} \left| \frac{\partial W}{\partial Q_{I}} \right|^2
   + \sum_i D^2_i
\end{equation}
o{\`u} la premi{\`e}re somme est effectu{\'e}e sur tous les champs chiraux
et la seconde sur les g{\'e}n{\'e}rateurs $i$ du groupe de jauge.
Dans le cas d'une th{\'e}orie $N=2$ minimalement coupl{\'e}e, le
superpotentiel et les D-termes s'{\'e}crivent 
en fonction des scalaires complexes $z$ 
et $(Q,\tilde Q)$ des multiplets vectoriels et hypermultiplets
respectivement selon
\begin{subequations}
\begin{align}
\label{superpot}
W &= z^i Q T_i \tilde Q \\
D_i = Q T_i Q^{\dagger} - \tilde Q T_i \tilde Q^{\dagger} + z T_i
z^{\dagger}\ ,
\end{align}
\end{subequations}
o{\`u} $T_i$ est la matrice hermitienne associ{\'e}e au g{\'e}n{\'e}rateur $i$.
Les vides supersym{\'e}triques sont donc obtenus pour des valeurs
des champs scalaires extr{\'e}misant le superpotentiel et annulant
le D-terme. En particulier, pour $z^i=0$, on est ramen{\'e} {\`a}
$N_V$ {\'e}quations alg{\'e}briques complexes et $N_V$ {\'e}quations
r{\'e}elles :
\begin{subequations}
\begin{eqnarray}
\label{moment}
Q T_i \tilde Q &=&0 \\
Q T_i Q^{\dagger} - \tilde Q T_i \tilde Q^{\dagger} &=& 0
\end{eqnarray}
\end{subequations}
Ces trois {\'e}quations r{\'e}elles se transforment comme un triplet sous la 
sym{\'e}trie $SU(2)_R$ agissant sur les structures complexes de la
vari{\'e}t{\'e} hyperk{\"a}hlerienne, et correspondent {\`a} 
l'annulation des trois {\it applications moment} correspondant
{\`a} l'action du groupe d'isom{\'e}trie sur la vari{\'e}t{\'e}.
Apr{\`e}s avoir quotient{\'e} l'espace des solutions de ces {\'e}quations
par l'action du groupe de jauge, on obtient une vari{\'e}t{\'e}
de dimension $4N_H - 3N_V - N_V$ dont on peut montrer qu'elle
est encore hyperk{\"a}hlerienne. Cette construction est connue
\index{vari{\'e}t{\'e}!hyperk{\"a}hlerienne}
\index{hyperk{\"a}hlerien, quotient}
sous le nom de {\it quotient hyperk{\"a}hl{\'e}rien} et g{\'e}n{\'e}ralise
les notions de quotients symplectique et k{\"a}hl{\'e}rien bien connues
en math{\'e}matiques. La vari{\'e}t{\'e} quotient d{\'e}crit alors les interactions
des hypermultiplets subsistant {\`a} basse {\'e}nergie apr{\`e}s 
le m{\'e}canisme de Higgs. Elle devient singuli{\`e}re lorsque 
\index{Higgs!m{\'e}canisme de}
l'action du groupe de jauge sur l'espace des solutions aux 
{\'e}quations (\ref{moment}) n'est plus libre, c'est-{\`a}-dire
lorsque la sym{\'e}trie non ab{\'e}lienne n'est pas totalement
bris{\'e}e, en accord avec l'id{\'e}e que les singularit{\'e}s de
l'action effective de basse {\'e}nergie co{\"\i}ncident avec
les points o{\`u} des particules de masse nulle apparaissent.
Notons finalement que cette construction peut {\^e}tre d{\'e}form{\'e}e
en introduisant des {\it termes de Fayet-Iliopoulos} dans l'action
\index{Fayet-Iliopoulos, terme de}
microscopique, dont l'effet est de remplacer le second membre
des {\'e}quations (\ref{moment}) par une constante non nulle pour
les g{\'e}n{\'e}rateurs ab{\'e}liens du groupe de jauge. La 
d{\'e}formation r{\'e}sultante de l'espace des solutions 
en g{\'e}n{\'e}ral restaure une action libre du groupe de jauge, 
et r{\'e}sout ainsi les singularit{\'e}s de l'espace quotient.
\index{singularit{\'e}!r{\'e}solution quantique}

Dans le travail annex{\'e} en appendice \ref{hk}, nous avons {\'e}tudi{\'e}
ces espaces quotients dans une vari{\'e}t{\'e} de situations, que
nous avons compar{\'e}s {\`a} la lumi{\`e}re des dualit{\'e}s 
{\'e}lectrique-magn{\'e}tique. Les branches de Higgs correspondant
{\`a} la th{\'e}orie microscopique $SU(2)$ avec $N_f$ saveurs
{\`a} faible couplage se sont
ainsi r{\'e}v{\'e}l{\'e}es co{\"\i}ncider avec les branches
de Higgs {\'e}manant des diff{\'e}rentes singularit{\'e}s {\`a}
fort couplage d{\'e}crites par des th{\'e}ories ab{\'e}liennes
en pr{\'e}sence d'hypermultiplets charg{\'e}s de masse nulle.
Nous avons {\'e}galement d{\'e}couvert l'{\'e}galit{\'e} de la branche
de Higgs d'une th{\'e}orie $U(N_c)$ avec $N_f$ saveurs
d'hypermultiplets dans la repr{\'e}sentation fondamentale,
avec la branche de Higgs d'une th{\'e}orie de groupe de
jauge $U(N_f-N_c)$ et autant de saveurs. Cette identification
rappelle la dualit{\'e} conjectur{\'e}e par Seiberg dans le 
\index{Seiberg, dualit{\'e} dans SYM $N=1$}
\index{dualit{\'e}!de Seiberg dans SYM $N=1$}
contexte des th{\'e}ories de jauge de supersym{\'e}trie $N=1$
entre ces m{\^e}mes th{\'e}ories, mais s'en distingue par la pr{\'e}sence
du multiplet chiral adjoint caract{\'e}ristique de la supersym{\'e}trie $N=2$.
Elle a par la suite
trouv{\'e} une justification dans le cadre de la th{\'e}orie
des D-branes. 

\section{Sym{\'e}tries cach{\'e}es des th{\'e}ories de supergravit{\'e}}
La r{\'e}solution par Seiberg et Witten des th{\'e}ories de jauge
de supersym{\'e}trie $N=2$ a consid{\'e}rablement modifi{\'e} notre
compr{\'e}hension des th{\'e}ories des champs supersym{\'e}triques,
en montrant comment l'information dans diff{\'e}rentes
limites perturbatives de l'espace des modules pouvait {\^e}tre 
combin{\'e}e avec les imp{\'e}ratifs de la supersym{\'e}trie 
gr{\^a}ce aux transformations de dualit{\'e} pour 
restituer la structure exacte globale de la th{\'e}orie de
basse {\'e}nergie. L'impact qu'elle a eu en th{\'e}orie des
supercordes ne s'expliquerait cependant pas sans le faisceau
d'indications issu des th{\'e}ories de supergravit{\'e},
qui repr{\'e}sentent les th{\'e}ories de supercordes dans la
limite de basse {\'e}nergie. Ces th{\'e}ories r{\'e}v{\`e}lent, tout au
moins au niveau de leur action classique, l'existence de
sym{\'e}tries {\it cach{\'e}es} de dualit{\'e} reliant des r{\'e}gions de l'espace
\index{sym{\'e}tries cach{\'e}es}
des modules faiblement coupl{\'e}es {\`a} des r{\'e}gions de
fort couplage, bien que le terme de dualit{\'e} {\'e}lectrique-magn{\'e}tique
ne soit plus ad{\'e}quat dans ce cas. Ces th{\'e}ories n'existent
au niveau quantique que par la th{\'e}orie des supercordes qui
les r{\'e}gularise et les prolonge {\`a} toute {\'e}nergie. 
Leurs sym{\'e}tries cach{\'e}es ne seront donc que le 
premier indice de la dualit{\'e} des th{\'e}ories
des supercordes que nous aborderons au chapitre suivant.

\subsection{Supersym{\'e}trie locale, gravitation et unification}
L'inclination naturelle du physicien contemporain le porte {\`a}
jauger toute nouvelle sym{\'e}trie interne qui lui {\'e}choit.
La supersym{\'e}trie globale ne fait pas exception, mais son jaugement
\index{supersym{\'e}trie!jaugement de la}
est loin d'{\^e}tre anodin : le commutateur de deux transformations
de supersym{\'e}trie g{\'e}n{\'e}rant une translation, toute th{\'e}orie
invariante sous les transformations locales de supersym{\'e}trie 
sera en particulier invariante sous le groupe des diff{\'e}omorphismes,
et incluera donc le graviton de spin 2, avec son partenaire
supersym{\'e}trique, le gravitino de spin 3/2. 
Cette complication cache en r{\'e}alit{\'e} une aubaine, puisqu'elle laisse 
esp{\'e}rer un meilleur comportement ultraviolet dans une th{\'e}orie
de la gravitation r{\'e}put{\'e}e non renormalisable. Qui plus est,
en pr{\'e}sence de supersym{\'e}trie {\'e}tendue $N\ge 2$, 
on unifie gravitation et interactions de jauge en un multiplet 
supersym{\'e}trique gravitationnel contenant simultan{\'e}ment le
graviton, les $N$ gravitini, et $N(N-1)/2$ particules de spin 1
dits {\it graviphotons}. L'absence de particules non massives de spin 
sup{\'e}rieur {\`a} 2 impose une borne sup{\'e}rieure $N\le 8$ sur le
nombre de supersym{\'e}tries autoris{\'e}es. L'unification maximale
est r{\'e}alis{\'e}e dans la {\it th{\'e}orie de supergravit{\'e}
maximalement supersym{\'e}trique},  qui n'autorise que
\index{supersym{\'e}trie!repr{\'e}sentations de la}
\index{supersym{\'e}trie!maximale}
\index{supergravit{\'e}}
le seul multiplet gravitationnel contenant tous les spins de 0 {\`a} 2.
Elle n'{\'e}chappe malheureusement pas {\`a} l'absence de chiralit{\'e}
caract{\'e}ristique des th{\'e}ories {\`a} supersym{\'e}trie {\'e}tendue.

D'abord d{\'e}velopp{\'e}es comme th{\'e}ories de champs {\`a} quatre dimensions,
les th{\'e}ories de supergravit{\'e} ont connu un regain d'int{\'e}r{\^e}t 
en relation avec le d{\'e}veloppement des th{\'e}ories des supercordes,
qu'elles d{\'e}crivent {\`a} basse {\'e}nergie. Les th{\'e}ories de supergravit{\'e}
en dimension dix ont ainsi pu {\^e}tre classifi{\'e}es, en parall{\`e}le avec 
la classification des supercordes critiques en dimension dix. La
supergravit{\'e} $N=1$ {\`a} onze dimensions est rest{\'e}e {\`a} l'{\'e}cart
de ce sch{\'e}ma avant d'{\^e}tre incorpor{\'e}e dans les d{\'e}veloppements
les plus r{\'e}cents.

\subsection{Supergravit{\'e} de type I {\`a} dix dimensions}
\index{supergravit{\'e}!de type I}
Les charges $Q_\alpha$ de la supersym{\'e}trie $N=1$ 
{\`a} 9+1 dimensions se r{\'e}partissent en un spineur de Majorana-Weyl
de 16 composantes, correspondant {\`a} $N=4$ charges supersym{\'e}triques
dans le d{\'e}compte en 3+1 dimensions. Ceci correspond {\`a} la supersym{\'e}trie
maximale, et donc la dimension maximale {\'e}galement, 
d'une th{\'e}orie de jauge {\`a} supersym{\'e}trie globale. 
Les $N_V$ {\it multiplets vectoriels}, contenant chacun un champ de
jauge et un fermion de Majorana-Weyl de chiralit{\'e} droite,
peuvent {\^e}tre coupl{\'e}s au {\it multiplet gravitationnel} 
$N=1$, contenant le graviton, 
le gravitino de chiralit{\'e} gauche, 
un {\it tenseur antisym{\'e}trique
de jauge}
\index{tenseur antisym{\'e}trique}
\footnote{On entend par tenseur antisym{\'e}trique de jauge, une
forme diff{\'e}rentielle $C$ de rang $n$ 
dont la dynamique pr{\'e}sente l'invariance
de jauge $C\rightarrow C+d\Lambda$, o{\`u} $\Lambda$ est toute forme
diff{\'e}rentielle de rang $n-1$. En d'autres termes, $C$ est
un {\'e}l{\'e}ment du groupe de cohomologie $H_n$ de l'espace-temps.
}
$B_{\mu\nu}$, un spineur droit et un scalaire $\phi$ dit
{\it dilaton}, pour donner la th{\'e}orie de supergravit{\'e} dite
de type I. Cet ensemble de champs correspond pr{\'e}cis{\'e}ment
aux modes de masse nulle {\it des th{\'e}ories de supercordes de type I
et h{\'e}t{\'e}rotiques}, dont la valeur moyenne du dilaton repr{\'e}sente
\index{supercordes, th{\'e}orie des!de type I}
\index{supercordes, th{\'e}orie des!h{\'e}t{\'e}rotiques}
la constante de couplage. La chiralit{\'e} de ce contenu en champs
entra{\^\i}ne l'existence d'{\it anomalies gravitationnelles
et de jauge}, par exemple visibles dans les diagrammes en 
hexagone, 
\index{anomalie!gravitationnelle}
qui ne peuvent {\^e}tre compens{\'e}es que pour les groupes
de jauge $SO(32)$ et $E_8\times E_8$.
La th{\'e}orie des cordes de type I fournit
une r{\'e}alisation de la premi{\`e}re possibilit{\'e}, o{\`u} la sym{\'e}trie
de jauge est r{\'e}alis{\'e}e gr{\^a}ce aux charges de Chan-Paton
\index{Chan-Paton, facteur de}
port{\'e}es par les extr{\'e}mit{\'e}s de la corde ouverte
\footnote{La restriction au groupe de jauge $SO(32)$ peut {\'e}galement
{\^e}tre comprise comme la condition de compensation des diagrammes
de tadpole.}.
La d{\'e}couverte du m{\'e}canisme de compensation d'anomalies
en supergravit{\'e} de type I par Green et Schwarz \cite{Green:1984sg}
\index{Green et Schwarz, m{\'e}canisme de}
a relanc{\'e} la recherche d'une th{\'e}orie
des cordes de sym{\'e}trie $E_8\times E_8$ et a conduit
{\`a} la d{\'e}couverte des cordes h{\'e}t{\'e}rotiques
$E_8\times E_8$ et $SO(32)$
\cite{Gross:1985dd}. L'unicit{\'e} de
la th{\'e}orie de supergravit{\'e} $N=1$ {\`a} dix dimensions
sera un {\'e}l{\'e}ment important en faveur de la dualit{\'e}
h{\'e}t{\'e}rotique-type I dont nous dirons un mot
\index{dualit{\'e}!h{\'e}t{\'e}rotique - type I}
au chapitre suivant.

\subsection{Supergravit{\'e} de type IIB}
\index{supergravit{\'e}!de type IIB}
La d{\'e}finition d'une th{\'e}orie de supergravit{\'e} $N=2$ {\`a} dix dimensions
requiert le choix de la chiralit{\'e} relative des deux charges spinorielles
de Majorana-Weyl d{\'e}finissant l'alg{\`e}bre de supersym{\'e}trie. 
Le choix de deux spineurs de chiralit{\'e}
oppos{\'e}e conduit {\`a} une th{\'e}orie non chirale, dite de type IIA, que
nous d{\'e}taillerons au paragraphe suivant. Dans le cas de deux spineurs
de m{\^e}me chiralit{\'e}, on obtient la 
{\it supergravit{\'e} chirale $N=2$ de type IIB},
dont le multiplet gravitationnel contient, en plus du graviton
et des deux gravitini, {\it deux} tenseurs antisym{\'e}triques de
jauge $B_{\mu\nu}$ et $\mathcal{B}_{\mu\nu}$, 
deux spineurs de m{\^e}me chiralit{\'e},
un scalaire {\it complexe} $\tau$, ainsi qu'un tenseur antisym{\'e}trique
de jauge {\`a} quatre indices $\mathcal{D}_{\mu\nu\rho\sigma}$
{\`a} courbure courbure auto-duale $d\mathcal{D}=*d\mathcal{D}$~;
\index{anomalie!gravitationnelle}
les anomalies gravitationnelles sont pr{\'e}cis{\'e}ment compens{\'e}es
pour ce multiplet \cite{Alvarez-Gaume:1984ig}, qui est le seul
multiplet de spin inf{\'e}rieur ou {\'e}gal {\`a} deux autoris{\'e}
par la supersym{\'e}trie maximale {\`a} 32 supercharges.
Ce contenu en champs correspond au spectre de masse nulle de la 
{\it th{\'e}orie des supercordes de type IIB}. En particulier,
la partie r{\'e}elle du scalaire complexe $\tau=\axion+i e^{-\phi}$,
la deux-forme $\mathcal{B}_{\mu\nu}$ et la quatre-forme 
$\mathcal{D}_{\mu\nu\rho\sigma}$
proviennent du {\it secteur de Ramond} de la corde de type IIB
\index{Ramond, secteur de}
\index{Neveu-Schwarz, secteur de}
\index{tenseur antisym{\'e}trique!de Ramond}
\footnote{Les notions de secteur de Ramond et de secteur de
Neveu-Schwarz seront introduites dans le chapitre 3, section
\ref{rns}. Nous noterons les champs de Ramond avec des
lettres rondes, pour les distinguer des champs de Neveu-Schwarz
en lettres droites.}, tandis
que la partie imaginaire de $\tau$ est identifi{\'e}e avec
$1/g$, l'inverse de la constante de couplage de la th{\'e}orie
des cordes. La supergravit{\'e} de type IIB se r{\'e}v{\`e}le {\^e}tre
invariante sous le groupe de sym{\'e}trie continue $Sl(2,\Real)$,
\index{S-dualit{\'e}!de la th{\'e}orie IIB}
\index{sym{\'e}tries cach{\'e}es}
agissant sur les champs bosoniques selon
\begin{equation}
\label{sl2b}
\tau \rightarrow \frac{a\tau + b}{c\tau + d} \sp
\begin{pmatrix} B\\ \mathcal{B} \end{pmatrix}
\rightarrow 
\begin{pmatrix} a & b \\ c & d \end{pmatrix}
\begin{pmatrix} B \\ \mathcal{B} \end{pmatrix}
\sp 
a,b,c,d\in\Real\ ,\quad ad-bc=1\ ,
\end{equation} 
et laissant invariants la m{\'e}trique et le tenseur antisym{\'e}trique
{\`a} quatre indices\cite{Schwarz:1983qr}. 
Cette sym{\'e}trie continue classique ne saurait
cependant survivre quantiquement en raison de la quantification de Dirac
\index{Dirac, condition de quantification de}
de la charge associ{\'e}e aux champs de jauges $B$ et $\mathcal{B}$.
Il peut cependant en subsister un sous-groupe discret $Sl(2,\Zint)$,
contenant en particulier l'inversion $\tau \rightarrow -1/\tau$
reliant le r{\'e}gime de faible couplage $\tau_2 \rightarrow \infty$
au r{\'e}gime de fort couplage $\tau_2 \rightarrow 0$
\cite{Hull:1995ys}. Cette sym{\'e}trie
jouera un r{\^o}le important dans la conjecture de dualit{\'e} des
supercordes, et central dans les d{\'e}veloppements dans lesquels
j'ai {\'e}t{\'e} impliqu{\'e}.

\subsection{Supergravit{\'e} de type IIA et supergravit{\'e} {\`a}
onze dimensions} 
\index{supergravit{\'e}!de type IIA}
\index{supergravit{\'e}!{\`a} onze dimensions}
L'autre option consiste {\`a} choisir deux spineurs de chiralit{\'e}
oppos{\'e}e pour d{\'e}finir une alg{\`e}bre de supersym{\'e}trie $N=2$
dite de type IIA. Le multiplet gravitationnel consiste alors,
en sus du graviton et de deux gravitini de chiralit{\'e} oppos{\'e}e,
de tenseurs de jauge $\mathcal{A}_\mu$, $B_{\mu\nu}$, 
$\mathcal{C}_{\mu\nu\rho}$, de deux spineurs de Weyl-Majorana
et d'un scalaire {\it r{\'e}el} correspondant au dilaton.
\index{Ramond, secteur de}
\index{Neveu-Schwarz, secteur de}
Les champs de jauge $\mathcal{A}$ et $\mathcal{C}$ correspondent
au secteur de Ramond de la supercorde de type IIA. Contrairement
{\`a} la supergravit{\'e} de type IIB, la supergravit{\'e} de type IIA
ne pr{\'e}sente pas de sym{\'e}trie reliant fort et faible couplage.
En revanche, son contenu en champs est identique {\`a} celui de
la r{\'e}duction dimensionnelle 
\index{Kaluza-Klein!r{\'e}duction de}
\index{Kaluza-Klein|textit}
{\it {\`a} la Kaluza-Klein} d'une {\it th{\'e}orie de supergravit{\'e}
{\`a} onze dimensions contenant}, en plus du graviton $g_{MN}$
et d'un gravitino de Majorana $\psi^{M}$, 
\index{tenseur antisym{\'e}trique}
une trois-forme de jauge $\mathcal{C}_{MNP}$. L'action de cette 
th{\'e}orie a {\'e}t{\'e} construite par Cremmer, Julia et Scherk
\cite{Cremmer:1978km}, et s'{\'e}crit,
en ne retenant que les termes bosoniques,
\begin{equation}
\label{sugra11}
\mathcal{S}=\int 
\frac{1}{2l_{11}^{9}}
\left( d^{11}x \sqrt{-g} R - \frac{1}{2\cdot 4!} d\mathcal{C} \wedge
  *d\mathcal{C} 
\right) + \int \frac{\sqrt{2}}{2^7\cdot 3^3} \mathcal{C}\wedge d\mathcal{C} 
\wedge d\mathcal{C} \ .
\end{equation}
La normalisation relative des termes cin{\'e}tiques et du terme
de Wess-Zumino $\mathcal{C}\wedge d\mathcal{C} \wedge d\mathcal{C}$ 
\index{Wess-Zumino, terme de!dans la SUGRA 11D}
est fix{\'e}e par l'alg{\`e}bre de supersym{\'e}trie.
La r{\'e}duction dimensionelle de Kaluza-Klein consiste {\`a} compactifier
la th{\'e}orie sur un cercle de rayon $R_{11}$ et {\`a} omettre la 
d{\'e}pendance des champs sur la dimension interne
\index{compactification!sur un cercle}
(voir par exemple \cite{Salam:1981xd}). Cette proc{\'e}dure
est valable aux {\'e}nergies inf{\'e}rieures {\`a} $1/R_{11}$, pour laquelle
les {\'e}tats de moment interne $p=N/R_{11}$ peuvent {\^e}tre ignor{\'e}s.
La m{\'e}trique {\`a} onze dimensions $g_{MN}$ se r{\'e}duit ainsi
en une m{\'e}trique {\`a} dix dimensions $g_{\mu\nu}$, un rayon
$R_{11}$ fluctuant dans l'espace non compact et un champ de
jauge $\mathcal{A}_\mu$, selon l'ansatz
\index{ansatz!de Kaluza-Klein}
\index{Kaluza-Klein!ansatz de}
\begin{equation}
ds^2 = R_{11}^2 
(dy - \mathcal{A}_\mu dx^\mu)^2 + g_{\mu\nu} dx^{\mu} dx^{\nu} \ .
\end{equation}
Le tenseur de jauge en onze dimensions $\mathcal{C}_{MNP}$
donne quant {\`a} lui apr{\`e}s r{\'e}duction dimensionelle
une trois-forme $\mathcal{C}_{\mu\nu\rho}$ et une
deux-forme $B_{\mu\nu}=\mathcal{C}_{\mu\nu 11}$. 
Dans le r{\'e}f{\'e}rentiel d'Einstein, c'est-{\`a}-dire apr{\`e}s 
dilatation de Weyl $g\rightarrow g (R_{11}/l_{11})^{-1/4}$
afin de d{\'e}coupler le rayon $R_{11}$
de l'action d'Einstein-Hilbert,
\index{Einstein, r{\'e}f{\'e}rentiel d'}
\index{Einstein-Hilbert, action d'}
\index{Weyl!dilatation de}
l'action de la supergravit{\'e}
{\`a} onze dimensions r{\'e}duite {\`a} dix dimensions s'{\'e}crit alors
\begin{align}
\label{sugraIIA}
\mathcal{S}=&\int  \frac{1}{2\alpha^{'4}}
\left(  d^{10}x \sqrt{-g} R + \frac{1}{2} d\phi\wedge *d\phi
-\frac{1}{4} e^{2\phi} F\wedge *F -\frac{1}{48} e^{-\phi} H \wedge
 *H\right)
\nonumber\\
&+ \int \frac{\sqrt{2}}{2^7} B\wedge d\mathcal{C}  \wedge d\mathcal{C} \ ,
\end{align}
o{\`u} on a identifi{\'e} le dilaton $\phi$ au rayon $R_{11}$ selon
$R_{11}/l_{11}=e^{2\phi/3}$.
Cette action reproduit pr{\'e}cis{\'e}ment l'action de la supergravit{\'e} de type IIA~;
les implications de cette co{\"\i}ncidence ne seront
pleinement r{\'e}alis{\'e}es
que dans le cadre de la th{\'e}orie {\it des supercordes} de type IIA, qui donne
une d{\'e}finition microscopique d'une th{\'e}orie des champs autrement
non renormalisable.

\subsection{Compactification toro{\"\i}dale et U-dualit{\'e}\label{sugracomp}}
Nous avons jusqu'{\`a} pr{\'e}sent d{\'e}crit les th{\'e}ories de
supergravit{\'e} en dimension 10, dimension critique des
th{\'e}ories de supercordes. Ces th{\'e}ories incorporent 
cependant la gravit{\'e}, et la g{\'e}om{\'e}trie
\footnote{Le terme {\it g{\'e}om{\'e}trie} inclut ici non seulement
la valeur moyenne de la m{\'e}trique $g_{\mu\nu}$, mais aussi
de tous les autres champs, scalaires, de jauge et fermioniques
inclus.} de l'espace-temps 
qu'elles choisissent est en principe une question d{\'e}termin{\'e}e
par la dynamique quantique. En raison de leur supersym{\'e}trie
\index{vide!s{\'e}lection du}
cependant, le potentiel effectif dans l'espace des g{\'e}om{\'e}tries
pr{\'e}sente des {\it directions plates}, et on est libre de
\index{directions plates!de la th{\'e}orie des cordes}
consid{\'e}rer la th{\'e}orie au voisinage d'un point quelconque
de ces g{\'e}om{\'e}tries, en particulier en un point o{\`u}
l'espace-temps se d{\'e}compose en une vari{\'e}t{\'e} plate 
non compacte correspondant {\`a} l'espace-temps ambiant,
et une vari{\'e}t{\'e} compacte de dimension $d$ et de
rayon tr{\`e}s petit
devant l'{\'e}chelle d'observation. Les fluctuations de
\index{compactification}
la g{\'e}om{\'e}trie le long des directions plates au
voisinage du vide choisi correspondent {\`a} des champs
scalaires de masse nulle dits {\it champs de modules}.
\index{modules, champs de}
La supersym{\'e}trie est en g{\'e}n{\'e}ral bris{\'e}e par la 
\index{supersym{\'e}trie!brisure par compactification}
compactification. La raison en est que les g{\'e}n{\'e}rateurs
de la supersym{\'e}trie correspondent aux configurations
covariantement constantes du spineur reli{\'e}
au gravitino par supersym{\'e}trie, lesquelles disparaissent
pour des vari{\'e}t{\'e}s de courbure quelconque, c'est-{\`a}-dire
de groupe d'holonomie $SO(d,\Real)$. Pour des
{\it vari{\'e}t{\'e}s {\`a} holonomie restreinte} cependant, il
\index{vari{\'e}t{\'e}!{\`a} holonomie restreinte}
\index{holonomie!de la m{\'e}trique}
existe des configurations de spineur invariantes
par transport parall{\`e}le, et donc une alg{\`e}bre de
supersym{\'e}trie {\'e}ventuellement {\'e}tendue. 
\index{compactification!sym{\'e}tries de jauge}
Les isom{\'e}tries de la vari{\'e}t{\'e} de compactification
apparaissent comme des {\it sym{\'e}tries de jauge} dans la
th{\'e}orie r{\'e}duite, ainsi que nous l'avons vu
dans l'exemple de compactification sur un
cercle du paragraphe pr{\'e}c{\'e}dent, et le groupe
des diff{\'e}omorphismes de la vari{\'e}t{\'e} interne 
peut {\'e}galement donner lieu {\`a} des 
{\it sym{\'e}tries globales} dans la th{\'e}orie r{\'e}duite. 
Le cas le plus simple est celui d'une vari{\'e}t{\'e} compacte
plate, c'est-{\`a}-dire d'un tore $T^{d}$, pour lequel
toutes les supersym{\'e}tries sont conserv{\'e}es
\index{compactification!toro{\"\i}{}dale}
\footnote{
En dimension paire, le groupe d'holonomie peut {\^e}tre
r{\'e}duit de $SO(d)$ {\`a} $SU(d/2)$, correspondant aux vari{\'e}t{\'e}s
k{\"a}hleriennes {\`a} courbure de Ricci nulle, dites vari{\'e}t{\'e}s
de Calabi-Yau, et il existe alors un spineur neutre sous
\index{Calabi-Yau, vari{\'e}t{\'e} de}
$SU(d/2)$ et donc covariantement constant. En dimension
multiple de 4, on peut encore r{\'e}duire le groupe d'holonomie
{\`a} $Sp(d/4)$, correspondant aux vari{\'e}t{\'e}s hyperk{\"a}hleriennes
\index{vari{\'e}t{\'e}!hyperk{\"a}hlerienne}
\index{hyperk{\"a}hlerienne, vari{\'e}t{\'e}}
que nous avons d{\'e}j{\`a} discut{\'e}es. On peut {\'e}galement
compactifier sur une sph{\`e}re $S^d$ sans briser aucune
supersym{\'e}trie, {\`a} condition de choisir un espace anti-de Sitter
\index{anti-de Sitter, espace}
pour les dimensions non compactes 
\cite{Freund:1980xh,Schwarz:1983qr,Duff:1998}}. 
Les $d$ isom{\'e}tries
de translation donnent lieu {\`a} une invariance de jauge
$U(1)^d$, {\'e}ventuellement {\'e}tendue {\`a} une sym{\'e}trie 
non ab{\'e}lienne pour certaines valeurs des modules
du tore, tandis que le groupe des diff{\'e}omorphismes 
du tore donne lieu {\`a} une sym{\'e}trie globale $Sl(d,\Real)$,
dont un sous-groupe $SO(d,\Real)$ agit trivialement.

Si l'int{\'e}r{\^e}t ph{\'e}nom{\'e}nologique de la compactification
est {\'e}vident, son int{\'e}r{\^e}t th{\'e}orique demande peut-{\^e}tre
plus de justification. On pourrait en effet argumenter que
la compactification d'une th{\'e}orie des champs ne fait que
quantifier l'impulsion des particules, entra{\^\i}nant l'existence
de tours d'{\'e}tats de Kaluza-Klein, et ne fait qu'ajouter au probl{\`e}me
\index{Kaluza-Klein!excitation de}
une complication inutile.
Cette vue oublie cependant de prendre en 
compte l'existence possible d'{\it {\'e}tats {\'e}tendus} 
dans la th{\'e}orie non compactifi{\'e}e, qui peuvent {\^e}tre stabilis{\'e}s
gr{\^a}ce {\`a} l'existence de cycles d'homologie non triviaux de
\index{cycle d'homologie!de la vari{\'e}t{\'e} de compactification}
l'espace compact. Il se peut {\'e}galement qu'une sym{\'e}trie 
<<cach{\'e}e>> dans la th{\'e}orie non compactifi{\'e}e se combine avec
les sym{\'e}tries inh{\'e}rentes {\`a} la compactification en un groupe
plus grand de sym{\'e}tries tout {\`a} fait visibles. C'est pr{\'e}cis{\'e}ment
\index{sym{\'e}tries cach{\'e}es|textit}
le cas des compactifications toro{\"\i}dales des th{\'e}ories de
supergravit{\'e} {\`a} supersym{\'e}trie maximale, ainsi que l'ont
reconnu Cremmer et Julia \cite{Cremmer:1979up}.

Ainsi, la compactification de la th{\'e}orie de type IIA sur un tore
$T^d$ montre une extension de la sym{\'e}trie $Sl(d,\Real)$ {\'e}vidente,
en un {\it groupe de sym{\'e}trie cach{\'e}} $Sl(d+1,\Real)$ m{\'e}langeant les 
modules de $T^d$ avec le dilaton et les lignes de Wilson du
\index{Wilson, ligne de}
champ de jauge {\it de Ramond} $\mathcal{A}$. Cette sym{\'e}trie
n'est autre que la manifestation de la filiation entre
la th{\'e}orie de supergravit{\'e} de type IIA et la supergravit{\'e}
{\`a} onze dimensions discut{\'e}e au paragraphe pr{\'e}c{\'e}dent.
\index{dualit{\'e}!des th{\'e}ories IIA et SUGRA 11D}
\index{Kaluza-Klein!r{\'e}duction de SUGRA 11D en IIA} 
Par ailleurs, la compactification de
la th{\'e}orie de type IIA sur un cercle de rayon $R_A$ 
(et {\it a fortiori} sur un tore $T^d$, $d>1$) 
et de couplage $e^{\phi_A}$, se r{\'e}v{\`e}le
strictement {\'e}quivalente {\`a} la compactification de la
supergravit{\'e} de type IIB 
de couplage $e^{\phi_B}=e^{\phi_A}/R_A$
sur un cercle de rayon $R_B=\alpha'/R_A$,
moyennant l'identification du scalaire de Ramond
$\axion$ de type IIB avec la ligne de Wilson $\mathcal{A}$
sur le cercle du champ de jauge $\mathcal{A}_\mu$ de 
la th{\'e}orie de type IIA.
\index{dualit{\'e}!des th{\'e}ories IIA et IIB}
Cette propri{\'e}t{\'e} n'est quant {\`a} elle que le reflet {\`a} basse {\'e}nergie
de la {\it T-dualit{\'e}} des th{\'e}ories des cordes perturbatives,
\index{T-dualit{\'e}}
que nous discuterons au chapitre suivant.
En utilisant cette {\'e}quivalence, la dualit{\'e} $Sl(2,\Zint)$ de
\index{S-dualit{\'e}!de la th{\'e}orie IIB}
la th{\'e}orie des supercordes de type IIB, apr{\`e}s  compactification
sur un cercle et T-dualit{\'e}, n'est autre que la sym{\'e}trie g{\'e}om{\'e}trique
$Sl(2,\Zint)$ de la supergravit{\'e} {\`a} onze dimensions compactifi{\'e}e
sur un tore $T^2$. 

Cette structure ne fait que s'{\'e}tendre pour des compactifications
sur des tores de dimension $d\ge 2$. L'espace des modules contient
en effet un {\it sous-espace homog{\`e}ne}  $Sl(d+1,\Real)/SO(d+1)$,
\index{vari{\'e}t{\'e}!homog{\`e}ne}
correspondant aux modules de la m{\'e}trique sur le tore $T^{d+1}$
d{\'e}finissant la compactification de la supergravit{\'e} {\`a} onze
dimensions, mais aussi un sous-espace 
$SO(d,d,\Real)/\left(SO(d)\times SO(d)\right)$, d'intersection
non vide avec le pr{\'e}c{\'e}dent, d{\'e}finissant les modules
de la m{\'e}trique et du tenseur antisym{\'e}trique $B_{\mu\nu}$
du secteur de Neveu-Schwarz sur le tore $T^d$. Du point
de vue de la th{\'e}orie des cordes, ces modules
co{\"\i}ncident avec les param{\`e}tres de la th{\'e}orie conforme
de $d$ bosons compactifi{\'e}s, et sont identifi{\'e}s sous l'action du
groupe de T-dualit{\'e} $SO(d,d,\Zint)$
\index{T-dualit{\'e}!sur un tore $T^d$}
\footnote{En effectuant deux
T-dualit{\'e}s successives, on obtient en effet une
sym{\'e}trie de la supergravit{\'e} de type IIA compactifi{\'e}e sur $T^d$ (et non
plus une dualit{\'e} avec la supergravit{\'e} de type IIB). L'ensemble
de ces transformations forme le groupe discret $SO(d,d,\Zint)$.}.
L'{\it espace des modules total} contenant ces deux sous-espaces 
s'{\'e}crit encore comme un quotient 
$E_d/H_d$, o{\`u} $E_d$ est le {\it groupe non compact g{\'e}n{\'e}r{\'e}
par $Sl(d+1,\Real)$ et $SO(d,d,\Real)$}, et $H_d$ son {\it sous-groupe
compact maximal}
\index{groupe!sous-groupe compact maximal}
\index{U-dualit{\'e}!de type II sur $T^d$|textit}
\footnote{Cette derni{\`e}re propri{\'e}t{\'e} est une cons{\'e}quence
de la positivit{\'e} de la m{\'e}trique sur l'espace $G_d/H_d$, 
c'est-{\`a}-dire de
\index{fant{\^o}me}
l'absence de fant{\^o}mes dans la dynamique des scalaires.}.
Pour $d\ge 6$, les groupes $E_d$ correspondent {\`a} des 
versions non compactes des {\it groupes de Lie exceptionnels}
\index{groupe!exceptionnel}
$E_6,E_7$ et $E_8$, tandis que pour $d\le 5$ ils se r{\'e}duisent
aux groupes classiques 
\begin{equation}
E_2=Sl(2,\Real)\ ,\quad
E_3=Sl(3,\Real)\times Sl(2,\Real)\ ,\quad
E_4=Sl(5,\Real)\ ,\quad 
E_5=SO(5,5,\Real)\ .
\end{equation}
Le groupe $E_d$ appara{\^\i}t comme une {\it sym{\'e}trie globale} de l'action
de la th{\'e}orie de supergravit{\'e} compactifi{\'e}e sur $T^d$ (et non
seulement de l'espace des modules), tandis que le groupe $H_d$
est un groupe de {\it sym{\'e}trie locale}. Il n'est cependant pas 
associ{\'e} {\`a} des champs de jauge fondamentaux mais {\`a} une
{\it connection composite} construite {\`a} partir des champs scalaires,
et l'invariance de jauge peut {\^e}tre fix{\'e}e en choisissant
un r{\'e}pr{\'e}sentant de ces scalaires dans l'orbite de $H_d$.

L'appellation historique de <<sym{\'e}trie cach{\'e}e>> peut sembler
\index{sym{\'e}tries cach{\'e}es}
{\`a} ce stade d{\'e}concertante : nous en avons attribu{\'e} une
partie {\`a} l'existence de la T-dualit{\'e}, et l'autre {\`a}
l'existence d'une th{\'e}orie de supergravit{\'e} {\`a} onze dimensions 
dont la th{\'e}orie de type IIA est la r{\'e}duction dimensionnelle.
Elle reste cependant justifi{\'e}e pour plusieurs raisons :
\begin{itemize}
\item la T-dualit{\'e} est visible dans la th{\'e}orie de
type IIA {\`a} dix dimensions, mais n'a {\`a} ce jour pas {\'e}t{\'e}
exhib{\'e}e comme sym{\'e}trie de la th{\'e}orie {\`a} onze dimensions.
En d'autres termes, l'action de la T-dualit{\'e} sur les
modes de Fourier d'impulsion non nulle autour du cercle
de rayon $R_{11}$ n'est pas connue.
\item la T-dualit{\'e} est une sym{\'e}trie perturbative
de la th{\'e}orie de cordes de type IIB, mais ce n'est pas
le cas de l'invariance de Lorentz {\`a} 11 dimensions.
\index{Lorentz, invariance {\`a} 11D}
En particulier, l'op{\'e}ration qui {\'e}change un rayon
de compactification avec le rayon de la onzi{\`e}me
dimension $R_{11}=e^{2\phi/3}$ ne pr{\'e}serve pas la 
s{\'e}rie de perturbation en $e^{\phi}$.
\item seul un sous groupe discret $SO(d,d,\Zint)$ 
de la sym{\'e}trie continue $SO(d,d,\Real)$ de l'action
effective de basse {\'e}nergie subsiste au niveau de la th{\'e}orie des cordes,
\index{action effective!sym{\'e}trie de}
en raison de la quantification des impulsions dans l'espace compact.
Existe-t-il donc un sous-groupe r{\'e}siduel $E_d(\Zint)$ du
groupe de sym{\'e}trie continue $E_d$,
dit {\it groupe de U-dualit{\'e}}, incluant des dualit{\'e}s
non perturbatives ?  
\end{itemize}
Ces questions ne peuvent avoir un sens que si une extension 
non perturbative des th{\'e}ories de supergravit{\'e} existe.
\index{supergravit{\'e}!r{\'e}gularisation ultraviolette}
La d{\'e}finition perturbative de ces th{\'e}ories non renormalisables
est d{\'e}j{\`a} un probl{\`e}me en soi, auxquelles les th{\'e}ories des
cordes, que nous introduirons  dans le chapitre suivant, fourniront
la solution, en offrant une r{\'e}gularisation finie dans l'ultraviolet.
La d{\'e}finition non perturbative quant {\`a} elle 
est partiellement r{\'e}solue par la conjecture de dualit{\'e} des
supercordes, mais n'a pas encore {\'e}t{\'e} totalement explicit{\'e}e.
Consid{\'e}r{\'e}es comme th{\'e}ories
effectives de basse {\'e}nergie des th{\'e}ories de supercordes, 
les th{\'e}ories de supergravit{\'e} fournissent 
n{\'e}anmoins des indications pr{\'e}cieuses sur 
le contenu de cette hypoth{\'e}tique th{\'e}orie des cordes non perturbative,
en particulier gr{\^a}ce {\`a} l'{\'e}tude de leurs solutions classiques
vers lesquelles nous nous tournons maintenant.

\section{Solitons de $p$-brane des th{\'e}ories de supergravit{\'e}}
L'{\'e}tude du spectre BPS des solutions classiques des th{\'e}ories
\index{etats BPS@{\'e}tats BPS!des th{\'e}ories de supergravit{\'e}}
de supergravit{\'e} r{\'e}v{\`e}le une diversit{\'e} 
d'objets solitoniques {\'e}tendus, en correspondance avec
la diversit{\'e} des champs de jauge pr{\'e}sents.
Avec les sym{\'e}tries cach{\'e}es d{\'e}crites dans la
section pr{\'e}c{\'e}dente, l'existence de ces {\'e}tats sera un argument
essentiel de la conjecture de dualit{\'e}s des th{\'e}ories de
supercordes. Nous en donnerons une br{\`e}ve description,
renvoyant le lecteur aux nombreux articles de revue pour
plus de d{\'e}tails (par exemple, 
\cite{Duff:1995an,Duff:1996zn,Stelle:1996tz,Townsend:1996xj,Townsend:1997wg}).

\subsection{Tenseurs antisym{\'e}triques de jauge et $p$-branes}
\index{tenseur antisym{\'e}trique!charge par rapport {\`a}}
\index{brane!charge {\'e}lectrique des}
\fig{5cm}{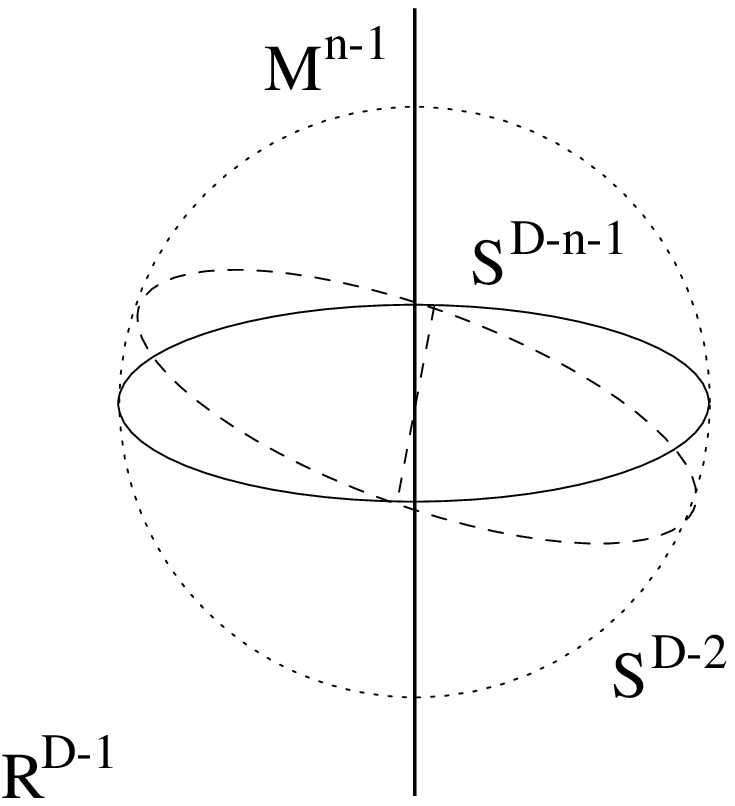}{La charge {\'e}lectrique sous une $n$-forme 
$\mathcal{C}_n$ dans un espace-temps de dimension $D$
est donn{\'e}e par l'int{\'e}grale de la
forme duale $*\mathcal{C}_n$ sur une section $S^{D-n-1}$ de la
sph{\`e}re {\`a} l'infini $S^{D-2}$.}{charge}
Avant de donner une description explicite de ces solutions
classiques, il est souhaitable d'identifier le type de configurations
solitoniques attendues. Le secteur bosonique des 
th{\'e}ories de supergravit{\'e} {\`a} dix dimensions
comprend la m{\'e}trique $g_{\mu\nu}$,
le dilaton $\phi$ et plusieurs tenseurs antisym{\'e}triques de
jauge $\mathcal{C}_n$, o{\`u} l'indice $n$ d{\'e}signe le rang
de la forme diff{\'e}rentielle. On s'int{\'e}resse en 
particulier aux solitons charg{\'e}s sous ces tenseurs de
jauge
\footnote{On peut {\'e}galement consid{\'e}rer des 
solutions purement gravitationnelles o{\`u} le potentiel
de jauge est nul et le dilaton constant. Except{\'e} pour le cas
de l'onde gravitationnelle, la g{\'e}om{\'e}trie
asymptotique de ces solutions n'est cependant pas minkovskienne,
et on n'obtient pas ainsi de solitons de la th{\'e}orie plate.
On peut cependant obtenir des solutions de g{\'e}om{\'e}trie
asymptotique $\RR^{D-1}\times S_1$ (en utilisant l'espace
d'Einstein self-dual de Taub-NUT, de g{\'e}om{\'e}trie
\index{Taub-NUT, instanton de}
\index{instanton!gravitationnel}
\index{Kaluza-Klein!monop{\^o}le de}
\index{monop{\^o}le magn{\'e}tique!de Kaluza-Klein}
asymptotique $\RR^3\times S_1$) d{\'e}crivant des 
{\it monop{\^o}les magn{\'e}tiques de Kaluza-Klein} de la
th{\'e}orie compactifi{\'e}e sur $S_1$
\cite{Sorkin:1983ns}, localis{\'e}s dans 4 dimensions
et donc assimilables {\`a} des $(D-5)$-branes.}. 
En dimension 3+1, la charge {\'e}lectrique associ{\'e}e
{\`a} un champ de jauge $A_\mu$ est mesur{\'e}e par le flux
\index{flux {\'e}lectrique} 
$\oint_{S^2} *dA$ du champ {\'e}lectrique {\`a} travers une {\it 2-sph{\`e}re}
entourant la configuration. La charge {\'e}lectrique est
donc essentiellement {\it ponctuelle}, en accord avec le fait
que la dynamique d'une particule ponctuelle peut {\^e}tre
naturellement coupl{\'e}e {\`a} la 1-forme $A_\mu$ par
un couplage $\int A$ dans l'action. La charge magn{\'e}tique
est d{\'e}finie de mani{\`e}re identique apr{\`e}s {\it dualit{\'e}
de Poincar{\'e}} $dA \rightarrow *dA$, 
\index{Poincar{\'e}!dualit{\'e} de}
\index{dualit{\'e}!de Poincar{\'e}}
soit par le flux $\oint_{S_2} dA$. Elle est conserv{\'e}e
en vertu de l'identit{\'e} de Bianchi $d(dA)=0$, tandis que
la conservation de la charge {\'e}lectrique requiert 
l'{\'e}quation du mouvement dans le vide $d(*dA)=0$.
De mani{\`e}re analogue, on d{\'e}finit {\it la charge {\'e}lectrique
associ{\'e}e {\`a} un tenseur de jauge de rang $n$} $\mathcal{C}_n$ dans un 
espace temps de dimension $D$ comme l'int{\'e}grale 
$\oint_{S^{D-n-1}} *d\mathcal{C}$ sur une sph{\`e}re de dimension $D-n-1$
{\`a} l'infini, sous-vari{\'e}t{\'e} de 
la sph{\`e}re $S^{D-2}$ {\`a} l'infini spatial (figure \ref{charge}).
La charge correspondante est donc port{\'e}e par un
objet transverse {\`a} $S^{D-n-1}$ dans $S^{D-2}$, soit
de dimension $n-1$. Le tenseur de jauge de rang $n$ se couple
d'ailleurs naturellement au volume d'univers d'un objet
{\'e}tendu de dimensionnalit{\'e} $n-1$ par le couplage
$\int \mathcal{C}_n$. Ainsi, en dimension 3+1, la charge associ{\'e}e {\`a} une
deux-forme $B_{\mu\nu}$ est port{\'e}e par une corde, et
mesur{\'e}e par le flux $\oint_{S^1} *dB$ sur un cercle {\`a} l'infini
transverse {\`a} la corde. Notons que la charge ne d{\'e}pend 
en g{\'e}n{\'e}ral pas des variations infinit{\'e}simales de
la section $S^{D-n-1}$ de la sph{\`e}re {\`a} l'infini
$S^{D-2}$, mais qu'elle subit une discontinuit{\'e} lorsque cette
section intersecte la corde, ou tout autre objet charg{\'e}.
En particulier, la charge s'annule si la corde 
ne s'{\'e}tend pas {\`a} l'infini.
La charge {\it magn{\'e}tique} sous $\mathcal{C}_n$
est bien entendu d{\'e}finie comme la charge {\'e}lectrique 
par rapport au {\it dual de Poincar{\'e}}
$\mathcal{C}_{D-n-2}$. Cette charge est donc port{\'e}e par des objets
{\'e}tendus dans $D-n-3$ directions.
Les objets charg{\'e}s par rapport au potentiel $\mathcal{C}_n$ correspondent
donc {\`a} des objets {\'e}tendus sur $p=n-1$ dimensions internes,
commun{\'e}ment baptis{\'e}s {\it $p$-branes}.
La {\it dualit{\'e} {\'e}lectrique-magn{\'e}tique} en dimension $D$
{\'e}change donc $p$-branes et $(D-p-4)$-branes
\index{dualit{\'e}!{\'e}lectrique-magn{\'e}tique}.

Les solutions des th{\'e}ories de supergravit{\'e} correspondant  
{\`a} ces objets {\'e}tendus peu\-vent {\^e}tre obtenues en d{\'e}composant
les $D$ coordonn{\'e}es de l'espace temps en $p+1$ coordonn{\'e}es
de {\it volume d'univers} $x^\mu$ et $D-p-1$ coordonn{\'e}es
{\it transverses} $y^m$. On consid{\`e}re alors un ansatz
\index{ansatz!des $p$-branes}
\index{brane!ansatz des $p$-branes}
\begin{subequations}
\begin{equation}
\label{pansatz}
ds^2=e^{2A(r)} \eta_{\mu\nu} dx^\mu dx^\nu
+ e^{2B(r)} \delta_{mn} dx^m dx^n 
\end{equation}
\begin{equation}
\phi = \phi(r) 
\end{equation}
\begin{equation}
F_{p+2} = d\mathcal{C}_{p+1}={\rm Vol}_{p+1} \wedge d(e^C(r)) 
\end{equation}
\end{subequations}
o{\`u} la m{\'e}trique et le dilaton,
invariants sous les translations le long du volume d'univers, 
ne d{\'e}pendent que du rayon transverse $r=\sqrt{y^2}$,
et o{\`u} ${\rm Vol}_{p+1}$ d{\'e}signe l'{\'e}l{\'e}ment de
volume du volume d'univers de la $p$-brane. De telles solutions
d{\'e}crivent en r{\'e}alit{\'e} des $p$-branes <<{\'e}paisses>>
dont la taille d{\'e}pend de la vitesse de d{\'e}croissance
de $A,B,C,\phi$. L'interpr{\'e}tation solitonique
requiert {\'e}galement que cette d{\'e}croissance soit assez
rapide pour que l'espace soit {\it asymptotiquement plat},
de sorte que de telles solutions puissent {\^e}tre mises
en pr{\'e}sence {\`a} grande distance.

Avant de pouvoir d{\'e}terminer 
les fonctions $A,B,C,\phi$ de l'ansatz pr{\'e}c{\'e}dent,
il est n{\'e}cessaire de pr{\'e}ciser le lagrangien
que l'on consid{\`e}re. Ne s'int{\'e}ressant qu'aux
solutions charg{\'e}es par rapport {\`a} un seul champ de jauge
$\mathcal{C}_{p+1}$, on peut omettre tous les autres potentiels
de jauge, et se restreindre {\`a}
\begin{equation}
\mathcal{S}=\int d^{D}x \sqrt{-g} 
\left( R -\frac{1}{2}\partial\phi\partial\phi
-\frac{1}{2n!} e^{a\phi} F_{p+1}^2 \right)\ .
\end{equation}
La valeur du couplage $a$ au dilaton d{\'e}pend du mod{\`e}le
consid{\'e}r{\'e}. Cette troncation pr{\'e}serve l'existence
d'une charge supersym{\'e}trique $Q$. L'invariance sous cette charge
implique l'invariance sous la moiti{\'e} des supersym{\'e}tries 
originelles.

\subsection{Membranes et 5-branes {\`a} onze dimensions}
Les fonctions $A,B,C,\phi$ peuvent maintenant
{\^e}tre d{\'e}termin{\'e}es par la condition BPS
d'invariance sous la supersym{\'e}trie $Q$.
Nous renvoyons le lecteur {\`a} l'article de revue
\cite{Stelle:1996tz} pour une pr{\'e}sentation compl{\`e}te,
et nous nous contentons ici de d{\'e}crire les caract{\'e}ristiques
g{\'e}n{\'e}rales des solutions obtenues, tout d'abord dans le
cas de la supergravit{\'e} {\`a} onze dimensions. On obtient 
dans ce cas des solutions solitoniques de {\it $2$-brane}, aussi
appel{\'e}es {\it membranes}, 
\index{membrane, soliton de SUGRA 11D}
\index{brane!membrane de SUGRA 11D}
charg{\'e}es {\'e}lectriquement sous la trois-forme $\mathcal{C}_3$, et des
solutions de {\it 5-brane} charg{\'e}es magn{\'e}tiquement
sous $\mathcal{C}_3$. Ces objets infiniment {\'e}tendus dans
2 (resp. 5) dimensions spatiales poss{\`e}dent une tension,
soit une masse par unit{\'e} de volume d'univers, de
$T_2=1/l_{11}^3$ (resp. $T_5=1/l_{11}^6$), o{\`u}
$l_{11}$ est l'unique {\'e}chelle de longueur de la supergravit{\'e}
{\`a} 11 dimensions. 
\index{tension!des membranes et 5-branes}
La g{\'e}om{\'e}trie de la 
membrane interpole entre l'espace asymptotiquement plat
{\`a} grande distance et un espace
\footnote{La notation $AdS_n$ d{\'e}signe un espace maximalement sym{\'e}trique
anti-de Sitter de dimension $n$.} 
$AdS_4\times S^7$
{\`a} l'horizon
\index{horizon, g{\'e}om{\'e}trie de l'}. 
\index{anti-de Sitter, espace}
Elle peut {\^e}tre continu{\'e}e analytiquement au-del{\`a}
de l'horizon jusqu'{\`a} une singularit{\'e} de genre temps
correspondant {\`a} une deux-brane infiniment fine.
\index{cinq-brane!soliton de SUGRA 11D}
\index{brane!cinq-brane de SUGRA 11D}
La 5-brane en revanche est r{\'e}guli{\`e}re partout, et interpole
entre l'espace asymptotiquement plat et $AdS_7 \times S^4$ {\`a}
l'horizon. La continuation au-del{\`a} de l'horizon montre
que l'espace {\it int{\'e}rieur} peut 
{\^e}tre identifi{\'e} avec l'espace {\it ext{\'e}rieur} {\`a} l'horizon .
Ces solutions correspondent {\`a} des configurations
{\it extr{\^e}mes} de solutions de <<p-branes noires>>
o{\`u} l'horizon interne co{\"\i}ncide avec l'horizon
externe. Ce cas est en effet le seul pr{\'e}servant la moiti{\'e} des
supersym{\'e}tries. En cons{\'e}quence, elles  saturent les in{\'e}galit{\'e}s
\index{Bogomolny, borne de}
de Bogomolny entre la masse par unit{\'e} de volume d'univers
et les charges {\'e}lectriques et magn{\'e}tiques sous le champ
de jauge $\mathcal{C}_3$~: ce sont donc des {\it {\'e}tats BPS
annihil{\'e}s par la moiti{\'e} des supersym{\'e}tries}.
\index{etats BPS@{\'e}tats BPS!de $p$-branes}

L'existence de ces solutions semble peser en faveur
d'une description microscopique de la th{\'e}orie
de supergravit{\'e} {\`a} onze dimensions en termes
d'une th{\'e}orie de supermembranes. Cette esp{\'e}rance
est malheureusement d{\'e}{\c c}ue par les difficult{\'e}s
de quantifier les non-lin{\'e}arit{\'e}s de la dynamique
des supermembranes. Nous reviendrons sur cette question
dans le dernier chapitre de ce m{\'e}moire.

\subsection{Solitons des supergravit{\'e}s {\`a} dix dimensions}
\index{brane|textit}
La m{\^e}me approche fournit des solutions de $p$-branes
pour toutes les th{\'e}ories de supergravit{\'e} en dimension 10,
associ{\'e}es {\`a} chaque tenseur de jauge. Dans le cas de
la supergravit{\'e} de type IIA, on peut de mani{\`e}re {\'e}quivalente
{\'e}tudier comment celle-ci h{\'e}rite des solutions de la
supergravit{\'e} {\`a} onze dimensions par compactification
sur le cercle de rayon $R_{11}$. Deux types de r{\'e}duction
sont en g{\'e}n{\'e}ral envisageables\footnote{On peut {\'e}galement r{\'e}duire
le long d'une isom{\'e}trie angulaire, mais le statut de 
cette proc{\'e}dure est encore peu clair \cite{Duff:1998}.}~:
\index{r{\'e}duction dimensionnelle!des solitons}
\begin{itemize}
\item Dans le cas o{\`u} le
volume d'univers de la $p$-brane contient la direction
compactifi{\'e}e, c'est-{\`a}-dire lorsque la $p$-brane
est {\it enroul{\'e}e} sur cette direction,
 on peut quotienter par l'isom{\'e}trie correspondante pour obtenir
une $(p-1)$-brane en dimension $D-1$. Cette op{\'e}ration est
dite {\it r{\'e}duction diagonale}. La tension de la membrane
r{\'e}duite est alors $T_{p-1}=R_{11} T_{p}$.
\index{tension!par r{\'e}duction}
\item Si au contraire le volume d'univers de la $p$-brane
est transverse {\`a}
la dimension compacte, la sym{\'e}trie de translation est bris{\'e}e
et on ne peut alors plus prendre le quotient. On peut en revanche
consid{\'e}rer un {\it empilement}
continu de $p$-branes dans la direction compacte, de mani{\`e}re
{\`a} restaurer la sym{\'e}trie de translation. La possibilit{\'e}
de superposer des configurations identiques de $p$-branes
est intimement li{\'e}e {\`a} l'absence d'interactions statiques
\index{etats BPS@{\'e}tats BPS!empilement}
entre configuration BPS. On obtient
ainsi une $p$-brane de m{\^e}me tension
en dimension inf{\'e}rieure, ce qui vaut
{\`a} ce processus le nom de {\it r{\'e}duction verticale}.
\end{itemize}
La membrane de la supergravit{\'e} {\`a} onze dimensions donne
ainsi par r{\'e}duction diagonale une 1-brane de tension
$R_{11}/l_{11}^3=1/\alpha'$ et charg{\'e}e par rapport au tenseur
de jauge de Neveu-Schwarz $C_{\mu\nu11}=B_{\mu\nu}$, 
qui n'est autre que la corde
fondamentale de la th{\'e}orie de type IIA ; 
et par r{\'e}duction verticale une D2-brane de tension
\index{D-brane!solitons de supergravit{\'e}}
\index{tension!des D-branes}
$1/l_{11}^3 = \alpha^{'-3/2}/g$ charg{\'e}e par rapport au
tenseur de jauge de Ramond $\mathcal{C}_3$~:
nous appellons {\it NS$p$-branes} les $p$-branes charg{\'e}es sous
les tenseurs de jauge de Neveu-Schwarz, et {\it D$p$-branes}
celles charg{\'e}es sous les champs de Ramond, pour des raisons
qui appara{\^\i}{}tront dans la section \ref{dbr}.

De m{\^e}me, la 5-brane de la supergravit{\'e} {\`a} onze dimensions
charg{\'e}e magn{\'e}tiquement sous $\mathcal{C}_3$ 
donne par r{\'e}duction la D4-brane de tension
$R_{11}/l_{11}^6=1/g\alpha^{'5/2}$, charg{\'e}e sous le tenseur de jauge 
de Ramond $\mathcal{R}_5$, ainsi que la NS5-brane de tension
$1/l_{11}^6=1/g^2\alpha^{'3}$,
charg{\'e}e {\it magn{\'e}tiquement} sous le tenseur de
Neveu-Schwarz $B_{\mu\nu}$. Le monop{\^o}le de Kaluza-Klein
de la supergravit{\'e} {\`a} onze dimensions appara{\^\i}{}t comme la
D6-brane de tension $R_{11}^2/l_{11}^9=1/g\alpha ^{'7/2}$
\footnote{Il donne {\'e}galement par r{\'e}duction diagonale le monop{\^o}le
de Kaluza-Klein de la supergravit{\'e} de type IIA compactifi{\'e}e sur
un cercle de rayon $R$, correspondant {\`a} une 5-brane de tension 
$R_{11} R^2/l_{11}^9=R^2/g^2 \alpha^{'4}$~; et par r{\'e}duction
verticale une 6-brane de tension $R^2/l_{11}^9=R^2 /g^3
\alpha^{'9/2}$, sur laquelle nous reviendrons dans le dernier chapitre.}. 
Il faut encore ajouter {\`a} ces objets {\'e}tendus les {\'e}tats issus de
l'onde gravitationnelle en dimension 11, correspondant au
supergraviton {\`a} dix dimensions ainsi qu'{\`a} ses excitations
\index{Kaluza-Klein!excitation de}
\index{D-brane!D0-branes}
de Kaluza-Klein, de masse $N/R_{11}=N/g\alpha^{'1/2}$, identifi{\'e}es
aux D0-branes de la th{\'e}orie de type IIA et {\`a} ses {\'e}tats li{\'e}s.

Dans le cas de la supergravit{\'e} de type IIB, cette proc{\'e}dure de r{\'e}duction
n'est bien entendu pas applicable et on doit reprendre 
l'approche pr{\'e}c{\'e}dente. On obtient ainsi une
D1-brane charg{\'e}e sous $\mathcal{B}_{\mu\nu}$ ; une
D3-brane charg{\'e}e sous le tenseur de jauge {\`a} courbure autoduale
$\mathcal{D}_{\mu\nu\rho\sigma}$ ; une D5-brane magn{\'e}tiquement
charg{\'e} sous $\mathcal{B}$, ainsi qu'une autre NS5-brane
charg{\'e}e sous le tenseur de jauge de Neveu-Schwarz $B_{\mu\nu}$.
La 1-brane solitonique est un candidat de corde duale {\`a}
la corde fondamentale sous la dualit{\'e}
$Sl(2,\Zint)$ de la th{\'e}orie de type IIB si cette dualit{\'e}
doit exister, et les deux 5-branes en formeraient alors
un doublet. On obtient {\'e}galement un cas d{\'e}g{\'e}n{\'e}r{\'e}
de solution avec un volume d'univers r{\'e}duit {\`a} un point,
soit une D$(-1)$-brane
\index{D-instanton!de type IIB}. L'espace correspondant {\'e}tant
de signature euclidienne, cette solution doit {\^e}tre identifi{\'e}e {\`a}
une {\it configuration instantonique} de la supergravit{\'e}
de type IIB. Cet objet sera essentiel aux d{\'e}veloppements 
trait{\'e}s au chapitre 4.

La th{\'e}orie de type I quant {\`a} elle comprend un tenseur de
jauge $\mathcal{B}_{\mu\nu}$ du secteur de Ramond en m{\^e}me
temps que les bosons de jauge de la sym{\'e}trie non ab{\'e}lienne
$SO(32)$. De fait, la supergravit{\'e} de type I montre l'existence
d'un soliton de D1-brane charg{\'e} {\'e}lectriquement sous
$\mathcal{B}_{\mu\nu}$, qu'il est tentant d'identifier {\`a} la
corde h{\'e}t{\'e}rotique de groupe de jauge $SO(32)$~;
ainsi que de trois types de solitons de 5-brane
\cite{Callan:1991dj}~: la 5-brane {\it neutre} est charg{\'e}e
sous le tenseur $B_{\mu\nu}$ de la corde h{\'e}t{\'e}rotique
ou le tenseur $\mathcal{B}_{\mu\nu}$ de la corde de type I
mais neutre sous le groupe de jauge, tandis que les 5-branes
{\it de jauge} et {\it sym{\'e}triques}, construites {\`a}
partir d'un instanton de Yang-Mills dans les directions transverses, 
sont charg{\'e}es {\`a} la fois sous $B$ et sous le groupe de jauge.
La 5-brane sym{\'e}trique pr{\'e}sente la particularit{\'e} d'{\^e}tre une
solution {\it exacte en $\alpha'$} des {\'e}quations du mouvement issues
de la th{\'e}orie des cordes, correspondant {\`a} une th{\'e}orie $(4,4)$
superconforme.

Par opposition aux particules ponctuelles, ces objets {\'e}tendus
poss{\`e}dent des degr{\'e}s de libert{\'e} externes correspondant
aux modes z{\'e}ro des champs de la supergravit{\'e} en leur pr{\'e}sence.
Ces modes sont associ{\'e}s aux sym{\'e}tries bris{\'e}es par la
pr{\'e}sence du soliton, en particulier la sym{\'e}trie de translation
dans les directions transverses, et correspondent {\`a} des 
\index{coordonn{\'e}e collective}
{\it coordonn{\'e}es collectives} g{\'e}n{\'e}ralis{\'e}es de la
$p$-brane \cite{Gervais:1975yg,Callan:1991ky}. Les excitations de ces modes 
modifient la position de la $p$-brane ainsi que sa forme, et
la $p$-brane doit bien plus {\^e}tre comprise comme un objet
compact {\'e}tendu que comme un plan rigide infini. 
La th{\'e}orie des champs (ou la m{\'e}canique quantique, pour $p=0$)
d{\'e}finissant la dynamique de
ces modes sur le volume d'univers peut en principe {\^e}tre
obtenue en {\'e}valuant l'action microscopique sur la
configuration des champs du soliton. Les supersym{\'e}tries
pr{\'e}serv{\'e}es par le soliton BPS correspondent {\`a} autant de
supersym{\'e}tries de la th{\'e}orie de volume d'univers, tandis
que les supersym{\'e}tries bris{\'e}es sont r{\'e}alis{\'e}es
non-lin{\'e}airement, c'est-{\`a}-dire {\it spontan{\'e}ment
bris{\'e}es}. Les premi{\`e}res peuvent {\^e}tre rendues 
\index{supersym{\'e}trie!sur le volume d'univers des branes}
manifestes gr{\^a}ce {\`a} l'introduction d'une supersym{\'e}trie
locale dite {\it kappa-sym{\'e}trie}, dont la fixation impose
l'{\'e}galit{\'e} des nombres de degr{\'e}s de libert{\'e}
\index{kappa-sym{\'e}trie}
\index{modes z{\'e}ros!des D-branes}
bosoniques et fermioniques. La d{\'e}termination
de la dynamique des modes z{\'e}ros d'un soliton est un probl{\`e}me
difficile~; Polchinski en a donn{\'e} la solution dans le cadre de la
th{\'e}orie des supercordes ouvertes, tout au moins pour les $p$-branes
charg{\'e}es sous les tenseurs de Ramond de la corde de type II.
\index{cinq-brane!h{\'e}t{\'e}rotique}
Dans le cas de la 5-brane h{\'e}t{\'e}rotique, certains modes
z{\'e}ros apparaissent non perturbativement et la th{\'e}orie
de volume d'univers est d{\'e}crite par une th{\'e}orie des cordes,
et non plus une th{\'e}orie des champs \cite{
Callan:1991dj,witten:1996,Seiberg:1997td}.

\subsection{Compactification et cycles supersym{\'e}triques
\label{susycyc}}
\index{cycle d'homologie!supersym{\'e}trique}
Etant donn{\'e} le spectre des configurations solitoniques en
dimension maximale, on peut en principe en d{\'e}duire le
spectre en dimension inf{\'e}rieure en {\it enroulant}
\index{enroulement!d'un soliton sur un cycle}
les directions internes des solitons autour des directions
compactes. Les r{\'e}ductions diagonale et verticale
que nous avons discut{\'e}es au paragraphe pr{\'e}c{\'e}dent fournissent 
l'exemple le plus simple de ce processus pour le cas
des compactifications toro{\"\i}{}dales, et n'affectent pas
le nombre de supersym{\'e}tries pr{\'e}serv{\'e}es par 
les {\'e}tats correspondants. Dans le cas de la compactification 
\index{Calabi-Yau, vari{\'e}t{\'e} de}
\index{cycle d'homologie!des espaces de Calabi-Yau}
sur des espaces courbes de Calabi-Yau, l'enroulement de l'{\'e}tat
sur un cycle quelconque brise toutes les supersym{\'e}tries, et
la stabilit{\'e} de l'{\'e}tat correspondant n'est pas garantie.
Pour certains cycles cependant, dits {\it supersym{\'e}triques},
la supersym{\'e}trie bris{\'e}e par la pr{\'e}sence du soliton
peut {\^e}tre compens{\'e}e par une {\it kappa-sym{\'e}trie}
sur le volume d'univers du soliton. Les conditions sur les
trois-cycles des espaces de Calabi-Yau de dimension 6 
ont {\'e}t{\'e} analys{\'e}es par K. et M. Becker et
A. Strominger \cite{Becker:1995kb}~; cette analyse a {\'e}t{\'e}
{\'e}tendue par Ooguri {\it et al.} par un argument bas{\'e} sur
la description microscopique des solitons en termes de D-branes,
et conduit {\`a} deux cat{\'e}gories de cycles
supersym{\'e}triques $C$ sur les espaces de Calabi-Yau $K$ de 
dimension complexe  $n$:
\cite{Ooguri:1996ck}~:
\begin{itemize}
\item les cycles de type A, de dimension {\it r{\'e}elle} $n$, 
sont tels que {\it le pull-back de la $n$-forme holomorphe 
$\Omega$ est proportionnel {\`a} la forme volume induite 
par la m{\'e}trique de $K$ sur le cycle $C$.}
\item les cycles de type B sont les {\it sous-vari{\'e}t{\'e}s
holomorphes} de la vari{\'e}t{\'e} $K$.
\end{itemize}
Les cycles supersym{\'e}triques sont donc en particulier de volume minimal
\footnote{En pr{\'e}sence de champs de jauge, cette d{\'e}finition doit
{\^e}tre {\'e}tendue, et c'est l'{\it action de Born-Infeld} du soliton qui doit {\^e}tre
extr{\'e}male.}.
Ainsi, dans le cas du tore $T^2$, les cycles de type B sont simplement
les points, $T^2$ et ses recouvrements, tandis que les cycles de type
A sont tels que $dz = e^{i\phi} \sqrt{|dz|^2}$ o{\`u} la
phase $\phi$ est constante~: il s'agit donc d'une ligne droite 
trac{\'e}e sur le tore. Dans le cas des vari{\'e}t{\'e}s de 
Calabi-Yau de dimension $4$, ou vari{\'e}t{\'e}s $K_3$,
les cycles de type A et B se confondent, gr{\^a}ce {\`a} la sym{\'e}trie
$SU(2)_R$ de la vari{\'e}t{\'e} hyperk{\"a}hlerienne qui {\'e}change
\index{vari{\'e}t{\'e}!hyperk{\"a}hlerienne}
\index{hyperk{\"a}hlerienne, vari{\'e}t{\'e}}
la deux-forme holomorphe $\Omega$ et la forme de K{\"a}hler $J$.



%% file: chap3.tex
\chapter{Pluralit{\'e}, dualit{\'e} et unicit{\'e} des th{\'e}ories de supercordes}
Le lecteur aura sans doute {\'e}t{\'e} g{\^e}n{\'e} par la discussion des
dualit{\'e}s non perturbatives
des th{\'e}ories de supergravit{\'e} au chapitre pr{\'e}c{\'e}dent,
alors que ces th{\'e}ories fonci{\`e}rement non renormalisables ne sont
\index{renormalisation!renormalisabilit{\'e} de la gravitation}
d{\'e}j{\`a} pas d{\'e}finies comme th{\'e}ories de perturbation. La
renormalisabilit{\'e} d'une th{\'e}orie des champs 
n'est cependant un crit{\`e}re important que lorsque cette th{\'e}orie
pr{\'e}tend {\`a} une description fondamentale {\`a} toute {\'e}nergie ;
et encore la non-renormalisabilit{\'e} au sens du comptage de
puissance n'implique-t-elle pas l'absence de sch{\'e}ma de renormalisation 
\index{action effective!renormalisabilit{\'e}}
\index{supergravit{\'e}!r{\'e}gularisation ultraviolette}
coh{\'e}rent. Les th{\'e}ories de supergravit{\'e} ne pr{\'e}tendent cependant
qu'{\`a} une description {\it effective} de la dynamique {\`a} basse {\'e}nergie
induite apr{\`e}s {\it int{\'e}gration sur les modes massifs} de fr{\'e}quence
sup{\'e}rieure {\`a} une {\'e}chelle $\Lambda$. Il n'est donc
pas question d'int{\'e}grer les diagrammes de boucles 
sur des moments infiniment grands, mais seulement 
sur les moments inf{\'e}rieurs {\`a} $\Lambda$.
La description {\it microscopique} est fournie par les th{\'e}ories de 
supercordes, qui contrairement {\`a} ces th{\'e}ories des champs
ne souffrent pas de divergences ultraviolettes. Les
contraintes de coh{\'e}rence restreignent ces th{\'e}ories
{\`a} cinq mod{\`e}les d{\'e}finis en dix dimensions, donnant
naissance {\`a} une vari{\'e}t{\'e} presque infinie de mod{\`e}les
en dimensions inf{\'e}rieures. Il existe de nombreux
ouvrages d'introduction {\`a} la th{\'e}orie des cordes
\cite{Green:1987sp,Green:1987mn,Lust:1989tj,Polchinski:1994,
Ooguri:1996ik,Kiritsis:1997hj,Vafa:1997pm}, 
dont le tr{\`e}s complet
livre vert de Green, Schwarz et Witten~; fid{\`e}les {\`a} notre
approche, nous nous contenterons d'en donner une introduction tr{\`e}s
condens{\'e}e dans les sections 1 et 2 de ce chapitre,
en insistant sur les points n{\'e}cessaires {\`a} la
compr{\'e}hension des travaux pr{\'e}sent{\'e}s dans cette th{\`e}se.

Une fois r{\'e}solu le probl{\`e}me de la d{\'e}finition perturbative des
th{\'e}ories de supergravit{\'e}, restera la question beaucoup plus
{\'e}pineuse de la d{\'e}finition non perturbative des th{\'e}ories
de supercordes. Cette d{\'e}finition n'existe pas encore explicitement.
Les relations de dualit{\'e} entre th{\'e}ories de
supercordes en donnent
une d{\'e}finition partielle, en identifiant le r{\'e}gime
de fort couplage d'une th{\'e}orie de supercordes au r{\'e}gime de faible
couplage de la th{\'e}orie duale. Nous d{\'e}crirons avec plus de d{\'e}tails
dans la section 3 de ce chapitre, apr{\`e}s en avoir donn{\'e} un avant-go{\^u}t au 
chapitre pr{\'e}c{\'e}dent dans le cadre des th{\'e}ories de supergravit{\'e}.
Ces dualit{\'e}s indiquent l'existence
d'une {\it th{\'e}orie non perturbative des supercordes}, dont
les cinq th{\'e}ories de supercordes perturbatives repr{\'e}sentent les
d{\'e}veloppement limit{\'e}s dans cinq limites de son espace 
des modules~; cette th{\'e}orie fondamentale fera l'objet des deux
derniers chapitres de ce m{\'e}moire.

Dans ce travail de th{\`e}se, nous nous sommes
particuli{\`e}rement int{\'e}ress{\'e}s {\`a} la dualit{\'e} des
th{\'e}ories h{\'e}t{\'e}rotique et de type II {\`a} seize charges 
supersym{\'e}triques, qui contient la richesse des effets 
non perturbatifs des th{\'e}ories des
supercordes sans requ{\'e}rir l'arsenal de la g{\'e}om{\'e}trie alg{\'e}brique des
espaces de Calabi-Yau. Nous avons en particulier donn{\'e} des
tests de cette dualit{\'e} dans les couplages gravitationnels
des compactifications {\it twist{\'e}es} de ces th{\'e}ories,
en collaboration avec A. Gregori, N. Obers, E. Kiritsis,
C. Kounnas et M. Petropoulos~; nous avons {\'e}galement 
v{\'e}rifi{\'e} cette dualit{\'e} dans les couplages scalaires
du dilaton de la th{\'e}orie de type II, au cours d'une
collaboration avec I. Antoniadis et T. Taylor. 
Ce chapitre constitue une introduction {\`a} ces contributions,
rassembl{\'e}es dans les appendices \ref{tt} et \ref{dds} de ce m{\'e}moire
Nous reviendrons sur ces travaux dans le chapitre 4, 
sous l'aspect des effets instantoniques de membrane.

\section{La corde bosonique}
\index{bosonique, th{\'e}orie des cordes}
L'id{\'e}e de r{\'e}soudre le probl{\`e}me des divergences ultraviolettes
des th{\'e}ories des champs en supposant une extension finie aux
particules {\'e}l{\'e}mentaires est aussi vieille que les th{\'e}ories
des champs elle-m{\^e}mes. Elle a cependant resurgi avec 
vigueur au d{\'e}but des ann{\'e}es 1970, apr{\`e}s la gestation
\index{duaux, mod{\`e}les}
des mod{\`e}les duaux, sous la forme de la th{\'e}orie
des cordes. Cette th{\'e}orie consid{\`e}re la dynamique relativiste
\index{corde relativiste}
d'un objet {\it unidimensionnel} filiforme. Tout comme une particule
ponctuelle de masse $m$ 
\index{ligne d'univers}
\index{particule relativiste}
d{\'e}crit une ligne d'univers de genre temps $X^{\mu}(\tau)$ dont l'action est
mesur{\'e}e par la longueur propre 
\begin{equation}
S\left(\{X^\mu(\tau)\} \right) =m \int d\tau 
\sqrt{ - \frac{dX^{\mu}}{d\tau}\frac{dX^{\nu}}{d\tau}
\eta_{\mu\nu}(X) }
\ ,
\end{equation}
la corde d{\'e}crit une surface d'univers $X^{\mu}(\sigma,\tau)$ dont
l'action est mesur{\'e}e par la surface propre
\index{Nambu-Goto, action de}
\begin{equation}
\label{ng}
S\left(\{X^\mu(\sigma,\tau)\} \right) =
T \int d\sigma^0 \wedge d\sigma^1 \sqrt{ - \epsilon_{\alpha\beta}
\frac{\partial X^{\mu}}{\partial\sigma_\alpha}
\frac{\partial X^{\nu}}{\partial\sigma_\beta}
\eta_{\mu\nu}(X) }
\ ,
\end{equation}
o{\`u} on a rassembl{\'e} le temps propre $\tau$ et l'abscisse curviligne
$\sigma$ sous la notation $\sigma^0=\tau,\sigma^1=\sigma$. Le param{\`e}tre
$T$, homog{\`e}ne {\`a} l'inverse d'une aire et souvent not{\'e}
$T=1/(2\pi\alpha')$, correspond {\`a} la {\it tension} de la corde.
\index{tension!de la corde fondamentale}
Tout comme la dynamique de la particule peut {\^e}tre quantifi{\'e}e {{\`a} la Feynman}
en sommant sur les lignes d'univers {\`a} extr{\'e}mit{\'e}s fix{\'e}es, la
corde relativiste quantique est obtenue en {\it sommant sur les
surfaces d'univers} tubulaires {\`a} bords fix{\'e}s
\cite{Gervais:1971}.
L'action (\ref{ng}), dite de Nambu-Goto,
est classiquement {\'e}quivalente {\`a} l'action de Polyakov
\cite{Brink:1976sc,Deser:1976rb}
\index{Polyakov, action de}
\begin{equation}
\label{polyakov}
S\left(\{X^\mu(\sigma^\gamma),g_{\alpha\beta}(\sigma^\gamma)\} \right) =
\frac{T}{2} \int d\sigma^0 \wedge d\sigma^1~g^{\alpha\beta}\eta_{\mu\nu} 
\partial_\alpha X^{\mu}
\partial_\beta  X^{\nu}
\ ,
\end{equation}
o{\`u} on introduit une {\it m{\'e}trique auxiliaire} $g_{\alpha\beta}$
fluctuante sur la surface d'univers~; les {\'e}quations du mouvement
identifient cette m{\'e}trique 
{\`a} la {\it m{\'e}trique induite} par le plongement $X^{\mu}$
dans l'espace-temps ambiant, {\`a} un {\it facteur conforme} non fix{\'e}
pr{\`e}s, en raison de la {\it sym{\'e}trie de Weyl}
\index{Weyl!sym{\'e}trie de}
\begin{equation}
g_{\alpha\beta}\rightarrow e^{f(\sigma,\tau)} g_{\alpha\beta}
\end{equation}
de l'action de Polyakov (\ref{polyakov}), 
propre au caract{\`e}re bidimensionnel de la surface d'univers.
L'invariance sous les diff{\'e}omorphismes de la surface
d'univers permet (localement) de ramener la
m{\'e}trique {\`a} la forme diagonale 
$g_{\alpha\beta}(\sigma)=\rho(\sigma) \eta_{\alpha\beta}$,
\index{Liouville, champ de}
Le facteur conforme $\rho$ dit {\it champ de Liouville}, se d{\'e}couple
gr{\^a}ce {\`a} la sym{\'e}trie de Weyl, et 
on se ram{\`e}ne ainsi {\`a} l'action {\it gaussienne}
\begin{equation}
\label{sgaussien}
S\left(\{X^\mu(\sigma)\} \right) =
\frac{T}{2} \int d\sigma^0 \wedge d\sigma^1 ~\eta_{\mu\nu}
\partial_\alpha X^{\mu}
\partial^\alpha  X^{\nu} \ ,
\end{equation}
La sym{\'e}trie de Weyl pr{\'e}sente cependant une {\it anomalie quantique},
\index{anomalie!conforme}
et Polyakov a pu prouver que cette anomalie s'annulait uniquement
en dimension 26, {\it dimension critique} de la th{\'e}orie des
\index{critique, dimension}
cordes bosoniques \cite{Polyakov:1981rd}. 

\subsection{Tachyon, spectre de masse nulle et {\'e}tats massifs}
Dans la jauge (\ref{sgaussien}), les 26 coordonn{\'e}es de plongement
$X^{\mu}$ se d{\'e}couplent, et v{\'e}rifient l'{\'e}quation des ondes
bidimensionnelle $\partial_\alpha\partial^\alpha X^\mu=0$~; chaque
coordonn{\'e}e
s'{\'e}crit donc comme somme de composantes droite et gauche,
\begin{equation}
X^{\mu}(\tau,\sigma)=X^{\mu}_L(\tau+\sigma) + X^{\mu}_R(\tau-\sigma)
\end{equation}
elles-m{\^e}mes d{\'e}coupl{\'e}es dans le cas de la corde ferm{\'e}e, auquel
nous nous restreignons ici
\footnote{Dans le cas de la corde ouverte, ces deux composantes
sont identifi{\'e}es par la condition de Neumann
\index{Neumann, condition de}
$\partial_{\sigma}X^{\mu}=0$ aux bords $\sigma=0,\pi$.}.
Il est commode d'effectuer une rotation de Wick sur la surface
d'univers, de sorte que les composantes gauches et droites
s'identifient aux fonctions {\it holomorphes} $X_L(z)$ et 
{\it anti-holomorphes} $X_R(\bar z)$ de la coordonn{\'e}e $z=\sigma+i\tau$.
Les excitations quantiques de la corde ferm{\'e}e sont donc
d{\'e}crites par deux s{\'e}ries d'oscillateurs $\alpha^\mu_n,\bar\alpha^\mu_n$,
\index{oscillateur}
correspondant aux modes de Fourier de $X_L^\mu(z)$ 
et $X_R^\mu(\bar z)$, v{\'e}rifiant les relations
\begin{align}
[\alpha^\mu_n,\alpha^\nu_m]&=[\bar\alpha^\mu_n,\bar\alpha^\nu_m]
=im\delta_{m+n}\eta^{\mu\nu}\\
\alpha^{\mu\dagger}_n&=\alpha^\mu_{-n}\ ,\quad
\bar\alpha^{\mu\dagger}_n=\bar\alpha^{\mu}_{-n}\ ,\quad
\end{align}
et agissant sur le vide de l'espace de Fock  $|p\rangle$ d'impulsion $p_\mu$.
Le mode z{\'e}ro $\alpha^{\mu}_0=\alpha_0^{\mu\dagger}=p^\mu/2$
correspond 
{\`a} l'impulsion totale transverse de la corde. 

L'invariance de jauge sous les diff{\'e}omorphismes de surface d'univers
impose l'annulation du {\it tenseur d'{\'e}nergie-impulsion} sur
la surface d'univers, de modes de Fourier
\begin{eqnarray}
L_0 &= -\frac{\alpha'}{4} p^2 + 
\sum_{n=1}^{\infty} \alpha_{-n}\alpha_{n} -a\ ,\quad
L_n &= \frac{1}{2}\sum_{m=-\infty}^{\infty} \alpha_{m-n}\alpha_{n}  \\
\bar L_0 &= -\frac{\alpha'}{4}p^2 + 
\sum_{n=1}^{\infty} \bar\alpha_{-n}\bar\alpha_{n} -\bar a\ ,\quad
\bar L_n &= \frac{1}{2}\sum_{m=-\infty}^{\infty} \bar\alpha_{m-n}
\bar\alpha_{n} 
\end{eqnarray}
o{\`u} les constantes  $a$ et $\bar a$, dites {\it intercepts},
\index{intercept|textit}
\index{fant{\^o}me}
sont fix{\'e}es {\`a} la valeur $a=\bar a=1$ par l'absence de
fant{\^o}mes. En particulier, l'annulation de $L_0$ et 
$\bar L_0$ implique la {\it formule de masse} et la
condition de {\it level matching}~:
\index{level matching, condition de}
\index{masse, formule de!de la corde bosonique}
\begin{align}
\label{levmat}
\mathcal{M}^2 &= \frac{2}{\alpha'} \left( N_L + N_R -2\right) \\
N_L &= N_R
\end{align}
o{\`u} $N_L=\sum_{n=1}^{\infty} \alpha_{-n}\alpha_{n}$
et $N_R=\sum_{n=1}^{\infty} \bar\alpha_{-n}\bar\alpha_{n}$
d{\'e}signent les nombres d'oscillateurs gauche et droit.

En particulier, l'{\'e}tat fondamental $N_L=N_R=0$
de l'espace de Fock correspond
{\`a} une particule de masse imaginaire dite {\it tachyon},
\index{tachyon}
r{\'e}v{\'e}latrice de l'instabilit{\'e} de la th{\'e}orie.
Le premier {\'e}tat excit{\'e} $N_L=N_R=1$, soit 
$\zeta_{\mu\nu}~\alpha^\mu_{-1}\bar\alpha^\nu_{-1}|p\rangle$, 
correspond {\`a} des particules de masse nulle, dont le
spin d{\'e}pend du choix du tenseur de polarisation $\zeta_{\mu\nu}$:
\begin{itemize}
\item $\zeta_{\mu\nu}=\zeta\eta_{\mu\nu}$ correspond au champ
scalaire $\phi$ du {\it dilaton} ;
\index{dilaton|textit}
\item $\zeta_{\mu\nu}$ antisym{\'e}trique correspond au tenseur
antisym{\'e}trique de jauge $B_{\mu\nu}$
\index{tenseur antisym{\'e}trique|textit}
\item $\zeta_{\mu\nu}$ sym{\'e}trique de trace nulle correspond au graviton
$g_{\mu\nu}$.
\index{graviton|textit}
\end{itemize}
Les {\'e}tats plus excit{\'e}s forment une tour d'{\'e}tats supermassifs
\index{supermassifs, {\'e}tats}
de masse $\mathcal{M}\sim 1/\alpha'$ de l'ordre de la masse de Planck,
et de spin arbitrairement {\'e}lev{\'e}.
La propagation de la corde dans les champs de fonds $g_{\mu\nu},B_{\mu\nu}(x)$
et $\phi(x)$ peut {\^e}tre d{\'e}crite en g{\'e}n{\'e}ralisant l'action
(\ref{polyakov}) {\`a}
\begin{equation}
\label{sigmod}
S=
\int d^2\sigma \sqrt{g}
\left( \frac{1}{4\pi\alpha'} g^{\alpha\beta}
\partial_\alpha X^{\mu}\partial_\beta  X^{\nu}
g_{\mu\nu}(X) + \frac{i}{2\pi} \epsilon^{\alpha\beta}
B_{\mu\nu}(X) \partial_\alpha X^{\mu}\partial_\beta  X^{\nu}
- \frac{1}{4\pi} \phi(X) R \right)
\end{equation}
Cette action n'est plus gaussienne que dans l'approximation
o{\`u} la taille de la corde $\sqrt{\alpha'}$ est tr{\`e}s
inf{\'e}rieure au rayon  de courbure caract{\'e}ristique de
l'espace-temps. L'{\it invariance conforme} de la th{\'e}orie des champs
\index{conforme, th{\'e}orie des champs!sur la surface d'univers}
sur la surface d'univers\footnote{On pourra se 
reporter {\`a} l'excellent cours \cite{Ginsparg:1988ui}
pour une introduction aux th{\'e}ories des champs conformes
bidimensionnelles.}, n{\'e}cessaire au d{\'e}couplage 
du champ de Liouville
\index{Liouville, champ de}
impose alors une dynamique sur les champs de fond de masse nulle, 
donn{\'e}e par l'annulation des fonctions beta
\index{beta, fonction!sur la surface d'univers}
relatives aux couplages $g_{\mu\nu}$, $B_{\mu\nu}$ et $\phi$~; celle-ci 
peut se r{\'e}sumer au premier ordre en $\alpha'$ en termes
de l'{\it action effective}
\begin{equation}
S=\frac{1}{\alpha^{'12}}\int d^{26}x \sqrt{-g} e^{-2\phi}
\left(R+4(\partial\phi)^2 - \frac{1}{2\cdot 3!} H^2 \right)
\end{equation}
o{\`u} la trois-forme $H=dB$ d{\'e}note la courbure du tenseur de jauge
$B_{\mu\nu}$. La gravitation d'Einstein-Hilbert se trouve
\index{Einstein-Hilbert, action d'}
ainsi incluse et g{\'e}n{\'e}ralis{\'e}e dans la th{\'e}orie des
cordes.

\subsection{Interactions, dilaton et s{\'e}rie de perturbation}
Les th{\'e}ories des champs associent une constante de couplage
{\`a} chaque type d'interaction selon la nature des particules
incidentes. La th{\'e}orie des cordes au contraire unifie tous
les types de particules comme diff{\'e}rentes excitations d'un 
m{\^e}me objet. Il n'existe alors plus qu'un seul type d'interaction
correspondant {\`a} l'ouverture de deux cordes ferm{\'e}es et leur 
recollement en une seule (interaction de {\it splitting and
joining}), soit l'ouverture d'un {\it manche} dans la surface
\index{splitting and joining}
d'univers. Par transformation conforme, une surface correspondant
{\`a} une interaction {\it {\`a} l'ordre des arbres} de plusieurs
\index{arbres, interaction {\`a} l'ordre des}
{\'e}tats asymptotiques peut {\^e}tre d{\'e}form{\'e}e en une 
{\it sph{\`e}re} ponctu{\'e}e d'autant de pattes ext{\'e}rieures
(figure \ref{arbres})\footnote{On pourra se reporter aux
r{\'e}f{\'e}rences \cite{Friedan:1986ge,D'Hoker:1988ta} 
pour un expos{\'e} d{\'e}taill{\'e}
des m{\'e}thodes de calcul d'amplitudes de diffusion en th{\'e}orie
des cordes~; on trouvera {\'e}galement {\`a} la fin de l'article
en annexe \ref{tt} un exemple d{\'e}taill{\'e} de calcul de diffusion
{\`a} une boucle.}.L'effet de chaque 
particule est incorpor{\'e} par l'insertion
d'un {\it op{\'e}rateur de vertex} local dans la th{\'e}orie
\index{op{\'e}rateur de vertex}
conforme bidimensionnelle, d{\'e}pendant de l'{\'e}tat d'oscillation
interne, de la polarisation et de l'impulsion de la particule.
Le tachyon et le graviton sont ainsi d{\'e}crits par les
insertions
\begin{equation}
\int d^2\sigma \sqrt{g} :e^{ik^\rho X_\rho(\sigma)}:
\sp
\int d^2\sigma \sqrt{g}~ \zeta_{\mu\nu} 
g^{\alpha\beta} :\partial_\alpha X^\mu\partial_\beta X^\nu
e^{ik^\rho X_\rho(\sigma)}: \ .
\end{equation}
La fonction de corr{\'e}lation de la th{\'e}orie conforme sur
la sph{\`e}re en pr{\'e}sence de ces insertions fournit
ainsi l'{\'e}l{\'e}ment de matrice S de diffusion entre ces 
particules.
\fig{3cm}{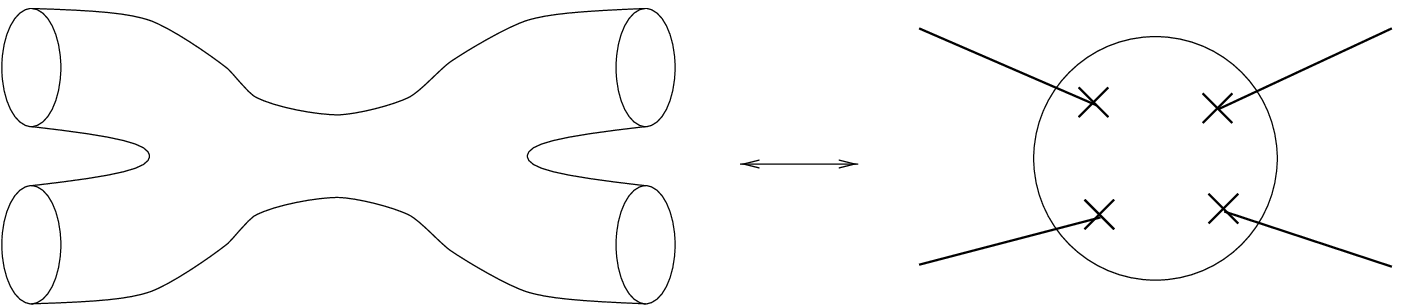}{Amplitude de diffusion {\`a} l'ordre des
arbres et repr{\'e}sentation par op{\'e}rateurs de vertex.}{arbres}

Les diagrammes {\`a} $n$ boucles conduisent {\`a} des
corr{\'e}lations d'op{\'e}rateurs de vertex sur des surfaces
de Riemann de genre $n$. Il ne suffit cependant pas d'int{\'e}grer
sur les champs $X^\mu$ de la th{\'e}orie conforme, mais il
faut aussi sommer sur les diff{\'e}rentes m{\'e}triques
$g_{\alpha\beta}$ sur la surface d'univers, ou plus exactement
sur les classes d'{\'e}quivalence de m{\'e}triques {\it modulo}
les diff{\'e}omorphismes et les transformations de Weyl. 
\index{Weyl!dilatation de}
Ces classes d'{\'e}quivalence sont param{\'e}tris{\'e}es par
\index{espace des modules!des surfaces de Riemann}
\index{surface de Riemann!espace des modules}
{\it un espace des modules}\footnote{
Il ne faut pas confondre cet espace des modules 
avec l'espace des modules
d{\'e}crivant les vides des th{\'e}ories de champs ou de cordes.
Cet espace correspond au quotient de l'{\it espace de Teichm{\"u}ller},
d{\'e}crivant les m{\'e}triques modulo les transformations de Weyl
et les diff{\'e}omorphismes connect{\'e}s {\`a} l'identit{\'e},
par le {\it groupe modulaire} d{\'e}crivant les classes de diff{\'e}omorphismes
\index{groupe! modulaire des surfaces de Riemann}
non connect{\'e}s {\`a} l'identit{\'e}.}
de dimension finie $6n-6$ (ou 0 pour $n=0$,et 2 pour $n=1$),
sur lequel on doit encore int{\'e}grer. Ainsi, dans le cas 
d'un diagramme {\`a} une boucle, la m{\'e}trique est d{\'e}finie aux
diff{\'e}omorphismes et transformations de Weyl pr{\`e}s par le 
param{\`e}tre de structure complexe $\tau$, tel que 
$ds^2\equiv |d\sigma^0 + \tau d\sigma^1|^2$. $\tau$ prend ses
valeurs dans {\it le demi-plan de Poincar{\'e} modulo le groupe
\index{Poincar{\'e}, demi-plan de}
modulaire $Sl(2,\Zint)$}. Ce quotient peut {\^e}tre repr{\'e}sent{\'e}
par le {\it domaine fondamental}
\index{domaine fondamental modulaire}
\begin{equation}
\mathcal{F}=\{ \tau\in\CC /\quad 
\tau_2 >0,\ |\tau|\ge 1 \mbox{ et } -\frac{1}{2}\le \tau_1 \le 
\frac{1}{2}\} \ .
\end{equation}
La partie imaginaire de $\tau_2$ peut {\^e}tre vue comme le {\it param{\`e}tre
de Schwinger} de la th{\'e}orie des champs ordinaire, tandis que
\index{Schwinger, param{\`e}tre de}
l'int{\'e}grale sur $\tau_1 \in [-1/2,1/2]$ impose la condition de
{\it level matching} (\ref{levmat}). En particulier, 
\index{level matching, condition de}
la r{\'e}gion de faible param{\`e}tre de Schwinger
$\tau_2 \rightarrow 0$,
correspondant {\`a} la r{\'e}gion ultraviolette, est exclue de ce
domaine (ou plus exactement {\'e}quivalente {\`a} la r{\'e}gion $\tau_2 
\rightarrow +\infty$ par transformation modulaire), ce qui explique
l'absence de divergences ultraviolettes en th{\'e}orie des cordes.
\index{divergence ultraviolette!absence en th{\'e}orie des cordes}
L'amplitude de transition vide-vide (ou, en d'autres termes,
la constante cosmologique)
\index{cosmologique, constante}
{\`a} une boucle s'{\'e}crit alors, pour la th{\'e}orie bosonique
en espace plat,
\begin{equation}
\label{Zboso}
\mathcal{A}=\int_{\mathcal{F}} \frac{d^2\tau}{\tau_2^2} Z(\tau,\bar\tau)\ ,
\quad Z(\tau,\bar\tau)=(\tau_2)^{-24/2} 
\frac{1}{\eta^{24}(\tau)\bar\eta^{24}(\bar\tau)}
\end{equation}
o{\`u} les fonctions modulaires
\index{modulaire, forme}
\footnote{L'article en appendice \ref{tt} rassemble les
identit{\'e}s utiles sur les fonctions modulaires, et
fournit une introduction d{\'e}taill{\'e}e au calcul
d'amplitudes {\`a} une boucle en th{\'e}orie des cordes.
On se reportera {\'e}galement avec profit aux r{\'e}f{\'e}rences
\cite{Ginsparg:1988ui,Kiritsis:1997hj}.}
 $\eta^{24}(\tau)$ 
et $\bar\eta^{24}(\bar\tau)$ de Dedekind 
correspondent aux contributions des 24 oscillateurs gauches
et droits, tandis que le pr{\'e}facteur $\tau_2^{-D/2}
\sim\int d^{D-2}p$ $e^{-\pi p^2\tau_2}$ correspond
{\`a} l'int{\'e}gration 
sur l'impulsion transverse des {\'e}tats dans la boucle.
L'int{\'e}grand $Z(\tau,\bar\tau)$ correspond {\`a} la {\it fonction de
  partition}
\index{partition, fonction de!des th{\'e}ories de cordes}
des modes de la corde. Son {\it invariance sous les transformations
modulaires de $\tau$} est
un pr{\'e}requis fondamental pour la coh{\'e}rence de la th{\'e}orie
\footnote{La fonction $\eta(\tau)$ n'est en effet invariante
modulaire qu'{\`a} une racine 24-i{\`e}me de l'unit{\'e} pr{\`e}s.
On retrouve ainsi la dimension critique $D=26$ de la 
\index{critique, dimension}
th{\'e}orie des cordes bosoniques, bien que cet argument
n'{\'e}limine pas les autres possibilit{\'e}s $D=24k+2, k\in\Zint$.}.

Le calcul d'une amplitude physique doit donc prendre en
compte les surfaces d'univers de {\it genre} $n$ arbitraire
pour d{\'e}crire les diagrammes de Feynman de nombre de boucles
quelconque, ce qui lui vaut le nom de 
\index{genre, d{\'e}veloppement en}
\index{serie de perturbation@s{\'e}rie de perturbation!en th{\'e}orie des cordes}
\index{surface de Riemann!genre}
\index{Euler, caract{\'e}ristique d'}
{\it d{\'e}veloppement en genre}. Le genre d'une surface de Riemann, reli{\'e}
{\`a} la {\it caract{\'e}ristique d'Euler} $\chi=2-2n$, peut s'{\'e}crire 
comme l'int{\'e}grale sur la surface d'univers d'un terme local
topologique
\begin{equation}
\chi = 2-2n = \frac{1}{4\pi} \int d^2\sigma~\sqrt{g} R 
\end{equation}
qui n'est autre que l'op{\'e}rateur 
d{\'e}crivant le couplage
du dilaton {\`a} la surface d'univers dans l'{\'e}quation (\ref{sigmod}). 
{\it Les surfaces de genre $n$
sont donc pond{\'e}r{\'e}es par un facteur $e^{(2n-2)\phi}$}, o{\`u}
$\phi$ d{\'e}signe ici la valeur moyenne du champ de dilaton,
que nous pouvons donc identifier avec {\it le couplage 
de la th{\'e}orie des cordes} :
\index{dilaton}
\begin{equation}
g=e^{\phi}\ .
\end{equation}
Cette formulation de la th{\'e}orie des cordes appara{\^\i}t donc
{\it essentiellement perturbative}. Parler d'effets 
non perturbatifs ne saurait avoir de sens que dans le
cadre d'une th{\'e}orie plus large dont le d{\'e}veloppement
en genre, ou {\it d{\'e}veloppement de cordes}, ne repr{\'e}sente 
qu'un sch{\'e}ma d'approximation.

\section{Les th{\'e}ories de supercordes}
La th{\'e}orie des cordes bosoniques d{\'e}crite {\`a} grands traits 
dans la section pr{\'e}c{\'e}dente pr{\'e}sente de nombreuses insuffisances.
L'{\it existence du tachyon} montre qu'elle n'est pas d{\'e}finie 
\index{tachyon}
au voisinage de son vide stable, et la th{\'e}orie de perturbation
n'a donc pas de sens. L'{\it absence de particules fermioniques} est
regrettable pour une th{\'e}orie
visant {\`a} l'unification des forces, bien qu'il ne soit pas exclu
que ces particules apparaissent autour du point stable de 
cette th{\'e}orie. Les {\it th{\'e}ories des supercordes} pallient
{\`a} ces insuffisances en introduisant des champs fermioniques
sur la surface d'univers, de mani{\`e}re {\`a} {\'e}tendre la sym{\'e}trie
conforme en une sym{\'e}trie {\it superconforme}. 
\index{conforme, th{\'e}orie des champs!superconforme}
Cette sym{\'e}trie locale contient en particulier
la supersym{\'e}trie locale de la supergravit{\'e} bidimensionnelle.
La contribution {\`a} la charge centrale de l'alg{\`e}bre conforme
des fant{\^o}mes de Faddeev-Popov correspondants abaisse la
\index{fant{\^o}me}
dimension critique des th{\'e}ories de supercordes {\`a} $D=10$.
\index{critique, dimension}
Tandis que les champs
bosoniques, repr{\'e}sentant les coordonn{\'e}es physiques de la
corde, doivent {\^e}tre d{\'e}finis globalement sur la surface
d'univers, les champs fermioniques peuvent {\^e}tre soit
p{\'e}riodiques soit antip{\'e}riodiques {\it le long de la corde},
d{\'e}finissant ainsi les secteurs de {\it Ramond} (R)
et de {\it Neveu-Schwarz} (NS). L'invariance modulaire 
\index{Ramond, secteur de|textit}
\index{Neveu-Schwarz, secteur de|textit}
impose alors l'introduction de p{\'e}riodicit{\'e}s d{\'e}finies
{\it le long de l'axe du temps propre}
de la surface d'univers, projetant effectivement
la moiti{\'e} de chaque secteur. Plus g{\'e}n{\'e}ralement, sur une surface
de Riemann de genre $n$, on est amen{\'e} {\`a} sommer sur tous les
choix de {\it structure de spin}
\index{spin, structure de|textit} 
$s=\ar{a_1\dots a_n}{b_1\dots b_n}$, 
soit $2n$ signes $ (-)^{a_i}, (-)^{b_i}$
correspondant aux p{\'e}riodicit{\'e}s autour de chacun des cycles 
d'homologie $A_i,B_i$ (figure \ref{riemann} page \pageref{riemann}), 
\index{surface de Riemann!homologie des}
avec une phase $(-)^{\alpha(s)}$ d{\'e}termin{\'e}e
par les imp{\'e}ratifs d'invariance modulaire et de factorisation
\cite{Alvarez-Gaume:1986es,Seiberg:1986by}.
Cette projection, dite
GSO du nom de ses inventeurs Gliozzi,
\index{GSO, projection|textit}
Scherk et Olive \cite{Gliozzi:1976jf}, {\'e}limine en particulier du spectre
le tachyon, et garantit l'existence d'un nombre {\'e}gal
de particules bosoniques et fermioniques {\`a} chaque
niveau d'excitation. La {\it supersym{\'e}trie d'espace-temps}
\index{supersym{\'e}trie!du spectre des supercordes}
n'est pas manifeste dans cette construction~;
Green et Schwarz en ont donn{\'e} une formulation {\'e}quivalente
\index{Green et Schwarz, corde de}
dans le c{\^o}ne de lumi{\`e}re o{\`u} cette supersym{\'e}trie est explicite 
\index{front de lumi{\`e}re}
\cite{Green:1981yb}. Ce gain est cependant obtenu 
au d{\'e}triment de l'invariance de Lorentz {\`a} dix dimensions.

\subsection{Cordes ferm{\'e}es de type II\label{rns}}

En associant {\`a} chaque boson $X^{\mu}(\sigma,\tau)$ son partenaire
supersym{\'e}trique $\psi^\mu(\sigma,\tau)$ et en appliquant la projection GSO
s{\'e}par{\'e}ment dans chacun des secteurs gauche et droit, on obtient
ainsi les supercordes de type IIA et IIB, selon le
\index{supercordes, th{\'e}orie des!de type IIA|textit}
\index{supercordes, th{\'e}orie des!de type IIB|textit}
choix du signe relatif de la projection GSO entre les deux secteurs.
Leur d{\'e}finition est r{\'e}sum{\'e}e succintement dans la fonction
\index{partition, fonction de!des th{\'e}ories de cordes}
de partition\footnote{C'est plut{\^o}t d'un {\it indice}
de partition qu'il faudrait parler, car les fermions sont
compt{\'e}s avec un signe oppos{\'e} {\`a} celui des bosons.}
\begin{equation}
\label{ztypeii}
Z_{\; \II}(\tau,\bar\tau)= \frac{1 }{ \t_2^4}
\cdot
\frac{1}{\eta^{8}\bar\eta^{8}}
\cdot
\frac{1}{2}\sum_{a,b=0}^1
(-1)^{a+b+ab} \left( \frac{\th\ss}{\eta} \right)^{8/2}
\cdot
\frac{1}{2}\sum_{\bar{a}, \bar{b} =0}^1
(-1)^{\ba + \bb + \epsilon \ba \bb }
\left(\frac{\thb\bss}{\bar\eta}\right)^{8/2}
\end{equation}
avec $\epsilon=1$ en type IIB et $\epsilon=0$ en type IIA.
L'interpr{\'e}tation de cette expression est transparente 
si l'on choisit la jauge du c{\^o}ne de lumi{\`e}re, 
\index{lumi{\`e}re, front de!th{\'e}orie des cordes}
o{\`u} $\mu$ est effectivement
restreint aux huit coordonn{\'e}es transverses.
On reconna{\^\i}t alors respectivement l'int{\'e}grale sur les
impulsions transverses, la contribution des huit oscillateurs 
bosoniques transverses, et celle des huits fermions
gauches et droits de structure de spin $\ss$ et $\bss$.
La fonction de partition comprend {\it quatre secteurs} selon la valeur
de $(a,\ba)$, et la sommation sur $(b,\bb)$ impose
les projections GSO dans chaque secteur.
Le caract{\`e}re bosonique ou fermionique
d{\'e}pend de la phase $(-)^{a+b+ab+\ba+\bb+\epsilon\ba\bb}$ 
{\'e}valu{\'e}e en  $(b,\bb)=0$. 
Apr{\`e}s projection, 
l'{\'e}tat fondamental d'un complexe de huit fermions
de Neveu-Schwarz ($a=1$) correspond {\`a} un vecteur 
${\bf 8_v}$ sous le groupe de Lorentz transverse $SO(8,\Real)$,
tandis que l'{\'e}tat fondamental d'un complexe de huit fermions
de Ramond ($a=0$) correspond {\`a} un spineur 
${\bf 8_s}$ ou ${\bf 8_c}$ selon le choix du signe de la projection.
Le spectre dans chaque secteur $(a,\ba)$ peut ainsi {\^e}tre obtenu 
par produit tensoriel des repr{\'e}sentations gauche et droite:

\begin{itemize}
\item le secteur $(1,1)$, dit {\it de Neveu-Schwarz} (ou NS-NS),
comprend 64 degr{\'e}s de libert{\'e}s dans les repr{\'e}sentations
$\mathbf{8_v} \otimes \mathbf{8_v} = 
\mathbf{1} \oplus \mathbf{28} \oplus\mathbf{35_v}$
correspondant au dilaton $\phi$, graviton 
\index{dilaton}
$g_{\mu\nu}$ et tenseur antisym{\'e}trique $B_{\mu\nu}$
respectivement ;
\index{tenseur antisym{\'e}trique}

\item le contenu du secteur $(0,0)$, dit {\it de Ramond} (ou R-R)
d{\'e}pend du choix de la chiralit{\'e} relative des spineurs
gauche et droit. Dans le cas de la th{\'e}orie de type IIA, le
produit des deux spineurs de chiralit{\'e} oppos{\'e}e donne un
contenu non-chiral $\mathbf{8_s} \otimes \mathbf{8_c} = 
\mathbf{8_v} \oplus \mathbf{56_v}$ correspondant aux tenseurs
antisym{\'e}triques de jauge $\mathcal{A}_\mu$ et $\mathcal{C}_{\mu\nu\rho}$
\footnote{On pourrait aussi bien consid{\'e}rer des potentiels
de jauge duaux de Poincar{\'e} {\`a} $\mathcal{A}_1$ et $\mathcal{C}_3$,
qui poss{\`e}deraient les m{\^e}mes degr{\'e}s de libert{\'e} sur la couche
de masse.}.
Dans le cas de type IIB, on obtient au contraire un contenu chiral
$\mathbf{8_s} \otimes \mathbf{8_s} = 
\mathbf{1} \oplus \mathbf{28} \oplus \mathbf{35_c}$
repr{\'e}sentable en termes d'un champ scalaire $\axion$,
d'un tenseur antisym{\'e}trique $\mathcal{B}_{\mu\nu}$
et d'un tenseur {\`a} courbure autoduale $\mathcal{D}_{\mu\nu\rho\sigma}$.
Le statut des champs de Ramond est tr{\`e}s diff{\'e}rent de celui
des champs de Neveu-Schwarz. La corde fondamentale n'est pas charg{\'e}e
par rapport {\`a} ces champs. En particulier, il n'existe pas de
mod{\`e}le sigma non-lin{\'e}aire g{\'e}n{\'e}ralisant le lagrangien
(\ref{sigmod}) en pr{\'e}sence de champs de fond de Ramond.

\item les deux secteurs fermioniques $(0,1)$ et $(1,0)$ (ou R-NS et NS-R)
sont identiques ({\`a} un changement de chiralit{\'e} pr{\`e}s en type IIA),
et contiennent chacun un gravitino et un fermion, selon
$\mathbf{8_v} \otimes \mathbf{8_s} =  
\mathbf{8_c} \oplus \mathbf{56_c}$.
Ils constituent les partenaires supersym{\'e}triques des bosons
des secteurs de Neveu-Schwarz et Ramond.

\end{itemize}
On obtient ainsi des th{\'e}ories de supersym{\'e}trie $N=2$ 
{\`a} dix dimensions (soit 32 supercharges), chirale
(type IIB) ou non chirale (type IIA), dont les descriptions
de basse {\'e}nergie sont donn{\'e}es par les th{\'e}ories de
supergravit{\'e} d{\'e}crites au chapitre pr{\'e}c{\'e}dent. Les
propri{\'e}t{\'e}s de supersym{\'e}trie se manifestent dans
le calcul des amplitudes physiques par diff{\'e}rentes annulations.
En particulier, la fonction de partition (\ref{ztypeii})
est identiquement nulle par l'identit{\'e} sur les fonctions
th{\'e}ta
\begin{equation}
\label{riem}
\sum_{a,b=0}^1
(-1)^{a+b+ab} \th^4 \ss = 0
\end{equation}
\index{Riemann, identit{\'e} de}
qui traduit la compensation entre les degr{\'e}s de libert{\'e} bosonique
et fermionique {\`a} chaque niveau. La constante cosmologique {\`a} une boucle,
\index{cosmologique, constante}
{\'e}gale {\`a} l'int{\'e}grale modulaire de la fonction de partition,
est donc aussi nulle\footnote{Cette annulation ne semble pas {\^e}tre
une cons{\'e}quence de la supersym{\'e}trie d'espace-temps, puisqu'il
existe des th{\'e}ories de supergravit{\'e} g{\'e}n{\'e}rant une constante
cosmologique non nulle {\`a} une boucle.}.

\subsection{Cordes de type I}
\index{supercordes, th{\'e}orie des!de type I|textit}
Ayant obtenu une version supersym{\'e}trique de la th{\'e}orie des
cordes ferm{\'e}es,
il est naturel de chercher une construction correspondante pour
les cordes ouvertes. Les {\it th{\'e}ories de supercordes ouvertes},
dites de type I, ont connu un
regain d'int{\'e}r{\^e}t r{\'e}cent en liaison avec la (re)d{\'e}couverte
des D-branes. Elles n'occupent cependant pas une position
\index{D-brane}
centrale pour ce travail de th{\`e}se, aussi serons-nous
plus brefs encore qu'{\`a} l'accoutum{\'e}e (le lecteur pourra se reporter
{\`a} \cite{Fabre:1997} pour plus de d{\'e}tails.).
La corde bosonique ferm{\'e}e pr{\'e}sentait deux s{\'e}ries d'oscillateurs
gauche et droits $\alpha_n$ et $\bar\alpha_n$. Les conditions aux
limites de Neumann des cordes ouvertes identifient ces op{\'e}rateurs
de mani{\`e}re {\`a} ne plus laisser que les modes stationnaires.
Cette op{\'e}ration revient {\`a} prendre le quotient par l'involution
$z\rightarrow \bar z$ {\it inversant l'orientation
de la surface d'univers}. Dans le cas de la th{\'e}orie des 
supercordes de type II, cette op{\'e}ration doit {\^e}tre combin{\'e}e
avec une involution $(-)^F$ sur l'espace de Fock fermionique
et un {\'e}change des fermions gauches et droits,
ce qui n'est possible qu'en type IIB lorsque ceux-ci sont
de m{\^e}me chiralit{\'e}. Cette identification {\'e}limine
la moiti{\'e} des supersym{\'e}tries de la th{\'e}orie de type IIB.
Elle introduit dans le d{\'e}veloppement en genre, 
\index{genre, d{\'e}veloppement en}
en plus des surfaces de Riemann orientables
de la th{\'e}orie IIB, des surfaces ferm{\'e}es {\it non-orientables}, ainsi
que des surfaces {\it avec bords}.
La caract{\'e}ristique d'Euler d'une surface de genre $n$ avec
\index{Euler, caract{\'e}ristique d'|textit}
$b$ bords et $c$ points de non-orientabilit{\'e} (dits
{\it crosscaps}) s'{\'e}crivant $\chi=2-2n-b-c$, on voit
\index{crosscap}
qu'{\`a} l'ordre d'une boucle il est n{\'e}cessaire d'introduire
la {\it bouteille de Klein} ($c=2,b=n=0$), l'{\it anneau}
($b=2,c=n=0$) et le {\it ruban de M{\"o}bius} ($b=c=1,n=0$).
L'existence d'un bord permet le couplage {\`a} un
champ de jauge par l'interm{\'e}diaire de charges ponctuelles
dites de {\it Chan-Paton}, de groupe $SO(n)$ ou $USp(n)$
\index{Chan-Paton, facteur de}
dans le cas de cordes ouvertes non orient{\'e}es, en accord
avec l'existence d'un champ de jauge dans le spectre de
masse nulle.  
Le crit{\`e}re d'invariance modulaire est beaucoup moins puissant dans le cas
\index{modulaire, invariance}
des th{\'e}ories de cordes ouvertes, et doit {\^e}tre compl{\'e}t{\'e} par le
crit{\`e}re de {\it compensation des  tadpoles} n{\'e}cessaire
\index{tadpoles, compensation de}
{\`a} l'absence de divergences ultraviolettes. Ce pr{\'e}requis
\index{divergence ultraviolette!absence en th{\'e}orie des cordes}
fixe les coefficients respectifs des diff{\'e}rents diagrammes,
et restreint le groupe de jauge {\`a} $SO(32)$, en accord
avec les restrictions de compensation d'anomalies
\index{anomalie!gravitationnelle}
dans la th{\'e}orie de supergravit{\'e} de type I
\index{supergravit{\'e}!de type I}
discut{\'e}es au chapitre pr{\'e}c{\'e}dent.
On obtient ainsi {\it la th{\'e}orie des supercordes de type I},
\index{supercordes, th{\'e}orie des!de type I}
dont le spectre de masse nulle comprend le dilaton
et graviton du secteur de Neveu-Schwarz des cordes ferm{\'e}es (mais pas 
le tenseur antisym{\'e}trique exclu par la projection),
le tenseur antisym{\'e}trique de Ramond $\mathcal{B}_{\mu\nu}$,
\index{secteur de Neveu-Schwarz}
\index{secteur de Ramond}
\index{tenseur antisym{\'e}trique}
ainsi que le champ de jauge $SO(32)$ du secteur des cordes ouvertes 
et leurs partenaires fermioniques sous la supersym{\'e}trie $N=1$
{\`a} dix dimensions.

\subsection{Cordes h{\'e}t{\'e}rotiques}
\index{supercordes, th{\'e}orie des!h{\'e}t{\'e}rotiques|textit}
L'introduction de supersym{\'e}trie sur la surface d'univers
et le choix d'une projection convenable 
a ainsi permis de r{\'e}soudre le probl{\`e}me du tachyon
et de l'absence de fermions de la  th{\'e}orie
des cordes bosonique. L'absence de sym{\'e}trie de
jauge non ab{\'e}lienne augure mal cependant de
l'avenir ph{\'e}nom{\'e}nologique de la th{\'e}orie des supercordes de type II.
La th{\'e}orie des cordes de type I pr{\'e}sente bien une
invariance de jauge non ab{\'e}lienne, mais la construction de mod{\`e}les
chiraux en dimension 4 est rest{\'e}e longtemps difficile en raison
de l'absence d'un crit{\`e}re d'invariance modulaire commode.
Gr{\^a}ce {\`a} la d{\'e}couverte de la th{\'e}orie des supercordes
h{\'e}t{\'e}rotiques, il est devenu possible de construire des mod{\`e}les
ph{\'e}nom{\'e}nologiques viables reproduisant les trois g{\'e}n{\'e}rations
de mati{\`e}re chirale, ouvrant ainsi l'{\`e}re de la 
\index{chiralit{\'e}}
<<ph{\'e}nom{\'e}nologie des supercordes>>.
\index{phenomenologie@ph{\'e}nom{\'e}nologie}
La construction des cordes h{\'e}t{\'e}rotiques 
repose sur le constat que les secteurs
gauche et droit des th{\'e}ories de cordes ferm{\'e}es ne sont coupl{\'e}s
que par les modes z{\'e}ro bosoniques et fermioniques. En revanche,
la d{\'e}finition d'une projection GSO ne requiert la supersym{\'e}trie
\index{GSO, projection}
que d'un seul c{\^o}t{\'e} et suffit {\`a} garantir l'absence de
tachyon et la pr{\'e}sence de fermions dans le spectre.
Gross, Harvey, Martinec et Rohm ont ainsi propos{\'e} en 1985
une th{\'e}orie des cordes combinant 
la th{\'e}orie superconforme $N=1$ d{\'e}finissant la corde
de type II dans le secteur gauche, et la th{\'e}orie conforme
non supersym{\'e}trique de la corde bosonique dans le 
secteur droit \cite{Gross:1985dd}. Les dix champs gauches $X_L^{\mu}(z)$
sont ainsi associ{\'e}s avec les dix
champs droits $X_R^{\mu}(\bar z)$ pour d{\'e}finir les
coordonn{\'e}es de plongement de la corde h{\'e}t{\'e}rotique,
tandis que les seize champs suppl{\'e}mentaires 
$X^I_R(\bar z)$, soit apr{\`e}s fermionisation les trente-deux
fermions droits $\psi^i(\bar z)$, sont utilis{\'e}s
pour r{\'e}aliser une alg{\`e}bre de courant de rang 16.
Les contraintes d'invariance modulaire restreignent alors
\index{modulaire, invariance}
cette alg{\`e}bre {\`a} $SO(32)$ et $E_8\times E_8$ comme attendu
d'apr{\`e}s l'analyse de la compensation d'anomalies par
Green et Schwarz. La fonction de partition de ces mod{\`e}les
\index{Green et Schwarz, m{\'e}canisme de}
s'{\'e}crit ainsi
\index{partition, fonction de!des th{\'e}ories de cordes}
\begin{equation}
\label{zhet}
Z_{\; \rm het}(\tau,\bar\tau)= \frac{1}{\t_2^4}
\cdot
\frac{1}{\eta^{8}\bar\eta^{8}}
\cdot
\frac{1}{2}\sum_{a,b=0}^1
(-1)^{a+b+ab} \left( \frac{\th\ss}{\eta} \right)^{8/2}
\cdot
\left[ \frac{1}{2}
\sum_{\bar{a}, \bar{b} =0}^1
\left(\frac{\thb\bss}{\bar\eta}\right)^{16/2}
\right]^2
\end{equation}
o{\`u} l'on reconna{\^\i}t {\`a} nouveau l'int{\'e}grale des modes z{\'e}ros,
les huit oscillateurs bosoniques transverses, leurs partenaires
supersym{\'e}triques gauches, et les 32 fermions droits r{\'e}partis
en deux groupes de 16. Le terme entre crochets reproduit
le caract{\`e}re de l'alg{\`e}bre affine $SO(32)$, ou
aussi bien celui de l'alg{\`e}bre affine $E_8\times E_8$~;
bien que les fonctions de partition de ces deux mod{\`e}les
soient identiques, les spectres
soient en effet distincts. A nouveau, la m{\^e}me identit{\'e}
(\ref{riem})
implique l'annulation de cette fonction de partition.

\section{Compactification et T-dualit{\'e}}
\index{compactification!des th{\'e}ories de cordes}
A ce stade, nous avons obtenu les cinq th{\'e}ories des
supercordes supersym{\'e}triques en dimension critique : 
cordes ferm{\'e}es de type IIA et IIB, cordes ouvertes
de type I, cordes h{\'e}t{\'e}rotiques $SO(32)$ et 
$E_8 \times E_8$. La simplicit{\'e} relative de cette
classification s'{\'e}vanouit d{\`e}s que l'on consid{\`e}re
ces th{\'e}ories en dimension inf{\'e}rieure, tout d'abord
en raison de la multiplicit{\'e} des compactifications
possibles, et ensuite de l'existence de constructions
non g{\'e}om{\'e}triques directement en dimension inf{\'e}rieure.
Nous nous contenterons ici de discuter les deux cas les
plus simples pr{\'e}servant respectivement tout ou moiti{\'e}
de la supersym{\'e}trie : compactifications toro{\"\i}dales
et sur vari{\'e}t{\'e} $K_3$. Ces deux cas suffiront aux besoins
de cette th{\`e}se, et {\`a} r{\'e}v{\'e}ler l'originalit{\'e}
de la perception de l'espace-temps par la th{\'e}orie
des cordes. Ils nous fourniront l'exemple de
dualit{\'e}s perturbatives, pr{\'e}curseurs des S-dualit{\'e}s
que nous consid{\`e}rerons dans la section suivante.

\subsection{Compactification toro{\"\i}dale de la corde bosonique ferm{\'e}e}
Nous avons d{\'e}ja discut{\'e} la compactification 
des th{\'e}ories de champs sur un cercle dans la section \ref{sugracomp}:
chaque {\'e}tat de la th{\'e}orie originale se scinde en une
tour d'{\'e}tats de Kaluza-Klein de masse $\mathcal{M}= m/R$.
\index{Kaluza-Klein!excitation de}
Dans le cas de la th{\'e}orie des cordes, la situation est qualitativement
diff{\'e}rente puisque l'existence d'un cycle non-trivial autorise
de nouvelles configurations, dites {\it instantons de surface d'univers},
\index{instanton!de surface d'univers}
o{\`u} la corde s'enroule $n$ fois autour de ce cercle. Les entiers
\index{enroulement!etat d'@{\'e}tat d'}
$m$ et $n$ correspondent aux {charges {\'e}lectriques} sous la
sym{\'e}trie de jauge $U(1)_L \times U(1)_R$ correspondant aux vecteurs
$g_{\mu1}$ et $B_{\mu1}$, o{\`u} l'indice $1$ d{\'e}signe la direction compacte.
La th{\'e}orie 
conforme d{\'e}crivant ces {\'e}tats est une th{\'e}orie de boson
compact libre soluble explicitement.
Chaque couple $(m,n)$
d{\'e}crit un {\'e}tat <<fondamental>> $|m,n\rangle$ de la th{\'e}orie conforme
d'{\'e}nergie $H= L_0 + \bar L_0$ et spin conforme $S=L_0 -\bar L_0$ avec
\begin{eqnarray}
L_0 &=& \frac{\alpha'}{4} \left( \frac{m}{R} + n \frac{R}{\alpha'} \right)^2 \\
\bar L_0 &=& \frac{\alpha'}{4} 
\left( \frac{m}{R} - n \frac{R}{\alpha'} \right)^2  \ ,
\end{eqnarray}
sur lequel est construite une tour d'{\'e}tats d'oscillations identique
{\`a} celle d'un boson non compact, soumise {\`a} la condition de 
{\it level matching}
\index{level matching, condition de}
\begin{equation}
N_R - N_L = mn\ .
\end{equation}
La fonction de partition 
s'{\'e}crit donc 
\begin{equation}
\label{Zr}
Z(R;\tau,\bar\tau) = \frac{1}{\eta\bar\eta}
\sum_{m,n=-\infty}^{\infty} q^{L_0} \bar q^{\bar L_0}
\end{equation}
o{\`u} $q=e^{2i\pi\tau}$. On constate en particulier que {\it cette
fonction de partition est invariante sous l'inversion du rayon}
$R\rightarrow 
\alpha'/R$.
\index{T-dualit{\'e}!sur un cercle}
\index{T-dualit{\'e}|textit}      
Cette sym{\'e}trie, dite {\it T-dualit{\'e}}\footnote{
La sym{\'e}trie de T-dualit{\'e} est d{\'e}crite en d{\'e}tail
dans l'article de revue \cite{Giveon:1994fu}.}, 
correspond {\`a} une dualit{\'e}
de Poincar{\'e} $\partial_\alpha X \rightarrow \epsilon_{\alpha\beta}
\index{dualit{\'e}!de Poincar{\'e}}
\partial^\beta X$ sur la surface d'univers,
soit $X(\bar z) \rightarrow -X(\bar z)$. Elle
agit sur le spectre des {\'e}tats en {\'e}changeant le
nombre classique d'enroulement $n$ avec le nombre quantique de moment $m$.
En particulier, lorsque $R\rightarrow 0$, les {\'e}tats enroul{\'e}s
autour du cercle $(m=0,n\ne 0)$ deviennent toujours plus l{\'e}gers,
tandis que les {\'e}tats de Kaluza-Klein deviennent plus massifs.
\index{decompactification@d{\'e}compactification, limite de}
De mani{\`e}re duale, lorsque le cercle est d{\'e}compactifi{\'e}, ce sont les
{\'e}tats d'enroulement qui deviennent supermassifs, tandis que les
{\'e}tats de moment approchent la masse nulle. Cette sym{\'e}trie,
que nous avons ici v{\'e}rifi{\'e}e  {\`a} l'ordre d'une boucle, est valide 
{\`a} tous les ordres en perturbation, moyennant la transformation
$e^{\phi}\rightarrow e^{\phi}\sqrt{\alpha'}/R$ du dilaton
\index{dilaton}
\footnote{Cette transformation pr{\'e}serve le dilaton
effectif $e^{-2\phi_{D-1}}=R e^{-2\phi_D}/\sqrt{\alpha'}$.}, 
et elle commute
donc avec le d{\'e}veloppement perturbatif. 
Au point {\it autodual} $R=\sqrt{\alpha'}$, les {\'e}tats
de charges $(m,n)=(\pm 1,\pm 1)$ deviennent de masse nulle
et correspondent aux {\it bosons de jauge}
associ{\'e}s {\`a} l'{\it extension} de la sym{\'e}trie de jauge
$U(1)_L\times U(1)_R$ en une {\it sym{\'e}trie non ab{\'e}lienne}
$SU(2)_L \times SU(2)_R$. Cette sym{\'e}trie de jauge est
\index{sym{\'e}trie de jauge!extension}
{\it spontan{\'e}ment bris{\'e}e} pour $R\ne 1$, et la T-dualit{\'e} 
\index{T-dualit{\'e}!comme sym{\'e}trie de jauge}
peut s'interpr{\'e}ter comme la {\it sym{\'e}trie de Weyl} r{\'e}siduelle
\index{Weyl!sym{\'e}trie de}
sous cette brisure. A ce titre,
elle doit donc {\^e}tre {\it exacte} dans une g{\'e}n{\'e}ralisation
non perturbative de la th{\'e}orie des cordes.

Cette sym{\'e}trie $\Zint_2$ admet une extension remarquable
dans le cas des compactifications toro{\"\i}dales sur un 
\index{T-dualit{\'e}!sur un tore $T^d$|textit}
tore $T^d$ de dimension sup{\'e}rieure. Il est alors n{\'e}cessaire
de pr{\'e}ciser {\`a} la fois la m{\'e}trique sur le tore $g_{IJ}$
et la valeur moyenne du tenseur antisym{\'e}trique $B_{IJ}$.
Ces $d^2$ champs scalaires correspondent {\`a} autant de
modules de la th{\'e}orie compactifi{\'e}e. Moments et enroulements
sont alors quantifi{\'e}s par des entiers $m_I$ et $n^I$ tels
que
\begin{equation}
X^{I}(\sigma,\tau) = g^{IJ} m_J \tau + n^I \sigma \ .
\end{equation}
L'{\'e}nergie et le spin conforme sont donn{\'e}s par
\begin{eqnarray}
L_0 + \bar L_0 &=& \frac{\alpha'}{2} \left( m_I +B_{IJ} n^J\right) g^{IK} 
\left( m_K +B_{KL} n^L\right) + \frac{1}{2\alpha'} n^{I} g_{IJ} n^J \\
L_0 - \bar L_0 &=& m_I n^I
\end{eqnarray}
Le vecteur $p=(m_I,n^I)$ {\`a} $2d$ composantes d{\'e}crit alors un
{\it r{\'e}seau} muni de la norme {\it paire} $2 m_I n^I$ de signature $(d,d)$,
et de la m{\'e}trique $L_0 + \bar L_0$ de volume {\it unit{\'e}}.
La fonction de partition associ{\'e}e {\`a} ce r{\'e}seau $\Gamma_{d,d}$
\index{partition, fonction de!d'un r{\'e}seau}
\index{auto-dual pair, r{\'e}seau}
{\it auto-dual pair} 
\begin{equation}
Z_{d,d}(g,B;\tau,\bar\tau) = 
\frac{1}{\eta ^d \bar\eta ^d} \sum_{m_I,n^I} q^{L_0} \bar q^{\bar L_0}
\ ,\quad q=e^{2\pi i \tau}
\end{equation}
est donc bien une fonction {\it invariante modulaire}, et repr{\'e}sente
\index{modulaire, invariance}
la fonction de partition des modes z{\'e}ro des $N$ bosons compacts,
ainsi que des oscillateurs gauches et droits associ{\'e}s.
L'espace des modules du r{\'e}seau
est un espace homog{\`e}ne 
\index{vari{\'e}t{\'e}!homog{\`e}ne}
\begin{equation}\frac{SO(d,d,\Real)}{SO(d)\times SO(d)}
\end{equation}
de dimension $d^2$, param{\'e}tr{\'e} par la matrice $M\in SO(d,d,\Real)$
de la forme quadratique
$L_0+\bar L_0$ :
\begin{equation}
L_0 +\bar L_0 = \frac{1}{2} p^t M p
,\quad
M=\begin{pmatrix}
\alpha' g^{-1} & \alpha' g^{-1}B \\ - \alpha' B g^{-1} &
\frac{g}{\alpha'} - \alpha' B g^{-1} B 
\end{pmatrix}
\ .
\end{equation}
L'action contragradiente $M\rightarrow \Omega^t M \Omega$
du groupe $O(d,d,\Real)$ sur l'espace homog{\`e}ne
peut {\^e}tre compens{\'e}e par une rotation du vecteur entier $p\rightarrow 
\Omega^{-1} p$ lorsque $\Omega$ est une matrice {\`a} coefficients
entiers, soit $\Omega\in O(d,d,\Zint)$. {\it La fonction
de partition du tore $T^d$ est donc {\it invariante 
sous le groupe de T-dualit{\'e}} $O(d,d,\Zint)$}. Un sous-groupe
\index{T-dualit{\'e}!sur un cercle}
g{\'e}n{\'e}rateur (mais non minimal) de ce groupe de T-dualit{\'e} 
consiste en les rotations
euclidiennes du tore $Sl(d,\Zint)$, les flots spectraux
$B_{IJ} \rightarrow B_{IJ}+1$ et les T-dualit{\'e}s {\'e}l{\'e}mentaires
\index{flot spectral!en th{\'e}orie des cordes}
sur chaque cercle du tore. 

Une simplification importante intervient dans le cas des compactifications
sur un tore $T^2$. L'espace homog{\`e}ne $O(2,2,\Real)/O(2)\times O(2)$ 
se scinde en effet en deux facteurs 
$Sl(2,\Real)/U(1) \times Sl(2,\Real)/U(1)$ correspondant {\`a} deux
param{\`e}tres complexes $T$ et $U$ :
\begin{equation}
T= B_{12} + i \sqrt{\det g}/\alpha', \qquad
U= g_{12}/g_{11} + i \sqrt{\det g}/g_{11}
\end{equation}
sur lequel le groupe de T-dualit{\'e} agit par {\it transformations modulaires}
ind{\'e}pendantes $Sl(2,\Zint)_T\times Sl(2,\Zint)_U$ et par {\it {\'e}change
de $T$ et $U$}
\cite{Dijkgraaf:1988vp}. 
Plus g{\'e}n{\'e}ralement, la sym{\'e}trie de T-dualit{\'e}
existe d{\`e}s lors que la vari{\'e}t{\'e} de compactification admet une 
isom{\'e}trie~; elle la transforme alors en une vari{\'e}t{\'e} de topologie
{\it classique} tout {\`a} fait distincte, mais {\'e}quivalente du point de 
vue de la propagation de la corde. Elle s'{\'e}tend {\'e}galement 
aux compactifications sur espaces de Calabi-Yau o{\`u} elle transforme
\index{Calabi-Yau, vari{\'e}t{\'e} de}
une vari{\'e}t{\'e} $K$ en sa vari{\'e}t{\'e} {\it miroir} $\tilde K$
\index{miroir, sym{\'e}trie}
(voir par exemple les cours \cite{Hosono:1994av,Greene:1996cy}). Cette
sym{\'e}trie de dualit{\'e} est donc de port{\'e}e tr{\`e}s g{\'e}n{\'e}rale
et correspond {\`a} une sym{\'e}trie de jauge de la th{\'e}orie des
cordes encore mal comprise.

\subsection{T-dualit{\'e} et supercordes ferm{\'e}es\label{Tramond}}
La discussion pr{\'e}c{\'e}dente s'appliquait {\`a} la corde bosonique et au secteur 
bosonique des supercordes ferm{\'e}es. Dans le cas de la supercorde de
type II, l'action de la T-dualit{\'e} sur les champs bosoniques
de la surface d'univers s'accompagne d'une action sur les champs
fermioniques renversant la chiralit{\'e} des spineurs 
du secteur droit. Les cordes de type IIA et de type IIB sont ainsi
\index{dualit{\'e}!des th{\'e}ories IIA et IIB|textit}
\index{T-dualit{\'e}!secteur de Ramond|textit}
\index{Ramond, secteur de}
{\it {\'e}chang{\'e}es} {\`a} chaque inversion de rayon, de sorte que le groupe
de T-dualit{\'e} se trouve r{\'e}duit {\`a} $SO(d,d,\Zint)$, le g{\'e}n{\'e}rateur
de d{\'e}terminant -1 reliant les deux th{\'e}ories IIA et IIB.
Les valeurs moyennes des tenseurs antisym{\'e}triques de jauge
du secteur de Ramond fournissent en outre des modules suppl{\'e}mentaires
de la th{\'e}orie compactifi{\'e}e, sur lesquels la T-dualit{\'e}
doit encore agir. Une analyse explicite de la g{\'e}om{\'e}trie 
de l'espace des modules de Ramond montre que la
T-dualit{\'e} sur la direction $I$ est repr{\'e}sent{\'e}e sur les
potentiels de Ramond par
\footnote{Cette {\'e}quation admet des corrections en pr{\'e}sence 
d'un champ $B_{\mu\nu}$ non nul (cf appendice \ref{dc}).}
\begin{equation}
\mathcal{R} \rightarrow dx^I \cdot \mathcal{R} + dx^I \wedge \mathcal{R} 
\end{equation}
o{\`u} $\mathcal{R}=\sum_p \mathcal{R}_p$ d{\'e}signe la somme des 
formes diff{\'e}rentielles de Ramond d'ordre pair (en type IIB)
ou impair (en type IIA), et o{\`u} les symboles $\cdot$ et $\wedge$
d{\'e}signent les produits int{\'e}rieurs et ext{\'e}rieurs de formes
diff{\'e}rentielles. L'ensemble de ces transformations
pour $I=1\dots d$ engendrent une {\it alg{\`e}bre de Clifford} qui n'est
\index{Clifford, alg{\`e}bre de}
autre que l'alg{\`e}bre associ{\'e}e {\`a} $SO(d,d,\Real)$. Les champs
scalaires de Ramond se transforment donc comme une repr{\'e}sentation
{\it spinorielle} de $SO(d,d,\Zint)$, et le passage de la th{\'e}orie
de type IIA {\`a} la th{\'e}orie de type IIB s'accompagne d'une inversion
de chiralit{\'e}. \index{chiralit{\'e}}

\index{supercordes, th{\'e}orie des!h{\'e}t{\'e}rotiques}
\index{dualit{\'e}!de Het $SO(32)$ et Het $E_8\times E_8$|textit}
Le cas de la th{\'e}orie h{\'e}t{\'e}rotique offre une particularit{\'e}
d'un autre ordre. Les bosons de jauge de la corde h{\'e}t{\'e}rotique
en dimensions 10 fournissent par compactification toro{\"\i}dale
$16d$ champs de modules correspondant aux {\it lignes de Wilson}
\index{Wilson, ligne de|textit}
$\oint A_I^a dx^I$ brisant la sym{\'e}trie de jauge de rang 16. 
Qui plus est, la
distinction entre coordonn{\'e}e de plongement et coordonn{\'e}e
d'alg{\`e}bre de courant dispara{\^\i}t du c{\^o}t{\'e} droit de
la corde, si bien qu'en r{\'e}alit{\'e} on a $d$ bosons
{\`a} gauche et $d+16$ {\`a} droite, compactifi{\'e}s
sur un {\it r{\'e}seau de Narain} $(P_L,P_R)\in \Gamma_{d,d+16}$,
\index{Narain, r{\'e}seau de}
param{\'e}tr{\'e} par l'espace homog{\`e}ne $O(d,d+16,\Real)/(SO(d)\times
SO(d+16)$ {\it modulo} le groupe de T-dualit{\'e} {\'e}tendu $O(d,d+16,\Zint)$
\cite{Narain:1986jj,Narain:1987am,Lerche:1989np}.
La formule de masse et la condition de {\it level matching}
\index{level matching, condition de}
\index{masse, formule de!de Het$/T^d$}
s'{\'e}crivent\footnote{$N_L$ contient ici les oscillateurs
fermioniques demi-entiers.} 
\begin{eqnarray}
N_L + \frac{P_L^2}{2} - \frac{1}{2} = N_R + \frac{P_R^2}{2} - 1 \ ,\\
\mathcal{M}^2 = \frac{4}{\alpha'} \left( N_R + \frac{P_R^2}{2} -
  1\right)\ .
\end{eqnarray}
Les {\'e}tats tels que $(P_L^2,P_R^2)=(0,2)$ et $(N_L,N_R)=(1/2,0)$
correspondent donc {\`a} des bosons de jauge de masse nulle
s'ajoutant aux 16 bosons $U(1)$ ($N_L,N_R=(1/2,1)$).
Ils signalent donc l'{\it extension de la sym{\'e}trie de 
jauge} h{\'e}t{\'e}rotique en certains points de l'espace des
\index{sym{\'e}trie de jauge!extension}
modules. Les racines ayant toutes la m{\^e}me longueur
$P_R^2-P_L^2=2$, on obtient ainsi les groupes de jauge
simplement lac{\'e}s A,D,E de la classification
\index{groupe!simplement lac{\'e}}
\index{Cartan, classification de}
de Cartan. En particulier, lorsque le r{\'e}seau $\Gamma_{d,d+16}$
est factoris{\'e} en $\Gamma_{d,d}\oplus \Gamma_{16}$,
soit pour des lignes de Wilson nulles, on retrouve
la sym{\'e}trie de jauge $SO(32)$ (ou $E_8\times E_8$) de la
th{\'e}orie {\`a} 10 dimensions. On peut cependant trouver une
valeur des lignes de Wilson restituant l'{\it autre} sym{\'e}trie
de jauge $E_8\times E_8$ (ou $SO(32)$)~: les deux th{\'e}ories
h{\'e}t{\'e}rotiques {\`a} dix dimensions sont ainsi 
contin{\^u}ment reli{\'e}es
par compactification sur un cercle.

\subsection{Cordes ouvertes et D-branes\label{dbr}}
\index{D-brane|textit}
\index{T-dualit{\'e}!en cordes ouvertes}
La T-dualit{\'e} que nous avons d{\'e}crite dans le cas des
cordes ferm{\'e}es permet ainsi de relier deux {\`a} deux les
th{\'e}ories des supercordes {\`a} dix dimensions que l'on
croyait distinctes. Son existence dans la th{\'e}orie
des cordes ouvertes de type I semble {\it a priori}
probl{\'e}matique, puisque les {\'e}tats d'enroulement
images des {\'e}tats de Kaluza-Klein sous la T-dualit{\'e}
\index{enroulement!etat d'@{\'e}tat d'}
\index{Kaluza-Klein!excitation de}
n'existent pas en th{\'e}orie des cordes ouvertes.
La r{\'e}solution de ce paradoxe a men{\'e} Horava, Polchinski, Dai, et
Leigh {\`a} la d{\'e}couverte des D-branes 
\cite{Dai:1989ua,Horava:1989ga} qui ont r{\'e}cemment pris
une importance consid{\'e}rable dans la compr{\'e}hension
des dualit{\'e}s des th{\'e}ories des cordes et des 
th{\'e}ories de jauge. Nous en rappellerons bri{\`e}vement
les points saillants indispensables {\`a} la compr{\'e}hension
des travaux de ce m{\'e}moire, renvoyant le lecteur
aux articles de revue pour plus de d{\'e}tails
\cite{Bachas:1996sc,Polchinski:1996fm,Polchinski:1996na,Taylor:1997dy}.

Revenant {\`a} l'interpr{\'e}tation
de la T-dualit{\'e} en termes de dualit{\'e} de Poincar{\'e} sur
la surface d'univers, on voit que dans le cas des cordes
ouvertes, cette op{\'e}ration remplace la 
condition de Neumann $\partial_\sigma X(\sigma=0,\pi)=0$ sur la 
coordonn{\'e}e compacte par la condition de {\it Dirichlet}
$\partial_\tau X(\sigma=0,\pi)=0$, traduisant le fait que les
\index{Neumann, condition de}
\index{Dirichlet, condition de}
extr{\'e}mit{\'e}s de la corde ouverte sont attach{\'e}es {\`a} un point
fixe dans la direction $X$, soit sur une {\it 8-brane} 
de l'espace-temps. Le moment n'est alors plus conserv{\'e},
mais l'enroulement autour de la direction compacte devient
un bon nombre quantique. La T-dualit{\'e} peut {\^e}tre appliqu{\'e}e
dans plusieurs directions distinctes successivement de mani{\`e}re
{\`a} g{\'e}n{\'e}rer des $p$-branes de toute dimension dites
{\it D-branes}. La 9-brane correspond {\`a} la propagation libre
des cordes ouvertes, mais la pr{\'e}sence de $N$
d'entre elles revient {\`a} attacher un facteur de Chan-Paton $U(N)$
\index{Chan-Paton, facteur de}
aux extr{\'e}mit{\'e}s des cordes ouvertes orient{\'e}es
\footnote{$SO(N)$ ou $Sp(N/2)$ dans le cas des cordes non
orient{\'e}es. En particulier, dans la th{\'e}orie de type I, la sym{\'e}trie
de jauge $SO(32)$ peut {\^e}tre interpr{\'e}t{\'e}e comme la
pr{\'e}sence de 16 D9-branes.}.
Ces D-branes apparaissent comme des objets infiniment massifs,
{\'e}tendus longitudinalement et localis{\'e}s transversalement
dans la th{\'e}orie de perturbation.
\fig{4cm}{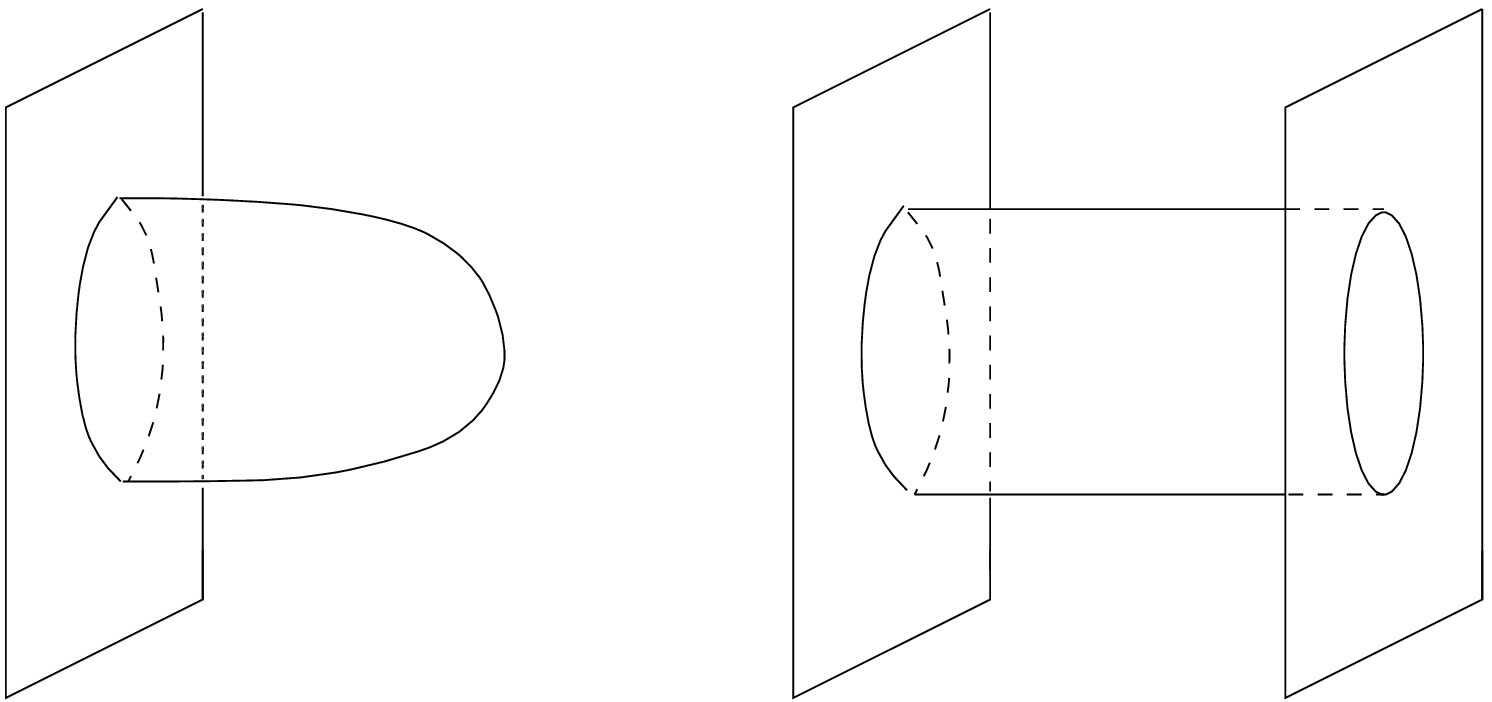}{A gauche~: fluctuation {\`a} l'ordre du disque du champ de
cordes ouvertes en pr{\'e}sence d'une D-brane~; {\`a} droite~: 
fluctuation {\`a} l'ordre de l'anneau du champ de
cordes ouvertes en pr{\'e}sence de deux D-branes, ou de mani{\`e}re
{\'e}quivalente, {\'e}change de cordes ferm{\'e}es entre celles-ci.}{1dbrane}

L'{\'e}change de moment avec les fluctuations du champ de cordes ouvertes
en sa pr{\'e}sence conf{\`e}re {\`a} la D-brane une dynamique 
justifiant cette appellation de membrane. Les {\it modes de masse nulle
des cordes ouvertes} attach{\'e}es {\`a} la D-brane 
correspondent en particulier aux {\it degr{\'e}s de libert{\'e} de la
D-brane}. Contrairement {\`a} la th{\'e}orie des cordes ouvertes <<libres>>
o{\`u} ces modes se propagent dans l'espace-temps {\`a} dix dimensions,
la brisure de la sym{\'e}trie de translation par la
D-brane r{\'e}duit leur d{\'e}pendance  aux coordonn{\'e}es
{\it longitudinales}, et ils ne propagent donc plus que {\it sur le
volume d'univers} de la D-brane. Ils se couplent n{\'e}anmoins
\index{volume d'univers}
aux modes des cordes {\it ferm{\'e}es} de l'espace-temps {\`a} dix 
dimensions, et en particulier au graviton. Le potentiel
vecteur $A_\mu$ de la th{\'e}orie des cordes ouvertes libres
donne ainsi naissance {\`a} $9-p$ champs $A_I$, {\it scalaires}
du point de vue du volume d'univers de la D-brane et
correspondant aux {\it fluctuations de position transverses}
{\`a} la D-brane, et un {\it champ de jauge} $A_{\mu}$ 
se propageant sur ce volume d'univers~; {\`a} ces fluctuations
marginales s'ajoutent les modes massifs des cordes ouvertes,
\index{supermassifs, {\'e}tats}
dont l'int{\'e}gration conduit {\`a} une action effective
pour les degr{\'e}s de libert{\'e} $A_I$ et $A_\mu$ de la D-brane.
Cette action peut {\^e}tre ais{\'e}ment d{\'e}termin{\'e}e par T-dualit{\'e}
{\`a} partir de l'action effective du champ de jauge $A_\mu$ de 
la th{\'e}orie des cordes ouvertes libres \cite{Leigh:1989jq}. 
Ce champ de jauge
couple au {\it bord} de la surface d'univers de la corde ouverte par
une ligne
de Wilson $\oint A_\mu \partial_\tau X^\mu$ ; l'amplitude 
\index{Wilson, ligne de}
{\`a} l'ordre du disque  de la th{\'e}orie
des cordes ouvertes en pr{\'e}sence du champ de fond $A_\mu$
(figure \ref{1dbrane})
peut {\^e}tre {\'e}valu{\'e}e explicitement\footnote{aux termes contenant
des d{\'e}riv{\'e}es de $F$ pr{\`e}s.} \cite{Abouelsaood:1987gd}
et conduit {\`a} l'action de l'
{\it {\'e}lectrodynamique de Born-Infeld}
\index{Born-Infeld, action de|textit}
\index{electrodynamique@{\'e}lectrodynamique!de Born-Infeld}
\cite{Born:1935}
\begin{equation}
\label{ebi}
\langle e^{\oint A_\mu \partial_\tau X^\mu} \rangle_{\mbox{disque}} = 
\int d^Dx~e^{-\phi} \sqrt{ \det(\eta_{\mu\nu} + \alpha' F_{\mu\nu} ) }
\end{equation}
o{\`u} $F=dA$ est la courbure du champ de jauge $A$. Cette action
restitue aux {\'e}nergies basses devant $1/\sqrt{\alpha'}$ l'action
de Maxwell ordinaire, mais la corrige par des interactions
{\`a} nombre de d{\'e}riv{\'e}es arbitraires\footnote{Cette s{\'e}rie
est g{\'e}n{\'e}r{\'e}e par le d{\'e}veloppement de la racine carr{\'e}e,
et non pas du d{\'e}terminant qui contient les termes $F^{2k}$,
$k=1\dots [D/2]$. La racine carr{\'e}e peut du reste {\^e}tre
{\'e}limin{\'e}e en introduisant un champ auxiliaire~:
$\sqrt{x}=\langle \frac{1}{2} (v + x/v) \rangle_v$.
}.
La T-dualit{\'e} dans la direction $I$
remplace le couplage de jauge $\int A_I \partial_\tau X^I$
par un couplage transverse $\int A_I \partial_\sigma X^I$
au {\it flux d'impulsion} $p^I$ traversant l'extr{\'e}mit{\'e}
de la corde vers la D-brane. Le champ $A_I$ appara{\^\i}{}t donc
bien comme la variable conjugu{\'e}e
au moment de la D-brane, soit comme la position  $x^I$
dans la direction $I$. L'{\'e}lectrodynamique
de Born-Infeld (\ref{ebi}) donne alors sous cette
r{\'e}interpr{\'e}tation l'{\it action de Dirac-Born-Infeld}
\cite{Dirac:1962}
d{\'e}crivant la dynamique de la D$p$-brane
\begin{equation}
\label{abi}
S=\int d^{p+1}x~e^{-\phi} \sqrt{ \det(\hat g_{\mu\nu} + \alpha' F_{\mu\nu} ) }
\end{equation}
o{\`u} $\hat g$ repr{\'e}sente la m{\'e}trique induite sur
le volume d'univers par le plongement $x^I$. En particulier,
la tension de la  D$p$-brane est donn{\'e}e par
\index{tension!des D-branes|textit}
\begin{equation}
T_{p}= e^{-\phi}/(\alpha')^\frac{p+1}{2}
\end{equation}
tr{\`e}s sup{\'e}rieure {\`a} celle de la corde fondamentale {\`a}
faible couplage, mais aussi tr{\`e}s inf{\'e}rieure {\`a} celle en $1/g^2$
des solitons des th{\'e}ories des champs habituelles. 

L'action de Born-Infeld peut {\^e}tre ais{\'e}ment g{\'e}n{\'e}ralis{\'e}e
en pr{\'e}sence d'un champ de fond de tenseur antisym{\'e}trique de Neveu-Schwarz
\index{tenseur antisym{\'e}trique}
$B_{\mu\nu}$. Celui-ci couple {\`a} la surface d'univers par 
un terme topologique $\int \hat B$, invariant sous la transformation
de jauge $B\rightarrow B+d\Lambda$ {\it {\`a} un terme de bord pr{\`e}s}
$\oint \Lambda$ qui peut {\^e}tre absorb{\'e} dans une transformation
du champ de jauge de volume d'univers  de
la D-brane $A \rightarrow A-\Lambda$. 
La courbure g{\'e}n{\'e}ralis{\'e}e $F_{\mu\nu}+\hat B_{\mu\nu}$
est alors invariante de jauge, et remplace $F$ dans l'{\'e}quation
(\ref{abi}). 

Nous avons jusqu'{\`a} pr{\'e}sent d{\'e}crit les D-branes dans le
cadre de la th{\'e}orie des cordes ouvertes. Elles existent
tout autant dans les th{\'e}ories des supercordes de type I et II, o{\`u}
elles d{\'e}crivent des {\'e}tats 1/2-{\it BPS satur{\'e}s}
\index{etats BPS@{\'e}tats BPS}
\footnote{La structure drastiquement diff{\'e}rente des modes gauches et
droits dans la corde h{\'e}t{\'e}rotique emp{\^e}che l'introduction de
bord sur la surface d'univers de la corde, et donc l'existence
de D-branes h{\'e}t{\'e}rotiques.}. 
Cette propri{\'e}t{\'e} appara{\^\i}{}t par exemple dans 
l'{\it annulation} de l'interaction {\`a} une boucle de 
deux D-branes parall{\`e}les {\it statiques} (figure \ref{1dbrane}), 
donn{\'e}e par la fonction de partition 
de la th{\'e}orie de supercordes sur l'anneau. 
L'action de Born-Infeld est
alors compl{\'e}t{\'e}e par des termes fermioniques
de mani{\`e}re {\`a} former une th{\'e}orie de volume
d'univers invariante sous {\it la moiti{\'e}} des
supersym{\'e}tries de la th{\'e}orie libre. L'autre moiti{\'e},
bris{\'e}e spontan{\'e}ment par la pr{\'e}sence de la
D-brane, est r{\'e}alis{\'e}e {\it non-lin{\'e}airement}
sur le volume d'univers.
\index{supersym{\'e}trie!sur le volume d'univers des branes}

Comme remarqu{\'e} par Polchinski, l'annulation du potentiel
d'interaction statique entre deux D-branes parall{\`e}les r{\'e}sulte
de la compensation entre les interactions gravitationnelles
\index{Ramond, secteur de}
\index{Neveu-Schwarz, secteur de}
du secteur de Neveu-Schwarz et {\it les interactions de
jauge du secteur de Ramond}. Elle implique en particulier
que la D$p$-brane {\it est charg{\'e}e sous le potentiel
de Ramond} $\mathcal{R}_{p+1}$, et porte la charge minimale
permise par la condition de quantification de Dirac. 
\index{Dirac, condition de quantification de}
L'action de la D-brane comprend donc, en sus de 
l'action de Born-Infeld (\ref{ebi}), le couplage topologique
de Wess-Zumino
\index{Wess-Zumino, terme de!sur le volume d'univers des D-branes}
\begin{equation}
\label{topo}
S = \int d^{p+1}x~e^{F+\hat B} \hat\mathcal{R}\ ,
\end{equation}
o{\`u} l'int{\'e}grale s{\'e}lectionne la $p+1$-forme apr{\`e}s
d{\'e}veloppement en s{\'e}rie de l'exponentielle
\cite{Green:1996bh,Green:1997dd}. Les
D$p$-branes peuvent ainsi {\^e}tre {\it identifi{\'e}es}
aux $p$-branes des th{\'e}ories de supergravit{\'e} de type I et II charg{\'e}es
sous ces m{\^e}mes champs, et en fournissent une description
en termes de th{\'e}orie conforme. Certaines configurations 
de D-branes non parall{\`e}les pr{\'e}servent quant {\`a}
elles une fraction inf{\'e}rieure de la supersym{\'e}trie,
et peuvent encore {\^e}tre identifi{\'e}es {\`a} des {\'e}tats
BPS des th{\'e}ories de supergravit{\'e}.

Si l'action de Born-Infeld (\ref{abi}) d{\'e}crit la dynamique
d'une D-brane, elle ne dit cependant rien de l'interaction entre 
plusieurs D-branes. Celle-ci peut {\`a} nouveau {\^e}tre d{\'e}termin{\'e}e
en consid{\'e}rant les fluctuations du champ de cordes entre elles.
Le champ de cordes ferm{\'e}es conduit aux interactions
gravitationnelles ordinaires, corrig{\'e}es par l'effet des modes
massifs. Le champ de corde ouvertes engendre cependant de nouvelles
interactions, puisque les cordes sont maintenant susceptibles 
d'avoir leurs deux extr{\'e}mit{\'e}s attach{\'e}es sur deux D-branes
diff{\'e}rentes. La masse des {\'e}tats fondamentaux de ces cordes
$\mathcal{M}= L/\alpha'$ est proportionnelle {\`a} leur 
{\'e}longation $L$, et ces modes deviennent de masse nulle lorsque 
les deux D-branes se touchent. Le cas de $N$ D-branes parall{\'e}les,
positionn{\'e}es {\`a} des abscisses $x^i$ le long d'une direction
compacte (figure \ref{Ndbrane}) est
particuli{\`e}rement simple~: sous la T-dualit{\'e} il correspond 
{\`a} $N$ 9-branes, soit {\`a} la pr{\'e}sence de facteurs de Chan-Paton
\index{Chan-Paton, facteur de}
\index{T-dualit{\'e}!en cordes ouvertes}
$U(N)$ aux extr{\'e}mit{\'e}s de la corde~; la sym{\'e}trie de jauge est
cependant bris{\'e}e en $U(1)^N$ par l'holonomie du champ de jauge
\index{holonomie!du champ de jauge}
autour du cercle
\begin{equation}
\begin{pmatrix} 
         e^{2\pi i x_1/R} &                &       & \\
                        & e^{2\pi ix_2/R} &       & \\
                        &                & \dots & \\
                        &                &       & e^{2\pi i x_N/R}
\end{pmatrix}
\in U(N)
\end{equation}
\fig{5cm}{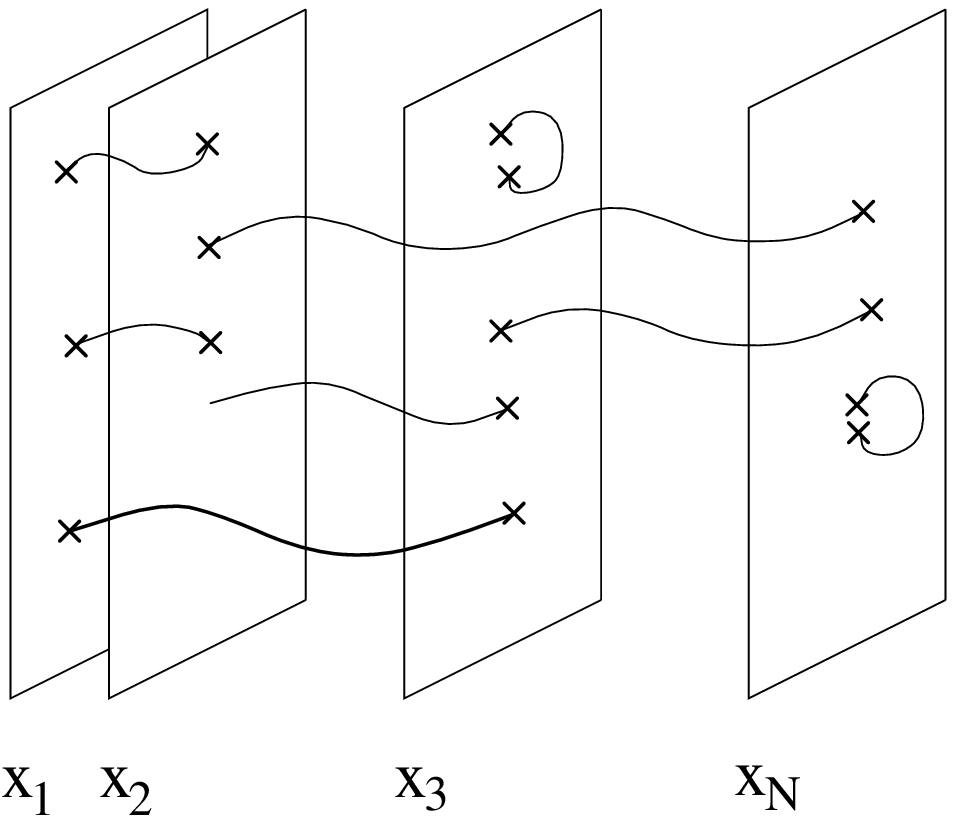}{Interactions de cordes ouvertes dans un
syst{\`e}me de $N$ D-branes parall{\`e}les.}{Ndbrane}
La co{\"\i}{}ncidence de $n$ D-branes s'accompagne donc
d'une {\it restauration de sym{\'e}trie non ab{\'e}lienne}
\index{sym{\'e}trie de jauge!restauration}
$U(n)\times U(1)^{N-n}$ dont les $n^2$ bosons vecteurs non ab{\'e}liens
correspondent aux cordes de masse nulle joignant les $n$ D-branes
deux {\`a} deux. La dynamique effective d{\'e}crivant ces modes peut
en principe encore {\^e}tre obtenue {\`a} partir de l'action effective
d'un champ de jauge non-ab{\'e}lien $U(N)$ en th{\'e}orie de type I~;
la non commutativit{\'e} des champs de jauges matriciels $A_\mu
\in su(N)$ emp{\^e}che une {\'e}valuation explicite de
l'amplitude {\`a} une boucle g{\'e}n{\'e}ralisant l'{\'e}quation
(\ref{ebi})\footnote{Dans le cas non ab{\'e}lien, en raison
de la relation $[D,D]F=[F,F]$, la distinction entre termes 
d{\'e}rivatifs et commutateurs n'existe plus, et on est donc
r{\'e}duit {\`a} garder ou omettre tous les deux. Dans le second
cas, on obtient une relation
analogue {\`a} (\ref{ebi}) o{\`u} la trace sym{\'e}tris{\'e}e 
dans la repr{\'e}sentation adjointe de $U(N)$ appara{\^\i}{}t devant 
la racine carr{\'e}e \cite{Tseytlin:1997cs}. 
Les effets conjugu{\'e}s de non lin{\'e}arit{\'e} et de non commutativit{\'e} 
sont cependant g{\^a}ch{\'e}s par cette approximation.
Le couplage topologique est quant
{\`a} lui donn{\'e} exactement par la trace de l'{\'e}quation
(\ref{topo}), o{\`u} la courbure g{\'e}n{\'e}ralis{\'e}e
doit {\^e}tre lue $F+\hat B\cdot \mathbb{I}$. 
}~;
la limite de basse {\'e}nergie est cependant non ambigu{\"e}, 
et correspond {\`a} une th{\'e}orie de Yang-Mills supersym{\'e}trique
\footnote{{\`a} 16 charges supersym{\'e}triques en type II,
ou 8 charges en type I.}
non ab{\'e}lienne de groupe de jauge $U(N)$.
Les champs de jauge $A_I$ deviennent par T-dualit{\'e} les
positions {\it non ab{\'e}liennes} transverses des D-branes
\cite{Witten:1996im},
sujettes au potentiel scalaire
\begin{equation}
V= \tr [X^i,X^j] [X_i,X_j]
\end{equation}
Dans la phase de Higgs, les matrices de positions commutent
\index{Higgs!phase de}
et peuvent {\^e}tre simultan{\'e}ment diagonalis{\'e}es~:
on recouvre ainsi la notion de position individuelle de chaque
D-brane. Au voisinage des points de sym{\'e}trie {\'e}tendue,
o{\`u} deux valeurs propres des matrices de position co{\"\i}{}ncident,
les fluctuations des degr{\'e}s de libert{\'e} non diagonaux 
induisent des effets de coh{\'e}rence et seule la position
du centre de masse, correspondant au facteur diagonal $U(1)\subset U(N)$,
est bien d{\'e}finie. Cette non-commutativit{\'e} de l'espace-temps
\index{geometrie non commutative@g{\'e}om{\'e}trie non commutative}
n'est pas sans rappeler la g{\'e}om{\'e}trie non commutative formul{\'e}e
par A. Connes \cite{Connes:1994}, 
mais la connection n'a jusqu'a pr{\'e}sent pas {\'e}t{\'e}
formul{\'e}e en toute g{\'e}n{\'e}ralit{\'e}.
Bien qu'incompl{\`e}te, cette formulation est particuli{\`e}rement 
adapt{\'e}e {\`a} l'{\'e}tude des {\it {\'e}tats li{\'e}s} de D-branes~:
l'approximation de basse {\'e}nergie {\`a} l'action de
$N$ D-branes est alors suffisante, et les {\'e}tats li{\'e}s
\index{li{\'e}, {\'e}tat li{\'e} de D-branes}
peuvent {\^e}tre identifi{\'e}s aux {\it {\'e}tats du vide
supersym{\'e}triques} de la th{\'e}orie de Yang-Mills
$U(N)$. L'absence de {\it mass gap} dans ces th{\'e}ories
en rend cependant l'{\'e}tude d{\'e}licate, et seule l'existence
d'{\'e}tats li{\'e}s pour $N=2$ a jusqu'{\`a} pr{\'e}sent pu {\^e}tre prouv{\'e}e~;
l'unicit{\'e} de l'etat li{\'e} de $N$ D0-branes, {\`a} la base
de nombreuses conjectures de dualit{\'e}, est encore incertaine.

\subsection{Orbifolds et compactification sur $K_3$}
\index{vari{\'e}t{\'e}!orbifold}
\index{compactification!sur orbifold}
La compactification toro{\"\i}dale que nous avons
d{\'e}crite jusqu'{\`a} pr{\'e}sent, n'agissant que sur
les modes z{\'e}ros bosoniques, pr{\'e}servait la 
totalit{\'e} des supersym{\'e}tries de la th{\'e}orie non
compactifi{\'e}e. La supersym{\'e}trie peut {\^e}tre r{\'e}duite
de mani{\`e}re commode en jaugeant une sym{\'e}trie discr{\`e}te
de l'espace-temps \cite{Dixon:1985jw,Dixon:1987qv}. 
L'espace r{\'e}sultant, dit {\it orbifold}, 
pr{\'e}sente des singularit{\'e}s aux {\it points fixes} de la sym{\'e}trie,
\index{singularit{\'e}!r{\'e}solution par la corde}
mais la th{\'e}orie conforme correspondante est parfaitement
r{\'e}guli{\`e}re, comme l'atteste l'{\'e}quivalence de la
compactification sur l'orbifold $S^1/\Zint_2$ de rayon
$R=\sqrt{\alpha'}$ 
avec la compactification sur un cercle de rayon $R=2\sqrt{\alpha'}$. 
\index{compactification!sur un cercle}
L'invariance modulaire impose l'inclusion
\index{modulaire, invariance}
d'{\'e}tats dits {\it twist{\'e}s} en sus des 
\index{twist{\'e}, {\'e}tat}
{\'e}tats de la th{\'e}orie originale
{\it invariants} sous la sym{\'e}trie. Les {\'e}tats twist{\'e}s
correspondent aux cordes ferm{\'e}es {\`a} une action de la 
sym{\'e}trie pr{\`e}s, soit aux {\'e}tats encerclant 
les singularit{\'e}s de l'orbifold. En particulier,
le secteur de masse nulle contient les modes non twist{\'e}s
correspondant aux modules de la th{\'e}orie initiale
pr{\'e}servant l'existence de la sym{\'e}trie, ainsi que les
{\it modes de r{\'e}solution} ({\it blow-up modes}) 
param{\`e}trant la r{\'e}solution de chaque singularit{\'e} de l'orbifold.
L'exemple le plus simple consiste {\`a} jauger la sym{\'e}trie
$\Zint_2: X^I\rightarrow -X^I$ renversant toutes
les coordonn{\'e}es $X^{i}$ du tore $T^d$. La fonction de partition
bosonique peut s'{\'e}crire en terme des blocs conformes {\it twist{\'e}s}
\index{partition, fonction de!d'un orbifold $T^d/\Zint_2$}
$Z_{d,d}\hg$ :
\begin{subequations}
\begin{equation}
Z_{d,d}^{orb} = \frac{1}{2} \sum_{g,h=0}^{1} Z_{d,d}\hg \ ,
\end{equation}
\begin{equation}
Z_{d,d} \ar{0}{0} =  Z_{d,d}
\sp
Z_{d,d} \hg  =2^d \frac{\vert \et \vert^{3d}}
{\left\vert  \th \ar{1 +h}{ 1 +g } \right\vert^d} \sp (h,g) \neq (0,0)
\ ,
\end{equation}
\end{subequations}
o{\`u} $h=0$ (resp. 1) dans le secteur non twist{\'e} (resp. twist{\'e}),
et o{\`u} $g=1$ correspond {\`a} l'insertion de l'op{\'e}rateur de sym{\'e}trie.
Les blocs twist{\'e}s $(h,g)\ne(0,0)$ ne d{\'e}pendent pas des modules
non twist{\'e}s $G_{IJ},B_{IJ}$, la sym{\'e}trie $\Zint_2$ ne laissant
que le secteur de charges nulles $m_I=n^I=0$. Ils d{\'e}pendraient en revanche
des modules de r{\'e}solution si on {\'e}cartait marginalement la
th{\'e}orie conforme du point d'orbifold $T^d/\Zint_2$.
Dans le cas des th{\'e}ories de supercordes, l'invariance du
courant de supersym{\'e}trie de surface d'univers impose que la
sym{\'e}trie agisse {\'e}galement sur les fermions selon 
$\psi^\mu\rightarrow -\psi^\mu$. Pour $d\le 3$, cette action brise la
supersym{\'e}trie d'espace-temps
\footnote{En effet, le spineur de 
Majorana-Weyl {\`a} 10 dimensions se r{\'e}duit en un
spineur de Majorana en dimension 9, un spineur de Weyl en
dimension 8, et un spineur de Dirac en dimension 7. Ce n'est qu'en
dimension 6 qu'il se scinde en {\it deux} spineurs dont l'un peut {\^e}tre
projet{\'e} sous la sym{\'e}trie $\Zint_2$.}.
Pour $d=4$, une moiti{\'e} de la supersym{\'e}trie est pr{\'e}serv{\'e}e
par la projection. L'orbifold $T^4/\Zint_2$ repr{\'e}sente en
effet le prototype
\footnote{Il existe trois autres mod{\`e}les d'orbifolds ab{\'e}liens
de $K_3$: $T^4/\Zint_3,T^4/\Zint_4,T^4/\Zint_6$. 
Le groupe discret $\Zint_n$ agit
dans tous ces cas sur les coordonn{\'e}es complexes du 4-tore
par $(z_1,z_2)\rightarrow (e^{2\pi i/n}z_1,e^{-2\pi i/n}z_2)$.
Cette action pr{\'e}serve en particulier la deux-forme holomorphe
$dz_1\wedge dz_2$.}
d'une {\it surface $K_3$}, c'est-{\`a}-dire
\index{compactification!sur $K_3$}
une vari{\'e}t{\'e} ({\'e}ventuellement singuli{\`e}re) k{\"a}hl{\'e}rienne 
compacte de dimension 4, simplement connexe,
et de courbure de Ricci triviale. Ces conditions
assurent l'existence d'un spineur covariantement constant,
et donc d'une supersym{\'e}trie r{\'e}siduelle. Compte tenu du
r{\^o}le central jou{\'e} par les compactifications sur $K_3$ dans
les conjectures de dualit{\'e} et dans les travaux pr{\'e}sent{\'e}s
en appendice \ref{tt} et \ref{dds} de ce m{\'e}moire, 
nous consacrerons le reste de cette section
{\`a} d{\'e}crire les aspects les plus importants de ces surfaces,
renvoyant {\`a} la r{\'e}f{\'e}rence \cite{aspinwall:1996mn} pour
une pr{\'e}sentation plus approfondie.

En r{\'e}alit{\'e}, c'est de {\it la } surface $K_3$ qu'il faudrait
parler : toutes les surfaces $K_3$ sont en effet 
{\it diff{\'e}omorphiquement} {\'e}quivalentes. Elles diff{\`e}rent
cependant dans le choix de la {\it 2-forme holomorphe} $\Omega$
d{\'e}finissant la {\it structure complexe},
et de la {\it classe de K{\"a}hler} $J$.
Le th{\'e}or{\`e}me de
\index{K3, surface@$K_3$, surface|textit}
Calabi et Yau assure l'existence et l'unicit{\'e} d'une m{\'e}trique
de courbure de Ricci nulle
(ou {\it m{\'e}trique d'Einstein}) ces deux structures {\'e}tant
fix{\'e}es \cite{Yau:1977ms}.
Toutes deux peuvent {\^e}tre sp{\'e}cifi{\'e}es en termes de vecteurs
dans le r{\'e}seau de cohomologie de $K_3$, qu'il est ais{\'e} de
d{\'e}terminer au point d'orbifold $T^4/\Zint_2$ : de la cohomologie
du tore $T^4$, seuls subsistent les formes {\it non twist{\'e}es}
$1,dx^i\wedge dx^j,dx_1\wedge dx_2\wedge dx_3\wedge dx_4$. 
La r{\'e}solution de chacune des 16 singularit{\'e}s 
de type $\CC^2/\Zint_2$ (soit $A_1$ dans la classification des
singularit{\'e}s des surfaces complexes) 
introduit autant de deux-cycles non triviaux~;
la cohomologie totale est r{\'e}sum{\'e}e dans le {\it diamant de Hodge}
\def\mb#1{\makebox[10pt]{$#1$}}
\begin{equation}
  {\arraycolsep=2pt
  \begin{array}{*{5}{c}}
    &&\mb{h^{0,0}}&& \\ &\mb{h^{1,0}}&&\mb{h^{0,1}}& \\
    \mb{h^{2,0}}&&\mb{h^{1,1}}&&\mb{h^{0,2}} \\
    &\mb{h^{2,1}}&&\mb{h^{1,2}}& \\ &&\mb{h^{2,2}}&&
  \end{array}} \;=\; 
  {\arraycolsep=2pt
  \begin{array}{*{5}{c}}
    &&\mb1&& \\ &\mb0&&\mb0& \\ \mb1&&\mb{(4+16)}&&\mb{1.} \\
    &\mb0&&\mb0& \\ &&\mb1&&
  \end{array}}
\end{equation}
pour une caract{\'e}ristique d'Euler $\chi=24$.
\index{Euler, caract{\'e}ristique d'}
Le second groupe de cohomologie $H_2(K_3,\Zint)$ est
particuli{\`e}rement important, puiqu'il peut {\^e}tre
munie d'une {\it forme d'intersection} 
$\alpha \# \beta = \int_{K_3} \alpha\wedge\beta$
de signature $(3,19)$, et {\`a} valeurs enti{\`e}res {\it paires}.
La dualit{\'e} de de Rham identifie en outre 
$H_2(K_3,\Zint)$ avec son dual $H^2(K_3\Zint)$.
$H_2(K_3,\Zint)$ constitue donc un {\it r{\'e}seau pair autodual},
analogue au r{\'e}seau de Narain d{\'e}finissant les compactifications
\index{Narain, r{\'e}seau de}
toro{\"\i}dales h{\'e}t{\'e}rotiques. Le choix de la 2-forme
holomorphe et de la
2-forme de K{\"a}hler d{\'e}finissent trois vecteurs $\Omega,\bar\Omega,J$ de
$H_2(K_3,\Zint)$. La surface $K_3$ {\'e}tant {\it hyperk{\"a}hl{\'e}rienne},
elle n'est cependant pas affect{\'e}e par une {\it rotation $SO(3)$}
de ces trois objets. Le choix de la m{\'e}trique d'Einstein sur
la vari{\'e}t{\'e} $K_3$ est donc param{\'e}tr{\'e} par l'espace 
des modules homog{\`e}ne
\index{espace des modules!de la vari{\'e}t{\'e} $K_3$}
\begin{equation}
O(3,19,\Zint) \backslash O(3,19,\Real)/(SO(3,\Real)\times
SO(19,\Real))\times \Real^+,
\end{equation}
correspondant au choix d'un 3-plan dans l'espace vectoriel
$H_2(K_3,\Real)\sim \Real^{3,19}$, aux diff{\'e}omorphismes
$SO(3,19,\Zint)$ du r{\'e}seau $H_2(K_3,\Zint)$ pr{\`e}s. Le facteur
$\Real^{+}$ correspond lui au volume de la m{\'e}trique d'Einstein. 

La donn{\'e}e de la m{\'e}trique d'Einstein ne suffit cependant pas
{\`a} d{\'e}finir la th{\'e}orie des cordes sur cet espace : il 
faut encore sp{\'e}cifier la valeur moyenne du tenseur
\index{tenseur antisym{\'e}trique}
antisym{\'e}trique $B_{\mu\nu}$, ainsi que des tenseurs
de jauge de Ramond dans les supercordes de type II, ou des
champs de jauge dans le cas des supercordes h{\'e}t{\'e}rotiques
et de type I. Nous laissons ce dernier cas de c{\^o}t{\'e} dans
cette th{\`e}se, puisque nous nous restreignons aux situations
avec au moins 16 supercharges de sym{\'e}trie. Les valeurs
moyennes du tenseur antisym{\'e}trique $B_{\mu\nu}$ peuvent {\^e}tre 
mesur{\'e}es par les 22 int{\'e}grales $\int_{\gamma^I} B$
sur les deux-cycles de $K_3$, et {\'e}tendent l'espace
des modules {\`a} \cite{Seiberg:1988pf}
\begin{equation}
O(4,20,\Zint) \backslash O(4,20,\Real)/(O(4,\Real)\times
O(20,\Real))
\end{equation}
auquel il faut encore ajouter un facteur $\Real^+$ correspondant 
au dilaton. La sym{\'e}trie $O(4,20,\Zint)$ contient la
sym{\'e}trie g{\'e}om{\'e}trique $O(3,19,\Zint)$ pr{\'e}c{\'e}dente, mais l'{\'e}tend
par des transformations suppl{\'e}mentaires 
{\it m{\'e}langeant la m{\'e}trique et le tenseur antisym{\'e}trique}.
\index{miroir, sym{\'e}trie}
Ces transformations correspondent aux {\it sym{\'e}tries miroir}
des mod{\`e}les sigma non-lin{\'e}aires sur les espaces de Calabi-Yau,
et correspondent {\`a} des dualit{\'e}s {\it perturbatives} des
th{\'e}ories des cordes analogues aux T-dualit{\'e}s des
compactifications toro{\"\i}dales. Dans le cas des supercordes
de type IIA compactifi{\'e}es sur $K_3$, on obtient ainsi
l'espace des modules complet, car les champs de Ramond,
correspondant {\`a} des formes diff{\'e}rentielles de degr{\'e}
{\it impair}, ne g{\'e}n{\`e}rent pas de degr{\'e}s de libert{\'e}
scalaires sur $K_3$. En revanche, dans le cas IIB,
il faut encore inclure la valeur moyenne du scalaire
de Ramond $\axion$, les 22 flux $\int_{\gamma^I} \mathcal{B}$
du tenseur antisym{\'e}trique de Ramond {\`a} travers les 
deux-cycles de $K_3$ ainsi que le flux total
$\int_{K_3} \mathcal{D}$ de la 4-forme $\mathcal{D}$. 
Avec le facteur du dilaton, ces champs se combinent en
l'espace des modules de type IIB
\index{espace des modules!de IIB sur $K_3$}
\begin{equation}
O(5,21,\Real)/(O(5,\Real)\times O(21,\Real))
\end{equation}
pour lequel il est naturel de conjecturer une identification sous le
groupe $O(5,21,\Zint)$. Cette dualit{\'e} 
\index{U-dualit{\'e}!de IIB sur $K_3$}
{\it est cependant non perturbative}, car m{\'e}langeant le dilaton
avec les param{\`e}tres g{\'e}om{\'e}triques de la compactification.
Notons qu'on ne sait d{\'e}finir la th{\'e}orie des
cordes perturbativement que pour une valeur moyenne nulle
des champs de Ramond.

Les compactifications sur $K_3$ permettent ainsi de r{\'e}duire
la dimension de l'espace-temps {\`a} $D=6$ tout en conservant la
moiti{\'e} des supersym{\'e}tries. La dimension
$D=4$ peut {\^e}tre atteinte par compactification suppl{\'e}mentaire
\footnote{On peut aussi envisager la compactification
de dix dimensions {\`a} quatre sur un espace de Calabi-Yau {\`a}
six dimensions \cite{Candelas:1985en}. 
\index{Calabi-Yau, vari{\'e}t{\'e} de}
Ce cas ne pr{\'e}serve cependant qu'un quart des
supersym{\'e}tries, aussi l'omettrons nous dans ce m{\'e}moire.}
sur un tore $T^2$.
On obtient ainsi, partant de la th{\'e}orie des cordes de type IIA ou IIB,
une th{\'e}orie supersym{\'e}trique $N=4$ en dimension 4. L'espace
des modules contient l'espace des modules 
\index{espace des modules!de Het/$T^4$ - IIA/$K_3$|textit}
$SO(4,20,\Real)/SO(4)\times SO(20) \times \Real^+$ de la
th{\'e}orie sur $K_3$ ainsi que les modules $T,U$ de la compactification
sur $T^2$. Il contient {\'e}galement les valeurs moyennes des
champs de Ramond sur les cycles de $K_3\times T^2$, ainsi que
\index{axion}
l'axion $a$ {\'e}quivalent au tenseur antisym{\'e}trique de Neveu-Schwarz
$B_{\mu\nu}$ sur la couche de masse. Avec le dilaton $\phi$,
\index{dilaton}
ce champ forme le scalaire complexe $S=a+i e^{-2\phi}$.
Dans le cas IIA, l'ensemble de ces champs 
s'organise dans l'espace homog{\`e}ne
\begin{equation}
\label{modii}
\frac{Sl(2,\Real)}{U(1)} \times 
\frac{O(6,22,\Real)}{O(6)\times O(22)}
\end{equation}
o{\`u} le premier facteur correspond au param{\`e}tre $T$ 
et le second rassemble $S,U$ et les modules de $K_3$ et de Ramond.
Le cas IIB s'obtient par T-dualit{\'e} $T \leftrightarrow U$. La encore,
il est naturel de conjecturer l'invariance de la th{\'e}orie
sur les transformations modulaires $Sl(2,\Zint)_T \times
SO(6,22,\Zint)$ de cet espace homog{\`e}ne.
{\it Cet espace des modules est identique {\`a}
celui de la corde h{\'e}t{\'e}rotique compactifi{\'e}e toro{\"\i}dalement
\index{dualit{\'e}!h{\'e}t{\'e}rotique - type II}
sur un tore $T^6$}~: le facteur $Sl(2,\Real)/U(1)$
est alors interpr{\'e}t{\'e} comme le champ complexe $S=a +ie^{-2\phi}$
{\it de la corde h{\'e}t{\'e}rotique}, tandis que le second
facteur n'est autre que l'espace des modules du r{\'e}seau
de Narain $\Gamma_{6,22}$. Cette co{\"\i}ncidence est une
\index{Narain, r{\'e}seau de}
des premi{\`e}res indications des dualit{\'e}s des supercordes,
vers lesquelles nous nous tournons dans la section suivante.
La S-dualit{\'e} de la th{\'e}orie 
\index{S-dualit{\'e}}
h{\'e}t{\'e}rotique est alors cons{\'e}quence de la T-dualit{\'e}
de la th{\'e}orie de type II \cite{Font:1990gx}.

La compactification sur $K_3\times T^2$ que nous venons de d{\'e}crire
n'est cependant pas la seule {\`a} fournir une th{\'e}orie de
supersym{\'e}trie $N=4$ en dimension 4. En particulier,
la vari{\'e}t{\'e} $K_3$ admet, sur une sous-vari{\'e}t{\'e} de
son espace des modules, une involution sans point fixe dite
{\it involution d'Enriques}\footnote{$T^4/\Zint_2$
admet ainsi les translations par une demi-p{\'e}riode 
$x^I\rightarrow x^I + 1/2$ comme involutions d'Enriques.
Ces involutions identifient les 16 points fixes de l'orbifold
deux {\`a} deux. Elles g{\'e}n{\`e}rent le groupe de sym{\'e}trie $(D_4)^4$
de la th{\'e}orie conforme associ{\'e}e {\`a} $T^4/\Zint_2$
\cite{Dijkgraaf:1988vp}.}. 
\index{Enriques, involution d'}
En combinant cette involution avec une
translation d'une demi-p{\'e}riode sur le tore, on obtient
une th{\'e}orie conforme sur $(K_3 \times T_2)$ qui pr{\'e}serve
la supersym{\'e}trie $N=4$. La projection {\'e}limine cependant
une partie des cycles de la vari{\'e}t{\'e} $K_3 \times T_2$, soit
une partie des bosons de jauge dans le spectre de basse {\'e}nergie.
On obtient ainsi des {\it th{\'e}ories de type II $N=4$ {\`a} rang r{\'e}duit},
\index{rang r{\'e}duit, th{\'e}orie des cordes {\`a}}
d'espace des modules $Sl(2,\Real)/U(1) \times O(6,6+N_V,\Real)/
(O(6)\times O(6+N_V))$. L'{\'e}tude des corrections de seuil
gravitationnelles dans ces th{\'e}ories et dans les th{\'e}ories
h{\'e}t{\'e}rotiques duales fait l'objet du travail en appendice \ref{tt}.

\section{La th{\'e}orie non perturbative des supercordes}
Apr{\`e}s le long chemin qui nous a conduit des dualit{\'e}s 
{\'e}lectriques-magn{\'e}tiques des th{\'e}ories de jauge supersym{\'e}triques 
aux dualit{\'e}s d'espace-cible des th{\'e}ories des supercordes
en passant par les sym{\'e}tries cach{\'e}es et le 
spectre des {\'e}tats solitoniques de membrane des th{\'e}ories
de supergravit{\'e}, nous avons maintenant en main les concepts
essentiels aboutissant {\`a} la {\it conjecture de dualit{\'e}
des th{\'e}ories de supercordes}. Nous nous sommes d'ailleurs
{\`a} plusieurs reprises approch{\'e}s de cette id{\'e}e, et
nous n'aurons ainsi qu'{\`a} rassembler les notions principales
apparues au cours de notre route.

\subsection{Dualit{\'e} des cordes $N=4$ h{\'e}t{\'e}rotique et de type II} 

Comme nous l'avons d{\'e}crit dans la section pr{\'e}c{\'e}dente,
les dualit{\'e}s d'espace-cible r{\'e}duisent {\`a} trois le
nombre de th{\'e}ories des supercordes perturbatives : 
les th{\'e}ories de type IIA et de type IIB sont identifi{\'e}es
par T-dualit{\'e} apr{\`e}s compactification sur un cercle,
de m{\^e}me que les cordes h{\'e}t{\'e}rotiques de groupe de jauge
$SO(32)$ et $E_8\times E_8$. Cette dualit{\'e} commute
\index{dualit{\'e}!des th{\'e}ories IIA et IIB}
\index{dualit{\'e}!de Het $SO(32)$ et Het $E_8\times E_8$}
avec le d{\'e}veloppement perturbatif en puissances
du couplage $g=e^\phi$, et peut {\^e}tre d{\'e}montr{\'e}e {\`a} tout ordre.
Elle peut en outre s'interpr{\'e}ter comme sym{\'e}trie de
jauge spontan{\'e}ment bris{\'e}e, et {\`a} ce titre doit
valoir {\'e}galement au niveau non perturbatif.

Nous avons {\'e}galement rencontr{\'e} de nombreux indices
de dualit{\'e} non perturbative entre ces trois th{\'e}ories.
Ainsi, la th{\'e}orie des cordes de type IIA compactifi{\'e}e
sur une vari{\'e}t{\'e} $K_3$ pr{\'e}sente {\it le m{\^e}me espace
des modules} que la th{\'e}orie h{\'e}t{\'e}rotique compactifi{\'e}e
sur un tore $T^4$. Cette indication peut {\^e}tre renforc{\'e}e
par l'examen des {\it actions effectives de basse {\'e}nergie}
\cite{Sen:1995cj}.
L'action {\`a} l'ordre des arbres de la th{\'e}orie h{\'e}t{\'e}rotique
compactifi{\'e}e sur $T^4$
s'obtient par r{\'e}duction de Kaluza-Klein de l'action
{\`a} dix dimensions :
\begin{equation}
S_{het}=\int d^6x~  e^{-2\phi_6}
\left[ \sqrt{-g} R + 4 (\nabla\phi_6)^2 - \frac{1}{12} \hat H\wedge *\hat H
+\frac{1}{ 8} \Tr (\partial M \partial M^{-1} ) 
-\frac{1}{ 4} M_{ij} F^i \wedge *F^{j} \right]  
\end{equation}
o{\`u} l'on a omis les champs fermioniques et les puissances de
$\alpha'$. Le couplage $g_6=e^\phi$ est 
reli{\'e} au dilaton de la th{\'e}orie {\`a} dix dimensions par
$e^{-2\phi_6}=V e^{-2\phi}$ o{\`u} $V$ est le volume de $T^4$~;
$F^i=dA^i$ est la courbure du champ de jauge $U(1)$ provenant
de la sym{\'e}trie de jauge {\`a} dix dimensions (bris{\'e}e par les
lignes de Wilson) ou de la sym{\'e}trie de jauge de Kaluza-Klein ; 
$\hat H=dB - L_{ij} A^i \wedge F^j$ contient
les corrections de Chern-Simons usuelles dans ces r{\'e}ductions ;
la matrice sym{\'e}trique $M$ param{\'e}trise l'espace des modules
homog{\`e}ne $SO(4,20,\Real)/SO(4)\times SO(20)$ et $L$ d{\'e}signe
la m{\'e}trique paire de signature (4,20). Le facteur $e^{-2\phi_6}$
correspond {\`a} une contribution {\`a} l'ordre de la sph{\`e}re.
D'un autre c{\^o}t{\'e}, l'action de la th{\'e}orie de type II r{\'e}duite
sur $K_3$ s'{\'e}crit
\begin{eqnarray}
S_{IIA}&=&\int d^6x~ e^{-2\phi_6}
\left[ \sqrt{-g} R +  4(\nabla\phi_6)^2 - \frac{1}{ 12} H\wedge *H
+\frac{1}{ 8} \Tr (\partial M \partial M^{-1} ) \right]   \nonumber\\
&& +\int d^6x \left[ -\frac{1}{ 4}
M_{ij} F^i \wedge *F^{j}  + L_{ij} B\wedge F^i\wedge F^j\right]
\end{eqnarray}
Les 24 champs de jauge $U(1)$ proviennent cette fois du
champ de jauge de Ramond  $\mathcal{A}$, 
de la r{\'e}duction du tenseur de jauge 
$\mathcal{C}=\sum_I \gamma_I \wedge \mathcal{A^I}$
sur les 2-cycles de $K_3$ et du tenseur $\mathcal{C}_3$
lui-m{\^e}me, {\'e}quivalent {\`a} un champ de jauge U(1) {\`a} 
six dimensions. Tous ces champs proviennent du secteur
de Ramond de la th{\'e}orie de type IIA, et pr{\'e}sentent
un couplage anormal au dilaton $e^{(\chi+2)\phi}$ sur
une surface de caract{\'e}ristique d'Euler $\chi$. 
On notera l'absence de
corrections de Chern-Simons dans la courbure $H=dB$,
et la pr{\'e}sence d'un terme topologique $B\wedge F\wedge F$
provenant d'un couplage $B \wedge d\mathcal{C} \wedge d\mathcal{C}$
{\`a} dix dimensions. L'identit{\'e} entre ces deux actions appara{\^\i}t
apr{\`e}s une transformation de Weyl $g\rightarrow e^{\phi_6/2} g_E$ 
{\'e}liminant le facteur du dilaton devant l'action
d'Einstein-Hilbert, et menant {\`a} l'action effective dans la
{\it m{\'e}trique d'Einstein}~:
\begin{eqnarray}
S_{het}&=&\int d^6x 
\left[ \sqrt{-g_E} R - (\nabla\phi_6)^2  - \frac{1}{ 12} e^{-2\phi_6}
\hat H\wedge *\hat H
+\frac{1}{ 8} \Tr (\partial M \partial M^{-1} ) \right. \nonumber\\
&& \left. -\frac{1}{ 4} e^{-\phi_6} M_{ij} F^i \wedge *F^{j} \right] \\
S_{IIA}&=&\int d^6x
\left[ \sqrt{-g_E} R - (\nabla\phi_6)^2 - \frac{1}{ 12} e^{-2\phi_6} H\wedge *H
+\frac{1}{ 8} \Tr (\partial M \partial M^{-1} ) \right.   \nonumber\\
&& \left. -\frac{1}{ 4} e^{\phi_6}
M_{ij} F^i \wedge *F^{j}  + L_{ij} B\wedge F^i\wedge F^j\right]
\end{eqnarray}
Ces deux actions sont pr{\'e}cis{\'e}ment {\'e}quivalentes sous
les identifications
\begin{equation}
\label{dual6d}
e^{\phi_{6;het}}=e^{-\phi_{6;IIA}} \ ,\quad
e^{-2\phi_{6;het}} \hat H_{het} = * H_{IIA}\ ,\quad
(g_E, A ^i, M)_{het} = (g_E, A ^i, M)_{IIA}\ .
\end{equation}
L'identification des m{\'e}triques d'Einstein peut alternativement
{\^e}tre prise en compte en transformant l'{\'e}chelle des cordes
$\alpha'$, de sorte que la dualit{\'e} h{\'e}t{\'e}rotique-type II
peut {\^e}tre commod{\'e}ment r{\'e}sum{\'e}e par les transformations
\begin{equation}
g_6 \rightarrow 1/g_6\ ,\quad
\alpha' \rightarrow \alpha' g_6^2\ .
\end{equation}
Cette dualit{\'e} en particulier inverse le couplage $g_6=e^{\phi_6}$, 
et {\'e}chappe donc {\`a} la th{\'e}orie de perturbations. 
Dans ces conditions, l'invariance
de l'action effective {\it {\`a} l'ordre des arbres} peut sembler une
b{\'e}n{\'e}diction exag{\'e}r{\'e}e : elle 
refl{\`e}te l'existence d'un {\it th{\'e}or{\`e}me de
non-renormalisation de l'action effective {\`a} deux d{\'e}riv{\'e}es}
dans les th{\'e}ories {\`a} 16 supersym{\'e}tries.
On ne saurait donc se contenter de l'examen de cette action
pour conclure {\`a} la dualit{\'e} exacte des th{\'e}ories h{\'e}t{\'e}rotique
et de type IIA {\`a} six dimensions.
Notons finalement qu'apr{\`e}s
compactification des deux th{\'e}ories sur un tore $T^2$,
l'identification (\ref{dual6d}) implique l'{\'e}quivalence
des deux th{\'e}ories de supersym{\'e}trie $N=4$
{\`a} quatre dimensions r{\'e}sultantes sous l'{\'e}change
$S \leftrightarrow T$ discut{\'e} {\`a} la fin de la section
pr{\'e}c{\'e}dente. 

Cette conjecture de dualit{\'e} re{\c c}oit une confirmation
suppl{\'e}mentaire par l'{\'e}tude du spectre des {\'e}tats BPS
dans chaque th{\'e}orie. 
Il existe en effet dans la th{\'e}orie de supergravit{\'e}
{\`a} six dimensions deux solitons de 1-brane : l'un,
r{\'e}gulier dans la m{\'e}trique h{\'e}t{\'e}rotique
et singulier dans la m{\'e}trique de type IIA, 
charg{\'e} sous le tenseur antisym{\'e}trique
de Neveu-Schwarz $B_{\mu\nu}$, peut {\^e}tre
interpr{\'e}t{\'e} comme la {\it corde fondamentale de type IIA},
apparaissant comme un soliton de la corde h{\'e}t{\'e}rotique ; 
son partenaire mag{\'e}tique est au contraire r{\'e}gulier
dans la m{\'e}trique de type IIA et singulier
dans la m{\'e}trique h{\'e}t{\'e}rotique, et admet
24 d{\'e}formations correspondant aux courants
{\'e}lectriques sous les 24 champs de jauge $U(1)$.
Il peut donc {\^e}tre interpr{\'e}t{\'e} comme {\it la corde
h{\'e}t{\'e}rotique fondamentale}, apparaissant comme
soliton dans la corde de type IIA \cite{Sen:1995cj}

Le spectre des {\'e}tats ponctuels peut aussi {\^e}tre
mis en correspondance pr{\'e}cise
\footnote{Le mat{\'e}riel pr{\'e}sent{\'e} dans cette section
appara{\^\i}t dans la seconde partie de l'article en
appendice \ref{tt}, o{\`u} il permet l'identification
des contributions non perturbatives
aux couplages scalaires {\`a} quatre d{\'e}riv{\'e}es 
dans la th{\'e}orie de type II compactifi{\'e}e
sur $K_3\times T^2$.}. Ces {\'e}tats BPS
correspondent aux excitations des oscillateurs
\index{oscillateur}
\index{etats BPS@{\'e}tats BPS!de het/$T^4$ - type II/$K_3$}
droits (non supersym{\'e}triques) de la corde
h{\'e}t{\'e}rotique\footnote{Les excitations BPS 
de la corde de type IIA se restreignent aux {\'e}tats
de masse nulle et n'apportent donc pas de nouvelle
confirmation de la dualit{\'e}.}, les oscillateurs gauches {\'e}tant 
dans leur {\'e}tat fondamental. La condition
de {\it level matching} pour $N_L=0$ impose
\index{level matching, condition de}
donc 
\begin{equation}
N_R -1 =\frac{1}{2} (P_L^2 - P_R^2) = \frac{1}{2} q^t L q \ ,
\end{equation}
o{\`u} $q^i$ est le vecteur des 24 charges enti{\`e}res sous les 
champs de jauge $U(1)$, et $L$ la m{\'e}trique de signature
(4,20). Cette {\'e}quation ne d{\'e}termine que le nombre
{\it total} d'excitations des oscillateurs 
bosoniques droits $\bar\alpha_{-n}$. Pour chaque valeur de $N_R$,
le nombre de possibilit{\'e}s $d(N_R)$, 
et donc la {\it d{\'e}g{\'e}n{\'e}rescence}
de l'{\'e}tat de charges $q^i$, est donn{\'e} par la fonction
de partition 
\index{partition, fonction de!des {\'e}tats BPS}
\begin{equation}
\label{etaexp}
\sum d(N) q^{N-1} = \frac{1}{q \prod_{n=1}^\infty (1-q^n)^{24}}
=\frac{1}{\eta^{24}(\tau)}\ ,\quad q=e^{2\pi i\tau}\ .
\end{equation}
La masse carr{\'e}e est quant {\`a} elle donn{\'e}e, dans la m{\'e}trique
h{\'e}t{\'e}rotique, par 
\begin{equation}
\mathcal{M}^2_{het} = \frac{1}{ 2} (P_L^2+P_R^2) + N_L +N_R -1 =P_L^2
=\frac{1}{2} q^t (M+L) q \ .
\end{equation}
\index{masse, formule de!des {\'e}tats BPS h{\'e}t{\'e}rotiques}
En particulier, ces {\'e}tats deviennent de masse nulle aux points o{\`u}
$P_L^2=0$ soit $N_R=0$ et $P_R^2=2$, correspondant aux {\it points
de sym{\'e}trie {\'e}tendue} de la corde h{\'e}t{\'e}rotique. 

Dans la m{\'e}trique de type IIA, la masse carr{\'e}e pr{\'e}sente le
comportement non perturbatif
en $\mathcal{M} \sim 1/g$ caract{\'e}ristique des D-branes :
\begin{equation}
\label{sol6d}
\mathcal{M}^2_{\II A} = \frac{1}{2} e^{-2\phi_{6;IIA}} q^t (M+L) q \ .
\end{equation}
\index{masse, formule de!des D-branes enroul{\'e}es sur $K_3$}
tandis que les charges $q^i$ correspondent maintenant aux charges
sous les 24 champs de jauge {\it du secteur de Ramond} de la th{\'e}orie
\index{Ramond, secteur de}
de type IIA. Ces {\'e}tats peuvent donc {\^e}tre identifi{\'e}s avec
les D0,D2,D4-branes de la th{\'e}orie de type IIA enroul{\'e}s autour
d'un cycle supersym{\'e}trique 
$\gamma$ de la vari{\'e}t{\'e} $K_3$. Les charges $q^i$
correspondent alors aux nombres d'enroulement de la D-brane autour de 
\index{enroulement!des solitons sur un cycle}
\index{cycle d'homologie!de $K_3$}
chacun des 24 cycles $\gamma^i$ de $K_3$, soit aux coefficients du 
d{\'e}veloppement $\gamma=\sum q^i \gamma_i$
du cycle $\gamma$ sur le r{\'e}seau d'homologie de $K_3$.
La norme enti{\`e}re $q^t L q$ correspond alors au nombre
d'auto-intersection $\gamma\#\gamma$ du cycle
\index{intersection}
\index{Euler, caract{\'e}ristique d'}
({\'e}gale {\`a} sa caract{\'e}ristique d'Euler dans le cas
d'un 2-cycle),et $q^t (M-L) q$ {\`a} son aire carr{\'e}e.
Pour {\^e}tre correcte, la dualit{\'e} h{\'e}t{\'e}rotique-type IIA
pr{\'e}voit donc {\it le nombre de cycles supersym{\'e}triques 
$N_s(q)$ homologiquement {\'e}quivalents {\`a}} $\sum q^i \gamma_i$:
\begin{equation}
N_s(q) = d\left( 1+\frac{1}{2}q^t L q \right) \ .
\end{equation}
Bershadsky {\it et al.} ont test{\'e} cette pr{\'e}diction en {\'e}tudiant
la th{\'e}orie topologiquement twist{\'e}e sur le volume d'univers
des D-branes \cite{Bershadsky:1996qy}~; Zaslow et Yau en ont
donn{\'e} une reformulation dans le cadre de la g{\'e}om{\'e}trie
alg{\'e}brique et ont pu en v{\'e}rifier l'exactitude pour
$\frac{1}{2}q^t L q <6$\cite{Yau:1996mv,Beauville:1997}. 
Cette pr{\'e}diction est r{\'e}alis{\'e}e
naturellement dans les corrections non perturbatives {\'e}tudi{\'e}es
dans l'appendice \ref{dds} de ce m{\'e}moire.

Les points de sym{\'e}trie {\'e}tendue de la corde h{\'e}t{\'e}rotique
correspondent aux points de l'espace des modules de $K_3$
o{\`u} la vari{\'e}t{\'e} devient singuli{\`e}re,
soit lorsqu'un ou plusieurs cycles deviennent d'aire nulle. 
Les D-branes enroul{\'e}es autour de ces cycles {\'e}vanescents
\index{cycle d'homologie!evanescent@{\'e}vanescent}
g{\'e}n{\`e}rent les particules de masse nulle d'une sym{\'e}trie
de jauge non ab{\'e}lienne d{\'e}termin{\'e}e par la matrice d'intersection
des cycles. Les singularit{\'e}s des surfaces $K_3$
ont {\'e}t{\'e} class{\'e}es dans la litt{\'e}rature math{\'e}matique
et se restreignent pr{\'e}cis{\'e}ment aux singularit{\'e}s
\index{singularit{\'e}!des surfaces $K_3$}
\index{sym{\'e}trie de jauge!extension}
de type ADE, en accord avec les sym{\'e}tries de jauge
observables du c{\^o}t{\'e} h{\'e}t{\'e}rotique. 

Cette pr{\'e}cise correspondance des spectres BPS des deux th{\'e}ories 
constitue l'argument le plus fort en faveur de la dualit{\'e}
des th{\'e}ories h{\'e}t{\'e}rotique et de type IIA compactifi{\'e}es
sur $T^4$ et sur $K_3$ respectivement. 
Le lecteur sceptique
pourrait cependant avec raison argumenter que toutes ces indications
ne concernent que le secteur BPS <<visible>> de la th{\'e}orie,
et ne pr{\'e}sument rien du secteur <<cach{\'e}>> de l'iceberg,
sur lequel se brisent toutes les approches.
Les contraintes de coh{\'e}rence de la th{\'e}orie des cordes
sugg{\`e}rent n{\'e}anmoins que l'identit{\'e} des secteurs BPS
puisse suffire {\`a} entra{\^\i}ner l'identit{\'e} compl{\`e}te.
On notera {\'e}galement que la th{\'e}orie h{\'e}t{\'e}rotique
{\it perturbative} appara{\^\i}{}t beaucoup plus compl{\`e}te
que la th{\'e}orie de type IIA, qui n{\'e}cessite l'introduction
des D-branes non perturbatives pour d{\'e}crire le spectre BPS.

\subsection{Dualit{\'e}s et unicit{\'e}}
Premi{\`e}re {\`a} {\^e}tre d{\'e}couverte, la dualit{\'e}
N=4 h{\'e}t{\'e}rotique-type II s'inscrit maintenant dans
un {\'e}cheveau complet de dualit{\'e}s qui a re{\c c}u
de nombreuses confirmations. Ce m{\'e}moire n'a pas vocation
{\`a} en donner une revue exhaustive, aussi nous bornerons
nous {\`a} en d{\'e}crire la structure g{\'e}n{\'e}rale sans entrer
dans tous les d{\'e}tails.
 
Des cinq th{\'e}ories des cordes perturbatives, la T-dualit{\'e}
apr{\`e}s compactification sur un cercle
ne laisse que trois th{\'e}ories apparamment distinctes:
la th{\'e}orie $SO(32)$ de type I, la th{\'e}orie h{\'e}t{\'e}rotique 
(de groupe de jauge $SO(32)$ ou $E_8\times E_8$) et
la th{\'e}orie de type II (A ou B). Ces th{\'e}ories sont
identifi{\'e}es par les dualit{\'e}s non perturbatives suivantes :

\begin{itemize}
\item La {\it dualit{\'e} h{\'e}t{\'e}rotique - type I} 
\index{dualit{\'e}!h{\'e}t{\'e}rotique - type I}
identifie les deux th{\'e}ories
des cordes de supersym{\'e}trie $N=1$ et groupe de jauge $SO(32)$
{\`a} dix dimensions. Les deux actions effectives de basse {\'e}nergie
sont identiques sous l'{\it inversion du couplage} $g=e^{\phi}$ et changement
\index{action effective!de type I et h{\'e}t{\'e}rotique $SO(32)$}
d'{\'e}chelle
\begin{equation}
g \rightarrow 1/g\ ,\quad
\alpha' \rightarrow \alpha' g 
\end{equation}
et l'identification du tenseur
$B_{\mu\nu}$ de la corde h{\'e}t{\'e}rotique avec
le tenseur $\mathcal{B}_{\mu\nu}$ du secteur de Ramond des
cordes ferm{\'e}es de la corde de type I. La corde h{\'e}t{\'e}rotique,
charg{\'e}e sous $B_{\mu\nu}$,
peut {\^e}tre identifi{\'e}e avec la {\it D1-brane} de la th{\'e}orie
de type I, charg{\'e}e sous $\mathcal{B}_{\mu\nu}$
\cite{Harvey:1995rn}.
Les  champs de la surface d'univers de la D1-brane, correspondant
aux modes de masse nulle des cordes ouvertes
dont la ou les extr{\'e}mit{\'e}s sont attach{\'e}es {\`a}
la D-brane, correspondent pr{\'e}cis{\'e}ment aux coordonn{\'e}es
de plongement et aux champs construisant l'alg{\`e}bre de
courant $SO(32)$ sur la surface d'univers de la corde h{\'e}t{\'e}rotique.
\index{cinq-brane!h{\'e}t{\'e}rotique}
La 5-brane h{\'e}t{\'e}rotique peut {\'e}galement {\^e}tre identifi{\'e}e
{\`a} la D5-brane de la th{\'e}orie de type I. 
Apr{\`e}s compactification sur un tore, les points
de sym{\'e}trie {\'e}tendue de la th{\'e}orie h{\'e}t{\'e}rotique
\index{sym{\'e}trie de jauge!extension}
sont associ{\'e}s {\`a} des points de fort couplage {\it local}
\footnote{Le champ de fond du dilaton de type I n'est en effet
pas uniforme sur le cercle.} de la
th{\'e}orie de type I \cite{Polchinski:1996df}.

\item L' {\it auto-dualit{\'e} de la th{\'e}orie de type IIB}
\index{S-dualit{\'e}!de la th{\'e}orie IIB}
identifie cette derni{\`e}re sous les transformations modulaires 
$Sl(2,\Zint)_B$
du scalaire complexe $\tau=\axion + ie^{-\phi}$. Cette sym{\'e}trie
associe {\`a} la corde fondamentale de charges {\'e}lectriques
(1,0) sous les
deux tenseurs de jauge $(B_{\mu\nu},\mathcal{B}_{\mu\nu})$
un {\it multiplet de cordes de charges} $(p,q)$, identiques
\index{pq, cordes de charge@$(p,q)$, cordes de charge}
en tout point {\`a} la corde fondamentale, et de tension
$1/\alpha'_{(p,q)} = |p+q\tau|/\alpha'$
\index{tension!de la corde de charge $(p,q)$}
\cite{Schwarz:1995dk}. La D3-brane est
invariante sous cette dualit{\'e}, mais celle-ci agit par {\it dualit{\'e}
{\'e}lectrique-magn{\'e}tique} dans la th{\'e}orie de jauge $N=4$
\index{dualit{\'e}!{\'e}lectrique-magn{\'e}tique}
d{\'e}crivant le volume d'univers de la D3-brane
\cite{Tseytlin:1996it}. Les solitons
de charge $(p,q)$ de cette th{\'e}orie de jauge sont interpr{\'e}t{\'e}s
\index{pq, dyons de charge@$(p,q)$, dyons de charge}
comme les {\it extr{\'e}mit{\'e}s} des cordes de type $(p,q)$ s'attachant
{\`a} la D3-brane. La NS5-brane
et la D5-brane forment quant {\`a} eux deux membres
d'un {\it multiplet de 5-branes} de charges magn{\'e}tiques $(p,q)$.
Apr{\`e}s compactification toro{\"\i}dale sur $T^d$,
la sym{\'e}trie non perturbative $Sl(2,\Zint)_B$ se combine avec la
sym{\'e}trie de T-dualit{\'e} $SO(d,d,\Zint)_T$ pour former
\index{T-dualit{\'e}}
\index{U-dualit{\'e}!de type II sur $T^d$}
la {\it sym{\'e}trie de U-dualit{\'e}} $E_{d+1}(\Zint)$
\cite{Hull:1995ys}. 
Cette sym{\'e}trie existe
aussi bien dans la {\it th{\'e}orie de type IIA compactifi{\'e}e
sur $T^d$}, que l'on obtient par une T-dualit{\'e} de d{\'e}terminant -1
arbitraire. Elle laisse invariante l'{\'e}chelle de Planck
\index{Planck, {\'e}chelle de}
\begin{equation}
l_P^{d-8} = \frac{ V e ^{-2\phi} }{\alpha^{'4}}\ .
\end{equation}

\item La {\it dualit{\'e} h{\'e}t{\'e}rotique-type II} relie les
\index{dualit{\'e}!h{\'e}t{\'e}rotique - type II|textit}
cordes h{\'e}t{\'e}rotiques et type IIA compactifi{\'e}es 
sur $T^4$ et $K_3$ respectivement. La sym{\'e}trie
$SO(4,20),\Zint)$ du r{\'e}seau de Narain h{\'e}t{\'e}rotique
appara{\^\i}t comme la {\it sym{\'e}trie miroir} 
\index{miroir, sym{\'e}trie}
\index{Narain, r{\'e}seau de}
associ{\'e}e {\`a} la vari{\'e}t{\'e} $K_3$. Les points
de sym{\'e}trie {\'e}tendue de la th{\'e}orie h{\'e}t{\'e}rotique
correspondent dans la th{\'e}orie de type IIA aux points
o{\`u} la vari{\'e}t{\'e} $K_3$ devient singuli{\`e}re. Les {\'e}tats
ponctuels non perturbatifs correspondant 
aux D0,2,4-branes enroul{\'e}es
autour du ou des cycles {\'e}vanescents de $K_3$ 
\index{cycle d'homologie!de $K_3$}
\index{cycle d'homologie!evanescent@{\'e}vanescent}
\index{sym{\'e}trie de jauge!extension}
deviennent alors de masse nulle et engendrent la sym{\'e}trie de
jauge.La dualit{\'e} h{\'e}t{\'e}rotique-type II
induit une
dualit{\'e} de ces m{\^e}mes th{\'e}ories compactifi{\'e}es
sur $T^4\times T^2$ et $K_3 \times T_2$ {\`a} quatre dimensions.
L'espace $K_3\times T^2$ du c{\^o}t{\'e} de type II
peut {\^e}tre fibr{\'e} non trivialement en un espace
de Calabi-Yau de dimension 6 de mani{\`e}re {\`a} obtenir
\index{Calabi-Yau, vari{\'e}t{\'e} de}
une compactification $N=2$ {\`a} quatre dimensions.
La dualit{\'e} h{\'e}t{\'e}rotique-type II continue {\`a}
valoir {\`a} condition de fibrer l'espace 
h{\'e}t{\'e}rotique $T^4\times T^2$
en une vari{\'e}t{\'e} $T^2 \times K_3$. 

\item La compactification de la th{\'e}orie de type IIB
sur une vari{\'e}t{\'e} $K_3$ conduit en revanche {\`a} une th{\'e}orie
{\it chirale} {\`a} seize charges supersym{\'e}triques, o{\`u} la dualit{\'e}
$Sl(2,\Zint)_B$ se combine avec la sym{\'e}trie miroir
\index{miroir, sym{\'e}trie}
$SO(4,20,\Zint)$  en
un {\it groupe de U-dua\-lit{\'e}} $SO(5,21,\Zint)$. La corde fondamentale
\index{U-dualit{\'e}!de IIB sur $K_3$}
est alors membre d'un multiplet contenant les D1,3,5-branes
enroul{\'e}es sur les 0,2,4-cycles de $K_3$, ainsi que leurs
partenaires sous la sym{\'e}trie $Sl(2,\Zint)_B$. Les points
singuliers de $K_3$ correspondent {\`a} l'apparition
de {\it cordes de tension nulle} et donc de tours d'{\'e}tats
\index{tension!cordes de tension nulle}
solitoniques non massifs.
\end{itemize}
Ces relations de dualit{\'e} d{\'e}terminent la
{\it dynamique {\`a} fort couplage} des th{\'e}ories de type IIB, type I
et h{\'e}t{\'e}rotique $SO(32)$. La dynamique {\`a} fort couplage
des th{\'e}ories de type IIA et h{\'e}t{\'e}rotique $E_8\times E_8$
peut {\it a priori} {\^e}tre obtenue
apr{\`e}s compactification sur un cercle de rayon
$R$ fixe, T-dualit{\'e} $(e^\phi,R) \rightarrow (e^{\phi'},R')=(e^\phi/R,1/R)$,
puis S-dualit{\'e} 
$(e^\phi/R,1/R)\rightarrow (e^{\phi''},R'')=(Re^{-\phi},\sqrt{R}e^{-\phi/2})$
vers la th{\'e}orie de type IIB ou I respectivement. Si cette
derni{\`e}re th{\'e}orie est faiblement coupl{\'e}e
($ e^{\phi''} \sim e^{-\phi} \rightarrow 0$), 
{\it elle ne se
r{\'e}duit pour autant pas {\`a} la th{\'e}orie perturbative
correspondante}, en raison du r{\'e}tr{\'e}cissement simultan{\'e}
du cercle de rayon $R'' \sim e^{-\phi/2} \rightarrow 0$.
En particulier, les D1-branes des th{\'e}ories de type I ou IIB
enroul{\'e}es autour du cercle de rayon $R''$
\index{enroulement!des solitons sur un cycle}
donnent lieu {\`a} des {\'e}tats de masse $e^{-\phi''}R''\sim e^{\phi/2}$
comparable {\`a} la masse des excitations de Kaluza-Klein 
\index{Kaluza-Klein!excitation de}
de masse $1/R'' \sim e^{\phi/2}$.

Le r{\'e}gime de fort couplage des th{\'e}ories de type IIA et
h{\'e}t{\'e}rotique $E_8\times E_8$ peut 
en revanche {\^e}tre d{\'e}termin{\'e}
{\it dans la limite de basse {\'e}nergie}. En effet,
la th{\'e}orie des cordes de type IIA est d{\'e}crite 
aux {\'e}nergies inf{\'e}rieures {\`a} l'{\'e}chelle des cordes
par la r{\'e}duction 
de la {\it supergravit{\'e} {\`a} onze dimensions} sur un
\index{supergravit{\'e}!{\`a} onze dimensions}
\index{Kaluza-Klein!r{\'e}duction de SUGRA 11D en IIA}
\index{dualit{\'e}!des th{\'e}ories IIA et SUGRA 11D}
cercle de rayon $R_{11}=l_{11} e^{2\phi/3}$. Cette
relation, observ{\'e}e sur l'action effective {\`a}
l'ordre des arbres de la th{\'e}orie de type IIA,
persiste pour toute valeur du couplage, en raison
de la non-renormalisation de l'action {\`a} deux
d{\'e}riv{\'e}es des th{\'e}ories {\`a} 32
supersym{\'e}tries. Elle s'{\'e}tend {\'e}galement aux
{\'e}nergies de l'ordre de $1/R_{11}$, gr{\^a}ce
{\`a} l'identification des {\'e}tats li{\'e}s de $N$ D0-branes 
de la th{\'e}orie de type IIA avec les {\'e}tats de Kaluza-Klein 
\index{Kaluza-Klein!excitation de la 11{\`e}me dimension}
de moment $N/R_{11}$ correspondant
{\`a} la r{\'e}duction du supergraviton {\`a} onze dimensions
\index{graviton}
\footnote{
La validit{\'e} de cette identification repose sur l'existence d'un 
unique {\'e}tat li{\'e} marginal de $N$ D0-branes, soit un unique
\index{li{\'e}, {\'e}tat li{\'e} de D-branes}
vide supersym{\'e}trique dans une th{\'e}orie de jauge $U(N)$ {\`a}
\index{vide!des th{\'e}ories supersym{\'e}triques}
seize supercharges. }\cite{Townsend:1995kk,Witten:1995ex}. 
A fort couplage, les D0-branes
deviennent de masse nulle et la th{\'e}orie de type IIA 
d{\'e}veloppe une onzi{\`e}me dimension non compacte.
La dimension critique $D=10$ de la
th{\'e}orie des supercordes de type IIA appara{\^\i}t donc
comme un {\it artefact} de la th{\'e}orie de perturbations.
\index{s{\'e}rie de perturbation!en th{\'e}orie des cordes}
Ce point de vue supprime {\'e}galement la distinction entre 
{\'e}tats fondamentaux et {\'e}tats solitoniques :
la corde fondamentale et la D2-brane ne sont que les
r{\'e}ductions diagonale et verticale d'une membrane
de la th{\'e}orie {\`a} onze dimensions, tandis que les
D4-branes et 5-brane de Neveu-Schwarz correspondent
aux descendants d'une unique 5-brane {\`a} onze dimensions.
\index{cinq-brane}\index{D-brane}
La D6-brane, charg{\'e}e magn{\'e}tiquement sous le
champ de jauge de Kaluza-Klein $\mathcal{A}_\mu$,
peut encore {\^e}tre identifi{\'e}e au {\it monop{\^o}le magn{\'e}tique
de Kaluza-Klein} sous la r{\'e}duction dimensionnelle
\cite{Townsend:1995kk}.
\index{Kaluza-Klein!monop{\^o}le de}
Enfin, la compactification de la th{\'e}orie {\`a} onze
dimensions sur un segment $S^1/\Zint_2$ peut {\^e}tre identifi{\'e}e
avec la th{\'e}orie des cordes h{\'e}t{\'e}rotiques $E_8 \times E_8$.
\index{compactification!de la M-th{\'e}orie sur un segment}
L'invariance de jauge appara{\^\i}t sous la forme de multiplets
vectoriels {\`a} dix dimensions se propageant sur les 9-branes
{\`a} chaque extr{\'e}mit{\'e} du segment.
Ces {\'e}tats sont analogues aux {\'e}tats {\it twist{\'e}s}
\index{twist{\'e}, {\'e}tat}
des constructions d'orbifold, bien qu'ils ne soient plus
\index{compactification!sur un orbifold}
impos{\'e}s par une condition d'invariance
modulaire, mais par la
\index{modulaire, invariance}
compensation {\it locale} de l'anomalie de cette th{\'e}orie
\index{anomalie!gravitationnelle}
chirale. Tout comme la onzi{\`e}me dimension dispara{\^\i}t
dans la th{\'e}orie de perturbation de type IIA, l'{\'e}cart
entre les deux 9-branes s'annule en th{\'e}orie des perturbations
h{\'e}t{\'e}rotique, menant {\`a} une th{\'e}orie des supercordes
en dimension critique. La corde fondamentale h{\'e}t{\'e}rotique
peut alors {\^e}tre interpr{\'e}t{\'e}e comme la membrane de la
th{\'e}orie {\`a} onze dimensions suspendue entre les deux 9-branes.

Les cinq th{\'e}ories de supercordes et la supergravit{\'e}
{\`a} onze dimensions apparaissent ainsi comme cinq diff{\'e}rentes
facettes d'une {\it th{\'e}orie non perturbative des
supercordes}, plus commun{\'e}ment nomm{\'e}e {\it M-th{\'e}orie}, 
\index{M-th{\'e}orie}
dont la formulation {\it ab initio} reste encore Myst{\'e}rieuse.
Les th{\'e}ories de cordes et leurs dualit{\'e}s d{\'e}finissent 
la M-th{\'e}orie comme les cartes et fonctions de transitions
d{\'e}finissent une vari{\'e}t{\'e} diff{\'e}rentielle, aux 
\index{vari{\'e}t{\'e}!diff{\'e}rentielle}
restrictions pr{\`e}s que la th{\'e}orie n'est d{\'e}finie
qu'{\it asymptotiquement} sur chaque carte, tandis que 
la compatibilit{\'e} des fonctions de transitions n'est
acquise que dans le secteur BPS.
Membranes et 5-branes semblent en tenir les r{\^o}les principaux,
mais les tentatives de quantification se heurtent aux
non-lin{\'e}arit{\'e}s de leur propagation libre. L'inclusion
non perturbative de leurs interactions semble du reste encore plus
inaccessible.
Plusieurs tentatives de d{\'e}finition totalement orthogonales {\`a}
cette approche
ont {\'e}t{\'e} propos{\'e}es, les unes bas{\'e}es sur des mod{\`e}les
de matrice supersym{\'e}triques, d'autres sur une th{\'e}orie
de Chern-Simons {\`a} onze dimensions \cite{Horava:1997dd}. La
\index{Chern-Simons, th{\'e}orie de}
formulation de Banks, Fischler, Shenker et Susskind
\index{Banks, Fischler, Shenker et Susskind, conjecture de}
semble pour le moment la plus prometteuse, et fera l'objet
du dernier chapitre de ce m{\'e}moire. 
En l'attente d'une formulation non perturbative {\it ab initio}
de la M-th{\'e}orie, une approche moins ambitieuse et plus
pragmatique consiste {\`a} {\'e}tudier dans quelle mesure les
{\it m{\'e}thodes semi-classiques} de la th{\'e}orie des champs peuvent
s'{\'e}tendre {\`a} la th{\'e}orie des cordes. Cette {\'e}tude
constitue le coeur de ce travail de th{\`e}se, et fait l'objet
du chapitre que nous abordons maintenant.


%% file: chap4.tex
\chapter{Approche semi-classique {\`a} la M-th{\'e}orie}
\index{semi-classique!calcul en th{\'e}orie des cordes}
Les m{\'e}thodes semi-classiques figurent parmi les rares outils
d'investigation des th{\'e}ories de champs non int{\'e}grables
au niveau non perturbatif. Elles sont particuli{\`e}rement
adapt{\'e}es aux situations o{\`u} les corrections perturbatives
sont nulles ou en nombre fini : les contributions des 
{\it points-selle} de l'action microscopique,
ou {\it instantons}, fournissent
\index{instanton}
les premi{\`e}res corrections non perturbatives {\`a} ces
amplitudes. Par chance, de nombreuses quantit{\'e}s physiques
des th{\'e}ories de supercordes sont prot{\'e}g{\'e}es des
corrections perturbatives par des {\it th{\'e}or{\`e}mes de 
non renormalisation}. Par malchance, l'action microscopique
de la th{\'e}orie non perturbative des supercordes n'est pas
connue, et les r{\`e}gles de somme sur les points-selle,
\index{instanton!mesure d'int{\'e}gration}
sinon les points-selle eux-m{\^e}mes, sont ind{\'e}termin{\'e}s.
Les th{\'e}ories de supergravit{\'e}, d{\'e}crivant les th{\'e}ories
de cordes {\`a} basse {\'e}nergie, fournissent une premi{\`e}re 
d{\'e}termination du spectre des instantons.
Les dualit{\'e}s des th{\'e}ories de cordes sont quant {\`a} elles
assez strictes pour autoriser la d{\'e}termination {\it exacte}
de certaines amplitudes physiques. Le d{\'e}veloppement
de ces amplitudes {\`a} faible couplage met en {\'e}vidence
une somme d'effets non perturbatifs que l'on peut chercher
{\`a} interpr{\'e}ter comme la contribution des points-selle
de la M-th{\'e}orie. On peut ainsi par recoupement d{\'e}terminer
les r{\`e}gles du calcul semi-classique dans une th{\'e}orie
dont on ne conna{\^\i}t pas la formulation microscopique,
et esp{\'e}rer les appliquer dans les situations o{\`u} les
contraintes de dualit{\'e} et de supersym{\'e}trie sont
insuffisantes. Cette ligne directrice m'a ainsi conduit {\`a}
{\'e}tudier les effets non-perturbatifs dans les couplages 
gravitationnels en $R^4$ des th{\'e}ories de type II {\`a} supersym{\'e}trie
maximale (annexes \ref{pq},\ref{dc} et \ref{nr4}) ; 
les couplages scalaires {\`a} quatre d{\'e}riv{\'e}es
dans les th{\'e}ories de type II {\`a} 16 charges supersym{\'e}triques
(annexe \ref{dds}) ;
et les couplages gravitationnels en $R^2$ dans ces m{\^e}mes th{\'e}ories
(annexe \ref{tt}).
Ces amplitudes physiques pr{\'e}sentent le d{\'e}nominateur commun
de ne recevoir de contributions que de configurations solitoniques
{\it pr{\'e}servant la moiti{\'e} des supersym{\'e}tries}. Cette propri{\'e}t{\'e}
de {\it saturation BPS} va de pair avec la propri{\'e}t{\'e} de 
non renormalisation et simplifie consid{\'e}rablement l'identification
des instantons contribuant {\`a} l'amplitude. Dans ce chapitre, 
nous introduirons et discuterons les r{\'e}sultats obtenus 
dans ces travaux, renvoyant aux articles eux-m{\^e}mes pour la
d{\'e}rivation.

\section{Calcul semi-classique en M-th{\'e}orie}

\subsection{Instantons de surface d'univers et instantons d'espace--temps}
\index{semi-classique!approximation}
La notion d'approximation semi-classique renvoie en g{\'e}n{\'e}ral
{\`a} la limite $\hbar\rightarrow 0$ et aux affres li{\'e}s au
r{\'e}tablissement d'un symbole si {\'e}troitement identifi{\'e} {\`a} 1 et
ses multiples. Dans le cas des th{\'e}ories de jauge, la normalisation
standard de l'action identifie la constante de couplage 
{\`a} $\hbar$, et la limite semi-classique est aussi une limite
de faible couplage. Dans le cas des th{\'e}ories de cordes,
il faut distinguer l'action de la corde, dite
{\it de premi{\`e}re quantification}, de l'action du
{\it champ de corde}, correspondant {\`a} la seconde quantification.
\index{champ de cordes, th{\'e}orie de}
$\hbar$ joue le m{\^e}me r{\^o}le que $\alpha'$ dans  le premier
cas, et le d{\'e}veloppement semi-classique correspond {\`a} un 
d{\'e}veloppement de {\it basse {\'e}nergie}. Il prend en compte
les configurations d'instantons du mod{\`e}le sigma bidimensionnel,
dits {\it instantons de feuille d'univers} 
({\it worldsheet instantons}), dans lesquels la surface
\index{instanton!de feuille d'univers}
d'univers s'enroule autour d'un cycle non trivial de l'espace-cible.
Ces instantons g{\'e}n{\`e}rent des effets d'ordre $e^{-A/\alpha'}$,
o{\`u} $A$ est l'aire du cycle, et peuvent appara{\^\i}tre 
{\it {\`a} tout ordre de la th{\'e}orie de perturbation en $g=e^{\phi}$.}
Le second cas est malheureusement moins explicite, l'action
du champ de corde n'{\'e}tant pas connue. La constante de Planck
doit cependant y appara{\^\i}tre comme diviseur commun, et le
d{\'e}veloppement semi-classique est alors un  d{\'e}veloppement 
autour des solutions classiques de l'{\'e}quation du mouvement
des {\it champs de cordes}. Ces {\'e}quations du mouvement 
se d{\'e}composent en autant d'{\'e}quations que de modes propres
de la corde, et les configurations d'instanton d{\'e}crivent
la configuration de tous ces modes. L'effet des modes massifs 
peut cependant {\^e}tre int{\'e}gr{\'e} de mani{\`e}re {\`a} ne laisser
\index{supermassifs, {\'e}tats}
que la dynamique effective des champs de masse nulle,
\index{action effective}
d{\'e}crite par une action o{\`u} ({\`a} l'ordre des arbres)
le dilaton appara{\^\i}t comme facteur global.  
L'approximation semi-classique correspond donc {\`a} toutes
fins utiles {\`a} une approximation de faible couplage.
Les instantons correspondants, dits {\it instantons d'espace-temps}, 
contribuent aux amplitudes par un facteur $e^{-S}$
o{\`u} $S$ repr{\'e}sente l'action de l'instanton.

\subsection{S{\'e}rie de perturbation et effets non-perturbatifs\label{bigord}}
Les instantons correspondent aux transitions
entre plusieurs vides perturbatifs d'une th{\'e}orie quantique.
\index{vide!transition entre}
Ils sont particuli{\`e}rement familiers dans les th{\'e}ories de Yang-Mills,
o{\`u} ils correspondent aux connexions
de jauge {\it {\`a} courbure auto-duales} $F=*F$ de la th{\'e}orie euclidienne. 
\index{euclidienne, action}
\index{instanton!des th{\'e}ories de jauge}
Ces configurations minimisent l'action euclidienne 
dans le secteur de seconde classe de Chern
\index{Chern, classe de} 
$\frac{1}{8\pi^2}\int \tr F\wedge F=n$,
et induisent des effets non perturbatifs d'ordre $e^{-8\pi ^2 n/g^2}$.
Des instantons analogues apparaissent {\'e}galement en pr{\'e}sence de
gravitation sous la forme d'{\it espaces asymptotiquement localement
euclidiens} ({\it ALE spaces}), de courbure de Riemann auto-duale
\index{ALE, espace|see{instanton gravitationnel}}
\index{instanton!gravitationnel}
(cf. \cite{Eguchi:1980jx} pour une revue d{\'e}taill{\'e}e).
Ils g{\'e}n{\`e}rent des effets du m{\^e}me ordre $e^{-1/g^2}$, o{\`u}
le couplage $1/g^2$ correspond cette fois au couplage gravitationnel
et multiplie l'action de Einstein-Hilbert. Ces effets sont comparables
\index{Einstein-Hilbert, action d'}
{\`a} l'{\it incertitude de la s{\'e}rie perturbative asymptotique}
\index{asymptotique, s{\'e}rie|textit}
aux grands ordres de perturbation. 
\index{serie de perturbation@s{\'e}rie de perturbation@incertitude de la}
Les diagrammes de Feynman {\`a} $l$ boucles 
\index{Feynman, diagramme de}
donnent en effet une contribution d'ordre $C^{-l}l! g^{2l}$
{\`a} l'amplitude totale, g{\'e}n{\'e}rant des p{\^o}les dans le
plan de Borel. Le choix de la prescription de contournement
\index{Borel, plan de}
des p{\^o}les introduit alors apr{\`e}s transformation de Borel 
inverse des incertitudes d'ordre $e^{-C/g^2}$
\footnote{Je remercie V. Rivasseau pour ses explications eclairantes
sur le sujet.}.

Les th{\'e}ories de cordes unifient en particulier les
interactions de jauge et la gravitation avec le m{\^e}me couplage
$g=e^{\phi}$, et doivent donc inclure de tels effets non perturbatifs.
Comme l'a remarqu{\'e} Shenker, le volume de l'espace des modules
des surfaces de Riemann de genre $l$, analogue aux facteurs
\index{espace des modules!des surfaces de Riemann}
\index{surface de Riemann!espace des modules}
combinatoires des diagrammes de Feynman de la th{\'e}orie des
champs, induit un comportement $C^{'-2l}(2l)!g^{2l}$
beaucoup plus explosif de la s{\'e}rie de perturbation
aux grands ordres, conduisant {\`a} une incertitude d'ordre
$e^{-C'/g}$ \cite{Shenker:1990}. 
Bien qu'observ{\'e}s explicitement dans les mod{\`e}les de matrice
des th{\'e}ories de cordes, ces effets instantoniques beaucoup plus importants
{\`a} faible couplage sont rest{\'e}s myst{\'e}rieux jusqu'{\`a} la 
d{\'e}couverte des D-branes
\index{D-brane}\index{D-instanton}
\footnote{L'explication de ces effets ne n{\'e}cessite en r{\'e}alit{\'e}
que le couplage anormal des champs de jauge de Ramond au dilaton,
conduisant aux solitons de supergravit{\'e} de masse $1/g$, et non
la formulation de ces D-branes en termes de th{\'e}ories conformes
bidimensionnelles {\`a} bord.}, 
dont la masse se comporte
pr{\'e}cis{\'e}ment comme $1/g$. Apr{\`e}s rotation de Wick, ces solitons
donnent ainsi lieu {\`a} des instantons dont l'action euclidienne
\index{Wick, rotation de}
\index{action euclidienne}
cro{\^\i}t comme $1/g$. La contribution en $e^{-1/g}$
peut {\'e}galement s'expliquer heuristiquement comme l'exponentiation
de diagrammes de disque d'ordre $g^{-\chi}=1/g$ correspondant
aux surfaces d'univers des cordes ouvertes attach{\'e}es {\`a}
la membrane\footnote{Ces diagrammes restes connexes
en raison du flux de moment d'un disque {\`a} l'autre
{\`a} travers la D-brane.} \cite{Polchinski:1994fq}. 

\subsection{M-th{\'e}orie et approximation semi-classique}
La M-th{\'e}orie est en l'{\'e}tat actuel d{\'e}finie par les th{\'e}ories 
de supercordes qui en donnent diff{\'e}rentes approximations perturbatives.
Elle pr{\'e}sente donc autant d'approximations
semi-classiques que de limites perturbatives, et le choix de telle
ou telle description d{\'e}pend du r{\'e}gime des param{\`e}tres consid{\'e}r{\'e}.
En particulier, les configurations d'instantons diff{\`e}rent 
d'une limite {\`a} l'autre. Elles peuvent {\^e}tre obtenues par 
l'{\'e}tude des solutions classiques des th{\'e}ories de supergravit{\'e}
{\it euclidiennes} d{\'e}crivant les modes de masse nulle de la M-th{\'e}orie
dans la limite consid{\'e}r{\'e}e, bien qu'on ne puisse exclure 
l'existence de solutions mettant en jeu les modes massifs.
Nous avons d{\'e}j{\`a} discut{\'e} les solutions des th{\'e}ories de
supergravit{\'e} {\it minkovskiennes}
du spectre 1/2-BPS des th{\'e}ories de cordes.
Apr{\`e}s rotation de Wick, ces solitons fournissent des
configurations instantoniques de $p$-brane pr{\'e}servant
la moiti{\'e} des supersym{\'e}tries. Ces configurations infiniment {\'e}tendues
poss{\`e}dent cependant une action euclidienne infinie, et ne d{\'e}crivent
donc pas des instantons {\`a} dix dimensions
\footnote{On pourrait imaginer de replier le 
volume d'univers de ces solutions en un volume fini, mais l'action
ne serait alors plus extr{\'e}male sous les contractions de ce volume.}.
Seule la supergravit{\'e} de type IIB pr{\'e}sente des configurations
instantoniques localis{\'e}es d'action euclidienne finie, correspondant
{\`a} une excitation de la m{\'e}trique et du scalaire de Ramond $\axion$.
Cette configuration est d{\'e}crite par la {\it D(-1)-brane}, ou 
{\it D-instanton}, de la th{\'e}orie de type IIB
\index{D-instanton!de type IIB}
\cite{Green:1991et}. Son action euclidienne
\begin{equation}
\label{sdi}
S_{D(-1)}=e^{-\phi} + i \axion
\end{equation}
peut {\^e}tre vue comme un cas d{\'e}g{\'e}n{\'e}r{\'e} de l'action de Born-Infeld
\index{Born-Infeld, action de}
(\ref{abi}).

Apr{\`e}s compactification en revanche, il devient possible de
stabiliser les configurations euclidiennes de $p$-branes
en les {\it enroulant} sur certains cycles non triviaux de la vari{\'e}t{\'e}
\index{cycle d'homologie!de la vari{\'e}t{\'e} de compactification}
\index{enroulement!des solitons en instantons|textit}
de compactification. On obtient ainsi des instantons 
d'action euclidienne finie en dimension inf{\'e}rieure. La pr{\'e}servation
d'une partie des supersym{\'e}tries impose les m{\^e}mes conditions 
g{\'e}om{\'e}triques sur le cycle que dans le cas des solitons BPS
\index{etats BPS@{\'e}tats BPS}
(section \ref{susycyc}).
Les D-branes donnent ainsi lieu {\`a} des configurations 
d'action euclidienne donn{\'e}e par l'action de Born-Infeld
\begin{equation}
S_{Dp}=\int d^{p+1}x~ e^{-\phi} \sqrt{\hat g +\hat B+ F}+ 
i \mathcal{R}e^{\hat B+F}\ ,
\end{equation}
faisant intervenir la courbure du champ de jauge du volume d'univers
de la D-brane, tandis que l'action associ{\'e}e {\`a} la 5-brane de Neveu-Schwarz 
\index{cinq-brane!de Neveu-Schwarz}
enroul{\'e}e sur un 6-cycle s'{\'e}crit
\begin{equation}
\label{actns5}
S_{NS5}=\int d^{6}x~e^{-2\phi} \sqrt{\hat g} +
i \hat B_6
\end{equation}
o{\`u} $B_6$ est le tenseur de jauge dual au tenseur de
Neveu-Schwarz $B_{\mu\nu}$ {\`a} dix dimensions sous la dualit{\'e} de
Poincar{\'e}.
\index{dualit{\'e}!de Poincar{\'e}}
La th{\'e}orie de type I recevra ainsi des contributions 
en $e^{-1/g}$ des D1- et D5-branes,
tandis que les th{\'e}ories de type IIA et IIB recevront des
contributions des D-branes de dimension paire et impaire respectivement ;
toutes les th{\'e}ories de supercordes pourront en outre recevoir des
contributions de NS5-brane
d'ordre $e^{-1/g^2}$. En particulier, les cordes h{\'e}t{\'e}rotiques
$SO(32)$ et $E_8\times E_8$
ne pr{\'e}sentant pas de D-brane dans leur spectre solitonique,
seules des corrections en $e^{-1/g^2}$ seraient attendues
\footnote{L'argument de Shenker de la 
section \ref{bigord} s'applique cependant tout autant {\`a} la corde
h{\'e}t{\'e}rotique, et de fait la dualit{\'e} h{\'e}t{\'e}rotique-type I
transforme les instantons de surface d'univers de type I
en des effets en $e ^{-1/g}$ de la corde h{\'e}t{\'e}rotique, dont
l'interpr{\'e}tation reste myst{\'e}rieuse. Il serait tr{\`e}s int{\'e}ressant
de disposer d'un exemple explicite de tels effets.}. La corde h{\'e}t{\'e}rotique
\index{supercordes, th{\'e}orie des!h{\'e}t{\'e}rotiques}
appara{\^\i}t ici encore comme la description perturbative <<la moins
incompl{\`e}te>> de la M-th{\'e}orie, si toutefois cette conjecture
s'av{\`e}re correcte. Encore ces instantons n'{\'e}puisent-ils
que le spectre des configurations pr{\'e}servant la moiti{\'e}
des supersym{\'e}tries, auxquelles il faut encore ajouter les configurations
de $p$-branes intersectantes, et nombre d'objets non d{\'e}crits
par l'ansatz (\ref{pansatz}). 

\subsection{Instantons et saturation BPS}
Les contraintes de supersym{\'e}trie permettent heureusement 
de contr{\^o}ler cette prolif{\'e}ration. Chaque classe de configuration
instantonique admet en effet des d{\'e}formations continues
param{\'e}tr{\'e}es par des {\it coordonn{\'e}es collectives},
\index{coordonn{\'e}e collective}
sur lesquelles les fonctions de corr{\'e}lation doivent {\^e}tre
int{\'e}gr{\'e}es \cite{Gervais:1975yg}:
\begin{equation}
\langle \prod{V_\phi} \rangle = 
\sum_{\mbox{secteurs}} \int d\mu 
~\langle \prod V_\phi \rangle_I~ \frac{1}{\sqrt{\det Q'}} e^{-S_I} 
\end{equation}
$S_I$ d{\'e}note l'action de l'instanton,
$Q$ la forme quadratique d{\'e}crivant les fluctuations gaussiennes
autour de l'instanton orthogonales aux modes z{\'e}ros ;
les $V_{\phi}$ repr{\'e}sentent les op{\'e}rateurs de vertex intervenant
dans la fonction de corr{\'e}lation, et d{\'e}pendent de l'amplitude
\index{op{\'e}rateur de vertex}
physique consid{\'e}ree. Une coordonn{\'e}e collective est en
particulier associ{\'e}e {\`a} chaque sym{\'e}trie bris{\'e}e
\index{instanton!sym{\'e}trie bris{\'e}e par}
par la configuration solitonique\footnote{Dans le cas
des th{\'e}ories de Yang-Mills, les instantons brisent
l'invariance par translation et par dilatation ;
ils pr{\'e}sentent par cons{\'e}quent des coordonn{\'e}es
collectives de position globale et de taille.}, bien que 
d'autres degr{\'e}s de libert{\'e} puissent {\^e}tre pr{\'e}sents.
En particulier, chaque charge supersym{\'e}trique bris{\'e}e
par l'instanton engendre une coordonn{\'e}e collective
{\it grassmannienne}, ou {\it mode z{\'e}ro fermionique}. 
\index{mode z{\'e}ro!fermionique}
L'int{\'e}gration conduit donc
{\`a} un r{\'e}sultat nul, {\`a} moins que les vertex
$V_\phi$ {\'e}valu{\'e}es autour de la configuration solitonique
ne {\it saturent} tous les modes z{\'e}ros fermioniques.  
Gr{\^a}ce {\`a} la supersym{\'e}trie, on peut toujours se ramener
au cas o{\`u} tous les op{\'e}rateurs de vertex sont fermioniques
\footnote{La supersym{\'e}trie $\delta_\epsilon \psi = \partial_\mu
\phi \gamma^\mu \epsilon$ convertit une d{\'e}riv{\'e}e bosonique
en un bilin{\'e}aire fermionique.}. Ces op{\'e}rateurs sont nuls 
dans la configuration bosonique de r{\'e}f{\'e}rence, mais 
d{\'e}pendent {\it lin{\'e}airement} au premier ordre
en les coordonn{\'e}es fermioniques grassmanniennes. 
{\it Le nombre d'insertions
$V_\phi$ doit donc {\it au moins} {\'e}galer le nombre de charges
supersym{\'e}triques bris{\'e}es par l'instanton.} 

Cette r{\`e}gle de s{\'e}lection est {\`a} la base de th{\'e}or{\`e}mes
de non renormalisation tr{\`e}s importants
\index{non renormalisation}
\footnote{Il s'agit ici de th{\'e}or{\`e}mes de non renormalisation
{\it semi-classique}. Ils ne pr{\'e}sagent pas de l'existence
de corrections non perturbatives d'autre nature, bien que
de telles corrections n'aient pas {\'e}t{\'e} observ{\'e}es.}.
En effet, les instantons
brisent au minimum la moiti{\'e} des $N$ charges supersym{\'e}triques.
Les interactions correspondant {\`a} $n_f < N/2$ 
vertex fermioniques ne re{\c c}oivent donc pas de contributions
instantoniques, tandis que celles correspondant {\`a} $n_f = N/2$
ne re{\c c}oivent de contributions {\it que des instantons BPS
pr{\'e}servant la moiti{\'e} des supersym{\'e}tries}, ou 1/2-BPS satur{\'e}s. 
Plus g{\'e}n{\'e}ralement, les interactions correspondant
{\`a} $n_f <N$ peuvent {\^e}tre corrig{\'e}s par des effets
d'instantons BPS pr{\'e}servant $N-n_f$ supercharges.
Cette supersym{\'e}trie r{\'e}siduelle suffit {\`a} garantir la compensation
entre les modes bosoniques et fermioniques autour de l'instanton,
soit $\det Q'=1$.
Les amplitudes v{\'e}rifiant $n_f<N$, ainsi que les termes de l'action effective
auxquels elles correspondent, sont dits {\it BPS satur{\'e}es}.
\index{saturation BPS|textit}
Cette propri{\'e}t{\'e} permet {\'e}galement de d{\'e}terminer le {\it nombre}
d'instantons contribuant. La stabilit{\'e} neutre des configurations
BPS permet en effet de mettre en pr{\'e}sence un nombre arbitraire
d'instantons BPS, dont les positions relatives sont d{\'e}crites
par des coordonn{\'e}es collectives suppl{\'e}mentaires, ainsi que
leurs partenaires supersym{\'e}triques grassmanniennes. Les modes
z{\'e}ros d'une telle configuration ne sont alors plus satur{\'e}s,
et la contribution {\`a} l'amplitude s'annule. Cet argument peut
{\^e}tre mis en d{\'e}faut aux points singuliers de l'espace des
\index{espace des modules!singularit{\'e}}
modules des instantons o{\`u} la dimension de l'espace tangent change,
et on ne peut donc exclure les contributions d'instantons BPS
co{\"\i}ncidant \cite{Bachas:1997xn}; 
ceux-ci sont cependant en g{\'e}n{\'e}ral indistinguables
des instantons de charge multiple.

Ainsi, {\it dans le cas des th{\'e}ories de supersym{\'e}trie maximale}
soit {\`a} $N=32$ supercharges, les premiers effets instantoniques
peuvent se manifester dans les couplages $1/2$-BPS satur{\'e}s {\`a}
16 fermions, reli{\'e}s par supersym{\'e}trie aux couplages {\`a} 
8 d{\'e}riv{\'e}es. Le seul multiplet {\'e}tant le multiplet gravitationnel,
ces couplages correspondent aux {\it interactions en $R^4$}, o{\`u} la 
contraction des indices de Lorentz n'est pas sp{\'e}cifi{\'e}e.
\index{non renormalisation!de l'action en $R^4$
dans type II $N=8$}
Ces couplages font l'objet des publications en appendices \ref{pq},
\ref{dc} et \ref{nr4}.
Les couplages en $R^6$ sont quant {\`a} eux $1/4$-BPS satur{\'e}s,
et les couplages en $R^8$ re{\c c}oivent {\it a priori} des
contributions des instantons non supersym{\'e}triques.
Dans le cas des th{\'e}ories {\`a} 16 supercharges, le r{\^o}le des
couplages $1/2$-BPS satur{\'e}s est jou{\'e} par les interactions
en $F^4$ et $R^2$, cette derni{\`e}re faisant l'objet de la 
publication en appendice \ref{dds}. Finalement, dans les th{\'e}ories
{\`a} 8 supercharges, l'action {\`a} deux d{\'e}riv{\'e}es elle-m{\^e}me
est corrig{\'e}e, et le potentiel scalaire dans les th{\'e}ories
{\`a} 4 supercharges. Dans ces deux derniers cas, la structure
holomorphe de la g{\'e}om{\'e}trie des multiplets vectoriels $N=2$
et du superpotentiel $N=1$ respectivement interdit 
les corrections en $e^{-1/g}=e^{-\sqrt{S-\bar S}/2i}$.
Cette restriction ne s'applique cependant pas {\`a}
la {\it g{\'e}om{\'e}trie des hypermultiplets} et au 
\index{hypermultiplet}
\index{Kahler, potentiel de@K{\"a}hler, potentiel de}
{\it potentiel de K{\"a}hler} respectivement. Elle ne limite
pas non plus les corrections de NS5-brane
en $e^{-1/g^2 + i a } = e^{-S}$ dans aucun de ces termes. 

\section{Instantons h{\'e}t{\'e}rotiques de 5-brane}
Les premiers exemples explicites de corrections non-perturbatives 
en th{\'e}orie des cordes ont {\'e}t{\'e} obtenus dans l'{\'e}tude de la m{\'e}trique
des multiplets vectoriels dans les compactifications $N=2$ de la 
corde h{\'e}t{\'e}rotique sur $K_3\times T^2$ ; ils sont bas{\'e}s sur la
conjecture de dualit{\'e} h{\'e}t{\'e}rotique-type IIA formul{\'e}e par Kachru
\index{dualit{\'e}!h{\'e}t{\'e}rotique - type II}
et Vafa dans le cadre des th{\'e}ories $N=2$ de type II
\cite{Kachru:1995wm}, et en fournissent
une v{\'e}rification cruciale (voir par exemple \cite{Partouche:1997}). 
Un ingr{\'e}dient essentiel {\`a} la 
d{\'e}monstration est la propri{\'e}t{\'e} de {\it d{\'e}couplage
des multiplets vectoriels et des hypermultiplets}. Le dilaton
\index{decouplage@d{\'e}couplage!des hypers et vecteurs}
appartient {\`a} un {\it hypermultiplet} dans la th{\'e}orie de type II, de sorte que
\index{dilaton}
\index{hypermultiplet}
{\it la m{\'e}trique des multiplets vectoriels ne peut recevoir
aucune correction perturbative ou non perturbative du cot{\'e}
de type IIA}. Elle est donc donn{\'e}e {\it exactement} par
la contribution {\`a} l'ordre des arbres
\footnote{En pratique, les instantons de surface d'univers
apparaissant dans la m{\'e}trique {\`a} l'ordre des arbres sont calcul{\'e}s
classiquement par {\it sym{\'e}trie miroir} {\`a} partir de la th{\'e}orie
\index{miroir, sym{\'e}trie}
de type IIB.}.
Ce th{\'e}or{\`e}me de non renormalisation ne s'applique
pas du c{\^o}t{\'e} h{\'e}t{\'e}rotique, o{\`u} le dilaton appartient {\`a} un
multiplet vectoriel. On peut appliquer la transformation de dualit{\'e}
et traduire la m{\'e}trique exacte des multiplets vectoriels de type II 
dans les variables
h{\'e}t{\'e}rotiques. Les effets d'{\it instantons de surface d'univers}
en $e^{-1/\alpha'}$ s'interpr{\`e}tent alors comme des 
{\it instantons d'espace-temps} en $e^{-1/g^2}$ de la th{\'e}orie
\index{instanton!de feuille d'univers}
h{\'e}t{\'e}rotique. Les complications g{\'e}om{\'e}triques des compactifications
de type II sur espace de Calabi-Yau et h{\'e}t{\'e}rotiques sur une
\index{Calabi-Yau, vari{\'e}t{\'e} de}
\index{K3, surface@$K_3$, surface}
vari{\'e}t{\'e} $K_3$ ont cependant emp{\^e}ch{\'e} jusqu'{\`a} pr{\'e}sent une
interpr{\'e}tation de ces corrections en termes de NS5-branes
enroul{\'e}es sur $K_3\times T^2$.

Les contributions des NS5-branes peuvent {\^e}tre plus facilement
analys{\'e}es dans le cas de la compactification toro{\"\i}dale
\index{compactification!toro{\"\i}{}dale}
de la corde h{\'e}t{\'e}rotique. On obtient ainsi une th{\'e}orie $N=4$
{\`a} quatre dimensions, dont l'action effective {\`a} deux d{\'e}riv{\'e}es
ne re{\c c}oit pas de corrections quantiques perturbatives ou
non perturbatives. En revanche, les corrections gravitationnelles
en $R^2$ correspondent {\`a} des couplages 1/2-BPS satur{\'e}s 
et peuvent recevoir des corrections de NS5-brane. Ces corrections
ont {\'e}t{\'e} d{\'e}termin{\'e}es par Harvey et Moore en utilisant
la dualit{\'e} entre la corde h{\'e}t{\'e}rotique compactifi{\'e}e sur
$T^6$ et la corde de type II compactifi{\'e}e sur $K_3\times T^2$
\cite{Harvey:1996ir}.
Le couplage en $R^2$ est en effet le premier 
d'une suite d'{\it amplitudes topologiques} $\mathcal{F}_g R^2 F^{2g-2}$
\index{amplitude de diffusion!topologique}
de la th{\'e}orie de type II,
o{\`u} $F$ represente ici la courbure d'un champ de jauge du
multiplet gravitationnel
\cite{Antoniadis:1994ze}. Ces amplitudes peuvent {\^e}tre {\'e}valu{\'e}es
dans le cadre de la {\it th{\'e}orie topologique} reli{\'e}e {\`a}
la th{\'e}orie conforme bidimensionnelle sur $K_3$ 
par un {\it twist}, et un comptage
des modes z{\'e}ros sur la surface d'univers montre qu'elles
\index{mode z{\'e}ro!sur la surface d'univers}
ne re{\c c}oivent de contributions que de surfaces de Riemann de 
genre $g$. Ce th{\'e}or{\`e}me de non renormalisation ne vaut
bien entendu qu'au niveau perturbatif. Il peut {\^e}tre {\'e}tendu
au niveau non perturbatif en notant que dans le cas des th{\'e}ories
$N=2$, les amplitudes $\F_g$ ne d{\'e}pendent que des multiplets
vectoriels, et ne re{\c c}oivent donc pas de corrections m{\^e}me
\index{vectoriel, multiplet}
non perturbatives. De fait, le calcul de l'amplitude 
$\F_1 R^2$ {\`a} une boucle en type IIA donne 
\begin{equation}
\label{r2iia}
\F_1^{IIA}= - \log \left( T_2 |\eta(T)|^4 \right)\ ,
\end{equation}
o{\`u} $T$ d{\'e}note le module de K{\"a}hler du tore $T^2$.
\index{Kahler, module de@K{\"a}hler, module de}
Cette expression est en elle-m{\^e}me invariante sous le groupe
de U-dualit{\'e} $Sl(2,\Zint)_T \times SO(6,22,\Zint)$, et
\index{U-dualit{\'e}!de Het/type II $N=4$}
ne requiert donc pas de contributions perturbatives ou 
non perturbatives suppl{\'e}mentaires. L'ind{\'e}pendance
de $\F_1^{IIA}$ sur les modules de $K_3$ traduit
le caract{\`e}re topologique de cette amplitude. Exprim{\'e}e
en terme des variables h{\'e}t{\'e}rotiques, elle devient
\begin{equation}
\F_1^{het}= - \log \left( S_2 |\eta(S)|^4 \right) \ ,
\end{equation}
o{\`u} $S=a + i V/g^2$ est le scalaire complexe d{\'e}crivant
le couplage de la th{\'e}orie h{\'e}t{\'e}rotique compactifi{\'e}e sur un tore
$T^6$ de volume $V$. Le d{\'e}veloppement {\`a} faible couplage
de cette amplitude 
\begin{equation}
\label{f1het}
\F_1^{het}= \frac{\pi}{3} S_2 + \sum_{N\ne 0} C(N)e^{iNS} - \log S_2
\end{equation}
reproduit la contribution {\`a} l'ordre des arbres 
$\F_{1;pert}^{het}=\frac{\pi}{3}S_2$, mais montre l'existence
d'une s{\'e}rie de contributions non perturbatives
\index{serie d'instantons@s{\'e}rie d'instantons!couplage $R^2$ de Het
  $N=4$}
d'ordre $e^{-1/g^2}$\footnote{L'{\'e}quation (\ref{f1het}) montre
{\'e}galement un terme $-\log S_2$ traduisant l'existence d'une
divergence logarithmique {\`a} fort couplage. Ce ph{\'e}nom{\`e}ne,
\index{divergence logarithmique en couplage}
que l'on retrouvera dans la suite, n'a pas encore 
re{\c c}u d'interpr{\'e}tation satisfaisante.}.
La comparaison du poids $e^{inS}$ de ces termes 
avec l'{\'e}quation (\ref{actns5}) permet d'identifier
ces contributions avec celle de $n$ NS5-branes euclidiennes
dont les six directions sont enroul{\'e}es autour
du tore\footnote{ou alternativement, d'une seule NS5-brane
enroul{\'e}e $n$ fois autour du tore. Le cas $n<0$ correspond {\`a} une
anti-brane, ou une brane d'orientation oppos{\'e}e {\`a} 
\index{anti-brane}
l'orientation de $T^6$.}. L'ind{\'e}pendance de $\F_1^{het}$
sur les modules du r{\'e}seau $\Gamma_{6,22}$ traduit
la neutralit{\'e} de la 5-brane vis-{\`a}-vis du groupe
de jauge h{\'e}t{\'e}rotique. L'extension de ce raisonnement
aux th{\'e}ories $N=4$ {\it de rang r{\'e}duit} fait
l'objet de la publication en appendice \ref{dds}.
Il montre en particulier l'existence d'une sym{\'e}trie $N=8$ 
restaur{\'e}e dans la limite de fort couplage et grand volume
de certaines th{\'e}ories h{\'e}t{\'e}rotiques (figure
\ref{tt:f1} page \pageref{tt:f1}).
Nous reviendrons {\'e}galement plus loin sur une autre
amplitude, cette fois exacte du c{\^o}t{\'e} h{\'e}t{\'e}rotique,
mais correspondant {\`a} des effets de D-brane 
du c{\^o}t{\'e} de type II.

\section{Instantons de type II et D-branes}
Dans la section pr{\'e}c{\'e}dente, nous avons d{\'e}crit des
effets non perturbatifs de la corde h{\'e}t{\'e}rotique
interpr{\'e}t{\'e}s en termes de NS5-branes neutres
enroul{\'e}es autour du tore $T^6$. La connaissance tr{\`e}s
imparfaite de la description des NS5-branes ne nous
a permis d'en donner une interpr{\'e}tation pr{\'e}cise
que dans le cas simple d'une compactification
toro{\"\i}dale sur $T^6$.  La dynamique des D-branes 
de type I et II est en revanche bien comprise
dans le cadre de l'action de Born-Infeld, et la
signature en $e^{-1/g}$ de ces effets en permet
une identification beaucoup plus s{\^u}re. 
La vari{\'e}t{\'e} de leurs dimensionnalit{\'e}s permet {\'e}galement
d'{\'e}tudier comment elles rentrent en jeu successivement
au cours de la compactification sur des tores de
dimension croissante. Avant de montrer ce point,
nous commencerons cependant par discuter bri{\`e}vement
le premier exemple explicite de corrections de 
D-branes construit par Ooguri et Vafa dans le cadre
des compactifications de type II sur espace de
Calabi-Yau. 

\section{D-instantons et g{\'e}om{\'e}trie des hypermultiplets}
La r{\'e}solution par Strominger
de la singularit{\'e} de {\it conifold}
\index{singularit{\'e}!de conifold}
a constitu{\'e} l'un des premiers succ{\`e}s de l'application
des id{\'e}es de Seiberg et Witten en th{\'e}orie des
cordes \cite{Strominger:1995qi}. 
Cette singularit{\'e} appara{\^\i}t dans la m{\'e}trique
des multiplets vectoriels d{\'e}crivant la structure
\index{vectoriel, multiplet}
complexe de l'espace de Calabi-Yau $K$ dans le cas
\index{Calabi-Yau, vari{\'e}t{\'e} de}
de la compactification $N=2$ de la th{\'e}orie de type IIB
{\`a} quatre dimensions, lorsqu'un 3-cycle $\gamma$ de
$K$ d{\'e}g{\'e}n{\`e}re. Elle se traduit par une divergence
logarithmique dans le pr{\'e}potentiel {\`a} l'ordre des arbres lorsque la 
\index{pr{\'e}potentiel}
p{\'e}riode $z=\int_\gamma \Omega$ s'annule. 
La m{\'e}trique des multiplets
vectoriels de type II {\'e}tant exacte {\`a} l'ordre des arbres,
cette singularit{\'e} ne peut {\^e}tre {\'e}limin{\'e}e par
les corrections quantiques, et doit donc avoir une origine
physique : Strominger a montr{\'e} qu'elle correspondait
{\`a} l'int{\'e}gration d'un {\'e}tat non perturbatif devenant
de masse nulle lorsque $z\rightarrow 0$ ; cet {\'e}tat n'est autre que
la D3-brane enroul{\'e}e autour du cycle {\'e}vanescent $\gamma$
\index{cycle d'homologie!evanescent@{\'e}vanescent}
\cite{Strominger:1995qi}.

Par sym{\'e}trie miroir, il existe donc {\'e}galement une singularit{\'e}
logarithmique dans la m{\'e}trique perturbative des {\it hypermultiplets}
de la th{\'e}orie de type IIA sur la vari{\'e}t{\'e} miroir
\index{miroir, sym{\'e}trie}
$\tilde K$. Il n'existe cependant pas de D3-brane
dans la th{\'e}orie de type IIA pour expliquer cette
singularit{\'e}, et du reste la m{\'e}trique des hypermultiplets
\index{hypermultiplet}
n'est plus prot{\'e}g{\'e}e des corrections quantiques.
Ooguri et Vafa ont montr{\'e} que les corrections instantoniques
de {\it D2-brane} enroul{\'e}es autour du cycle {\'e}vanescent
$\gamma$ pouvaient r{\'e}soudre la singularit{\'e},
et ont donn{\'e} une expression explicite pour ces corrections
dans le cas id{\'e}alis{\'e} o{\`u} la vari{\'e}t{\'e} des hypermultiplets
se restreignait {\`a} une vari{\'e}t{\'e} hyperk{\"a}hl{\'e}rienne de
dimension 4 \cite{Ooguri:1996me}. 
\index{vari{\'e}t{\'e}!hyperk{\"a}hlerienne}
Ce cas correspond {\`a} la limite de {\it double
scaling} $z\rightarrow 0, g\rightarrow 0, z/g=\mbox{cte}$.
Les quatre scalaires de l'hypermultiplet correspondent
alors {\`a} la p{\'e}riode $z\in \CC$ et deux variables p{\'e}riodiques r{\'e}elles
$x,t$  d{\'e}crivant la valeur moyenne
du tenseur de Ramond $\mathcal{C}_3$ sur $\gamma$ et sur
son dual. La m{\'e}trique est classiquement invariante
par translation selon $t$ et $x$, mais les instantons 
enroul{\'e}s sur $\gamma$ brisent l'invariance de translation selon $x$
{\`a} un sous groupe discret $\Zint$.
Une m{\'e}trique hyperk{\"a}hlerienne de dimension 4 avec une isom{\'e}trie
admet la forme g{\'e}n{\'e}rale\footnote{Cet ansatz appara{\^\i}{}t {\'e}galement
dans l'{\'e}tude des instantons gravitationnels, et d{\'e}finit les
\index{instanton!gravitationnel}
\index{Eguchi-Hanson, espace d'}
espaces d'Eguchi-Hanson, $V=\sum_k 1/4\pi\|x-x_k\|$ et de Taub-NUT,
$V=1+\sum_k 1/4\pi\|x-x_k\|$ (voir \cite{Eguchi:1980jx}). La solution de
Ooguri et Vafa resurgira dans le chapitre 5.}
\begin{equation}
ds^2 = \frac{1}{V} (dt - A_i dy^i)^2 + V (dy^i)^2
\end{equation}
o{\`u} le potentiel $V(y)$ et le vecteur $A$ v{\'e}rifient
les conditions
\begin{equation}
V^{-1} \Delta V =0, \nabla V = \nabla \wedge A
\end{equation}
$V$ et $A_i$ peuvent donc {\^e}tre interpr{\'e}t{\'e}s comme des potentiels
{\'e}lectrostatique et magn{\'e}tostatique respectivement, cr{\'e}es par la m{\^e}me
distribution de charges, axisym{\'e}trique puisque la phase de
$z$ ne doit pas intervenir.
Les effets instantoniques {\'e}tant {\'e}limin{\'e}s
pour $|z|\rightarrow \infty$, $V$ v{\'e}rifie la condition
aux limites
\begin{equation}
V\rightarrow  -\frac{1}{4\pi} \log |z|^2
\end{equation}
Ces conditions d{\'e}terminent alors $V$ {\`a} toute distance par
\begin{equation}
V=\frac{1}{4\pi} \sum_{n=-\infty}^{\infty}
\left( \frac{1}{\sqrt{(x-n)^2+|z|^2/g}} - \frac{1}{|n|}\right)~+\mbox{cte}
\end{equation}
correspondant {\`a} une distribution de charges ponctuelles
r{\'e}guli{\`e}rement espac{\'e}es le long de l'axe $Ox$.
Les p{\^o}les simples en $(z=0,x=n)$ ne correspondent {\`a} aucune
singularit{\'e} de la m{\'e}trique
\footnote{Ce ne serait plus le cas si les charges ponctuelles
avaient une valeur double, tout comme la m{\'e}trique en
coordonn{\'e}es polaires $ds^2=dr^2 + \alpha r^2 d\theta^2$
pr{\'e}sente une singularit{\'e} conique lorsque $\alpha\ne 1$.}
, et la singularit{\'e} de
conifold dispara{\^\i}t donc sous les corrections d'instantons.
Le comportement {\`a} faible couplage peut {\^e}tre exhib{\'e}
par resommation de Poisson sur l'entier $n$ :
\index{Poisson, resommation de}
\begin{equation}
\label{vof}
V = -\frac{1}{4\pi} \log \frac{|z|^2}{\mu^2}
+ \sum_{m\ne 0} \frac{1}{2\pi} 
K_0\left(2\pi e^{-\phi} |mz| \right)
e ^{2\pi i m x}
\end{equation}
o{\`u} $\mu$ est une constante arbitraire,
et redonne la contribution logarithmique perturbative, 
ainsi qu'une {\it somme d'effets non perturbatifs} exprim{\'e}s
en terme de la fonction de Bessel $K_0$~; les fonctions de Bessel
\index{Bessel, fonction de|textit}
$K_\nu$
se comportent {\`a}  grand argument comme
\footnote{On trouvera dans l'appendice de la publication
en annexe \ref{pq} un rappel des propri{\'e}t{\'e}s des fonctions
$K_\nu(z)$, ainsi qu'une m{\'e}thode commode pour
effectuer ce type de resommation de Poisson~; la s{\'e}rie
dans l'{\'e}quation (\ref{k0}) se termine {\`a} $k=\nu-\frac{1}{2}$
lorsque $\nu$ est demi-entier.}
\begin{equation}
\label{k0}
K_{\nu}(z)= \sqrt{\frac{\pi}{2z}} e^{-z} \left(1
+ \sum_{k=1}^{\infty} \frac{1}{(2z)^k} 
\frac{\Gamma\left(\nu+k+\frac{1}{2}\right)}
{k!\Gamma\left(\nu-k+\frac{1}{2}\right)}  
\right)\ .
\end{equation}
\index{serie d'instantons@s{\'e}rie d'instantons!dans la g{\'e}om{\'e}trie
  des hypers}
L'action euclidienne associ{\'e}e {\`a} ces effets
s'{\'e}crit donc $e^{-2\pi m \left( \frac{|z|}{g} \pm i x\right)}$,
soit pr{\'e}cis{\'e}ment l'action euclidienne associ{\'e}e {\`a} une
D3-brane enroul{\'e}e $m$ fois autour du 3-cycle de volume $|z|$ et
de flux de Ramond $x$. Les corrections $O(1/z)$ {\`a} l'approximation
(\ref{k0}) traduisent l'existence de corrections perturbatives
autour de la configuration d'instanton. L'absence de corrections
perturbatives au terme logarithmique dans le vide trivial 
suppose en revanche un {\it th{\'e}or{\`e}me de non renormalisation
perturbative au-del{\`a} d'une boucle} de la m{\'e}trique des hypermultiplets.

Ce r{\'e}sultat 
constitue {\`a} ma connaissance le premier exemple de calcul
explicite de corrections de D-branes {\`a} une amplitude physique.
Il pr{\'e}sente bon nombre de ph{\'e}nom{\`e}nes que nous retrouverons
{\`a} l'oeuvre dans la suite de ce m{\'e}moire. Il ne s'applique
cependant qu'{\`a} une situation <<id{\'e}alis{\'e}e>>, 
et fait une utilisation cruciale des
contraintes g{\'e}om{\'e}triques des vari{\'e}t{\'e}s hyperk{\"a}hl{\'e}riennes
de dimension 4, lesquelles n'ont pas d'{\'e}quivalent en dimension
sup{\'e}rieure. Dans les sections suivantes, nous consid{\`e}rerons des
amplitudes physiques pour lesquelles les contraintes de dualit{\'e}
permettent de donner des r{\'e}sultats {\it exacts}, et {\'e}tudierons
leur interpr{\'e}tation en termes de somme d'instantons.

\section{D-instantons et compactification toro{\"\i}dale}
L'{\'e}tude des compactifications toro{\"\i}{}dales de la M-th{\'e}orie
permet de s'affranchir des complications alg{\'e}briques li{\'e}es {\`a}
la d{\'e}termination des cycles supersym{\'e}triques des espaces
de Calabi-Yau pour se concentrer sur les contributions propres
des D-branes. Le groupe de U-dualit{\'e} apparaissant dans ces
th{\'e}ories {\`a} supersym{\'e}trie maximale est en outre assez
contraignant pour permettre une d{\'e}termination exacte de
certains couplages, dont on peut alors analyser les contributions
non perturbatives. Nous nous int{\'e}resserons en particulier
aux couplages en $R^4$, qui sont 1/2-BPS satur{\'e}s dans ces
\index{saturation BPS}
th{\'e}ories {\`a} 32 charges supersym{\'e}triques\footnote{Des 
couplages en $R^4$ ont {\'e}galement {\'e}t{\'e} {\'e}tudi{\'e}s
dans les th{\'e}ories h{\'e}t{\'e}rotiques et de type I,
o{\`u} ils sont reli{\'e}s par supersym{\'e}trie {\`a} des termes 
d'anomalie et donc non renormalis{\'e}s au-del{\`a} d'une 
boucle \cite{Bachas:1997mc,Kiritsis:1997hf}~; des couplages reli{\'e}s
par supersym{\'e}trie aux couplages en $R^4$ ont {\'e}t{\'e}
consid{\'e}r{\'e}s par \cite{Kehagias:1997cq,Green:1997me}.}, et qui sont
en outre invariants sous la U-dualit{\'e}, cette
derni{\`e}re pr{\'e}servant la m{\'e}trique dans le r{\'e}f{\'e}rentiel
d'Einstein. La M-th{\'e}orie compactifi{\'e}e sur un cercle est d{\'e}crite
{\`a} faible rayon par la th{\'e}orie des cordes de type IIA
{\`a} dix dimensions. Cette derni{\`e}re n'admet pas de
configurations d'instantons 1/2-BPS, et on s'attend donc 
{\`a} ce que le couplage en $R^4$ soit donn{\'e} par sa
valeur perturbative. Le groupe de U-dualit{\'e} de la th{\'e}orie de type
IIA non compactifi{\'e}e
est cependant trivial, et il n'est pas clair {\`a} ce stade 
quels ordres de la th{\'e}orie de perturbation contribueront
{\`a} ce couplage. La th{\'e}orie de type IIB {\`a} dix dimensions
pr{\'e}sente en revanche une sym{\'e}trie non perturbative 
$Sl(2,\Zint)_B$ qui nous permettra, {\`a} la suite de Green
et Gutperle, de d{\'e}terminer les contributions perturbatives
ainsi que la contribution des D(-1)-instantons~; nous pourrons alors
en d{\'e}duire le couplage de type IIA apr{\`e}s compactification
{\`a} 9 dimensions sur un cercle suivie d'une T-dualit{\'e}.
La poursuite de ce raisonnement pour des compactifications
nous permettra de d{\'e}terminer les contributions des
D-branes de toutes dimensions {\`a} ces couplages.

\subsection{Couplages $R^4$ dans la th{\'e}orie de type IIB non compactifi{\'e}e}
Le couplage $R^4$ {\`a} l'ordre des arbres et {\`a} une boucle
dans l'action effective peut
{\^e}tre ais{\'e}ment d{\'e}termin{\'e} en calculant l'amplitude de
diffusion {\`a} quatre gravitons
\index{amplitude de diffusion!{\`a} quatre gravitons en type II}
\footnote{Il est techniquement plus commode de calculer
l'amplitude {\`a} quatre gravitons et un module dans la
th{\'e}orie compactifi{\'e}e sur $T^2$.} et en isolant la contribution
dominante {\`a} basse {\'e}nergie, soit huit puissances des moments
externes. On obtient ainsi, dans la m{\'e}trique des cordes,
\begin{equation}
  \label{r410d}
S_{R^4} =\int d^{10}x\sqrt{-g} \left[ 2\zeta(3)e^{-2\phi} 
(t_8t_8 + \frac{1}{8} \epsilon_{10}\epsilon_{10}) R^4
+ \frac{2\pi^2}{3} (t_8t_8 + (-)^{\mu}
\frac{1}{8} \epsilon_{10}\epsilon_{10}) R^4
\right]
\end{equation}
o{\`u} $\mu=1$ (resp. $0$) en type IIB (resp. IIA). Les tenseurs
$t_8$ et $\epsilon_{10}$ d{\'e}signe les contractions particuli{\`e}res
des indices du tenseur de Riemann provenant des contractions
des fermions de surface d'univers dans les structures de spin
impaires et paires~:
\begin{align}
t_8 t_8 R^4 &= 
 t^{\bal_1 \bbe_1 \bal_2 \bbe_2 \bal_3 \bbe_3 \bal_4 \bbe_4}
 t^{\a_1 \b_1 \a_2 \b_2 \a_3 \b_3 \a_4 \b_4}
R_{\bal_1 \bbe_1 \a_1 \b_1}
\dots
R_{\bal_4 \bbe_4 \a_4 \b_4} \\
\epsilon_{10} \epsilon_{10} R^4
&=
\epsilon^{\bal_1 \bbe_1 \bal_2 \bbe_2 
          \bal_3 \bbe_3 \bal_4 \bbe_4 \bar\mu\bar\nu}
\epsilon^{\a_1 \b_1 \a_2 \b_2 
          \a_3 \b_3 \a_4 \b_4 \mu \nu}
g_{\bar\mu\mu}g_{\bar\nu\nu}
R_{\bal_1 \bbe_1 \a_1 \b_1}
\dots
R_{\bal_4 \bbe_4 \a_4 \b_4} 
\end{align}
En particulier, les diagrammes {\`a} l'ordre des arbres et {\`a} une
boucle contribuent en type IIB au m{\^e}me couplage dans la m{\'e}trique
d'Einstein
\begin{equation}
S_{R^4}= \int d^{10}x\sqrt{-g_{E}}
~f_{10}^B(\tau,\bar\tau)~
(t_8t_8 + \frac{1}{8} \epsilon_{10}\epsilon_{10}) R^4
\end{equation}
o{\`u}
\begin{equation}
f_{10}^B(\tau,\bar\tau)= 
e^{\phi/2}~\left( 2\zeta(3) e^{-2\phi} + \frac{2\pi^2}{3} 
+ O(e^{2\phi}) \right)
\end{equation}
Le couplage complet $f_{10}^B$, invariant sous les transformations
modulaires du param{\`e}tre $\tau = \axion + i e^{-\phi}$,
ne saurait donc se restreindre aux seules contributions perturbatives
\index{modulaire, forme}
de la sph{\`e}re et du tore. D'un autre c{\^o}t{\'e}, on ne peut utiliser
les r{\'e}sultats usuels sur les fonctions modulaires, puisque 
$f_{10}^B$ n'est manifestement ni holomorphe, ni harmonique.
Il existe cependant une classe de fonctions {\it r{\'e}elles}
invariantes sous le groupe modulaire $Sl(2,\Zint)_B$, correspondant
\index{Eisenstein, s{\'e}rie d'}
aux {\it s{\'e}ries d'Eisenstein} d'ordre $s$
\begin{equation}
\label{eis}
E_s(\tau,\bar\tau)
=\sum_{(m,n)\ne 0} \left( \frac{\tau_2}{|m+n\tau|^2} \right)^{s}
\end{equation}
dont le comportement {\`a} faible couplage, soit $\tau_2\rightarrow
\infty$, peut s'obtenir par resommation de Poisson sur l'entier $m$~:
\index{Poisson, resommation de}
\begin{align}
\label{eisfour}
E_s(\tau,\bar\tau)
=&2\zeta(2s)\tau_2^s + 2\sqrt{\pi} \tau_2^{1-s} 
\frac{\Gamma(s-1/2)}{\Gamma(s)} \zeta(2s-1) \nonumber\\
&+ \frac{\pi ^s \sqrt{\tau_2}}{\Gamma(s)}
\sum_{m\ne 0} \sum_{n\ne 0}
\left| \frac{m}{n} \right|^{s-1/2} 
K_{s-1/2} \left(2\pi \tau_2 |mn|\right) 
e^{2\pi i m n \tau_1}
\end{align}
Green et Gutperle ont en particulier
\index{Green et Gutperle, conjecture de|textit}
remarqu{\'e} que la s{\'e}rie d'Eisenstein $E_{s=3/2}$ reproduisait pr{\'e}cis{\'e}ment
les deux termes perturbatifs de $f_{10}^B$~; elle les compl{\`e}te en
une fonction invariante sous la U-dualit{\'e} par
une {\it somme infinie de termes non perturbatifs}
\cite{Green:1997tv}. En utilisant
le comportement asymptotique de la fonction de Bessel
\index{Bessel, fonction de}
(\ref{k0}), on obtient
\begin{equation}
\label{di}
E_{3/2}= 2\zeta(3) \e ^{-3\phi/2} + \frac{2\pi ^2}{3} e ^{\phi/2} +
2\pi \sum_{m\ne 0}  \sum_{n\ne 0} \frac{ |mn|^{1/2} }{n^2}
e^{-2\pi |mn| (e^{-\phi} \pm i a)} \left( 1+ O(e^{\phi}) \right) \nonumber
\end{equation}
La comparaison avec l'action euclidienne (\ref{sdi}) montre que chaque 
terme non perturbatif peut s'interpr{\'e}ter comme
\index{D-instanton!de type IIB}
\index{serie d'instantons@s{\'e}rie d'instantons!couplage $R^4$ de IIB}
la contribution de {\it $N=mn$ D-instantons de la th{\'e}orie
de type IIB} {\`a} l'amplitude $R^4$, ou alternativement d'un instanton
de charge $N=mn$. Les corrections $O(e ^{\phi})$ correspondent {\`a}
des corrections perturbatives dans le champ de fond de l'instanton.
Il est donc naturel de conjecturer, {\`a} la suite de Green et
Gutperle, que {\it le couplage $R^4$ exact dans la th{\'e}orie
de type IIB est donn{\'e} par la s{\'e}rie d'Eisenstein $E_{3/2}$}.
Cette conjecture suppose en particulier que {\it toutes les
corrections perturbatives {\`a} plus d'une boucle s'annulent}.
Ce th{\'e}or{\`e}me de non-renormalisation perturbative pourrait en principe
{\^e}tre d{\'e}montr{\'e} par un comptage soigneux des modes
z{\'e}ros fermioniques correspondant {\`a} l'amplitude {\`a}
\index{mode z{\'e}ro!sur la surface d'univers}
quatre gravitons sur une surface de genre arbitraire~;
Berkovits a pu en donner une preuve bas{\'e}e sur les contraintes
de supersym{\'e}trie apr{\`e}s compactification sur un tore
\cite{Berkovits:1997pj}.
Il est particuli{\`e}rement int{\'e}ressant de noter la {\it mesure 
de sommation} sur ces instantons~:
\index{instanton!mesure d'int{\'e}gration|textit}
\begin{equation}
\mu(N)=  \sum_{m\ne 0,n\ne 0,mn=N} \frac{ |mn|^{1/2}}{n^2}
= \sqrt{N} \sum_{n|N} \frac{1}{n^2}
\end{equation}
Cette mesure a pu {\^e}tre d{\'e}termin{\'e}e par Sethi et Stern
dans le cas $N=2$ dans le cadre du mod{\`e}le de matrice d{\'e}crivant
les D-instantons \cite{Sethi:1997pa}, mais la d{\'e}monstration de cette formule
pour tout $N$ reste un probl{\`e}me ouvert
\cite{Green:1997tn}. Avant d'{\'e}tudier
les cons{\'e}quences de cette conjecture pour les D-branes
de dimension plus {\'e}lev{\'e}e, nous discutons dans la section
suivante quelques propri{\'e}t{\'e}s des s{\'e}ries de Eisenstein
et leur pertinence pour les couplages de la th{\'e}orie de
type IIB.

\subsection{Fonctions modulaires r{\'e}elles et s{\'e}ries de Eisenstein}
\index{Eisenstein, s{\'e}rie d'|textit}
En comparaison avec les formes modulaires holomorphes, les fonctions
modulaires r{\'e}elles, c'est-{\`a}-dire les fonctions d{\'e}finies
sur le demi-plan de Poincar{\'e} $\{\tau \in \CC~;\tau_2>0\}$ et
\index{Poincar{\'e}, demi-plan de}
invariantes sous les transformations modulaires de $\tau$,
sont encore largement m{\'e}connues. Leur
importance pour certains probl{\`e}mes de th{\'e}orie des nombres a 
cependant aiguis{\'e} l'int{\'e}r{\^e}t des math{\'e}maticiens, et on
trouvera une pr{\'e}sentation abordable des principaux r{\'e}sultats
{\`a} leur sujet dans l'ouvrage de A. Terras
\cite{Terras:1985}.
Les s{\'e}ries de Eisenstein constituent en r{\'e}alit{\'e} les seules
fonctions modulaires r{\'e}elles connues explicitement. La s{\'e}rie 
(\ref{eisfour}) en fournit une d{\'e}finition pour $\Re s>1$, qui
peut {\^e}tre prolong{\'e}e analytiquement dans tout le plan complexe $s$
priv{\'e} de $s=1$\footnote{Les p{\^o}les de la fonction $\Gamma(s-1/2)$
en $s-1/2\in -\NN$ sont en effets compens{\'e}s par les z{\'e}ros de 
la fonction $\zeta(2s-1)$, et le p{\^o}le en $s=1/2$ de la fonction $\zeta(2s)$
est compens{\'e} par le p{\^o}le de la fonction $\Gamma(s-1/2)$
dans le second terme.}, et la fonction $E_1$ peut {\^e}tre
r{\'e}gularis{\'e}e en soustrayant le p{\^o}le simple 
\begin{equation}
\hat E_1(\tau,\bar\tau) = \lim_{s\rightarrow 1} \left(E_s
(\tau,\bar\tau)  - \frac{\pi}{s-1} \right)
=-\pi \log \tau_2 |\eta(\tau)|^4
\end{equation}
o{\`u} l'on reconna{\^\i}{}t la contribution de l'orbite d{\'e}g{\'e}n{\'e}r{\'e}e
{\`a} la fonction de partition de $T^2$. L'existence du p{\^o}le
\index{partition, fonction de!du tore $T^2$}
en $s=1$ peut {\'e}galement {\^e}tre observ{\'e}e {\`a} partir
de la relation de sym{\'e}trie
\begin{equation}
\label{symeis}
\frac{\Gamma(s) E_s(\tau,\bar\tau)}{\pi ^s}=
\frac{\Gamma(1-s) E_{1-s}(\tau,\bar\tau)}{\pi ^{1-s}}\ .
\end{equation}
Les fonctions de Eisenstein sont des fonctions propres du
Laplacien $\Delta=\tau_2^2(\partial_{\tau_1}^2+\partial_{\tau_2}^2)$
\index{Laplacien}
sur le demi-plan de Poincar{\'e}~:
\begin{equation}
\Delta E_s = s(s-1) E_s
\end{equation}
{\`a} l'exception de la s{\'e}rie r{\'e}gularis{\'e}e $\hat E_1$~:
\begin{equation}
\Delta \hat E_1 = \pi
\end{equation}
qui souffre de l'{\it anomalie holomorphique} usuelle, correspondant
{\`a} la divergence logarithmique en $\tau_2\rightarrow 0$.
\index{divergence logarithmique en couplage}
La relation (\ref{symeis}) traduit la non d{\'e}g{\'e}n{\'e}rescence
des valeurs propres du Laplacien. En sus des s{\'e}ries d'Eisenstein,
le Laplacien admet une infinit{\'e} discr{\`e}te de 
{\it formes cuspo{\"\i}{}dales} ({\it cusp forms}) $v_n(\tau,\bar\tau)$,
\index{cuspoidale, forme@cuspo{\"\i}{}dale, forme}
\index{modulaire, forme}
de valeurs propres $s_n(s_n-1)$ localis{\'e}es sur l'axe
$\Re s_n= 1/2$~; ces formes sont d{\'e}finies par l'annulation
des coefficients $a$ et $b$
dans le d{\'e}veloppement de Fourier g{\'e}n{\'e}rale des modes propres
\begin{equation}
f(\tau,\bar\tau) = a \tau_2^s + b \tau_2^{1-s} + \sqrt{\tau_2} 
\sum_{n\ne 0} a_n K_{s-1/2}(2\pi|n|\tau_2) e^{2\pi i n \tau_1}
\end{equation}
et d{\'e}croissent donc exponentiellement
en $\tau_2\rightarrow\infty$. Aucune d'entre elles n'est malheureusement
connue explicitement. Etant donn{\'e} le spectre du Laplacien, on 
peut maintenant {\'e}noncer le th{\'e}or{\`e}me de d{\'e}composition
spectrale de Roelcke-Selberg : {\it toute fonction modulaire $f$
\index{Roelcke-Selberg, th{\'e}or{\`e}me de}
de carr{\'e} int{\'e}grable\footnote{pour la mesure invariante
$d^2\tau/\tau_2^2$. Noter que les s{\'e}ries d'Eisenstein $E_s$ ne
sont jamais de carr{\'e} int{\'e}grable, et sont int{\'e}grables pour
$s\in]0,1[$.} se d{\'e}compose sur les modes propres 
du Laplacien selon}
\begin{equation}
f(\tau,\bar\tau) = \sum_{n\ge 0} (f,v_n) v_n(\tau,\bar\tau) 
+\frac{1}{4\pi i} \int_{Re s=1/2} (f,E_s) E_s(\tau,\bar\tau) ds
\end{equation}

Apr{\`e}s avoir rappel{\'e} cet arri{\`e}re-plan math{\'e}matique,
nous pouvons maintenant discuter la relevance des
s{\'e}ries de Eisenstein pour la d{\'e}termination des
couplages exacts de la th{\'e}orie de type IIB. Les deux
termes d'ordre $\tau_2^s$ et $\tau_2^{1-s}$ dans
le  d{\'e}veloppement (\ref{eisfour}) peuvent s'interpr{\'e}ter
comme deux corrections perturbatives de cordes ferm{\'e}es
s'ils sont s{\'e}par{\'e}s d'une puissance {\it paire} de $e^{\phi}$~:
$s=n+1/2$ doit alors {\^e}tre demi-entier
\footnote{Les fonctions de Bessel
\index{Bessel, fonction de}
$K_{n}$ contiennent alors une s{\'e}rie {\it infinie} de corrections
perturbatives, par opposition aux fonctions $K_{n+1/2}$.}.
L'entier $|n|$ correspond au nombre
d'ordres de perturbations s{\'e}parant ces deux contributions.
En particulier, dans le cas des couplages en $R^4$ que nous
avons consid{\'e}r{\'e}, l'ordre $s=3/2$ de la s{\'e}rie {\'e}tait
impos{\'e} par l'existence de corrections {\`a} l'ordre des
arbres et {\`a} une boucle. Les puissances exactes du dilaton
\index{dilaton}
\index{Weyl!dilatation de}
provenant de l'ordre de perturbation et de la transformation
de Weyl $g_{str}=e^{\phi/2}g_E$ doivent {\'e}galement {\^e}tre
reproduites~: si on consid{\`e}re une interaction de dimension
\footnote{Dans la convention pr{\'e}sente, $L$ d{\'e}signe la
dimension de longueur de l'op{\'e}rateur, soit
$L=8$ pour le terme d'Einstein-Hilbert {\`a} dix dimensions
\index{Einstein-Hilbert, action d'}
$\int d^{10}x\sqrt{-g}R$, ou $L=2$ pour le couplage $\int d^{10}x\sqrt{-g}R^4$.}
$L$ recevant des contributions de genre $n$ et $n'$,
les deux conditions s'{\'e}crivent
\begin{equation}
e ^{\frac{L}{4}\phi} e ^{(2n-2)\phi} = \tau_2^{s} \ ,\quad
e ^{\frac{L}{4}\phi} e ^{(2n'-2)\phi} = \tau_2^{1-s}
\end{equation}
dont la compatibilit{\'e} impose 
\begin{equation}
L+4(n+n')=6
\end{equation}
Cette condition est en particulier v{\'e}rifi{\'e}e pour 
le cas des couplages en $R^4$, mais aussi pour
les couplages $R^4 H^{4g-4}$, dont on peut donc imaginer
qu'ils soient donn{\'e}s par $E_{{1/2}+g}$,
correspondant {\`a} deux termes perturbatifs
{\`a} $0$ et $g$ boucles respectivement. Cette conjecture
vient d'{\^e}tre renforc{\'e}e par Vafa et Berkovits, qui 
ont reli{\'e} ce couplage {\`a} une amplitude topologique
\index{amplitude de diffusion!topologique}
similaire aux amplitudes $\mathcal{F}_g R^2 F^{2g-2}$ des
th{\'e}ories $N=2$ \cite{Berkovits:1998}.

Cette discussion ne fixe cependant que les contributions 
des s{\'e}ries d'Eisenstein au couplage. Les formes
\index{cuspoidale, forme@cuspo{\"\i}{}dale, forme}
cuspo{\"\i}{}dales ne sont pas visibles en s{\'e}rie de
perturbation, et ne contributent qu'aux effets
instantoniques. Il serait particuli{\`e}rement int{\'e}ressant de mettre
en {\'e}vidence par un raisonnement de dualit{\'e} de
telles contributions, compte tenu de l'int{\'e}r{\^e}t
math{\'e}matique de ces formes. Dans le cas des couplages
en $R^4$ de la th{\'e}orie de type IIB, Green et Vanhove
ont conjectur{\'e} que la condition $\Delta E_{3/2} = \frac{3}{4} E_{3/2}$
\index{Laplacien}
pourrait {\^e}tre une cons{\'e}quence de la supersym{\'e}trie\cite{Green:1997di}.
Gr{\^a}ce aux techniques de superespace d{\'e}velopp{\'e}es par Berkovits,
j'ai pu prouver dans la note en annexe \ref{nr4} \cite{Pioline:1998} que 
cela {\'e}tait bien le cas~:
ce th{\'e}or{\`e}me suffit donc 
{\`a} disqualifier les contributions de formes
cuspo{\"\i}{}dales {\`a} $f_{10}^B$. Nous obtiendrons
ce m{\^e}me r{\'e}sultat par un argument tr{\`e}s
diff{\'e}rent dans la section \ref{ddsdec}.

\subsection{Couplages $R^4$ et compactification de la M-th{\'e}orie}
Connaissant l'expression exacte des couplages en $R^4$ dans la
th{\'e}orie de type IIB {\`a} dix dimensions, on en d{\'e}duit 
imm{\'e}diatement le r{\'e}sultat apr{\`e}s compactification {\`a}
\index{compactification!sur un cercle}
neuf dimensions sur un cercle~: la th{\'e}orie de type IIB
{\`a} neuf dimensions ne poss{\`e}de toujours que les 
D(-1)-branes comme configurations instantoniques 1/2-BPS,
\index{D-instanton!de type IIB}
et leur contribution est simplement multipli{\'e}e par la
circonf{\'e}rence du cercle, correspondant {\`a} l'int{\'e}gration
sur la coordonn{\'e}e collective de translation le long de ce cercle.
\index{coordonn{\'e}e collective}
L'amplitude de 4 gravitons {\`a} une boucle pr{\'e}sente la
contribution suppl{\'e}mentaire des {\it instantons de 
surface d'univers} correspondant aux modes d'enroulement
\index{enroulement!etat d'@{\'e}tat d'}
de la corde de type IIB (ou les {\'e}tats de moment de la
corde de type IIA), mais on peut v{\'e}rifier que cette contribution
correspond {\`a} une structure tensorielle distincte
$(t_8 t_8 - \frac{1}{8} \epsilon_9 \epsilon_9) R^4$.
On obtient ainsi un r{\'e}sultat invariant sous la U-dualit{\'e}
{\`a} dix dimensions $Sl(2,\Zint)_U$, qui n'est autre que la
dualit{\'e} $Sl(2,\Zint)_B$ de la th{\'e}orie de type IIB.
\index{U-dualit{\'e}!de type II sur $S^1$}
Apr{\`e}s traduction dans les variables de la th{\'e}orie de type IIA
\index{dualit{\'e}!des th{\'e}ories IIA et IIB}
gr{\^a}ce {\`a} la T-dualit{\'e}, on obtient le couplage
\begin{equation}
f_{9}^A=2\zeta(3) R_A e^{-2\phi} +  
         \frac{2\pi^2}{3 R_A} + 0
+~ 4\pi e^{-\phi} \sum_{m\ne 0} \sum_{n\ne 0}
\left|\frac{m}{n}\right| K_1\left(2\pi R_A e^{-\phi} |mn|\right) 
          e^{2\pi i m n {\cal A}}
\end{equation}
o{\`u} $\mathcal{A}$ d{\'e}note la valeur moyenne du champ de Ramond
$\mathcal{A}_\mu$ le long du cercle. Cette expression
reproduit pr{\'e}cis{\'e}ment les contributions 
{\`a} l'ordre des arbres et {\`a} une 
boucle, et exclue toute contribution perturbative d'ordre plus {\'e}lev{\'e}.
Elle inclut en outre une somme d'effets non perturbatifs
\begin{equation}
f_9^{A}=f_{9;{\rm pert}}^A+
2\pi \sum_{m\ne 0}  \sum_{n\ne 0} \frac{ |mn|^{1/2}}{n^2}
e^{-2\pi |mn| (e^{-\phi} R_A \pm i \mathcal{A})} 
\left( 1+ O(e^{\phi}) \right) \nonumber
\end{equation}
\index{serie d'instantons@s{\'e}rie d'instantons!couplage $R^4$ de IIB}
pouvant s'interpr{\'e}ter comme la contribution des {\it instantons
de D0-branes} dont la ligne d'univers s'enroule autour du cercle :
\index{ligne d'univers}
l'action euclidienne d'une telle configuration s'{\'e}crit en effet
\begin{equation}
S_{D0} = e^{-\phi} R_A + i \mathcal{A}
\end{equation}
La signification des entiers $m$ et $n$ {\`a} ce stade n'est pas
claire, mais nous d{\'e}couvrirons bient{\^o}t que $m$ correspond
{\`a} la {\it charge} de la D0-brane (c'est-{\`a}-dire au nombre
de D0-branes {\'e}l{\'e}mentaires li{\'e}es ensemble), tandis
que $n$ correspond au {\it nombre d'enroulement} de la ligne
d'univers autour de la direction compacte. 

La {\it limite de d{\'e}compactification} $R_A \rightarrow \infty$
\index{decompactification@d{\'e}compactification, limite de}
{\`a} dilaton $e^{\phi}$ constant supprime les contributions 
des {\'e}tats d'enroulement et des instantons de D0-branes.
Le couplage $(t_8 t_8 +\frac{1}{8} \epsilon_{10}\epsilon_{10})R^4$ 
de la th{\'e}orie de type IIA {\`a} dix dimensions
se r{\'e}duit donc {\it exactement} {\`a} la contribution {\`a} l'ordre 
des arbres ; s'y ajoute le couplage 
$(t_8 t_8 -\frac{1}{8} \epsilon_{10}\epsilon_{10})R^4$ 
correspondant {\`a} la d{\'e}compactification de la contribution 
\index{Kaluza-Klein!excitation de}
des {\'e}tats de Kaluza-Klein {\`a} neuf dimensions. On obtient
ainsi l'action effective en $R^4$ de la th{\'e}orie de type IIA:
\begin{equation}
\label{r4iia}
S^{A}_{R^4}= \int d^{10}x\sqrt{-g}
~2\zeta(3) e^{-2\phi}
(t_8t_8 + \frac{1}{8} \epsilon_{10}\epsilon_{10}) R^4
+
\frac{2\pi^2}{3} 
(t_8t_8 - \frac{1}{8} \epsilon_{10}\epsilon_{10}) R^4
\end{equation}
Contrairement {\`a} la th{\'e}orie de type IIB, {\it l'action
en $R^4$ de la th{\'e}orie de type IIA ne re{\c c}oit donc
pas de corrections radiatives au-del{\`a} d'une boucle}.

Si la th{\'e}orie de type IIA d{\'e}crit en effet la M-th{\'e}orie
compactifi{\'e}e sur un cercle, la limite de d{\'e}compactification
$R_{11}=e^{2\phi/3}l_{11}\rightarrow\infty$ 
doit reproduire un couplage {\it invariant
sous le groupe de Poincar{\'e} {\`a} onze dimensions}. Le couplage
\index{Lorentz, invariance {\`a} 11D}
 (\ref{r4iia}) v{\'e}rifie bien cette condition, puisque sous
cette limite, le terme {\`a} l'ordre des arbres est supprim{\'e},
et seul subsiste le couplage
\begin{equation}
\label{r4m}
S^{M}_{R^4}= \int d^{11}x\sqrt{-g}
\frac{2\pi^2}{3} 
(t_8t_8 - \frac{1}{2\cdot 3!} \epsilon_{11}\epsilon_{11}) R^4\ .
\end{equation}
\index{supergravit{\'e}!{\`a} onze dimensions}
La supergravit{\'e} {\`a} onze dimensions d{\'e}crivant la M-th{\'e}orie
ne peut restituer cette pr{\'e}diction~: le calcul de l'amplitude
de diffusion de quatre gravitons {\`a} une boucle dans cette
th{\'e}orie non-renormalisable souffre d'une
\index{amplitude de diffusion!en SUGRA 11D}
\index{divergence ultraviolette!en supergravit{\'e}}
divergence {\it cubique}. La reproduction de ce r{\'e}sultat 
constituerait cependant un test crucial des propositions de d{\'e}finition
microscopique de la M-th{\'e}orie.

La restauration de l'invariance de Poincar{\'e} {\`a} onze dimensions
dans la limite de fort couplage ne contraint pas seulement la
limite $e^{\phi}\rightarrow \infty$, mais implique {\'e}galement 
l'invariance des couplages de la th{\'e}orie de type IIA compactifi{\'e}e
sur $T^d$ sous les sym{\'e}tries du tore {\'e}tendu
$T^{d+1}$~; ce groupe de sym{\'e}trie $Sl(d+1,\Zint)$ avec le
groupe de T-dualit{\'e} $SO(d,d,\Zint)$ g{\'e}n{\`e}re en effet
\index{T-dualit{\'e}}\index{U-dualit{\'e}!de type II sur $T^d$}
le groupe de U-dualit{\'e}. L'invariance  
du couplage $R^4$ de la th{\'e}orie de type IIA {\`a} neuf dimensions
sous ces sym{\'e}tries peut {\^e}tre rendue manifeste en l'exprimant
en termes des modules $g_{IJ}$
de la M-th{\'e}orie compactifi{\'e}e sur $T^2$~:
\begin{equation}
\label{eis11d}
f_9^A = V_{11} \sum_{(n^1,n^2)\ne (0,0)} 
\left( n^{I} g_{IJ} n^{J} \right)^{-3/2}\ ,\quad V_{11}=\sqrt{\det g_{IJ}}
\end{equation}
en parfaite analogie avec le couplage de type IIB (\ref{eis}).
Comme l'ont remarqu{\'e} Green, Gutperle et Vanhove, cette
expression et sa g{\'e}n{\'e}ralisation en dimension $d\ge 2$
peut {\^e}tre reproduite,
{\it {\`a} la divergence cubique mentionn{\'e}e ci-dessus pr{\`e}s},  
par un calcul
de diffusion {\`a} une boucle de quatre gravitons de
\index{amplitude de diffusion!en SUGRA 11D}
la supergravit{\'e} {\`a} onze dimensions
\cite{Green:1997as}. Les nombres quantiques
$n^I$ apparaissent alors comme les charges {\it duales
\index{Poisson, resommation de}
par resommation de Poisson} aux moments de Kaluza-Klein $n_I$
du graviton <<tournant>> dans la boucle.
Une telle connection n'est pas {\'e}tonnante, les D0-branes 
{\'e}tant identifi{\'e}es aux modes de Kaluza-Klein
\index{Kaluza-Klein!excitation de}
du supergraviton {\`a} onze dimensions. Elle suppose n{\'e}anmoins
des propri{\'e}t{\'e}s de non-renormalisation des couplages $R^4$
de la M-th{\'e}orie. On en d{\'e}duit les contributions du supergraviton
{\`a} 10 dimensions ($n_I=0$) et des D0-branes ($n_I\ne 0$)
au couplage $R^4$ {\it en toutes dimensions}~:
\begin{align}
\label{d0all}
f_{SUGRA}^A =& V_{11} \sum_{n^I\ne 0} 
\left( n^{I} g_{IJ} n^{J} \right)^{-3/2} \nonumber\\
=&
2\zeta(3) V_A e^{-2\phi} +  2V \sum_{n^i\ne 0} 
\frac{1}{n^i g_{ij}n^j} \\
&+~ 4\pi V e^{-\phi} \sum_{m\ne 0} \sum_{n\ne 0}
\frac{|m|}{\sqrt{n^i g_{ij} n^j} }
K_1\left(2\pi |m| e^{-\phi} \sqrt{n^i g_{ij} n^j}  \right) 
          e^{2\pi i m n^i {\cal A}_i} \nonumber
\end{align}
o{\`u} la deuxi{\`e}me ligne s'obtient par resommation de
Poisson sur l'entier $n^{11} \rightarrow m$. Cette
description unifie ainsi {\it en un r{\'e}sultat
$Sl(d+1,\Zint)$-invariant} l'amplitude {\`a} l'ordre
des arbres, la contribution {\`a} l'ordre d'une boucle
de l'{\it orbite d{\'e}g{\'e}n{\'e}r{\'e}e}
\footnote{L'orbite d{\'e}g{\'e}n{\'e}r{\'e}e d{\'e}signe la contribution
des cordes non excit{\'e}es de nombres d'enroulement nuls {\`a} l'amplitude {\`a}
une boucle, et correspond donc, tout comme l'amplitude
{\`a} l'ordre des arbres, {\`a} la contribution du supergraviton
{\`a} dix dimensions.} et la contribution de $m$ D0-branes,
d'action classique
\begin{equation}
\label{d0class}
S_{D0}=e^{-\phi} sqrt{n^i g_{ij} n^j} + i n^i {\cal A}_i \ .
\end{equation}
Elle justifie en outre l'interpr{\'e}tation que nous
avons donn{\'e}e dans le cas $d=1$ des entiers 
\index{D-brane!D0-branes}
$m$ et $n^i$.

\subsection{Cordes de charges $(p,q)$ et somme solitonique}

Nous avons vu que ce r{\'e}sultat reproduisait la
totalit{\'e} du couplage 
$(t_8 t_8 + \frac{1}{8} \epsilon\epsilon) R^4$ en dimensions
9 et 10, en accord avec l'{\'e}galit{\'e} de groupe
de U-dualit{\'e} $E_{d+1}(\Zint)$ et de son sous-groupe $Sl(d+1,\Zint)$
\index{U-dualit{\'e}!de type II sur $T^d$}
dans ces dimensions. En dimension 8, les D0-branes restent
les seules configurations instantoniques de la th{\'e}orie
de type IIA compactifi{\'e}e sur un tore $T^2$, et on
s'attend donc {\`a} ce que ce r{\'e}sultat reste valide.
De fait, le groupe de U-dualit{\'e} en dimension 8 
se s{\'e}pare en deux facteurs :
\begin{itemize}
\item $Sl(3,\Zint)$ correspond au groupe modulaire du tore $T^3$
du point de vue de la M-th{\'e}orie. Il contient en particulier
le groupe de dualit{\'e} non perturbative
$Sl(2,\Zint)$ de la th{\'e}orie de type
IIB, et le sous-groupe $Sl(2,\Zint)_U$ de la T-dualit{\'e}
$SO(2,2,\Zint)$ correspondant aux transformations modulaires
de la structure complexe $U$ du tore de la th{\'e}orie de type IIA
(ou aux transformations de la structure de K{\"a}hler $T$ du tore
de type IIB). Il agit sur l'espace des modules homog{\`e}ne
$Sl(3,\Real)/SO(3)$ comprenant en particulier le dilaton.
\item $Sl(2,\Zint)$ correspond inversement aux transformations modulaires
de la structure complexe $U$ du tore de la th{\'e}orie de type IIB
et aux transformations de la structure de K{\"a}hler $T$ du tore
de type IIA. Du point de vue de la M-th{\'e}orie, il correspond
aux transformations modulaires du module complexe
$\mathcal{C}_{123} + i V_{11}$ param{\'e}trant
$Sl(2,\Real)/U(1)$. Dans tous les cas il correspond
{\`a} des transformations perturbatives.
\end{itemize}
Le calcul de l'amplitude {\`a} quatre gravitons {\`a} genre 0 et 1
montre que le couplage $(t_8 t_8 + \frac{1}{8} \epsilon_8\epsilon_8)
R^4$, ici {\'e}crit dans les variables de type IIB,
\begin{equation}
f_8^A = 2\zeta(3) T_2 e^{-2\phi} - 2\pi\log T_2 |\eta(T)|^4
\end{equation}
ne d{\'e}pend que des modules de $Sl(3,\Real)/Sl(3,\Zint)$, 
tandis que le couplage $t_8 t_8 - \frac{1}{8} \epsilon_8\epsilon_8 R^4$
\begin{equation}
f_8^{A'} = - 2\pi\log U_2 |\eta(U)|^4
\end{equation}
est donn{\'e} {\`a} une boucle et ne d{\'e}pend que des 
modules du second facteur $Sl(2,\Real)/U(1)$
 (cf. annexe \ref{pq}). Cette analyse a {\'e}t{\'e} {\'e}tendue {\`a} tous
les ordres en perturbations par Berkovits \cite{Berkovits:1997pj}.
En collaboration avec E. Kiritsis, nous avons conjectur{\'e} que
le premier couplage {\'e}tait donc donn{\'e} {\it exactement} par
\index{Eisenstein, s{\'e}rie d'}
la s{\'e}rie d'Eisenstein d'ordre 3/2 (\ref{d0all}) pour le groupe 
$Sl(3,\Zint)$ , tandis que le second couplage {\'e}tait exact 
{\`a} une boucle, et donc donn{\'e} par la s{\'e}rie d'Eisenstein d'ordre 1
$E_1(U)$. Cette conjecture contient les r{\'e}sultats pr{\'e}c{\'e}dents
par d{\'e}compactification du tore $T^2$, et restitue les 
contributions attendues des D0-branes de la th{\'e}orie de type IIA.
La situation est cependant beaucoup plus int{\'e}ressante du c{\^o}t{\'e}
de type IIB, o{\`u} l'on attend, en plus des D-instantons, 
les contributions des {\it D1-branes}
\index{pq, cordes de charge@$(p,q)$, cordes de charge}
ainsi que des {\it cordes de charges $(p,q)$} qui leur sont reli{\'e}es
par S-dualit{\'e}. De fait, l'expansion (\ref{d0all}), en termes
des variables de type IIB, donne
\begin{align}
\label{f8b}
f_8^B =& 2\zeta(3)T_2 e^{-2\phi} - 2\pi\log T_2 |\eta(T)|^4\nonumber\\
+& 4\pi V e^{-\phi} \sum_{m\ne 0} \sum_{(n,q)\ne 0}
\frac{|m|}{\sqrt{(n+qB)^2+V^2 q^2} }
K_1\left(2\pi |m| e^{-\phi}\sqrt{ (n+qB)^2 + V^2 q^2}  \right) \nonumber\\
&\cdot e^{2\pi i m\left(n\axion + q (\mathcal{B}+\axion B)\right)}
\end{align}
Cette expression contient pour $q=0$ la contribution des 
D-instantons (\ref{di}) d{\'e}j{\`a} observ{\'e}e en dimension sup{\'e}rieure.
Lorsque $n\ne 0$ cependant, l'action euclidienne des effets
non perturbatifs 
\begin{equation}
\label{sd1}
S_{D1} = e^{-\phi}\sqrt{ (n+qB)^2 + V^2 q^2}  
+i~ \left[ n\axion + q (\mathcal{B}+\axion B) \right]
\end{equation}
correspond pr{\'e}cis{\'e}ment {\`a} l'action de Born-Infeld d'une
\index{Born-Infeld, action de}
{\it D1-brane enroul{\'e}e $q$ fois autour du tore et portant un
flux {\'e}lectrique $n$}. Une telle configuration 
\index{flux {\'e}lectrique}
correspond en effet {\`a} une configuration des champs de plongement 
$X^i$ et un champ {\'e}lectrique $F_{\alpha\beta}$ 
\begin{equation}
X^i(\sigma^\alpha) = N^i_\alpha \sigma^\alpha, \quad
F_{\alpha\beta}=\epsilon_{\alpha\beta} n
\end{equation}
sur la surface d'univers D1-brane
\footnote{Si la quantification des nombres d'enroulement 
$N^i_\alpha$ est clairement impos{\'e}e par la d{\'e}finition
univoque du plongement $\sigma^\alpha \rightarrow X^i$, 
la quantification du flux $F$ r{\'e}sulte quant {\`a} elle 
de la compacit{\'e} du groupe de jauge $U(1)$ support{\'e}
sur la D-brane.}, et l'{\'e}valuation de la
m{\'e}trique et des tenseurs de jauge induits par ce
plongement conduit {\`a}
\begin{eqnarray}
\det(\hat G + \hat B  + F) &=& (n+ \frac{1}{2}n^{ij}B_{ij})^2
+\frac{1}{2} n^{ij}~g_{ik} g_{jl}~n^{kl} \ , \\
e^{\hat B+F}\mathcal{R}&=&n\axion + 
\frac{1}{2}n^{ij} (\mathcal{B}_{ij}+\axion B_{ij})
\end{eqnarray}
o{\`u} la {\it deux-forme d'enroulement} $n^{ij}$
\index{enroulement!charge d'}
\begin{equation}
n^{ij}=\epsilon^{\alpha\beta} N^i_\alpha N^j_\beta\ ,
\end{equation}
invariante  sous les reparam{\'e}trisations de la surface d'univers, 
sp{\'e}cifie le 2-cycle sur lequel la D1-brane s'enroule. Dans le
cas pr{\'e}sent d'une compactification sur un tore $T^2$, seule
la composante $n^{12}=-n^{21}$ intervient et s'identifie
{\`a} la charge $q$ de l'{\'e}quation (\ref{sd1}). L'identification
du flux {\'e}lectrique $n$ avec la charge de D-instanton
r{\'e}alise explicitement une proposition de Douglas et Witten
sur la description d'{\'e}tats li{\'e}s de solitons de D-branes,
ici transpos{\'e}e dans le contexte des D-instantons
\cite{Douglas:1995bn,Witten:1996im}.

Si les D1-branes et les D-instantons sont apparus naturellement
dans l'interpr{\'e}tation de la s{\'e}rie d'instantons, le r{\^o}le
du multiplet des cordes de charges $(p,q)$ de la th{\'e}orie
de type IIB est jusqu'{\`a} pr{\'e}sent rest{\'e} obscur.
Ce n'est qu'en choisissant une repr{\'e}sentation o{\`u} la
S-dualit{\'e} $Sl(2,\Zint)$ {\it commute avec le d{\'e}veloppement
en s{\'e}rie} que leurs effets deviennent manifestes.
Il suffit pour cela d'effectuer une resommation de Poisson
\index{Poisson, resommation de}
{\it sur la charge de D-instanton}\footnote{une resommation
\index{D-instanton!de type IIB}
de Poisson sur l'entier $m$ ram{\`e}nerait au contraire
{\`a} la s{\'e}rie d'Eisenstein (\ref{eis11d})},
apr{\`e}s avoir toutefois s{\'e}par{\'e} la contribution des
D-instantons $q=0,n\ne 0$. La fonction de Bessel $K_1(z)$
\index{Bessel, fonction de}
donne alors apr{\`e}s resommation de Poisson une 
fonction exponentielle $K_{1/2}(z)=\sqrt{\frac{\pi}{2z}}e^{-z}$,
et la s{\'e}rie peut alors {\^e}tre r{\'e}{\'e}crite sous la forme
\begin{align}
\label{solsum}
f_8^B =& 2\zeta(3) V\sum_{p\wedge q=1} 
\left(\frac{|p+q\tau|}{\alpha'}\right) 
\left(\frac{\tau_2}{|p+q\tau|^2}\right)^2
- 2\pi \sum_{p \wedge q =1}\log T_{(p,q);2} |\eta(T_{(p,q)})|^4
\end{align}
o{\`u} la somme est effectu{\'e}e sur les couples d'entiers
$(p,q)$ sans diviseurs communs, et o{\`u} le param{\`e}tre
\begin{equation}
T_{(p,q)}= p B_{12} - q\mathcal{B}_{12} + i~|p+q\tau| V/\alpha'
\end{equation}
n'est autre que le module de K{\"a}hler du tore $T^2$ 
{\it mesur{\'e} par une corde de charges $(p,q)$ sous les tenseurs
$(B,\mathcal{B})$ et de tension $|p+q\tau|/\alpha'$}. Le couplage
en $R^4$ de la th{\'e}orie de type IIB s'interpr{\`e}te alors comme 
la somme des amplitudes {\`a} l'ordre des arbres et {\`a} une boucle
des cordes de type $(p,q)$
c'est-{\`a}-dire comme une {\it somme sur les solitons de la th{\'e}orie de
type IIB} (Ce point de vue a {\'e}galement {\'e}t{\'e} discut{\'e}
par Kehagias et Partouche \cite{Kehagias:1997jg}). 
Le d{\'e}veloppement (\ref{solsum}) commute
cependant avec la S-dualit{\'e}, et ne correspond donc plus {\`a} un
d{\'e}veloppement semi-classique.

\subsection{Instantons de D-branes et T-dualit{\'e}}
\index{U-dualit{\'e}!de type II sur $T^3$}
Apr{\`e}s compactification sur un cercle suppl{\'e}mentaire, soit
dans le cas de la M-th{\'e}orie compactifi{\'e}e sur $T^4$,
le groupe de U-dualit{\'e} $Sl(5,\Zint)$ ne se r{\'e}duit plus 
au sous-groupe g{\'e}om{\'e}trique $Sl(4,\Zint)$. Cette extension
co{\"\i}{}ncide avec l'apparition des D2-branes de la th{\'e}orie
de type IIA, qui peuvent maintenant s'enrouler autour du tore
$T^3$. Elle co{\"\i}{}ncide {\'e}galement avec la fusion des
deux amplitudes $(t_8t_8\pm \epsilon\epsilon)R^4$, en
raison de l'insuffisance des modes z{\'e}ros fermioniques
\index{mode z{\'e}ro!fermionique}
\footnote{alternativement, la densit{\'e} d'Euler {\`a}
\index{Euler, densit{\'e} d'}
huit dimensions $\epsilon_8\epsilon_8 R^4$ s'annule 
en dimensions inf{\'e}rieures.}. La s{\'e}rie d'Eisenstein
d'ordre  $3/2$ pour le groupe $Sl(5,\Zint)$ constitue
un candidat naturel pour le couplage en $t_8 t_8 R^4$
exact : nous avons pu prouver, en collaboration avec
Elias Kiritsis, qu'elle restituait en effet
\footnote{gr{\^a}ce {\`a} une propri{\'e}t{\'e}
arithm{\'e}tique assez miraculeuse.} les
amplitudes {\`a} l'ordre des arbres et {\`a} une boucle
(toutes orbites confondues). Du point de vue de type IIB,
elle reproduit {\'e}galement la s{\'e}rie instantonique
des D-instantons et D-cordes, ou la s{\'e}rie solitonique
des cordes de charge $(p,q)$. Du point de vue de type IIA,
elle montre au contraire les contributions des D2-branes
d'action de Born-Infeld euclidienne
\index{Born-Infeld, action de}
\begin{eqnarray}
S_{D2}&=&e^{-\phi} \left[
\left( n^i + \frac{1}{2} n^{ijk} B_{jk} \right) g_{il}
\left( n^l + \frac{1}{2} n^{lmn} B_{mn} \right)
+ \frac{1}{6} n^{ijk}~g_{il}g_{jm}g_{kn}~n^{lmn} \right]^{1/2}\nonumber\\
&&+ i~ \left( n^i {\cal A}_i + \frac{1}{6} n^{ijk} {\cal C}_{ijk} \right)
\end{eqnarray}
o{\`u} les formes $n^i$ et $n^{ijk}$ s'obtiennent en termes des
configurations des champs sur la surface d'univers par
\begin{equation}
n^{ijk}=\epsilon^{\alpha\beta\gamma} N^i_\alpha N^j_\beta N^k_\gamma, \quad
n^i=\frac{1}{2}\epsilon_{\alpha\beta\gamma} N^i_\alpha F_{\beta\gamma}
\end{equation} 
Les D0-branes apparaissent donc comme des {\it flux {\'e}lectriques
et magn{\'e}tiques} sur le volume d'univers des D2-branes, en parfaite
analogie avec le cas des D-instantons et des D1-branes.

La d{\'e}termination de la param{\'e}trisation
de l'espace homog{\`e}nes des modules n{\'e}cessaire {\`a} l'application
de cette m{\'e}thode devient assez rapidement prohibitive
pour des compactifications en dimensions inf{\'e}rieures. 
Dans la publication annex{\'e}e en appendice \ref{pq}, Elias Kiritsis
et moi-m{\^e}me avons alors adopt{\'e} une autre approche,
et utilis{\'e} une s{\'e}quence de {\it T-dualit{\'e}s}
\index{T-dualit{\'e}}
\index{Lorentz, transformation de}
et {\it transformations de Lorentz}  pour fixer les configurations
des D-branes de dimension arbitraire. Appliqu{\'e}e sur l'action
(\ref{d0class}) d'une 
configuration de D0-branes enroul{\'e}es autour
d'un cycle $n^i$ du tore $T^d$, la T-dualit{\'e} selon la direction 1
g{\'e}n{\`e}re une configuration d'action\footnote{Le tilde d{\'e}note des
corrections aux champs de Ramond proportionnelles {\`a} $B$, soit ici
$\tilde{\cal B}_{ij}={\cal B}_{ij} + \axion B_{ij}$, d{\'e}taill{\'e}es
dans l'appendice \ref{dc}.}
\begin{equation}
 S_{cl} \to e^{-\phi} \sqrt{ ( n^1 + B_{1a} n^a )^2 + n^a~g_{11} g_{ab}~n^b }
+ i~( n^1 \axion + n^a \tilde{\cal B}_{1a} ) \ .
\end{equation}
Pour satisfaire l'invariance sous les transformations de Lorentz internes
$Sl(d,\Zint)$, il est n{\'e}cessaire de r{\'e}interpr{\'e}ter la premi{\`e}re
composante $n^1$ de la charge de la D0-brane comme un singlet $n$,
et les composantes suivantes $n^{a}$ comme les composantes $n^{1a}$
d'un tenseur de charges $n^{ij}$. On d{\'e}couvre ainsi les charges
\index{D-brane!charges de}
$(m,m^{ij})$ et l'action euclidienne (\ref{sd1}) de la D1-brane et des
D-instantons de type IIB. Cette proc{\'e}dure peut {\^e}tre r{\'e}it{\'e}r{\'e}e
pour obtenir les charges et l'action des D-branes paires ou impaires
de dimension arbitraire. On obtient ainsi pour la D3-brane
\begin{eqnarray}
S_{D3}^B=e^{-\phi} &\left[ 
\left(n + \frac{1}{2}n^{ij} B_{ij} +\frac{1}{8} n^{ijkl} B_{ij} B_{kl} 
\right)^2
+ \frac{1}{2} \left(n^{ij} + \frac{1}{2} n^{ijkl} B_{kl} \right)
g_{im}g_{jn}  \left(n^{mn} + \frac{1}{2} n^{mnpq} B_{pq} \right) \right. 
\nonumber\\
&\left.  
+ \frac{1}{24} n^{ijkl} g_{im}g_{jn}g_{kp}g_{lq} n^{mnpq}  \right]^{1/2}
+ i~\left( n \axion + \frac{1}{2}
 n^{ij} \tilde{\cal B}_{ij} 
+ \frac{1}{24} n^{ijkl} \tilde{{\cal D}}_{ijkl}  \right)
\end{eqnarray}
que l'on peut encore relier {\`a} l'action de Born-Infeld de la
D3-brane sous les identifications
\begin{equation}
n^{ijkl} = \epsilon^{\alpha\beta\gamma\delta} 
N^i_\alpha N^j_\beta N^k_\gamma N^l_\delta
\ ,\quad
n^{ij} = \frac{1}{2} \epsilon^{\alpha\beta\gamma\delta} 
N^i_\alpha N^j_\beta F_{\gamma\delta}
\ ,\quad
n = \frac{1}{8} \epsilon^{\alpha\beta\gamma\delta} 
F_{\alpha\beta} F_{\gamma\delta}\ ,
\end{equation}
En particulier, les D1-branes de charges d'enroulement
$n^{ij}$ apparaissent comme les
\index{flux {\'e}lectrique}\index{enroulement!charge d'}
flux electriques et magn{\'e}tiques sur la D3-brane de charge
d'enroulement $n^{ijkl}$, tandis que les D-instantons
de charge $n$ apparaissent comme {\it des instantons} 
de la th{\'e}orie de jauge $U(1)$ sur le volume d'univers
de la 3-brane\footnote{Ces instantons, absents des th{\'e}ories
\index{instanton!des th{\'e}ories de jauge}
de jauge $U(1)$ {\it dans l'espace non-compact}, sont 
stabilis{\'e}s par la compacit{\'e} du volume d'univers
de la 3-brane.}. 

Comme nous l'avons remarqu{\'e} dans la section (\ref{Tramond}),
les champs de Ramond se transforment dans une repr{\'e}sentation
{\it spinorielle} $\mathcal{R}$ du groupe de T-dualit{\'e} $SO(d,d,\Zint)$.
\index{Ramond, secteur de}
\index{T-dualit{\'e}!secteur de Ramond}
\index{spinorielle, repr{\'e}sentation}
La repr{\'e}sentation de ces champs comme formes de degr{\'e} pair
$(\axion,\mathcal{B},\mathcal{D},\dots)$ en type IIB ou impair
$(\mathcal{A},\mathcal{C},\dots)$ en type IIA ne fait que
refl{\'e}ter la d{\'e}composition d'un spineur de $SO(d,d)$
de chiralit{\'e} fix{\'e}e en repr{\'e}sentations irr{\'e}ductibles
de $Sl(d)$. Cette propri{\'e}t{\'e} est tout aussi est tout
aussi valable des charges $\mathcal{N}=(m,m^{ij},m^{ijkl},\dots)$
ou $(m^i,m^{ijk},\dots)$ des D-branes de type IIB ou IIA
sous ces champs. La partie imaginaire de l'action de Born-Infeld
\index{Born-Infeld, action de}
correspond alors {\`a} la contraction des spineurs conjugu{\'e}s
$\mathcal{R}$ et $\mathcal{N}$. La partie r{\'e}elle d{\'e}pend en
revanche des modules $g_{ij}, B_{ij}$ d{\'e}finissant l'{\'e}lement
de l'espace homog{\`e}ne $SO(d,d)/(O(d)\times SO(d))$, soit
la m{\'e}trique dans le r{\'e}seau de compactification
\begin{equation}
\| p \|^2_V = (p_i + B_{ik} p^k)g^{ij}(p_j+B_{jl}p^l)
+ p^i g_{ij} p^j
\end{equation}
o{\`u} $(p_i,p^i)$ se transforme dans la repr{\'e}sentation
fondamentale de $SO(d,d)$. Cette m{\'e}trique induit une
m{\'e}trique sur toutes les repr{\'e}sentations de $SO(d,d)$,
et en particulier sur la repr{\'e}sentation spinorielle 
$\mathcal{M}$. L'action de Born-Infeld peut alors s'{\'e}crire
\begin{equation}
S_{D} = e^{-\phi} \sqrt{\| \mathcal{N} \|^2} 
+ i \langle \mathcal{R} ,\mathcal{N} \rangle
\end{equation}
o{\`u} la T-dualit{\'e} est maintenant manifeste.
Connaissant l'action g{\'e}n{\'e}rale des D-branes, on peut alors
l'ins{\'e}rer dans la s{\'e}rie d'instantons
\index{serie d'instantons@s{\'e}rie d'instantons!couplage $R^4$ de II sur $T^d$}
\begin{equation}
f = f_{\rm pert} +  4\pi V e^{-\phi} \sum_{m\ne 0} \sum_{\mathcal{N}}
\frac{|m|}{\|\mathcal{N}\| }
K_1\left(2\pi |m| e^{-\phi} \| \mathcal{N}\| \right)
e^{2\pi i m \langle\mathcal{R} ,\mathcal{N} \rangle}
\end{equation}
et essayer de reproduire les amplitudes en $R^4$ {\`a}
l'ordre des arbres et {\`a} une boucle, tout au moins
en dimension $d>4$ o{\`u} les contributions de NS5-brane
\index{cinq-brane!de Neveu-Schwarz}
ne sont pas attendues. La r{\`e}gle de sommation
sur les {\'e}tats de D(-1), D0-,D1- et D2-branes est d{\'e}termin{\'e}e par
l'exactitude des couplages en $R^4$ en dimensions 7,8,9,10
\footnote{On peut n{\'e}anmoins concevoir de modifier cette r{\`e}gle
de sommation tout en pr{\'e}servant l'invariance sous la T-dualit{\'e},
par exemple en introduisant une contrainte 
$\langle\mathcal{R} ,\mathcal{R} \rangle=0$.}. 

\subsection{Instantons de D-branes et U-dualit{\'e}}
Si l'invariance sous la T-dualit{\'e} est ainsi garantie, ce n'est pas
le cas de l'invariance sous la U-dualit{\'e}. Tout comme l'invariance
sous la S-dualit{\'e} de la th{\'e}orie de type IIB, celle-ci peut
{\^e}tre {\'e}tudi{\'e}e par resommation de Poisson sur l'entier $m$.
On obtient ainsi une s{\'e}rie d'Eisenstein dont l'invariance
\index{Eisenstein, s{\'e}rie d'}
\index{Poisson, resommation de}
\index{U-dualit{\'e}!de type II sur $T^d$}
sous la sym{\'e}trie $Sl(d+1,\Zint)$ du tore {\'e}tendu doit
{\^e}tre manifeste. De mani{\`e}re inattendue, {\it cette invariance
n'est pas satisfaite\footnote{Corr{\'e}lativement, la s{\'e}rie
d'instantons de D-branes ne reproduit pas les contributions
perturbatives au couplage en $R^4$.} 
par les contributions de D-branes en dimension $d\le 6$}.
La resommation de Poisson conduit en effet {\`a} une s{\'e}rie
d'Eisenstein de terme g{\'e}n{\'e}ral $\mathcal{M}^{-3/2}$, o{\`u}
\begin{eqnarray}
\mathcal{M}^2 = &R_{11}^2 \left( n^{11} + {\cal A}_i n^i + \frac{1}{6} n^{ijk} 
\tilde{\cal C}_{ijk} \right)^2 
+\left(n^i + \frac{1}{2} n^{ijk} B_{jk} \right)\frac{g_{il}}{R_{11} }
\left(n^l + \frac{1}{2} n^{lmn} B_{mn} \right) \nonumber\\
&+ \frac{R_{11}^2}{6}
n^{ijk}~\frac{g_{il}g_{jm}g_{kn}}{R_{11}^3} ~n^{lmn} 
\end{eqnarray}
L'invariance sous les transformations de Lorentz de $T^{d+1}$
\index{D-brane!charges de}
requiert alors l'introduction d'une {\it quatre-forme} $n^{IJKL}$
aux c{\^o}t{\'e}s de la charge de D2-brane $n^{ijk}=n^{ijk11}$.
Les charges $n^I$ et $n^{IJKL}$ forment alors une repr{\'e}sentation
fondamentale du groupe de U-dualit{\'e} $SO(5,5,\Zint)$, de norme
carr{\'e}e
\begin{align}
\mathcal{M}^2 =&
\left( n^I + \frac{1}{6} n^{IJKL} {\cal C}_{JKL} \right)
g_{IM} \left( n^M + \frac{1}{6} n^{MNPQ} {\cal C}_{NPQ} \right)\nonumber\\
&+ \frac{1}{24} n^{IJKL} g_{IM}~g_{JN}~g_{KP}~g_{LQ}~n^{MNPQ}\ .
\end{align}
Cette extension est loin d'{\^e}tre anodine : la norme de la
charge $n^{ijkl}$ cro{\^\i}{}t en effet en $1/g^2$ par contraste
avec la norme des charges de D-branes en $1/g$, ce qui implique
l'existence d'effets en $e^{-1/g^2}$ l{\`a} o{\`u} seuls
des effets en $e^{-1/g}$ seraient attendus. L'existence
de ces effets n'est cependant pas av{\'e}r{\'e}e dans la mesure
o{\`u} toutes les tentatives pour reproduire les amplitudes
perturbatives {\`a} partir de cette s{\'e}rie d'Eisenstein ont
{\'e}chou{\'e}.

\section{D-instantons et la g{\'e}om{\'e}trie de $K_3$}
\index{K3, surface@$K_3$, surface}
\index{compactification!sur $K_3$}
L'{\'e}tude des couplages en $R^4$ dans les compactifications
toro{\"\i}{}dales de la M-th{\'e}orie nous a permis
d'identifier en d{\'e}tail les effets non perturbatifs 
intervenant dans ces couplages en termes 
d'instantons de D-branes enroul{\'e}es sur les cycles
non triviaux de la vari{\'e}t{\'e} de compactification.
Le succ{\'e}s de cette identification repose largement
sur la param{\'e}trisation explicite de l'espace des
modules, et sur la connaissance du spectre des cycles
supersym{\'e}triques du tore de dimension arbitraire.
Il est cependant souhaitable de d{\'e}terminer si
cette image persiste dans des situations plus r{\'e}alistes
et donc moins supersym{\'e}triques. L'{\'e}tude de
la compactification {\`a} 16 charges de supersym{\'e}trie sur $K_3$ repr{\'e}sente un
premier pas dans cette direction. La
dualit{\'e} h{\'e}t{\'e}rotique-type II  permet
\index{dualit{\'e}!h{\'e}t{\'e}rotique-type II}
encore de contr{\^o}ler les effets non perturbatifs
dans les couplages 1/2-BPS satur{\'e}s, et la simplicit{\'e}
\index{saturation BPS}
g{\'e}om{\'e}trique de la vari{\'e}t{\'e} $K_3$ permet encore
d'en donner une interpr{\'e}tation en termes de D-branes.

Nous avons d{\'e}j{\`a} discut{\'e} les cas de la th{\'e}orie de type
II compactifi{\'e}e sur $K_3\times T^2$ et de la th{\'e}orie h{\'e}t{\'e}rotique
sur $T^6$ dans le cadre des corrections gravitationnelles
en $R^2$~: ces derni{\`e}res sont calculables {\it exactement}
{\`a} une boucle dans la th{\'e}orie de type II, en raison du
d{\'e}couplage des multiplets vectoriels et des hypermultiplets
\index{decouplage@d{\'e}couplage!des hypers et vecteurs}
dans la th{\'e}orie $N=2$ sous-jacente~;
elles s'interpr{\`e}tent alors du c{\^o}t{\'e} h{\'e}t{\'e}rotique 
comme les contributions des instantons de NS5-brane enroul{\'e}es
sur le tore $T^6$. Il existe cependant d'autres couplages 
1/2-BPS satur{\'e}s pour lesquels cette restriction ne s'applique pas.
\index{saturation BPS}
C'est le cas du couplage {\`a} quatre d{\'e}riv{\'e}es $\tilde\mathcal{F}_1$
\begin{equation}
\label{ddss}
\frac{\tilde\mathcal{F}_1}{2S_2^2}
\left(\partial_\mu \partial_\nu S \partial^\mu \partial^\nu S
+\partial_\mu \partial_\nu \bar S \partial^\mu \partial^\nu \bar S\right)
\end{equation}
o{\`u} $S$ d{\'e}signe le scalaire complexe du dilaton dans la th{\'e}orie
\index{dilaton}
de type II (soit le module de K{\"a}hler $T$ de la th{\'e}orie
h{\'e}t{\'e}rotique duale). La comparaison des op{\'e}rateurs de vertex
du dilaton $S$ et du graviton montre que $\tilde\mathcal{F}_1$ est reli{\'e}
{\it perturbativement} au couplage gravitationnel $R^2$ par la
{\it sym{\'e}trie miroir} . Il se r{\'e}duit donc 
\index{miroir, sym{\'e}trie}
perturbativement {\`a} la contribution 
{\`a} une boucle donn{\'e}e par la fonction modulaire
\index{modulaire, forme}
\begin{equation}
\label{ddsiia}
\Ftilde_1^{IIA}= - 24 \log \left(U_2 |\eta(U)|^4 \right) \ ,
\end{equation}
obtenue {\`a} partir du couplage $\mathcal{F}_1$ (\ref{r2iia})
par {\'e}change des modules $T$ et $U$. Du point de vue de la
supersym{\'e}trie $N=2$, $U$ est cependant comme le dilaton membre d'un 
\index{hypermultiplet}
hypermultiplet, et les corrections non perturbatives ne sont
donc pas exclues. Elles sont du reste n{\'e}cessaires {\`a} l'invariance
de l'amplitude (\ref{ddss}) sous le groupe de U-dualit{\'e}
\index{U-dualit{\'e}!de Het/type II $N=4$}
$SO(6,22,\Zint)$ m{\'e}langeant le module $U$ aux modules de $K_3$
et au dilaton. On peut cependant les calculer perturbativement
du point de vue h{\'e}t{\'e}rotique, o{\`u} le d{\'e}couplage des
\index{decouplage@d{\'e}couplage!des hypers et vecteurs}
hypermultiplets au multiplet vectoriel du dilaton implique 
\index{vectoriel, multiplet}
que l'amplitude image $\partial\partial T\partial\partial T$
soit {\it exacte} {\`a} une boucle. Le d{\'e}veloppement {\`a} faible
couplage de type IIA, soit {\it {\`a} grand volume} $T_2\rightarrow\infty$
du c{\^o}t{\'e} h{\'e}t{\'e}rotique, permet alors d'analyser les
contributions du c{\^o}t{\'e} de type II.
Cette approche a {\'e}t{\'e} d{\'e}velopp{\'e}e conjointement avec
I. Antoniadis et T. Taylor, et a fait l'objet de la publication
annex{\'e}e en appendice \ref{dds}. Nous y r{\'e}f{\`e}rons le lecteur pour
les d{\'e}tails de la d{\'e}rivation, et nous contentons d'en 
d{\'e}crire le r{\'e}sultat.

\subsection{Couplage exact $\tilde\mathcal{F}_1$ {\`a} quatre dimensions}
Le calcul de l'amplitude de diffusion {\`a} une boucle h{\'e}t{\'e}rotique
\index{amplitude de diffusion!de quatre modules h{\'e}t{\'e}rotiques}
de deux modules $T$ avec deux autres modules quelconques $\phi_1$ et 
$\phi_2$ conduit {\`a} l'int{\'e}grale de la fonction de partition du
r{\'e}seau de Narain $\Gamma_{6,22}$ sur le domaine fondamental
\index{Narain, r{\'e}seau de}
\index{partition, fonction de!d'un r{\'e}seau}
\index{domaine fondamental modulaire}
du tore, en pr{\'e}sence d'insertions de moments $P_R$~:
\begin{equation}
\A_{\phi_1\phi_2}\!=\frac{\p^2}{T_2^2}
\int_{\F} d^2 \tau \,\tau_2\!\!\!  \sum_{P_L,P_R \in\Gamma_{6,22}}
\left[ P_R^I v_{IJ}(\phi_1) P_R^J\right]
\left[ P_R^I v_{IJ}(\phi_2) P_R^J \right]\;
e^{i\pi\tau P_L^2} e^{-i\pi\taubar
P_R^2}\frac{1}{\bar\eta^{24}(\bar\tau)}
\ .
\end{equation}
Le tenseur $v_{IJ}(\phi_i)$ d{\'e}crit la polarisation associ{\'e}e
au module $\phi_i$, et la fonction $1/\bar\eta^{24}$ repr{\'e}sente
la contribution des 24 oscillateurs du c{\^o}t{\'e} droit. 
\index{oscillateur}
\index{saturation BPS}
L'absence d'oscillateurs du c{\^o}t{\'e} supersym{\'e}trique gauche
traduit la saturation BPS de ce couplage. 

\index{decompactification@d{\'e}compactification, limite de}
Dans la limite de grand volume du deux-tore $T_2\rightarrow\infty$, le 
r{\'e}seau de Narain $\Gamma_{6,22}$ se factorise en deux r{\'e}seaux
$\Gamma_{2,2}\times\Gamma_{4,20}$~; les {\'e}tats de nombre
\index{enroulement!etat d'@{\'e}tat d'}
d'enroulement non nul autour de $T^2$ donnent lieu {\`a} des corrections
d'ordre $e^{-T_2/\alpha'}\sim e^{-1/g_{\rm IIA}^2}$ en principe
identifiables {\`a} des configurations instantoniques de NS5-branes 
\index{cinq-brane!de Neveu-Schwarz}
de type II, n{\'e}gligeables devant les corrections qui nous
int{\'e}ressent ici. Les moments droits $P_R$ s'identifient alors aux
moments gauches $P_L$, et l'amplitude {\`a} quatre modules
devient alors {\it int{\'e}grable} par rapport aux modules $\phi_1$
et $\phi_2$~:
\begin{equation}
\A_{\phi_1\phi_2} ~\approx~ \frac{1}{T_2^2}D_{\phi_1}D_{\phi_2} 
\tilde\mathcal{F}_1
\end{equation}
Le couplage $\tilde\mathcal{F}_1$ 
s'{\'e}crit alors, apr{\`e}s resommation de Poisson sur les
\index{Poisson, resommation de}
charges de moments autour de $T^2$ et int{\'e}gration sur $\tau_1$,
\begin{equation}
\Ftilde_1 ~=~ T_2
\int_0^{\infty} 
\frac{d\tau_2}{(\tau_2)^2}  
{\sum_{n^I,q^i}}'
d\left(\frac{q^t L q}{2}\right)
e^{-\frac{\pi}{\t_2}n^t G n~-~\p\t_2 q^t(M+L) q
~-~2\p i n^t Y^t q}
\end{equation}
Les deux charges $n^{I=1,2}$ sont duales aux charges
de moments autour du tore $T^2$ de m{\'e}trique $G_{IJ}$, 
tandis que les 24 charges $q^i$
indexent les vecteurs du r{\'e}seau $\Gamma_{4,20}$, de
norme carr{\'e}e $q^t M q$ et produit pair $q^t L q$. Les
$2\times 24$ modules $Y_{iI}$ d{\'e}signent les lignes de Wilson
\index{Wilson, ligne de}
des champs de jauge du r{\'e}seau $\Gamma_{4,20}$ autour
des deux cycles de $T^2$. Finalement, les coefficients
$d(N)$ correspondent aux coefficients de Fourier de 
\index{modulaire, forme}
la fonction modulaire $1/\eta ^{24}$ comme dans l'{\'e}quation 
(\ref{etaexp}). Les {\'e}tats de charge $q=0$ engendrent la contribution
dominante {\`a} cette expression~:
\begin{equation}
\Ftilde_1 = -24 \log \left(U_2 |\eta(U)|^4 \right) + \delta\Ftilde_1\ ,
\end{equation}
laquelle reproduit pr{\'e}cis{\'e}ment
l'amplitude {\`a} une boucle dans la th{\'e}orie de type IIA
duale (\ref{ddsiia}), donnant un test suppl{\'e}mentaire de la
dualit{\'e} $N=4$ h{\'e}t{\'e}rotique-type II. Les {\'e}tats de charge $q$ non nulle
\index{dualit{\'e}!h{\'e}t{\'e}rotique - type II}
induisent quant {\`a} eux des contributions 
d'ordre $e ^{-\sqrt{T_2}}\sim e^{-1/\sqrt{\alpha'}}$~:
\begin{equation}
\label{dft1}
\delta\Ftilde_1 = 2 T_2
{\sum_{n^I,q^i}}'
d\left(\frac{q^t L q}{2}\right)
\left[ \frac{q^t (M + L) q}{n^t G n} \right]^{1/2}
K_1 \left( 2\pi \sqrt{ n^t G n \cdot 
\frac{q^t (M+L) q}{2}}\right)
e^{-~2\pi i n^t Y^t q}\ , 
\end{equation}
\subsection{Instantons et l'homologie de $K_3$}  
Dans les variables de type IIA, ces contributions correspondent
alors {\`a} des effets d'ordre $e ^{-1/g_{\rm IIA}}$~:
\begin{equation}
\delta\Ftilde_1 = 2 T_2 e^{-\phi_6}
{\sum_{n^I,q^i}}'
d\left(\frac{q^t L q}{2}\right)
\left[  \frac{q^t (M + L) q}{n^t G n} \right]^{1/2}
K_1 \left( 2\p e^{-\phi_6}\sqrt{ n^t G n \cdot 
 \frac{q^t (M+L) q}{2}}\right)
e^{-~2\pi i n^t Y^t q}\ , 
\end{equation}
\index{serie d'instantons@s{\'e}rie d'instantons!couplage
  $(\partial\partial\phi)^2$ de II sur $K_3\times T^2$}
correspondant {\`a} une action euclidienne
\begin{equation}
\label{sdk3}
S=\sqrt{ n^t G n } \cdot e^{-\phi_6} \sqrt{ \frac{q^t (M+L) q}{2}}
- i n^t Y^t q
\end{equation}
o{\`u} $e^{-2\phi_6}=S_2/T_2$ d{\'e}signe le dilaton de la th{\'e}orie
de type IIA {\`a} six dimensions. Ces effets s'interpr{\`e}tent
\index{D-instanton!de type II sur $K_3\times T^2$}
\index{cycle d'homologie!de $K_3$}
naturellement en termes de {\it D-branes de la th{\'e}orie de
type IIA s'enroulant sur les cycles impairs de la vari{\'e}t{\'e}
de compactification} $K_3\times T^2$~: on reconna{\^\i}{}t en
effet dans l'action (\ref{sdk3}) le produit de la masse 
d'un {\'e}tat solitonique {\`a} six dimensions (\ref{sol6d}) par
la longueur $\sqrt{n^t G n}$
d'un cycle $S_1$ de nombres d'enroulement $(n^1,n^2)$
autour des deux cercles du tore $T^2$. L'{\'e}tat solitonique
{\`a} six dimensions correspondant lui m{\^e}me {\`a} une D-brane
enroul{\'e}e sur un cycle pair $\gamma$ de $K_3$, on obtient
bien ainsi l'action euclidienne de la D-brane enroul{\'e}e
sur le produit $S_1\times \gamma$. L'identification des
lignes de Wilson $Y$ h{\'e}t{\'e}rotiques avec les valeurs de fond
$\int_{\gamma^i} \mathcal{R}$ des potentiels de Ramond
permet {\'e}galement d'interpr{\'e}ter la partie imaginaire de 
\index{Wess-Zumino, terme de!sur le volume d'univers des D-branes}
l'action effective comme le couplage de la D-brane aux
champs de Ramond~:
\begin{equation}
n^t Y^t q =
\int_{S_1} q^i \mathcal{R}_i =
\int_{S_1\ti K_3} \gamma \w \gamma_i \w \mathcal{R}_i =
\int_{S_1\ti K_3} \gamma \w \mathcal{R} =
\int_{S_1\ti \gamma} \mathcal{R}\ . 
\end{equation}

Du point de vue de la th{\'e}orie de type IIB
\footnote{Ce paragraphe constitue une addition par rapport
{\`a} la publication en annexe \ref{dds}.}, 
l'action euclidienne des effets non perturbatifs devient 
\begin{equation}
\label{sdk3b}
S=\sqrt{ (n+m B)^2 +  V^2 m^2} \cdot \sqrt{e^{-2\phi_6} 
\frac{q^t (M+L) q}{2}}
- i (n Y^t + m \tilde Y^t)q
\end{equation}
o{\`u} on a renomm{\'e} les charges $n^1,n^2$ en $m,n$ et s{\'e}par{\'e}
la matrice des champs de Ramond en deux vecteurs $Y$,
correspondant aux valeurs de fond des champs de Ramond 
sur les 24 cycles de $K_3$, et $\tilde Y$, correspondant
aux valeurs de fond sur $T^2$ des 24 tenseurs antisym{\'e}triques
{\`a} six dimensions obtenus en r{\'e}duisant les champs de Ramond sur $K_3$.
Les entiers $n$ et $m$ diff{\'e}rencient les deux types de
configurations instantoniques susceptibles d'appara{\^\i}{}tre
dans la compactification de la th{\'e}orie de type IIB sur 
$K_3 \times T^2$~: 
\begin{itemize}
\item Les D(-1)-,D1- et D3-branes enti{\`e}rement enroul{\'e}es sur 
un cycle $\gamma=q^i\gamma_i$ de $K_3$ 
donnent lieu {\`a} des {\it D-instantons} dans
la th{\'e}orie {\`a} six dimensions, et {\it a fortiori} {\`a} quatre
dimensions.  Leur action euclidienne est
proportionnelle {\`a} l'aire du cycle $\gamma$ et reproduit l'{\'e}quation
(\ref{sdk3b}) pour $(m,n)=(0,1)$. La phase $i n Y^t q$
est {\'e}galement en accord avec cette interpr{\'e}tation.
\item Les m{\^e}mes D-branes peuvent {\'e}galement {\^e}tre enroul{\'e}es
partiellement sur le cycle $\gamma$, et donner lieu {\`a} des
{\it D-cordes} {\`a} six dimensions, de tension
\index{tension!des D-cordes sur $K_3$}
\begin{equation}
T=\sqrt{e^{-2\phi_6} \frac{q^t (M+L) q}{2}}\ ;
\end{equation}
celles-ci peuvent encore {\^e}tre enroul{\'e}es sur le tore $T^2$, 
pour former des D-instantons de la th{\'e}orie {\`a} quatre dimensions.
L'action euclidienne associ{\'e}e {\`a} ces effets est alors le
produit de l'action de Born-Infeld $\sqrt{V^2+B^2}$ par
\index{Born-Infeld, action de}
la tension, reproduisant l'{\'e}quation (\ref{sdk3b}) dans le cas 
$(m,n)=(1,0)$.
\end{itemize}
Les configurations solitoniques pour $(m,n)$ quelconques
correspondent {\`a} des superpositions de D-instantons et
de D-cordes, soit, en analogie avec le cas de la compactification 
toro{\"\i}{}dale, {\`a} des D-cordes de flux {\'e}lectrique non nul.

Si l'interpr{\'e}tation du poids semi-classique $e^{-S}$ en termes
de configurations de D-bra\-nes ne pose pas de difficult{\'e},
les coefficients apparaissant devant
ce poids dans l'{\'e}quation (\ref{dft1}) sont moins clairs
mais devraient en principe r{\'e}sulter de l'expression des op{\'e}rateurs de vertex
\index{op{\'e}rateur de vertex}
du dilaton dans le champ de fond de ces instantons, ainsi que de 
l'int{\'e}gration sur les modes z{\'e}ro. Le coefficient entier
$d(q^t L q/2)$ peut en revanche {\^e}tre interpr{\'e}t{\'e}
comme le {\it nombre de cycles supersym{\'e}triques de $K_3$}
dans la classe d'homologie $\sum \gamma_i q^i$
\footnote{$q^t L q$ est alors {\'e}gale {\`a} l'auto-intersection
du cycle, c'est-{\`a}-dire dans le cas des surfaces de Riemann
\index{intersection}
\index{Euler, caract{\'e}ristique d'}
sa caract{\'e}ristique d'Euler.}. Nous avons d{\'e}ja mentionn{\'e}
cette conjecture dans le cadre du comptage
des {\'e}tats BPS de la th{\'e}orie de type IIA, et nous la retrouvons
\index{etats BPS@{\'e}tats BPS!de Het$/T^4$ - type IIA$/K_3$}
ici de mani{\`e}re naturelle et ind{\'e}pendante.

\subsection{D{\'e}compactification et couplages en $R^4$\label{ddsdec}}
Le couplage $\Ftilde_1$ exact {\`a} quatre dimensions {\'e}tant ainsi
obtenu, il est tentant d'en {\'e}tudier les cons{\'e}quences en
dimensions sup{\'e}rieures par simple d{\'e}compactification.
\index{decompactification@d{\'e}compactification, limite de}
Du point de vue de type IIA, la d{\'e}compactification
$T_2\rightarrow\infty$ {\`a} dilaton $e^{\phi_6}$ fix{\'e} {\'e}limine
toutes les contributions de D-branes, puisque celles ci
s'enroulent autour d'un des cercles du tore. La contribution
{\`a} une boucle, ind{\'e}pendante du volume du tore, dispara{\^\i}t
{\'e}galement, et le couplage 
$\partial\partial\phi_6\partial\partial\phi_6$
est donc nul en type IIA {\`a} six dimensions. La situation
est plus int{\'e}ressante du c{\^o}t{\'e} de type IIB, o{\`u} les
D-instantons $(m,0)$ subsistent en dimension 6, ainsi
\index{D-instanton!de type IIB sur $K_3$}
que la contribution {\`a} une boucle~:
\begin{equation}
\label{amplB6}
\Ftilde_1^{(6)} = 8\pi
~+~2 e^{-\varphi_6}\! 
{\sum_{m\ne 0 \atop q^i\ne 0}}
d\!\left( \frac{q^t L q}{2}\right)
\frac{ \sqrt{ q^t (M + L) q }}{|m|}
~K_1 \left( 2\pi |m| e^{-\varphi_6} 
\sqrt{ \frac{q^t (M+L) q}{2}}  \right)
e^{-2\pi i m Y_i q^i} 
\end{equation}
Dans la limite d{\'e}cadimensionnelle o{\`u} le volume de $K_3$ tend
lui m{\^e}me vers l'infini, seuls les D(-1)-instantons subsistent
dans la somme, qui se r{\'e}duit alors {\`a} 
\begin{equation}
\Ftilde_1^{(10)} = \frac{24}{2\pi} \left[ 
\frac{2\pi ^2}{3} + 4\pi e^{-\varphi_{10}} 
{\sum_{m\ne 0, q\ne 0}}
\left| \frac{q}{m} \right|
~K_1 \left( 2\p |q m| e^{-\varphi_{10}} \right)
e^{-2\pi i m q a} \right]
\end{equation}
De mani{\`e}re inattendue, {\it la limite {\`a} dix dimensions des couplages
$\partial\partial\phi\partial\partial\phi$ de la th{\'e}orie de
type IIB compactifi{\'e}e sur $K_3$ reproduit ainsi le couplage
en $R^4$ exact de Green et 
\index{Green et Gutperle, conjecture de}
Gutperle\footnote{au terme {\`a} l'ordre des arbres pr{\`e}s. 
On montre dans l'appendice de l'article en annexe \ref{dds} que ce terme
est en r{\'e}alit{\'e} pr{\'e}sent, mais qu'il n'affecte pas les
amplitudes physiques {\`a} quatre modules.} de la th{\'e}orie de
type IIB {\`a} dix dimensions~!}
Cette co{\"\i}{}ncidence peut cependant s'expliquer de la mani{\`e}re
suivante~: les couplages en $R^4$ de la th{\'e}orie de type IIB
sont reli{\'e}s par supersym{\'e}trie {\`a} des couplages {\`a}
huit d{\'e}riv{\'e}es $R^2\partial\partial\phi\partial\partial\phi$,
dont la r{\'e}duction sur $K_3$, de nombre de Pontryagin
$\int R^2 =24$, g{\'e}n{\`e}re l'interaction
$\Ftilde_1$. Inversement, {\it on d{\'e}montre ainsi} la validit{\'e}
de la conjecture de Green et Gutperle, et en particulier 
\index{cuspoidale, forme@cuspo{\"\i}{}dale, forme}
l'absence de contributions des formes cuspo{\"\i}{}dales de 
$Sl(2,\Zint)$ aux couplages en $R^4$.


%% file: chap5.tex
\chapter{M comme Matrice~?}
L'{\'e}cheveau des dualit{\'e}s des th{\'e}ories de cordes
que nous avons d{\'e}crit au chapitre 3 de ce m{\'e}moire nous
a conduit {\`a} l'id{\'e}e que les cinq th{\'e}ories des
supercordes {\`a} dix dimensions n'{\'e}taient que cinq regards
perturbatifs sur une th{\'e}orie fondamentale encore
myst{\'e}rieuse, dite M-th{\'e}orie. Si le spectre perturbatif de ces 
th{\'e}ories des cordes correspond {\`a} celui
d'une th{\'e}orie {\`a} nombre infini de champs en dix dimensions
le spectre BPS non perturbatif signale l'existence 
d'une onzi{\`e}me dimension {\it compacte}~: les D0-branes
repr{\'e}sentent les modes de Kaluza-Klein du supergraviton, et
les cordes perturbatives apparaissent alors comme les membranes enroul{\'e}es
selon cette direction. Il est donc naturel de rechercher une formulation de 
la th{\'e}orie fondamentale dans l'espace-temps {\it non compact}
{\`a} onze dimensions, reproduisant la dynamique perturbative 
des th{\'e}ories de supercordes apr{\'e}s compactification.
La th{\'e}orie des matrices, propos{\'e}e par Banks, Fischler, Shenker et
\index{matrices, th{\'e}orie des}
\index{Banks, Fischler, Shenker et Susskind, conjecture de}
Susskind en 1996 et revisit{\'e}e par Susskind en 1997, 
constitue une tentative de d{\'e}finition
{\it ab initio} de cette th{\'e}orie
\cite{Banks:1997vh,Susskind:1997cw}. Nous donnerons
une br{\`e}ve introduction {\`a} la th{\'e}orie des matrices
\footnote{Le lecteur pourra {\'e}galement se reporter aux articles de revue
\cite{Banks:1997mn,Bigatti:1997jy,Bilal:1997fy} pour plus de
d{\'e}tails.} et discuterons
plus particuli{\`e}rement ses compactifications toro{\"\i}{}dales,
dans le but de comprendre les U-dualit{\'e}s des th{\'e}ories
des cordes maximalement supersym{\'e}triques. Ce chapitre visera
en m{\^e}me temps {\`a} introduire au travail effectu{\'e} en 
collaboration avec Niels Obers et Eliezer Rabinovici, annex{\'e}
en appendice \ref{mu}.

\section{Quantification sur le front de lumi{\`e}re}

\subsection{Cin{\'e}matique sur le front de lumi{\`e}re}
Si l'invariance de Poincar{\'e} est un pr{\'e}requis de toute th{\'e}orie
physique, sa manifestation explicite dans les th{\'e}ories de
jauge oblige cependant {\`a} l'introduction de degr{\'e}s 
de libert{\'e} <<fant{\^o}matiques>> qui obscurcissent la physique.
La d{\'e}finition de l'espace de Hilbert d'une th{\'e}orie
des champs n{\'e}cessite du reste une foliation ({\it slicing})
\index{lumi{\`e}re, front de|textit}\index{foliation}
de l'espace-temps qui brise l'invariance de Poincar{\'e}
(figure \ref{front}).
La foliation par {\it surfaces de temps $t=x^0$ {\'e}gal} est g{\'e}n{\'e}ralement
choisie, et est invariante sous les translations et rotations 
spatiales, correspondant aux g{\'e}n{\'e}rateurs {\it cin{\'e}matiques}
du groupe de Poincar{\'e}. Les g{\'e}n{\'e}rateurs  de translation
selon le temps et de {\it boost} font quant {\`a} eux explicitement intervenir les
complications de la dynamique. 
\fig{5cm}{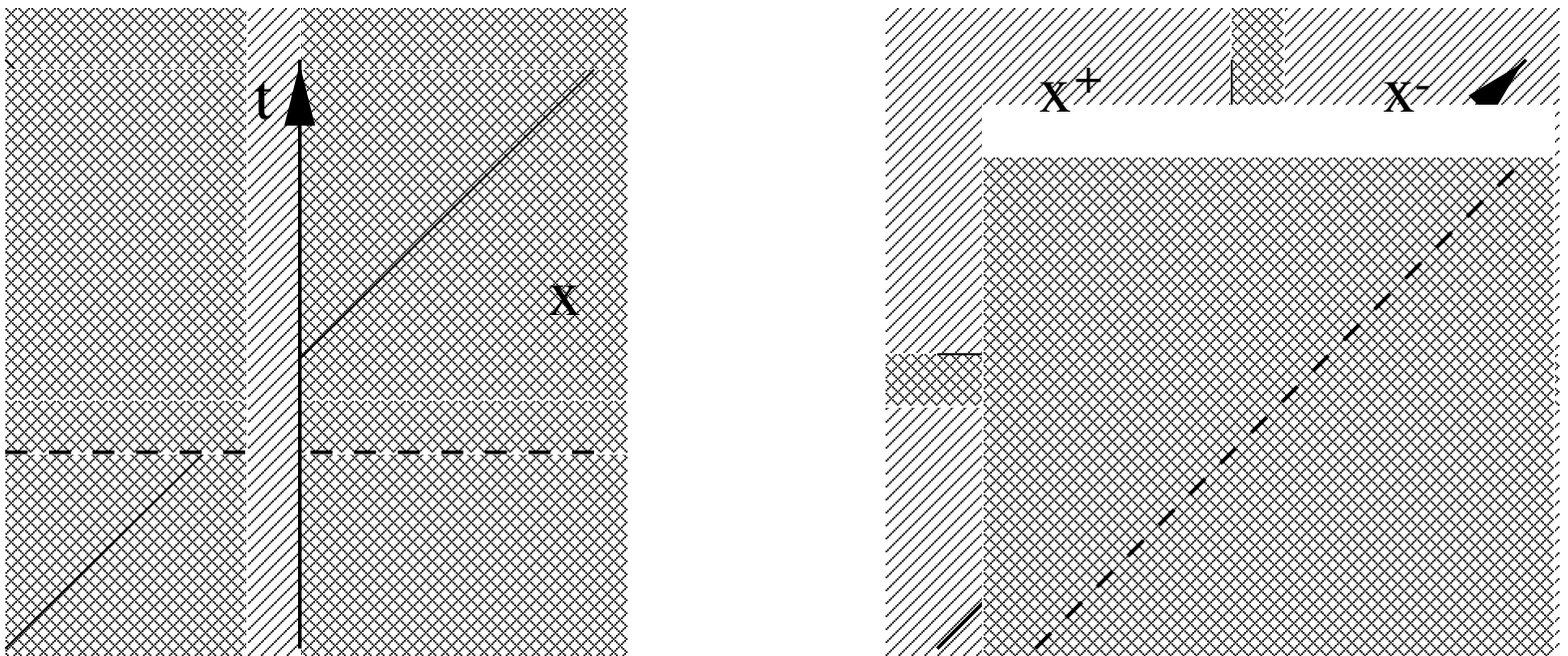}{Quantification {\`a} temps {\'e}gal (gauche) et
sur le front de lumi{\`e}re (droit).}{front}
La foliation par {\it surfaces de <<temps de lumi{\`e}re>>} 
$x^+=(x^0+x^1)/\sqrt{2}$, correspondant {\`a} la {\it quantification
sur le front de lumi{\`e}re}\footnote{Le terme <<c{\^o}ne de lumi{\`e}re>>
est inappropri{\'e}, car seule la moiti{\'e} du c{\^o}ne $(x^0)^2-(x^1)^2=0$
supporte la fonction d'onde. On utilise {\'e}galement la 
d{\'e}nomination de <<r{\'e}f{\'e}rentiel de moment infini>>, qui n'est
gu{\`e}re plus adapt{\'e}e.},
est avantageuse de ce point de vue, puisque les translations
et rotations transverses $P^i$ et $L^{ij}$ ainsi que 
le {\it moment longitudinal} $P^+$ et les boosts $L^{-i}$ et
$L^{+-}$ sont des g{\'e}n{\'e}rateurs cin{\'e}matiques\footnote{La
quantification {\`a} $x^-=(x^0-x^1)/\sqrt{2}$ constant pr{\'e}sente ainsi 
un g{\'e}n{\'e}rateur cin{\'e}matique suppl{\'e}mentaire sur
la quantification {\`a} $t$ constant, quelque soit la dimension
de l'espace.}. $P^-$ engendre les translations selon $x^-$
et joue le r{\^o}le du Hamiltonien.
Par contraste avec la relation de dispersion
non polynomiale de la quantification {\`a} temps {\'e}gal
(figure \ref{cone})
\begin{equation}
\label{etdr}
H = \sqrt{ P^i P_i + \mathcal{M}^2} \ ,
\end{equation}
la relation de dispersion prend sur le
front de lumi{\`e}re la forme
\index{lumi{\`e}re, front de!relation de dispersion}
\begin{equation}
\label{lcdr}
P^- = \frac{P^iP_i + \mathcal{M}^2}{2P^+}\ .
\end{equation}
La similitude de cette relation avec la relation de dispersion
non relativiste $H=\mathcal{M}+(p^i)^2/2\mathcal{M}$ r{\'e}sulte de l'invariance 
sous le {\it groupe de Galil{\'e}e} de l'espace transverse.
\fig{5cm}{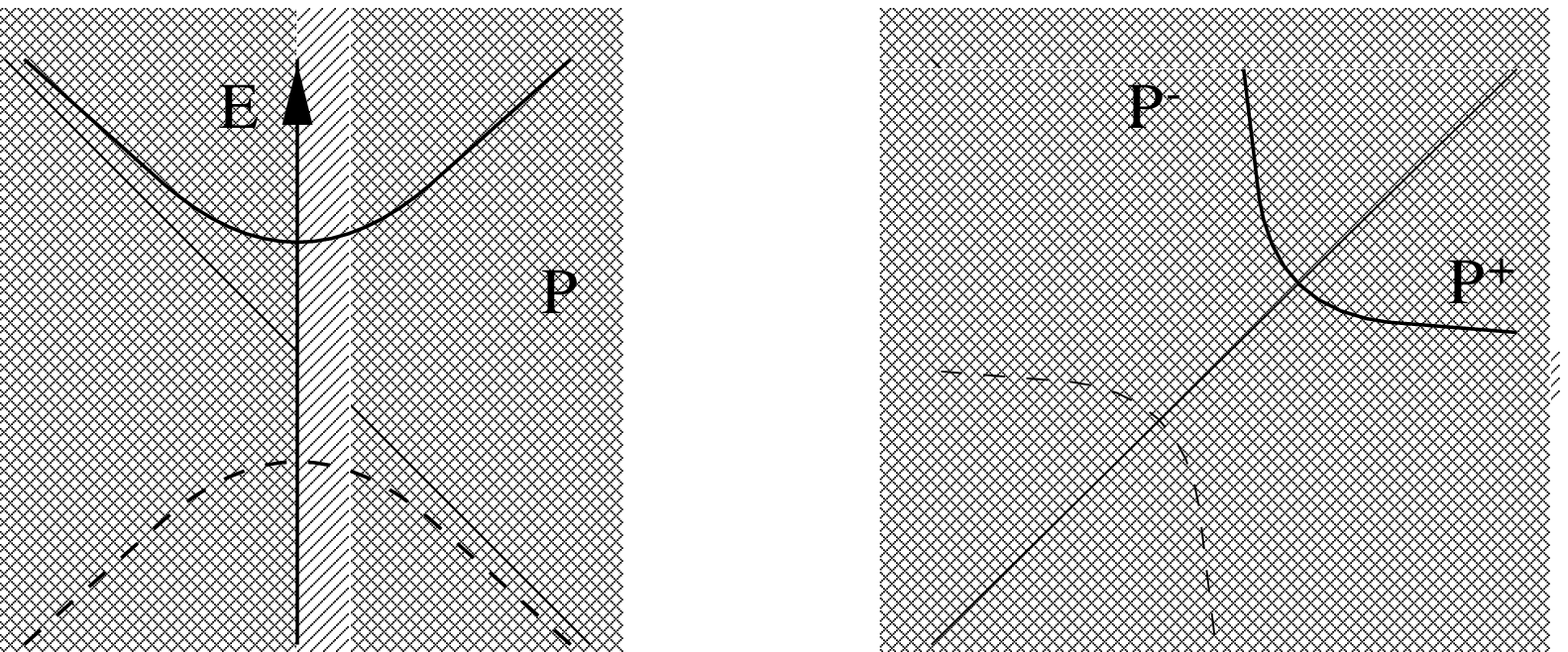}{Relation de dispersion {\`a} temps {\'e}gal (gauche) et
sur le front de lumi{\`e}re (droit).}{cone}
L'{\'e}quation (\ref{lcdr}) implique en particulier
que les particules, correspondant {\`a} $P^->0$, 
\index{Galil{\'e}e, invariance de}
ont un {\it moment longitudinal $P^+$ positif}, tandis  que les 
antiparticules ont un moment {\it n{\'e}gatif} (figure \ref{cone}). 
L'{\'e}tat du vide
de $P^-$ se r{\'e}duit donc {\`a} l'{\'e}tat fondamental
$|0\rangle$ de l'espace de Fock
\footnote{{\`a} la contribution des modes de moment $P^+$ nul pr{\`e}s.
Ces derniers sont exclus par la relation de dispersion
(\ref{lcdr}), sauf lorsque la masse $m$ s'annule.}.
Les {\'e}tats fant{\^o}matiques de norme n{\'e}gative sont {\'e}galement
exclus dans cette formulation.
En contrepartie de ces simplifications, le hamiltonien sur
le front de lumi{\`e}re pr{\'e}sente des interactions non locales
{\it instantan{\'e}es} correspondant {\`a} la pr{\'e}sence du p{\^o}le
en $P^+=0$ dans la relation (\ref{lcdr}), et donnant
naissance aux diagrammes de <<mouettes>> caract{\'e}ristiques
\index{mouette, diagramme de}
de la quantification sur le front de lumi{\`e}re. Le traitement
des modes de moment $P^+\rightarrow 0$ constitue le point le plus
d{\'e}licat de ce sch{\'e}ma de quantification, sous lequel se
r{\'e}fugient 
les
\index{vide!sur le front de lumi{\`e}re}
complications de la structure du vide dans l'approche
ordinaire (voir par exemple \cite{Pinsky:1993ey}).

\subsection{Front de lumi{\`e}re discret et r{\'e}solution spectrale}
La th{\'e}orie des champs peut {\^e}tre r{\'e}gularis{\'e}e {\`a} la fois dans
l'infrarouge et dans l'ultraviolet en supposant la direction
$x^-$ compacte de <<longueur>> $2\pi L$
\cite{Maskawa:1976,Pauli:1985pv}. Le moment longitudinal 
de la particule $i$ est
alors quantifi{\'e} selon
\begin{equation}
P^+_i = \frac{n}{L}  \ ,\quad n=1,2,\dots\ , 
\end{equation}
ce qui vaut {\`a} cette proc{\'e}dure le nom de 
{\it quantification sur le front de lumi{\`e}re discret} 
({\it discrete light cone quantization}, ou DLCQ). 
\index{lumi{\`e}re, front de!discret|textit}
Le moment total $P^+$ {\'e}tant conserv{\'e}, l'espace de Hilbert
se d{\'e}compose alors en secteurs de supers{\'e}lection 
$\mathcal{H}_N$ de moment 
$P^+ =N/L$, de dimension finie\footnote{dans le
cas d'une th{\'e}orie en dimension $1+1$. En dimension
sup{\'e}rieure, le continuum des impulsions transverses
subsiste, et peut {\^e}tre lev{\'e} par une compactification
appropri{\'e}e.}
engendr{\'e} par les {\'e}tats de l'espace de Fock 
$\alpha^\dagger_{n_1}\alpha^\dagger_{n_2}\dots
\alpha^\dagger_{n_k} |0\rangle$  tels que
$n_1+n_2+\dots+n_k=N$~; la diagonalisation du
hamiltonien $P^-$ peut alors {\^e}tre effectu{\'e}e
num{\'e}riquement dans chaque secteur.
Contrairement au cas de la quantification
{\`a} temps {\'e}gal ordinaire, cette finitude ne requiert pas de 
{\it cut-off} ultraviolet $n\le \Lambda$ sur les impulsions
longitudinales $P^+_i$~: la dimension finie
est assur{\'e}e par la condition $n_i>0$.
Il faut cependant prendre garde au fait que la direction $x^-$ est 
{\it de genre lumi{\`e}re}, et que $L$ n'est pas une longueur
invariante. Elle peut {\^e}tre modifi{\'e}e {\`a} volont{\'e} par
un boost de Lorentz $L^{+-}$~:
\begin{equation}
\begin{pmatrix} x^0 \\ x^1 \end{pmatrix}
\rightarrow
\begin{pmatrix} \cosh \beta & -\sinh\beta \\
                -\sinh \beta & \cosh\beta \end{pmatrix}
\begin{pmatrix} x^0 \\ x^1 \end{pmatrix}\ ,\quad
L \rightarrow e^{\beta} L\ ,\quad
P^- \rightarrow e^{\beta} P^-\ ,\quad
P^+ \rightarrow e^{-\beta} P^+\ .
\end{equation}
La d{\'e}pendance en $L$ du moment $P^+$ et du hamiltonien
$P^-$ est donc particuli{\`e}rement simple~:
\begin{equation}
P^+ = \frac{N}{L} \ , P^- = L H_N
\end{equation}
et la masse $\mathcal{M}^2 = 2 P^+ P^-$ est en particulier 
ind{\'e}pendante de $L$. 
La compacit{\'e} de la direction $x^-$ n'est cependant qu'un
artifice de calcul, et les r{\'e}sultats physiques sont
obtenus {\it dans la limite double} 
$L\rightarrow\infty$, $N\rightarrow\infty$
{\`a} moment longitudinal $P^+$ fix{\'e}. 
Le spectre de masse, correspondant
aux valeurs propres de la matrice $H_N$, d{\'e}pend
de $N$ et se {\it pr{\'e}cise} au fur et {\`a} mesure que $N$ augmente.
L'entier $N$ correspond donc en r{\'e}alit{\'e} {\`a} la {\it r{\'e}solution
spectrale} bien plus qu'au moment longitudinal. On peut {\'e}valuer
\index{spectrale, r{\'e}solution|textit}
 la valeur de $N$ minimale pour repr{\'e}senter
un {\'e}tat au repos de masse $\mathcal{M}$ et de taille caract{\'e}ristique
$r$ en demandant que cette derni{\`e}re soit tr{\`e}s
inf{\'e}rieure au rayon du cercle de genre lumi{\`e}re $L$
\cite{Bigatti:1997gm}.
En tenant compte de l'{\'e}galit{\'e} au repos $\mathcal{M}=P^-=P^+$,
on obtient ainsi la condition 
\begin{equation}
N \gg r~\mathcal{M}\ .
\end{equation}

\subsection{La th{\'e}orie des cordes sur le front de lumi{\`e}re
discret\label{lcst}}
La quantification sur le front de lumi{\`e}re a jou{\'e} un r{\^o}le historique
tr{\`e}s important dans la compr{\'e}hension des mod{\`e}les duaux et
\index{duaux, mod{\`e}les}
des th{\'e}ories de cordes \cite{Goddard:1973qh}. 
Elle permet en effet de fixer compl{\`e}tement
les reparam{\'e}trisations de la surface d'univers, et
ne laisse que les degr{\'e}s de libert{\'e} physiques des fluctuations transverses
de la corde. Nous avons pu appr{\'e}cier l'efficacit{\'e} de cette
proc{\'e}dure au chapitre 3 pour la d{\'e}termination des fonctions
de partition des th{\'e}ories des cordes. Elle peut {\^e}tre ais{\'e}ment
adapt{\'e}e au cas de la compactification sur un cercle de genre
lumi{\`e}re $x^-\equiv x^-+2\pi L$, comme nous le rappelons
maintenant.
\index{lumi{\`e}re, front de!th{\'e}orie des cordes}

La jauge du front de lumi{\`e}re consiste {\`a} identifier le
temps de la surface d'univers avec le {\it temps propre}
de l'espace cible~:
\begin{equation}
X^+(\sigma,\tau) = x^{+} + \alpha' P^+ \tau
\end{equation}
La condition de Virasoro associ{\'e}e {\`a} l'invariance sur les
reparam{\'e}trisations
\begin{equation}
0=\partial_\sigma X^\mu \partial_\tau X_\mu
= P^+ \partial_\sigma X^- - \partial_\sigma X^i \partial_\tau X^i
\end{equation}
permet d'{\'e}liminer la coordonn{\'e}e $X^-$ au profit
des coordonn{\'e}es transverses. Les coefficients de Fourier
de $X^-$ s'expriment alors en fonction des oscillateurs 
transverses~:\index{oscillateur}
\begin{eqnarray}
\label{modeex}
\alpha_n^- &=& \frac{1}{P^+} \left( 
\frac{1}{2}\sum_{m=-\infty}^\infty 
~: \alpha^i_{n-m}\alpha^i_m~: - a\delta_{n} \right) \\
\tilde\alpha_n^- &=& \frac{1}{P^+} \left( 
\frac{1}{2}\sum_{m=-\infty}^\infty 
~: \tilde\alpha^i_{n-m}\tilde\alpha^i_m~: - \tilde a\delta_{n} \right) 
\end{eqnarray}
La compactification $x^-\equiv x^-+2\pi L$ peut {\^e}tre construite
comme une construction d'orbifold habituelle~: elle projette
donc sur les {\'e}tats invariants de moment $P^+=N/L$,
et introduit des {\'e}tats {\it twist{\'e}s} s'enroulant
\index{twist{\'e}, {\'e}tat}
$n$ fois autour de la direction compacte~:
\begin{equation}
X^-(\sigma+2\pi) = X^- +2\pi n L
\end{equation}
En ins{\'e}rant les modes z{\'e}ros gauches et droits de $X^-$
\begin{equation}
\alpha_0^-=  \frac{P^-}{2} + \frac{nL}{2\alpha'} \ ,\quad
\tilde\alpha_0^-=  \frac{P^-}{2} - \frac{nL}{2\alpha'}  \ ,\quad
\alpha_0^i= \bar\alpha_0^i= \frac{P^i}{2} \ ,\quad
\end{equation} 
la somme et la diff{\'e}rence des {\'e}quations
(\ref{modeex}) conduisent {\`a} la formule de masse et la
condition de {\it level matching}~:
\index{masse, formule de!sur le front de lumi{\`e}re}
\index{level matching, condition de}
\begin{eqnarray}
 P^- &=& \frac{1}{2P^+} \left( P^i P^i + 2\frac{N_L + N_R -a -\bar a}
{\alpha'} \right) \\
 n N &=& N_L - N_R - a +\bar a
\end{eqnarray}
Ces relations peuvent {\^e}tre {\'e}tendues au cas des supercordes
en utilisant les valeurs appropri{\'e}es des {\it intercepts}
\index{intercept}
$a$ et $\bar a$, et incluant les contributions fermioniques
dans les nombres d'excitations $N_L$ et $N_R$.
La limite de grand $N$ est triviale dans le cas de la th{\'e}orie
des cordes libres, mais devient int{\'e}ressante en pr{\'e}sence 
d'interactions~; celles-ci n{\'e}cessitent cependant 
l'introduction d'une th{\'e}orie de champs de cordes,
\index{champ de cordes, th{\'e}orie de}
qui n'a pas encore {\'e}t{\'e} pleinement d{\'e}velopp{\'e}e.
La quantification du moment $P^-$ peut {\^e}tre commod{\'e}ment
incorpor{\'e}e en red{\'e}finissant la p{\'e}riodicit{\'e} de la 
coordonn{\'e}e $\sigma$ {\`a} $\sigma\equiv\sigma+N/L$. 
Chaque corde consiste alors en un nombre entier de
<<partons>> {\'e}l{\'e}mentaires de longueur $1/L$,
et la conservation du moment $P^-$ au cours des 
interactions entre cordes correspond
{\`a} la conservation du nombre de partons. 
\index{parton}

\section{M-th{\'e}orie sur le front de lumi{\`e}re}
Selon la conjecture BFSS, dans sa formulation renforc{\'e}e
par Susskind, la M-th{\'e}orie est d{\'e}crite {\it sur le 
front de lumi{\`e}re discret} dans le secteur de
moment longitudinal $P^+=N/L$ par {\it la m{\'e}canique quantique
de $N$ D0-branes}. Cette conjecture peut {\^e}tre heuristiquement
\index{D-brane!D0-branes}
justifi{\'e}e en identifiant le cercle de genre {\it lumi{\`e}re}
de <<rayon de lumi{\`e}re>> 
$L$ d{\'e}finissant le front de lumi{\`e}re discret {\`a} un cercle
spatial ordinaire de rayon $R_s$ {\it infiniment boost{\'e}}~:
\begin{equation}
\begin{pmatrix} \cosh \beta  & -\sinh\beta \\
                -\sinh \beta & \cosh\beta \end{pmatrix}
\begin{pmatrix} 0 \\ R_s \end{pmatrix}
= \frac{L}{\sqrt{2}}
\begin{pmatrix} -1+e ^{-2\beta} \\ 1+e^{-2\beta} \end{pmatrix}
\rightarrow \frac{1}{\sqrt{2}}
\begin{pmatrix} -L \\ L \end{pmatrix}
\end{equation}
A rayon de lumi{\`e}re $L$ fix{\'e}, le rayon $R_s= \sqrt{2}L e ^{-\beta}$ 
tend donc vers 0 dans la limite de boost infini
$\beta\rightarrow\infty$. Le secteur 
$(P^+= N/L,P^-=L H_N)$ du front
de lumi{\`e}re discret correspond donc au secteur
$(P^+_s=N/R_s,P^-=R_s H_N)$,
soit $(P\sim N/R_s, H= P+ R_s H_N)$  de la M-th{\'e}orie
{\it compactifi{\'e}e sur un cercle spatial de rayon}
$R_s \rightarrow 0$. Cette th{\'e}orie correspond pr{\'e}cis{\'e}ment
{\`a} la th{\'e}orie des cordes de type IIA {\it en pr{\'e}sence de
$N$ D0-branes}.
\index{D-brane!D0-branes}
Pour d{\'e}finir sans ambigu{\"\i}{}t{\'e} la limite {\`a} consid{\'e}rer,
il est n{\'e}cessaire de d{\'e}terminer le r{\'e}gime d'{\'e}nergie
pertinent pour la description de la M-th{\'e}orie sur le 
front de lumi{\`e}re discret.
Le hamiltonien sur le front de lumi{\`e}re discret $P^-${\'e}tant proportionnel
au <<rayon de lumi{\`e}re>> $L$, il s'{\'e}crit donc, pour raison de dimension,
\begin{equation}
P^- = \frac{L}{l_{11}^2} h_N
\end{equation}
o{\`u} le spectre de $h_N$ est d'ordre unit{\'e}.
Apr{\`e}s le boost, on obtient ainsi une {\'e}nergie
\begin{equation}
H-P = \frac{R_s}{l_{11}^2} h_N
\end{equation}
et il est donc commode d'effectuer un changement d'{\'e}chelle
$l_{11}\rightarrow l_{11s} = l_{11} e^{2\beta}$ de mani{\`e}re {\`a} maintenir
cette {\'e}nergie finie. Les degr{\'e}s de libert{\'e} pertinents 
seront s{\'e}lectionn{\'e}s dans la limite de {\it scaling}
\index{scaling, limite de|textit}
\begin{equation}
\label{scaling}
R_s\rightarrow 0\ , \quad M=\frac{R_s}{l_{11s}^2}= \frac{L}{l_{11}^2}
\ \mbox{~fix{\'e}}
\end{equation}
Exprim{\'e}s en termes de $R_s$ et $M$, les param{\`e}tres de la
th{\'e}orie de type IIA s'{\'e}crivent alors
\begin{equation}
g = (R_s M)^{3/4}\ ,\quad
\alpha' = \frac{ R_s^{1/2}}{ M^{3/2}}
\end{equation}
La th{\'e}orie des cordes de type IIA est donc faiblement coupl{\'e}e
et {\`a} basse {\'e}nergie dans la limite. Les corrections
de boucles et de d{\'e}riv{\'e}es sup{\'e}rieures {\`a} la dynamique des 
D0-branes sont donc supprim{\'e}es, et celle-ci est 
{\it exactement} d{\'e}crite par l'approximation de Yang-Mills 
{\`a} l'action de Born-Infeld supersym{\'e}trique~:
\index{Born-Infeld, action de}
\begin{equation}
\mathcal{L}=\int d\tau \frac{(\alpha')^{3/2}}{2g} \tr \left(
(\nabla_\tau X^i)^2
+ 2 \theta^t\nabla_\tau\theta -\frac{1}{2}  [X^i,X^j]^2
-2 \theta ^t\gamma ^i [\theta,X^i] \right)
\end{equation}
o{\`u} les neuf coordonn{\'e}es $X^i$, maintenant interpr{\'e}t{\'e}es comme
les coordonn{\'e}es transverses de la M-th{\'e}orie sur le front de 
lumi{\`e}re discret, prennent leurs valeurs dans l'adjoint de $U(N)$.
Apr{\`e}s red{\'e}finition du champ $X^i\rightarrow g^{1/3} X^i$,
on obtient en unit{\'e}s de $l_{11}$
\begin{equation}
\mathcal{L}=\int d\tau \tr \left(
\frac{1}{2R_s} (\nabla_\tau X^i)^2
+ 2 \theta^t\nabla_\tau \theta -\frac{R_s}{4} \tr [X^i,X^j]^2
-R_s \theta ^t\gamma ^i [\theta,X^i] \right)
\end{equation}
avec $\nabla_\tau = \partial_\tau + i [A,\cdot]$.
Le hamiltonien associ{\'e} {\`a} cette th{\'e}orie des champs en
dimensions 0+1 s'{\'e}crit alors en termes du moment
$\Pi_i=\nabla_\tau X^i$ conjugu{\'e} {\`a} $X^i$~:
\begin{equation}
\label{hbfss}
H=R_s \tr\left( \frac{1}{2} (\Pi^i )^2 
+\frac{1}{4} ([X^i,X^j])^2
+ \theta ^t\gamma ^i [\theta,X^i] \right) + \frac{N}{R_s}\ ,
\end{equation}
auquel on a ajout{\'e} l'{\'e}nergie au repos des
$N$ D0-branes {\`a} grande s{\'e}paration.
L'invariance de jauge permet de fixer $A=0$, mais il faut encore
restreindre l'espace de Hilbert aux {\'e}tats invariants de jauge
en imposant la contrainte de Gauss 
\index{Gauss, contrainte de}
\begin{equation}
[X^i,\Pi ^i]+ [\theta ,\theta ^t]\equiv 0
\end{equation}
Cet argument nous conduit donc {\`a} la conjecture BFSS~:
{\it la M-th{\'e}orie est d{\'e}crite sur le front de lumi{\`e}re
\index{matrices, th{\'e}orie des|textit}
\index{Banks, Fischler, Shenker et Susskind, conjecture de|textit}
discret dans le secteur de moment longitudinal $P^+=N/L$
par la m{\'e}canique quantique supersym{\'e}trique
de 9 matrices hermitiennes $N\times N$ $X^i$, de hamiltonien}
\begin{equation}
\label{hsuss}
P^-=L \tr\left( \frac{1}{2} (p^i )^2 +\frac{1}{2} (\Pi^i )^2 
+\frac{1}{4} ([X^i,X^j])^2
+ \theta ^t\gamma ^i [\theta,X^i] \right) \ ,
\end{equation}
o{\`u} on a d{\'e}coupl{\'e} le facteur ab{\'e}lien de moment $p^i$ et
le facteur $SU(N)$ de moment $\Pi^i$~; le premier d{\'e}crit le
mouvement transverse du centre de masse du syst{\`e}me,
tandis que le second d{\'e}crit les interactions des D0-branes,
identifi{\'e}es aux {\it partons} \index{parton}
de la M-th{\'e}orie sur le front de lumi{\`e}re discret. La M-th{\'e}orie
sur le front de lumi{\`e}re non compact est obtenue dans la
limite de {\it r{\'e}solution spectrale infinie}~:
\index{spectrale, r{\'e}solution}
\begin{equation}
N \rightarrow \infty\ ,\quad L\rightarrow\infty\ ,\quad
P^+=\frac{N}{L}\ \mbox{fix{\'e}.}
\end{equation}
Cette conjecture appelle plusieurs commentaires~:
\begin{itemize}
\item le hamiltonien (\ref{hsuss}) remplit tout d'abord les
conditions de sym{\'e}trie sous le {\it groupe de Galil{\'e}e
supersym{\'e}trique de l'espace transverse}
\index{Galil{\'e}e, invariance de}
\begin{equation}
\{q_\alpha,q_\beta\} = \delta_{\alpha\beta} P^+ \ ,\qquad
\{Q_\alpha,Q_\beta\} = \delta_{\alpha\beta} P^- \ ,\qquad
\left[Q_\alpha,q_\beta\right] = \gamma_{\alpha\beta}^i P^i
\end{equation}
engendr{\'e} par les 16+16 charges supersym{\'e}triques
\begin{eqnarray}
q_\alpha = \frac{1}{\sqrt{L}} \tr \theta \\
Q_\alpha = \sqrt{L} \tr( \gamma_{\alpha\beta}^i P^i
+i [X^i,X^j]\gamma_{\alpha\beta}^{ij} ) \theta_\beta
\end{eqnarray}
\index{supersym{\'e}trie!en th{\'e}orie des matrices}
Les 16 charges $q_\alpha$ sont r{\'e}alis{\'e}es non lin{\'e}airement
et n'agissent que sur le facteur ab{\'e}lien du groupe de jauge $U(N)$.
Elles correspondent aux charges spontan{\'e}ment bris{\'e}es par la
pr{\'e}sence des D0-branes dans le langage de la th{\'e}orie de type
IIA. Seules les charges $Q_\alpha$, correspondant aux 16 charges
de la th{\'e}orie de Yang-Mills {\`a} dix dimensions, contraignent
la dynamique relative des D0-branes. 
\index{D-brane!D0-branes}

\item la d{\'e}pendance du hamiltonien en le rayon de lumi{\`e}re $L$
est compatible avec l'invariance de Lorentz selon
$L^{+-}$. L'invariance sous les g{\'e}n{\'e}rateurs dynamiques
$L^{+i}$ est en revanche loin d'{\^e}tre {\'e}vidente, et
constitue l'essence de la conjecture. Elle est du reste
bris{\'e}e par la compacit{\'e} de la direction $x^-$, et n'a lieu
d'{\^e}tre que dans la limite $L\rightarrow \infty,
N\rightarrow\infty$. La construction d'op{\'e}rateurs $L^{+i}$
dans le secteur de supers{\'e}lection 
$\mathcal{H}_N$ g{\'e}n{\'e}rant le groupe de Poincar{\'e} {\`a} grand $N$
reste un probl{\`e}me ouvert de premi{\`e}re importance.
\index{Lorentz, invariance {\`a} 11D}

\item pour pr{\'e}tendre {\`a} d{\'e}crire la M-th{\'e}orie, la moindre
des exigences est que la th{\'e}orie des matrices inclue
le supergraviton de masse nulle et de moment longitudinal
\index{graviton}
$N$. Un tel {\'e}tat doit correspondre
{\`a} un {\it {\'e}tat fondamental supersym{\'e}trique} du hamiltonien $SU(N)$
\index{li{\'e}, {\'e}tat li{\'e} de D-branes}
d{\'e}crivant le mouvement relatif des $N$ partons. La
d{\'e}monstration de l'existence et l'unicit{\'e} d'un tel {\'e}tat 
constitue un second probl{\`e}me non r{\'e}solu {\`a} ce jour. Les 
{\'e}tats {\`a} plusieurs gravitons de moments $N_i/L$
peuvent {\^e}tre d{\'e}crits
asymptotiquement en d{\'e}composant la matrice $N\times N$
en blocs diagonaux~; un calcul de diffusion dans 
l'approximation de Born-Oppenheimer peut alors {\^e}tre
effectu{\'e} et compar{\'e} {\`a} la pr{\'e}diction de la
th{\'e}orie de supergravit{\'e}. K. et M. Becker ont
ainsi observ{\'e} l'accord au second ordre
dans le d{\'e}veloppement en boucles de la 
m{\'e}canique quantique des D0-branes \cite{Becker:1997wh}.

\item les membranes et cinq-branes de la M-th{\'e}orie
ne jouent aucun r{\^o}le dans cette formulation. Elles doivent
donc appara{\^\i}{}tre comme {\it {\'e}tats li{\'e}s de D0-branes}.
\index{brane!membrane et cinq-brane en th{\'e}orie des matrices}
De fait, le hamiltonien (\ref{hsuss}) est identique au
hamiltonien de la supermembrane apr{\`e}s remplacement
du crochet de Lie de l'alg{\`e}bre $su(N)$ par le crochet
de Poisson sur l'alg{\`e}bre des fonctions sur 
la surface de la membrane \cite{Banks:1997nn}. Dans le cas d'une membrane
toro{\"\i}{}dale, la correspondance
peut {\^e}tre pr{\'e}cis{\'e}e en associant {\`a} chaque fonction
$X^i(\sigma_1,\sigma_2)$
la matrice de ses coefficients $X^i_{mn}$ sur la base de Fourier
$e^{i(m\sigma_1+n\sigma_2)}$. Le statut des 5-branes est
moins clair {\`a} ce jour \cite{Berkooz:1997is}.

\item ayant ramen{\'e} l'{\'e}tude de la gravit{\'e} quantique {\`a}
onze dimensions {\`a} un probl{\`e}me de m{\'e}canique quantique,
on pourrait pousser le raisonnement plus loin et consid{\'e}rer le
mod{\`e}le de matrice correspondant {\`a} la m{\'e}canique
statistique de $N$ D-instantons de la th{\'e}orie de type IIB,
d{\'e}crite par la r{\'e}duction dimensionnelle {\it totale}
de la th{\'e}orie de Yang-Mills $U(N)$ {\`a} 10 dimensions~:
\begin{equation}
S= \frac{1}{4} \tr \left( [X^i,X^j]^2 + \theta ^t \gamma ^i
  [\theta,X^i] \right)
\end{equation}
o{\`u} l'indice $i$ va maintenant de 0 {\`a} 9. Cette approche
a {\'e}t{\'e} initi{\'e}e par Ishibashi, Kawai, Kitazawa et
Tsuchiya \cite{Ishibashi:1996xs}, et n'a pas re{\c c}u la m{\^e}me attention que
la proposition concurrente. La construction des membranes
mentionn{\'e}e au paragraphe pr{\'e}c{\'e}dent peut cependant
{\^e}tre transpos{\'e}e dans ce formalisme pour obtenir
la corde de type IIB \cite{Fukuma:1997en}. 
Comme nous le verrons dans la section
suivante, la th{\'e}orie des matrices peut {\^e}tre consid{\'e}r{\'e}e
\index{D-instanton!th{\'e}orie des matrices des}
comme une compactification de cette th{\'e}orie des D-instantons.
\end{itemize}

\section{Compactification de la th{\'e}orie des matrices}
\index{compactification!de la th{\'e}orie des matrices}
La th{\'e}orie des matrices pr{\'e}tend d{\'e}crire la M-th{\'e}orie
{\it non compactifi{\'e}e} sur le front de lumi{\`e}re. Dans une
th{\'e}orie contenant des objets {\'e}tendus, la compactification
est une op{\'e}ration non triviale qui peut changer
drastiquement les degr{\'e}s de libert{\'e} pertinents.
La prescription pour la compactification de la th{\'e}orie
des matrices est {\`a} l'heure actuelle incompl{\`e}te~:
elle fournit une formulation acceptable pour les compactifications
sur un tore $T^d$ de dimension $d\le 3$, mais ind{\'e}finie 
en dimension inf{\'e}rieure ou sur des vari{\'e}t{\'e}s courbes.
Elle montre toutefois une extension dramatique des degr{\'e}s
de libert{\'e} {\`a} prendre en compte, puisque la m{\'e}canique
quantique doit c{\'e}der la place {\`a} une authentique th{\'e}orie
des champs en dimension $d+1$.

\subsection{Compactification toro{\"\i}{}dale et th{\'e}ories de jauge}

La compactification toro{\"\i}{}dale d'une th{\'e}orie des champs
proc{\`e}de en g{\'e}n{\'e}ral en imposant l'invariance sous la
sym{\'e}trie discr{\`e}te $X^9 \rightarrow X^9 + 2\pi R$. Dans le cas de
la th{\'e}orie des matrices, cette identification doit {\^e}tre
prise {\it {\`a} une transformation de jauge pr{\`e}s}, et on
est donc conduit {\`a} restreindre notre attention aux matrices $X$
telles que 
\begin{equation}
\label{ccomp}
U X^9 U^{-1} = X^9 + 2\pi R \mathbb{I} 
\ ,\quad U X^i U^{-1}=X^i\ ,\quad i=1\dots 8,
\end{equation}
o{\`u} la matrice $U$ est unitaire et
la translation agit sur la coordonn{\'e}e du centre de masse $\Tr X^9$
uniquement. Cette condition n'admet de solution qu'{\`a} $N$ infini,
soit en pr{\'e}sence d'un nombre infini de D0-branes. En choisissant
pour $U$ la permutation $i\rightarrow i+M$, elle traduit le
fait que le groupe de $M$ D0-branes, d{\'e}crit par les matrices
hermitiennes $M\times M$ $(X_0^i,X_0^9)$, est dupliqu{\'e} autour de chaque 
point $(\tr X_0^i,\tr X^9_0 + 2\pi R \Zint)$. 
Les matrices infinies $X$ prennent donc la
forme
\begin{equation}
X^i=
\begin{pmatrix}
\ddots & \ddots  & \ddots  &        &     \\
\ddots & X^i_0    & X^i_1    & X^i_2      &     \\
\ddots & X^i_{-1} & X^i_0    & X^i_1      & \ddots \\
       & X^i_{-2} & X^i_{-1} & X^i_0      & \ddots  \\
       &        & \ddots & \ddots   & \ddots 
\end{pmatrix}\ ,
\qquad
X^9=
\begin{pmatrix}
\ddots & \ddots  & \ddots  &        &     \\
\ddots & X^9_0 -2\pi R & X^9_1    & X^9_2      &     \\
\ddots & X^9_{-1} & X^9_0    & X^9_1      & \ddots \\
       & X^9_{-2} & X^9_{-1} & X^9_0+2\pi R      & \ddots  \\
       &        & \ddots & \ddots   & \ddots 
\end{pmatrix}
\end{equation}
Les blocs {\'e}l{\'e}mentaires $X_{n}^I$ peuvent {\^e}tre commod{\'e}ment
regroup{\'e}s en un seul en introduisant une nouvelle coordonn{\'e}e
compacte $\sigma$ de p{\'e}riode $1/R$, et en d{\'e}finissant
\begin{equation}
X^i(\sigma,\tau)=\sum_{n=-\infty}^{\infty} X_n^i(\tau) 
e^{2\pi i n R\sigma}\ ,\quad
X^9(\sigma,\tau)=\sum_{n=-\infty}^{\infty} X_n^9(\tau) 
e^{2\pi i n R\sigma}\ ,\quad
\end{equation}
Les commutateurs des matrices $U(N)$ infinies s'expriment alors
simplement en termes des fonctions  $X(\sigma,\tau)${\`a} valeurs dans
$su(M)$ par
\begin{eqnarray}
\left[X^i,X^j\right] (\sigma) &=& \left[X^i(\sigma),X^j(\sigma)\right] \\
\left[X^9,X^i\right] (\sigma) &=& 
i\partial_\sigma X^i(\sigma) + \left[X^9(\sigma),X^j(\sigma)\right]
\end{eqnarray}
et la trace dans $U(N)$ est simplement remplac{\'e}e par
$\frac{1}{2\pi R} \int d\sigma \tr$. Le mod{\`e}le de 
matrices $U(N)$ {\`a} 0+1 dimensions, dans le secteur
des matrices infinies satisfaisant aux conditions (\ref{ccomp}),
est donc remplac{\'e} par une {\it th{\'e}orie
de jauge supersym{\'e}trique $U(M)$ {\`a} 1+1 dimensions}
\footnote{Par la m{\^e}me construction,
le mod{\`e}le de matrices de BFSS 
appara{\^\i}{}t donc comme la compactification 
du mod{\`e}le des instantons de 
Ishibashi {\it et al.}}
\index{D-instanton!th{\'e}orie des matrices des}
\begin{equation}
\mathcal{L}=\int d\tau~d\sigma \frac{1}{R} \tr \left(
\frac{1}{2L} \nabla_\alpha X^i \nabla ^\alpha X^i
+ 2 \theta^t \gamma ^\alpha\nabla_\alpha\theta -\frac{L}{4} 
\tr \left[X^i,X^j\right]^2
-L \theta ^t\gamma ^i 
\left[\theta,X^i\right] \right)
\end{equation}
o{\`u} la coordonn{\'e}e compacte $X^9$ est interpr{\'e}t{\'e}e comme
la connexion de jauge $A_\sigma$.
Les transformations de jauge $X\rightarrow \Omega X \Omega^{-1}$ 
de la th{\'e}orie initiale, commutant avec la matrice unitaire $U$,
engendrent ainsi les transformations de jauge de la th{\'e}orie
compactifi{\'e}e~:
\begin{eqnarray}
(\Omega X^i \Omega^{-1} ) (\sigma) &=& 
\Omega(\sigma) X^i(\sigma) \Omega^{-1} (\sigma) \\
(\Omega X^9 \Omega^{-1} ) (\sigma) &=& 
\Omega(\sigma) X^i(\sigma) \Omega^{-1} (\sigma) -i 
\Omega(\sigma) \partial_\sigma \Omega ^{-1} (\sigma)
\end{eqnarray}
Cette th{\'e}orie de jauge {\`a} 1+1 dimensions peut {\^e}tre
interpr{\'e}t{\'e}e comme {\it la d{\'e}finition non perturbative de
la th{\'e}orie de type IIA} sur le front de lumi{\`e}re discret
\cite{Banks:1997my,Dijkgraaf:1997vv,Motl:1997th}.
A grande distance, les 8 coordonn{\'e}es matricielles $X^i(\sigma,\tau)$
deviennent en effet simultan{\'e}ment diagonalisables, et d{\'e}crivent 
les positions transverses de $N$ cordes {\'e}l{\'e}mentaires. 
La diagonalisation n'est cependant
pas globalement d{\'e}finie autour de la direction $\sigma$, mais 
seulement {\`a} une permutation des N valeurs propres pr{\`e}s.
Les $N$ cordes {\'e}l{\'e}mentaires s'arrangent alors 
selon les $k$ cycles intervenant dans la permutation
en $k$ cordes de longueur $n_i$, et de moment longitudinal 
$P^+_i=n_i/L$. On retrouve ainsi la formulation de la th{\'e}orie
des cordes perturbatives sur le front de lumi{\`e}re discret,
{\it dans un formalisme de seconde quantification}.
Les interactions sont induites par l'extension de la sym{\'e}trie de
jauge lorsque deux valeurs propres co{\"\i}{}ncident.
La corde h{\'e}t{\'e}rotique non perturbative est quant {\`a} elle 
obtenue par projection $\Zint_2$, et d{\'e}crite par une th{\'e}orie
de jauge $SO(N)$ {\`a} 1+1 dimensions et 8 charges supersym{\'e}triques
\cite{Banks:1997it}

Cette construction peut ais{\'e}ment {\^e}tre r{\'e}p{\'e}t{\'e}e pour
des compactifications sur un tore $T^d$, et conduit,
{\it dans le cas o{\`u} les matrices unitaires de compactification $U_i$ 
commutent},
{\`a} une th{\'e}orie de jauge {\`a} 16 charges supersym{\'e}triques en 
$d+1$ dimensions, d{\'e}finie sur l'espace r{\'e}ciproque du tore de
compactification. Cette th{\'e}orie n'est autre que la
r{\'e}duction dimensionnelle de la th{\'e}orie de Yang-Mills
\index{reduction dim@r{\'e}duction dimensionnelle!de SYM $10D$}
supersym{\'e}trique {\`a} dix dimensions. La simplicit{\'e}
de la th{\'e}orie des matrices dispara{\^\i}{}t donc rapidement
par compactification, et sa signification devient m{\^e}me 
probl{\'e}matique pour $d>3$, puisque la th{\'e}orie de
jauge perd sa libert{\'e} asymptotique et devient ind{\'e}finie {\`a}
courte distance. {\it Si en revanche les matrices $U_i$ ne
commutent pas}, la th{\'e}orie de jauge ordinaire doit {\^e}tre
remplac{\'e}e par une {\it th{\'e}orie de jauge sur le tore
non commutatif} \cite{Connes:1997cr,Connes:1994}. 
\index{geometrie non commutative@g{\'e}om{\'e}trie non commutative}
Le statut de ces th{\'e}ories et leurs propri{\'e}t{\'e}s
sont encore largement m{\'e}connues.

\subsection{Compactification et limite de scaling}
Le sch{\'e}ma de compactification discut{\'e} ci-dessus peut {\^e}tre
compris plus g{\'e}n{\'e}ralement en revenant {\`a} la limite 
de scaling qui nous a conduit au mod{\`e}le de BFSS
\cite{Seiberg:1997ad,Sen:1997we}. 
En identifiant le cercle de genre lumi{\`e}re {\`a} un cercle
spatial infiniment boost{\'e}, on peut obtenir la
quantification sur le front de lumi{\`e}re discret
de la M-th{\'e}orie compactifi{\'e}e sur le tore $T^d$
comme {\it la th{\'e}orie de type IIA compactifi{\'e}e
sur $T^d$ en pr{\'e}sence de $N$ D0-branes.} dans la limite
de scaling
\index{scaling, limite de}
\begin{equation}
g = (R_s M)^{3/4}\ ,\quad
\alpha' = \frac{R_s^{1/2}}{M^{3/2}} \ ,\quad
R_i = r_i \left(\frac{R_s}{M}\right)^{1/2}\ ,\quad 
R_s\rightarrow 0\ ;\ M=\frac{L}{l_{11}^2},~r_i=\frac{R_i}{l_{11}} 
\mbox{~fix{\'e}s}
\end{equation}
o{\`u} les rayons $R_i$ du tore sont gard{\'e}s constants 
{\it en unit{\'e}s de Planck $l_{11}$}. Le tore devient
donc de taille nulle, et il est alors commode d'effectuer 
une T-dualit{\'e} sur toutes les directions de $T^d$~: on obtient
ainsi la th{\'e}orie des cordes de type IIA ($d$ pair)
ou IIB ($d$ impair) en pr{\'e}sence de $N$ D$p$-branes,
de param{\`e}tres
\begin{equation}
\tilde g = \frac{ (R_s M)^{\frac{3-d}{4}} }{\prod r_i}\ ,\quad
\tilde \alpha' = \frac{R_s^{1/2}}{M^{3/2}} \ ,\quad
\tilde R_i = \frac{1}{r_i M} \ ,\quad 
R_s\rightarrow 0\ ;\ M,r_i \mbox{~fix{\'e}s}\ .
\end{equation}
Le tore est maintenant de taille fix{\'e}e, et l'{\'e}chelle des cordes
tend vers $0$, de sorte que les modes massifs se d{\'e}couplent toujours.
\index{decouplage@d{\'e}couplage!de la gravitation des D-branes}
Pour $d\le 2$, la th{\'e}orie de type II reste faiblement coupl{\'e}e,
et la M-th{\'e}orie peut donc {\^e}tre d{\'e}crite par la th{\'e}orie
de Yang-Mills {\`a} 16 charges supersym{\'e}triques
sur le volume d'univers de la D$p$-brane. Le couplage
et les param{\`e}tres g{\'e}om{\'e}triques 
de cette th{\'e}orie de jauge sont donn{\'e}s par
\begin{equation}
\frac{1}{g_{\rm YM}^2}= \frac{\tilde\alpha^{'\frac{3-d}{2}}}{\tilde g}
= M^{d-3} \prod r_i\ ,\qquad
s_i = \tilde R_i = \frac{1}{r_i M}
\end{equation}
et sont donc gard{\'e}s fixes dans la limite de scaling. La M-th{\'e}orie
sur le front de lumi{\`e}re non compact est alors obtenue dans
la limite de r{\'e}solution spectrale infinie $M\sim L\sim
N\rightarrow\infty$, soit
\begin{equation}
g_{\rm YM}^2 N^{d-3} = \mbox{cte}\ ,\quad s_i N =\mbox{cte}\ .
\end{equation}
Le couplage de Yang-Mills {\'e}tant de dimension $d-3$, il est plus
int{\'e}ressant de le ramener {\`a} la taille du tore~:
\begin{equation}
g_{\rm YM}^2 \left(\prod s_i\right)
^{\frac{3-d}{d}} = \left(\prod s_i\right)^\frac{-3}{d}\ ,
\end{equation}
et le couplage sans dimension est donc gard{\'e} 
fixe dans les limites de scaling et de r{\'e}solution spectrale infinie.

Cette description reste correcte dans le cas $d=3$ bien que le couplage
de la th{\'e}orie des cordes soit maintenant fini, 
car les corrections de boucles sont toujours
supprim{\'e}es par la limite $\alpha'\rightarrow 0$. 
On note en particulier que la limite $N\rightarrow\infty$
est diff{\'e}rente de la limite de grand $N$ de 't Hooft 
\index{tHooft@'t Hooft, limite de}
couramment consid{\'e}r{\'e}e
dans les th{\'e}ories de jauge {\`a} 4 dimensions \cite{'tHooft:1974jz}.
Pour $d=4$ cependant, la th{\'e}orie de type IIA devient fortement
coupl{\'e}e et g{\'e}n{\`e}re une nouvelle <<onzi{\`e}me dimension>> 
\footnote{{\`a} ne pas confondre avec la dimension d{\'e}finissant le
front de lumi{\`e}re discret.}, de rayon et d'{\'e}chelle de Planck
\begin{equation}
\tilde R= \tilde g \sqrt{\tilde\alpha'}= \frac{1}{M\prod r_i}\ ,\quad
\tilde l_{11} = \tilde g^{1/3} \sqrt{\tilde\alpha'} = R_s^{1/6} M^{-5/6} 
\prod r^{-1/3}_i\ . 
\end{equation}
Les $N$ D4-branes de la th{\'e}orie de type IIA correspondent maintenant {\`a} 
$N$ 5-branes enroul{\'e}es sur la direction de rayon $\tilde R$.
La masse de Planck $1/\tilde l_{11}$ divergeant dans la 
limite de scaling, la gravitation se d{\'e}couple des degr{\'e}s de
libert{\'e} sur le volume d'univers de la 5-brane, et 
la M-th{\'e}orie compactifi{\'e}e sur $T^4$ peut donc {\^e}tre d{\'e}crite
sur le front de lumi{\`e}re discret par {\it la th{\'e}orie de
volume d'univers de $N$ 5-branes}
\index{cinq-brane!de Neveu-Schwarz}
\cite{Berkooz:1997cq}. Cette th{\'e}orie est malheureusement
trop mal connue {\`a} l'heure actuelle pour fournir une d{\'e}finition 
utilisable de la M-th{\'e}orie compactifi{\'e}e sur $T^4$.

Dans le cas $d=5$, le fort couplage de la th{\'e}orie de type IIB
invite {\`a} effectuer une transformation de S-dualit{\'e}~; les $N$
D5-branes deviennent alors $N$ NS5-branes de la th{\'e}orie de
type IIB de param{\`e}tres
\begin{equation}
\hat g= \frac{1}{\tilde g} =  (R_s M)^{1/2} \prod r_i\ ,\quad
\hat\alpha'= \tilde g \tilde \alpha' =\frac{\prod r_i}{M^2} \ ,\quad
\hat R_i = \tilde R_i\ .
\end{equation}
Dans la limite de scaling, les modes se propageant dans l'espace-temps
ambiant peuvent encore {\^e}tre d{\'e}coupl{\'e}s des modes localis{\'e}s
sur les NS5-branes, laissant une th{\'e}orie des cordes
de tension finie $1/\hat\alpha'$ se propageant sur la NS5-brane.
L'existence de cette th{\'e}orie des cordes non critiques est
hautement conjecturale, et ne permet pas de donner une 
d{\'e}finition effective de la M-th{\'e}orie compactifi{\'e}e
sur $T^5$. La situation est encore plus s{\'e}rieuse
dans le cas de la compactification sur $T^6$, o{\`u} le
rayon de la onzi{\`e}me dimension g{\'e}n{\'e}r{\'e}e {\`a} fort couplage
diverge et o{\`u} l'{\'e}chelle de Planck reste finie dans la
limite de scaling.

A d{\'e}faut de fournir une d{\'e}finition non perturbative
de la M-th{\'e}orie, ces descriptions
justifient au moins heuristiquement les groupes
de U-dualit{\'e} observ{\'e}s dans les compactifications
\index{U-dualit{\'e}!et th{\'e}orie des matrices}
de la M-th{\'e}orie sur $T^4$ et $T^5$~: le groupe
$E_4(\Zint)=Sl(5,\Zint)$ peut {\^e}tre interpr{\'e}t{\'e} comme
le groupe modulaire du volume d'univers $T^5$ de la
5-brane dans le cas $d=4$, tandis que le groupe
$E_5(\Zint)=SO(5,5,\Zint)$ appara{\^\i}{}t comme le
groupe de T-dualit{\'e} de la th{\'e}orie des cordes 
hypoth{\'e}tique se propageant dans le volume d'univers des NS5-branes de la 
th{\'e}orie de type IIB pour $d=5$. Dans le paragraphe suivant,
nous obtiendrons une meilleure compr{\'e}hension de ces
sym{\'e}tries de U-dualit{\'e} en les traduisant dans le langage
des th{\'e}ories de jauge.

\subsection{Dualit{\'e} {\'e}lectrique-magn{\'e}tique et U-dualit{\'e}}
La prescription de BFSS pour la d{\'e}finition de la M-th{\'e}orie
demeure donc pour l'heure s{\'e}rieusement incompl{\`e}te, et
pour les compactifications sur des vari{\'e}t{\'e}s de dimension
$d>3$, se heurte {\`a} la difficult{\'e} d'{\'e}tendre dans l'ultraviolet
les th{\'e}ories de jauge en dimension $d+1$. Certaines propri{\'e}t{\'e}s
de cette extension peuvent cependant {\^e}tre fix{\'e}es en traduisant
les contraintes de la U-dualit{\'e} de la M-th{\'e}orie compactifi{\'e}e
toro{\"\i}{}dalement en termes de cette th{\'e}orie de jauge. Cette approche
\index{U-dualit{\'e}!et th{\'e}orie des matrices}
\index{U-dualit{\'e}!de type II sur $T^d$}
a {\'e}t{\'e} suivie par Elitzur, Giveon, Kutasov et Rabinovici 
et prolong{\'e}e en collaboration avec Obers et Rabinovici,
dans le travail pr{\'e}sent{\'e} en appendice \ref{mu}
\cite{Elitzur:1997zn,Obers:1997kk}.

Le groupe de U-dualit{\'e} de la M-th{\'e}orie compactifi{\'e}e sur $T^{d}$
est engendr{\'e} par deux sous-groupes discrets~: le {\it groupe modulaire}
$Sl(d,\Zint)$ du tore de compactification, et le {\it groupe
de T-dualit{\'e}} $SO(d-1,d-1,\Zint)$. L'action du premier est manifeste
\index{T-dualit{\'e}!sur un tore $T^d$}
dans la th{\'e}orie de jauge d{\'e}crivant la M-th{\'e}orie sur le
front de lumi{\`e}re discret, et s'identifie au groupe
modulaire du tore r{\'e}ciproque sur lequel se propage la th{\'e}orie
de jauge. Si on se restreint pour simplicit{\'e} aux tores
rectangulaires en l'absence de champ de fond pour le tenseur
de jauge $C_{\mu\nu\rho}$\footnote{L'extension de
ce formalisme aux compactifications toro{\"\i}{}dales g{\'e}n{\'e}rales
fait l'objet de l'article en appendice \ref{mu}.}, ce groupe 
modulaire se restreint au 
{\it groupe de permutations}
\begin{equation}
\label{uperm}
S_{ij}\ :\  R_i \leftrightarrow R_j
\end{equation}des $d$ rayons,
qui n'est autre que le {\it groupe de Weyl} de $Sl(d,\Zint)$.
\index{Weyl, groupe de Weyl de la U-dualit{\'e}}
\index{groupe!de Weyl de la U-dualit{\'e}}

L'action du second peut {\^e}tre d{\'e}termin{\'e}e en choisissant
une dimension $k$ parmi les $d$ dimensions du tore
(not{\'e}e 11 dans les chapitres pr{\'e}c{\'e}dents), et en
identifiant la M-th{\'e}orie compactifi{\'e}e sur $T^{d}$ {\`a} la
th{\'e}orie de type IIA compactifi{\'e}e sur $T^{d-1}$, de
couplage et d'{\'e}chelle des cordes
\begin{equation}
g=\left(\frac{R_k}{l_{11}}\right)^{3/2}\ ,\quad
\alpha'=\frac{l_{11}^3}{R_k}\ .
\end{equation}
L'inversion simultan{\'e}e de deux rayons $(R_i,R_j,g) \rightarrow
(\alpha'/R_j,\alpha'/R_i,g\ \alpha'/R_i R_j)$ s'{\'e}crit alors,
en termes des variables de la M-th{\'e}orie
\footnote{$l_{11}$ peut {\^e}tre rendue invariante par un changement
d'{\'e}chelle, mais les transformations des rayons $R_{i,j,k}$ seraient
alors moins {\'e}l{\'e}gantes.}
\begin{equation}
T_{ijk}\ :\quad
R_i \rightarrow \frac{l_{11}^3}{R_j R_k}\ ,\quad
R_j \rightarrow \frac{l_{11}^3}{R_i R_k}\ ,\quad
R_k \rightarrow \frac{l_{11}^3}{R_i R_j}\ ,\quad
l_{11}^3 \rightarrow \frac{l_{11}^6}{R_i R_j R_k}\ .
\end{equation}
La transformation de T-dualit{\'e} $T_{ijk}$ est donc
\index{T-dualit{\'e}!en M-th{\'e}orie}
invariante sous les permutations des trois rayons $R_{i,j,k}$,
qui apparaissent ainsi sur un pied d'{\'e}galit{\'e} du point de vue
de la M-th{\'e}orie. Elle laisse en particulier invariante
l'{\'e}chelle de Planck {\`a} $11-d$ dimensions
$\prod R_i /l_{11}^9$. En termes des param{\`e}tres de
\index{Planck, {\'e}chelle de}
la th{\'e}orie de jauge, la transformation $T_{ijk}$ s'{\'e}crit
\begin{equation}
\label{ymdual}
S_{ijk}\ :\ 
\begin{cases}
g_{\rm YM}^2 & \rightarrow \frac{g_{\rm YM}^{2(2-4)}}{W^{d-5}}\ ,
\quad W=\prod_{a\ne i,j,k} s_a \\
s_\alpha & \rightarrow s_\alpha\ ,\quad \alpha=i,j,k \\
s_a & \rightarrow \frac{g_{\rm YM}^2}{W} s_a\ ,\quad a\ne i,j,k\ .
\end{cases}
\end{equation}
Dans le cas de la compactification sur $T^3$, cette transformation
se r{\'e}duit {\`a} la {\it sym{\'e}trie de dualit{\'e} {\'e}lectrique-
magn{\'e}tique} $g^2 \rightarrow 1/g^2$ de la th{\'e}orie de jauge
\index{dualit{\'e}!{\'e}lectrique-magn{\'e}tique}
$N=4$ {\`a} 3+1 dimensions \cite{Susskind:1996uh,Fischler:1997kp}. 
Pour $d> 3$, la transformation (\ref{ymdual}) peut encore
s'interpr{\'e}ter comme S-dualit{\'e} en interpr{\'e}tant le
tore $T^{d}$ comme {\it fibration} de tores $T^3$~:
la th{\'e}orie de jauge sur $T^{d}$ peut alors {\^e}tre r{\'e}duite
en une th{\'e}orie de Yang-Mills sur le 3-tore g{\'e}n{\'e}r{\'e} par
les directions $i,j,k$, de couplage
\begin{equation}
\frac{1}{g_{\rm eff}^2}=\frac{W}{g_{\rm YM}^2}
\end{equation}
La transformation (\ref{ymdual}) correspond alors simplement {\`a} la
S-dualit{\'e} $(g^2_{\rm eff},s_\alpha,s_a)\rightarrow (1/g^2_{\rm eff},s_\alpha,
g_{\rm eff}^2 s_a)$ dans la th{\'e}orie de Yang-Mills sur $T^3$.
Comme l'ont montr{\'e} Elitzur {\it et al}, le groupe g{\'e}n{\'e}r{\'e} par
les transformations $S_{ij}$ et $S_{ijk}$ n'est autre que
{\it le groupe de Weyl} $\mathcal{W}(E_{d})$ 
\index{Weyl, groupe de Weyl de la U-dualit{\'e}}
du groupe de U-dualit{\'e} $E_{d}$.
Les g{\'e}n{\'e}rateurs manquants de $E_{d}(\Zint)$, que nous 
appellerons {\it g{\'e}n{\'e}rateurs de Borel}, 
\index{Borel, g{\'e}n{\'e}rateur de}
correspondent aux transformations
modulaires $\gamma_i\rightarrow\gamma_i+\gamma_j$ du tore
$T^{d}$, ainsi qu'aux transformations de T-dualit{\'e}
$\mathcal{C}_{ijk}\rightarrow \mathcal{C}_{ijk}+1$~;
ces derni{\`e}res {\'e}tendent en particulier la transformation $S_{ijk}$
au groupe de S-dualit{\'e} $Sl(2,\Zint)$ de la th{\'e}orie de Yang-Mills
sur la fibre $T^3$.
Pour observer l'invariance sous le groupe de U-dualit{\'e}
total, il est donc n{\'e}cessaire d'autoriser les compactifications sur
des tores de m{\'e}trique non rectangulaire
en pr{\'e}sence de champ de fond pour le tenseur $\mathcal{C}_{\mu\nu\rho}$.

\subsection{Etats BPS et multiplets de U-dualit{\'e}}
Ayant d{\'e}crit comment le groupe de U-dualit{\'e} {\'e}mergeait
du point de vue de la th{\'e}orie de jauge d{\'e}crivant la
M-th{\'e}orie compactifi{\'e}e sur le front de lumi{\`e}re
discret, il nous reste encore {\`a} comprendre comment le
spectre BPS s'organise en repr{\'e}sentations
de ce groupe. Etant donn{\'e} les incertitudes de la
d{\'e}finition de la M-th{\'e}orie, on ne peut en l'{\'e}tat 
actuel d{\'e}river ce spectre~; l'existence de certains
{\'e}tats est cependant requise, et donc {\'e}galement
celle de leurs images sous la U-dualit{\'e}.
\index{etats BPS@{\'e}tats BPS!en th{\'e}orie des matrices}

Ainsi, la M-th{\'e}orie devant inclure le supergraviton de masse nulle dans
son spectre, elle doit apr{\`e}s compactification contenir
les {\'e}tats de Kaluza-Klein de masse $\mathcal{M}=1/R_i$ et de moment 
\index{Kaluza-Klein!excitation de}
$P^+=N/L$. Ceux-ci correspondent {\`a} des {\'e}tats d'{\'e}nergie
\begin{equation}
\label{f0}
P^- = \frac{\mathcal{M}^2}{P^+} = \frac{L}{N R^2_i} = 
\frac{g_{\rm YM}^2 s^2_i}{N V_s}
\end{equation}
et s'identifient donc dans la th{\'e}orie de jauge {\`a} l'{\'e}tat portant
un flux {\'e}lectrique dans la direction $i$. Par dualit{\'e}
\index{flux {\'e}lectrique}
\index{flux, multiplet de}
$S_{ijk}$, cet {\'e}tat donne naissance {\`a} un {\it multiplet de flux},
d'{\'e}nergies et de masses\footnote{Les indices $i,j,\dots$ sont
distincts, except{\'e}s lorsqu'ils sont s{\'e}par{\'e}s par un point-virgule.}
\begin{align}
\label{f1}
P^- = \frac{V_s}{Ng_{\rm YM}^2 (s_i s_j)^2} &\rightarrow 
\mathcal{M}= \frac{R_i R_j}{l_{11}^3} \\
\label{f2}
P^- = \frac{V_s^3}{N g_{\rm YM}^6(s_i s_j s_k s_l s_m)^2}&\rightarrow 
\mathcal{M}= \frac{R_i R_j R_k R_l R_m}{l_{11}^6} \\
\label{f3}
P^- = \frac{V_s^5}{N g_{\rm YM}^{10}(s_i; s_j s_k s_l s_m s_n s_p s_q)^2}&\rightarrow 
\mathcal{M}= \frac{R_i; R_j R_k R_l R_m R_n R_p R_q }{l_{11}^9}
\end{align}
o{\`u} l'on s'est restreint aux {\'e}tats apparaissant pour $d\le 7$.
$V_s=\prod s_i$ d{\'e}signe le volume du tore dual.
L'{\'e}tat (\ref{f1}) n'est autre que le flux magn{\'e}tique selon les
directions $i$ et $j$, et d{\'e}crit la membrane de la M-th{\'e}orie 
enroul{\'e}e sur un deux-cycle du tore $T^d$.
L'{\'e}tat (\ref{f2}) appara{\^\i}{}t pour $d\ge 5$, et,
comme le montre sa masse, correspond {\`a} la
5-brane de la M-th{\'e}orie enroul{\'e}e sur un 5-cycle du tore~;
son {\'e}nergie en $1/g_{\rm YM}^6$ ne permet cependant de l'identifier
{\`a} aucun {\'e}tat connu de la th{\'e}orie de Yang-Mills ordinaire.
Le troisi{\`e}me {\'e}tat, apparaissant pour $d\ge 7$, correspond quant {\`a}
lui au monop{\^o}le magn{\'e}tique de Kaluza-Klein~: lorsque $i=j$,
\index{Kaluza-Klein!monop{\^o}le de}
il pr{\'e}sente une masse en $(R_i)^2/l_{11}^9$ caract{\'e}ristique
d'un monop{\^o}le charg{\'e} magn{\'e}tiquement sous le champ de
Kaluza-Klein $g_{i\mu}$. Si on choisit la onzi{\`e}me direction 
orthogonalement aux indices $i$ {\`a} $q$, on obtient un {\'e}tat
de masse $\mathcal{M}\sim 1/g^3$ myst{\'e}rieux du point de la th{\'e}orie
des cordes.  Comme l'ont remarqu{\'e} Blau et O'Loughlin
\cite{Blau:1997du}, ceci revient {\`a} consid{\'e}rer 
un {\it r{\'e}seau de monop{\^o}les Taub-NUT}
\index{Taub-NUT, instanton de}
dans la direction $s$, dont la m{\'e}trique est donn{\'e}e pr{\'e}cis{\'e}ment
par l'espace de Ooguri et Vafa ({\'e}quation \ref{vof} page
\pageref{vof}) apparu dans l'{\'e}tude de 
la singularit{\'e} de conifold dans l'espace des hypermultiplets.
\index{singularit{\'e}!de conifold}
La divergence logarithmique pour $|z|\rightarrow\infty$ implique
que cet espace n'est pas asymptotiquement plat, et ne d{\'e}crit donc
pas un soliton au sens propre. De mani{\`e}re g{\'e}n{\'e}rale, une
d{\'e}pendance en $1/g^n$ de la <<masse>> implique  un champ
gravitationnel en $g^{2-n}$, et le traitement {\`a} faible couplage
n'est justifi{\'e} que pour $n\le 2$ \cite{Elitzur:1997zn}.
L'existence de ces {\'e}tats ne pr{\'e}sente cependant pas d'obstruction 
de principe au niveau de la M-th{\'e}orie\footnote{Un {\'e}tat singulier
du point de vue de la th{\'e}orie des cordes appara{\^\i}{}t donc
r{\'e}gulier dans la M-th{\'e}orie, de mani{\`e}re tr{\'e}s analogue au cas
du monop{\^o}le $U(1)$ de Kaluza-Klein construit par Sorkin dans le
cas de la r{\'e}duction de 5 {\`a} 4 dimensions, qui  appara{\^\i}{}t 
singulier en dimension 4 mais r{\'e}gulier lorsqu'interpr{\'e}t{\'e} dans
les variables appropri{\'e}es {\`a} la dimension 5\cite{Sorkin:1983ns}.}, 
et semble requise par la 
U-dualit{\'e}~;
on note en particulier qu'elle ne pr{\'e}serve pas la structure
asymptotique de l'espace-temps \cite{Sfetsos:1997xs}. 
Notons
finalement que pour $d\ge 8$, de nouveaux {\'e}tats apparaissent,
de masse en $R^3_i/l_{11}^{18}$ qui n'admettent
pas m{\^e}me l'interpr{\'e}tation heuristique que nous avons donn{\'e}e
ici. 

Comme nous l'avons vu dans la section (\ref{lcst}), la quantification
sur le front de lumi{\`e}re discret de la th{\'e}orie des cordes
introduit des {\'e}tats enroul{\'e}s sur le cercle compact de genre
lumi{\`e}re. Dans la M-th{\'e}orie, ils correspondent donc {\`a} des
membranes enroul{\'e}es selon le cercle de genre lumi{\`e}re et
un cercle $i$ du tore de compactification, de masse 
$\mathcal{M}=R_i L/l_{11}^3$. 
Ces {\'e}tats correspondent donc dans la th{\'e}orie de jauge
{\`a} des excitations d'{\'e}nergie
\begin{equation}
\label{m0}
P^- = \mathcal{M}= \frac{L R_i}{l_{11}^3} = \frac{1}{s_i}
\end{equation}
qui peuvent donc {\^e}tre identifi{\'e}s aux excitations de Kaluza-Klein
de la th{\'e}orie de jauge sur le tore $T^d$. Sous la U-dualit{\'e}
$S_{ijk}$, ces {\'e}tats donnent naissance au multiplet
dit {\it de moment}~:
\index{moment, multiplet de}
\begin{align}
\label{m1}
P^- = \frac{V_s}{g^2 s_i s_j s_k s_l} &\rightarrow 
\mathcal{M}= \frac{L R_i R_j R_k R_l }{l_{11}^6} \\
\label{m2}
P^- = \frac{V_s^2}{g^4 s_i; s_j s_k s_l s_m s_n s_p}&\rightarrow 
\mathcal{M}= \frac{L R_i; R_j R_k R_l R_m R_n R_p}{l_{11}^9} 
\end{align}
o{\`u} l'on s'est cette fois restreint aux {\'e}tats apparaissant en
dimension $d\le 6$. L'{\'e}tat (\ref{m1}) s'interpr{\`e}te comme
la 5-brane de la M-th{\'e}orie enroul{\'e}e sur le cercle de
genre lumi{\`e}re et sur un 4-cycle du tore $T^d$~; en termes
de la th{\'e}orie de jauge, il peut {\^e}tre identifi{\'e} 
{\`a} un instanton de la th{\'e}orie de Yang-Mills en dimension 3+1,
\index{instanton!des th{\'e}ories de jauge}
<<relev{\'e}>> en dimension d+1. L'{\'e}tat (\ref{m2}) correspond quant {\`a}
lui aux monop{\^o}les de Kaluza-Klein.

Nous avons d{\'e}crit jusqu'{\`a} pr{\'e}sent l'orbite des {\'e}tats
de flux et de moment sous le groupe de Weyl $\mathcal{W}(E_{d})$,
et donn{\'e} leur masse pour des compactifications de la M-th{\'e}orie
sur un tore rectangulaire en l'absence de champ de fond 
$\mathcal{C}_{\mu\nu\rho}$. Comme nous l'avons montr{\'e} dans
le travail en appendice \ref{mu}, les g{\'e}n{\'e}rateurs
de Borel $\gamma_i \rightarrow \gamma_i+\gamma_j$ et
$\mathcal{C}\rightarrow \mathcal{C}_{ijk}+1$ g{\'e}n{\`e}rent cependant
\index{flot spectral!en M-th{\'e}orie}
un {\it flot spectral} imposant de prendre en compte les
{\it superpositions arbitraires} des {\'e}tats {\'e}l{\'e}mentaires
pr{\'e}c{\'e}dents. Les {\'e}tats du multiplet de flux peuvent ainsi
{\^e}tre d{\'e}crit par un ensemble de {\it charges enti{\`e}res}
$m_I,m^{IJ},m^{IJKLM},m^{I;JKLMNPQ},\dots$, totalement
antisym{\'e}triques correspondant aux {\'e}tats (\ref{f0},\ref{f1},
\ref{f2},\ref{f3}) respectivement. La formule de masse
d'une telle combinaison d'{\'e}tats, {\it invariante sous
le groupe de U-dualit{\'e} total $E_d(\Zint)$}, a {\'e}t{\'e}
obtenue dans la r{\'e}f{\'e}rence \cite{Obers:1997kk} 
(annexe \ref{mu}) par analyse des
transformations de ces charges sous la T-dualit{\'e}.
Les {\'e}tats (\ref{m0},\ref{m1},\ref{m2}) peuvent pareillement
{\^e}tre d{\'e}crits en termes de charges enti{\`e}res
$m^I,m^{IJKL},m^{I~;JKLMNP}$.

\subsection{Invariance de Lorentz et dualit{\'e} de Nahm}
Les multiplets de flux et de moment {\'e}puisent donc l'ensemble des
{\'e}tats 1/2-BPS attendus de la M-th{\'e}orie compactifi{\'e}e sur $T^d$,
tout en fournissant une vari{\'e}t{\'e} d'{\'e}tats plus exotiques
\index{Lorentz, invariance {\`a} 11D}
que l'extension microscopique de la th{\'e}orie de jauge sur $T^d$
se doit de reproduire. Ensemble, ces {\'e}tats forment deux repr{\'e}sentations
du groupe de U-dualit{\'e}, r{\'e}sum{\'e}es dans la table suivante, dont les
deux derni{\`e}res lignes deviendront claires incessamment~:
\begin{equation*}
\begin{array}{|c|c||c|c|c|c|c|c|c|c|}
\hline
\multicolumn{2}{|c||}{d} &  1  & 2 & 3      & 4 & 5 & 6   & 7 & 8 \\ \hline
\hline  
\multicolumn{2}{|c||}{E_d(\Zint) } &   
 1  & Sl(2)  & Sl(3) \times Sl(2)   & 
Sl(5) & SO(5,5) &   E_6    & E_7  & E_8   \\ 
\hline  
{\rm Flux}   & \{m\}   & 1 & 3 & (3,2)  &10 &16 &27   &56 &248\\
{\rm Moment} & \{n\} & 1  & 2 & (3,1)  &5  &10 &27   &133&3875\\
\hline
{\rm Rang}   &  \{ N \}  & 1 &  1 & 1      & 1 & 1 & 1+1 &56+1+1+1&\infty \\
\hline
{\rm Total}  & \{ M\}   & 3 & 6 & 10     &16 &27 &56   &248&\infty\\
\hline
\end{array}
\end{equation*}
On a ainsi v{\'e}rifi{\'e} l'invariance du spectre (convenablement
{\'e}tendu) sous le groupe de U-dualit{\'e}. L'invariance de Lorentz
{\`a} onze dimensions de la th{\'e}orie des matrices impose cependant
l'existence d'une sym{\'e}trie suppl{\'e}mentaire dans le spectre~:
la quantification {\it sur le front de lumi{\`e}re discret}
de la M-th{\'e}orie compactifi{\'e}e sur $T^d$ peut en effet {\^e}tre
consid{\'e}r{\'e}e comme une compactification {\`a} part enti{\`e}re
de la M-th{\'e}orie sur $T^{d+1}$, et il doit donc exister
{\it une action du groupe de U-dualit{\'e} {\'e}tendu $E_{d+1}(\Zint)$
sur le spectre}\cite{Hacquebord:1997nq}. 
\index{U-dualit{\'e}!sur le front de lumi{\`e}re}
Le g{\'e}n{\'e}rateur de Weyl manquant correspond
naturellement {\`a} l'{\'e}change $R_i \leftrightarrow L$
d'une direction spatiale avec le cercle de genre lumi{\`e}re~;
il {\'e}change donc les {\'e}tats du multiplet de flux
(\ref{f1},\ref{f2}) avec ceux du multiplet de moment 
(\ref{m0},\ref{m1}). Comme on le montre dans l'appendice \ref{mu}
et le repr{\'e}sente dans les deux derni{\`e}res lignes de la
table ci-dessus,
les multiplets de flux et de moment d{\'e}crivant les {\'e}tats
1/2-BPS de la M-th{\'e}orie compactifi{\'e}e sur $T^d$ peuvent
{\^e}tre assembl{\'e}s en un {\it multiplet de flux} du
groupe {\'e}tendu $E_{d+1}(\Zint)$, {\it moyennant l'introduction
d'une charge suppl{\'e}mentaire}
\footnote{trois charges dans le cas $d=7$, et une infinit{\'e} dans
le cas $d=8$.} $N$ qui n'est autre que le moment
selon la direction compacte du front de lumi{\`e}re
\footnote{des conclusions identiques ont {\'e}t{\'e} atteintes
par \cite{Blau:1997du,Hull:1997kb}.}. $N$ repr{\'e}sentant
{\'e}galement le rang de la th{\'e}orie de jauge d{\'e}crivant la
th{\'e}orie des matrices compactifi{\'e}e, cette extension suppose
donc l'existence de sym{\'e}trie {\it m{\'e}langeant le rang $N$ 
et les charges {\'e}lectriques et magn{\'e}tiques}. Cette sym{\'e}trie
rappelle fortement la sym{\'e}trie {\it classique} de Nahm \cite{Nahm:1982}
\index{dualit{\'e}!de Nahm}
{\'e}changeant le rang et le flux {\'e}lectrique dans les th{\'e}ories
de jauge {\`a} deux dimensions sur $T^2$.
La mise en {\'e}vidence des g{\'e}n{\'e}rateurs de Borel suppl{\'e}mentaires
requiert l'{\'e}tude de la compactification sur le front de lumi{\`e}re
discret en pr{\'e}sence de lignes de Wilson pour le champ de jauge
de Kaluza-Klein et de champs de fond pour le tenseur de jauge
$\mathcal{C}_{\pm IJ}$, qui semble n{\'e}cessiter une description dans le cadre
de la g{\'e}om{\'e}trie non commutative \cite{Connes:1997cr,Douglas:1997fm}.
\index{geometrie non commutative@g{\'e}om{\'e}trie non commutative}


%% file: thesebbl.tex
\newcommand{\etalchar}[1]{$^{#1}$}

%% file: resume.tex


\vspace*{-1.8cm}
\thispagestyle{empty}
\enlargethispage{1.5cm}
\small \centerline{Aspects non perturbatifs de la th{\'e}orie des supercordes}
\vskip .3cm
Les th{\'e}ories de supercordes sont {\`a} l'heure actuelle le seul
candidat {\`a} l'unification quantique des interactions de jauge
et de la gravit{\'e}. Ces th{\'e}ories n'etaient jusqu'{\`a} r{\'e}cemment
d{\'e}finies que dans le r{\'e}gime de {\it faible couplage} par 
leur s{\'e}rie de perturbation. La d{\'e}couverte r{\'e}cente des sym{\'e}tries
de dualit{\'e}, prolongeant la dualit{\'e} {\'e}lectrique-magn{\'e}tique des
{\'e}quations de Maxwell, permet maintenant d'identifier ces th{\'e}ories
perturbatives comme {\it diff{\'e}rentes 
approximations} d'une th{\'e}orie fondamentale, la {\it M-th{\'e}orie}. 
Les sym{\'e}tries de dualit{\'e}s donnent acc{\`e}s aux effets
{\it non perturbatifs} d'une th{\'e}orie des cordes donn{\'e}e 
{\`a} partir de calculs {\it perturbatifs} dans une 
th{\'e}orie duale.

Le premier chapitre de ce m{\'e}moire fournit une introduction
non technique {\`a} ces d{\'e}veloppements. Le second introduit
les dualit{\'e}s non perturbatives observ{\'e}es en th{\'e}ories
de jauge et de supergravit{\'e}, le spectre BPS non-perturbatif
dans ces th{\'e}ories, et pr{\'e}sente une nouvelle
dualit{\'e} reliant les branches de Higgs de certaines th{\'e}ories de jauge
supersym{\'e}triques $N=2$. Dans le troisi{\`e}me chapitre,
nous introduisons bri{\`e}vement les th{\'e}ories des
cordes perturbatives et v{\'e}rifions explicitement les
conjectures de dualit{\'e}s dans les th{\'e}ories des cordes
de supersym{\'e}trie $N=4$ {\`a} quatre dimensions. 
Gr{\^a}ce aux dualit{\'e}s, nous obtenons
des r{\'e}sultats {\it exacts} non perturbativement pour certains
couplages dans l'action effective de basse {\'e}nergie,
et interpr{\'e}tons dans le chapitre 4 les effets
non perturbatifs ainsi obtenus
en termes de {\it configurations instantoniques de $p$-branes}
enroul{\'e}es sur les cycles supersym{\'e}triques de la vari{\'e}t{\'e}
de compactification. Enfin, nous discutons une proposition
r{\'e}cente de d{\'e}finition {\it a priori} de la M-th{\'e}orie
en termes de th{\'e}ories de jauge supersym{\'e}triques $U(N)$
{\`a} grand $N$, l'{\'e}tendons {\`a} des compactifications toro{\"\i}{}dales
en pr{\'e}sence de champs de fond constants, et interpr{\'e}tons
le spectre d'{\'e}tats BPS en termes d'excitations de la th{\'e}orie
de jauge. Les publications originales d{\'e}crivant ces
travaux sont reproduites en appendice.

\vskip .3cm
\noindent
{\sc mots-cl{\'e}s}~: {\it action effective, instantons, M-th{\'e}orie, p-branes, solitons,
sym{\'e}tries de dualit{\'e}, supergravit{\'e}, th{\'e}orie de jauge
supersym{\'e}trique}
\vfill
\centerline{Non-perturbative Aspects of Superstring Theory}
\vskip .3cm
 
Superstring theories are to date the only viable candidate 
for quantum unification of gauge interactions with gravity.
Until recently, these theories were only defined in the
{\it weak coupling} regime by their perturbative series.
Thanks to the recent discovery of duality symmetries, extending
the electric-magnetic duality of Maxwell equations,
these theories can now be identified as distinct approximations
to a fundamental theory known as {\it M-theory}. Duality
symmetries give access to {\it non-perturbative} effects
in one string theory from {\it perturbative} computations
in a dual theory.

The first chapter of this thesis gives a non-technical
introduction to these developments. The second chapter
introduces the non-perturbative dualities that have
been observed in gauge and supergravity theories as well as the 
non-perturbative BPS spectrum, and presents a novel duality
relating the Higgs branches of certain supersymmetric N=2
gauge theories. In the third chapter, we briefly introduce
the perturbative string theories and give some checks on the
duality between four-dimensional N=4 string theories. From
duality arguments, we obtain {\it exact} non-perturbative results
for various couplings in the low energy effective action, and
interpret the corresponding non-perturbative effects in terms
of {\it instantonic configurations of $p$-branes} wrapped
on supersymmetric cycles of the compactification manifold.
Finally, we discuss a recent proposal of definition
of M-theory in terms of large $N$ supersymmetric $U(N)$
gauge theories~; we extend it to toroidal compactifications 
in constant background fields, and interpret the resulting
spectrum of BPS states in terms of excitations in the
gauge theory. Original publications for these results 
are included in the Appendices.

\vskip .3cm
\noindent
{\sc keywords}: {\it duality symmetries, effective action, instantons,
  M-theory, p-branes, solitons, supergravity, supersymmetric gauge theory}

